\documentclass[onecolumn]{aastex62}

\newcommand{\chn}{{\it Chandra}}
\newcommand{\xmm}{XMM-{\it Newton}}
\newcommand{\swf}{{\it Swift}}
\newcommand{\cs}{G4Jy-3CRE}

\usepackage{multirow}
\usepackage{array}
\usepackage{bbding}
\usepackage{graphicx}
\usepackage{longtable}
\usepackage{mdwlist}
\usepackage{color}
\usepackage{enumitem}
\usepackage{lipsum}
\usepackage{amsmath}
\shorttitle{Powerful radio galaxies in the Southern Hemisphere II}
\shortauthors{F. Massaro et al.}

\begin{document}
\title{Powerful Radio Sources in the Southern Sky. II. A {\it SWIFT} X-Ray Perspective}

\author[0000-0002-1704-9850]{F. Massaro}
\affiliation{Dipartimento di Fisica, Universit\`a degli Studi di Torino, via Pietro Giuria 1, I-10125 Torino, Italy.}
\affiliation{Istituto Nazionale di Astrofisica (INAF) - Osservatorio Astrofisico di Torino, via Osservatorio 20, 10025 Pino Torinese, Italy.}
\affiliation{Istituto Nazionale di Fisica Nucleare (INFN) - Sezione di Torino, via Pietro Giuria 1, I-10125 Torino, Italy.}
\affiliation{Consorzio Interuniversitario per la Fisica Spaziale, via Pietro Giuria 1, I-10125 Torino, Italy.}

\author[0000-0002-2340-8303]{S. V. White}
\affiliation{Department of Physics and Electronics, Rhodes University, PO Box 94, Grahamstown, 6140, South Africa.}

\author[0000-0002-5646-2410]{A. Paggi}
\affiliation{Dipartimento di Fisica, Universit\`a degli Studi di Torino (UniTO), via Pietro Giuria 1, I-10125 Torino, Italy.}
\affiliation{Istituto Nazionale di Astrofisica - Osservatorio Astrofisico di Torino, via Osservatorio 20, 10025 Pino Torinese, Italy.}
\affiliation{Istituto Nazionale di Fisica Nucleare - Sezione di Torino, via Pietro Giuria 1, I-10125 Torino, Italy.}

\author[0000-0003-4413-7722]{A. Jimenez-Gallardo}
\affiliation{Dipartimento di Fisica, Universit\`a degli Studi di Torino, via Pietro Giuria 1, I-10125 Torino, Italy.}
\affiliation{Istituto Nazionale di Astrofisica (INAF) - Osservatorio Astrofisico di Torino, via Osservatorio 20, 10025 Pino Torinese, Italy.}
\affiliation{Dipartimento di Fisica e Astronomia, Universit\`a di Bologna, via P. Gobetti 93/2, 40129 Bologna, Italy.}

\author{J. P. Madrid}
\affiliation{Department of Physics and Astronomy, The University of Texas Rio Grande Valley, Brownsville, TX 78520, USA.}

\author[0000-0002-5941-5214]{C. Mazzucchelli}
\affiliation{European Southern Observatory, Alonso de C\'ordova 3107, Vitacura, Regi\'on Metropolitana, Chile.}
\affiliation{Instituto de Estudios Astrof\'{\i}sicos, Facultad de Ingenier\'{\i}a y Ciencias, Universidad Diego Portales, Avenida Ejercito Libertador 441, Santiago, Chile.}

\author[0000-0002-9478-1682]{W. R. Forman}
\affiliation{Center for Astrophysics $\mid$ Harvard \& Smithsonian, 60 Garden Street, Cambridge, MA 02138, USA.}

\author[0000-0003-3684-4275]{A. Capetti}
\affiliation{Istituto Nazionale di Astrofisica (INAF) - Osservatorio Astrofisico di Torino, via Osservatorio 20, 10025 Pino Torinese, Italy.}

\author{C. Leto}
\affiliation{ASI - Agenzia Spaziale Italiana, Via del Politecnico snc, 00133 Roma, Italy.}

\author[0000-0002-9896-6430]{A. Garc\'ia-P\'erez}
\affiliation{Dipartimento di Fisica, Universit\`a degli Studi di Torino, via Pietro Giuria 1, I-10125 Torino, Italy.}
\affiliation{Instituto Nacional de Astrof\'isica, \'Optica y Electr\'onica, Luis Enrique Erro 1, Tonantzintla, Puebla 72840, México.}

\author[0000-0002-4377-0174]{C.C. Cheung}
\affiliation{Space Science Division, Naval Research Laboratory, Washington, DC 20375, USA.}

\author[0000-0002-2558-0967]{V. Chavushyan}
\affiliation{Instituto Nacional de Astrof\'isica, \'Optica y Electr\'onica, Luis Enrique Erro 1, Tonantzintla, Puebla 72840, México.}
\affiliation{Center for Astrophysics $\mid$ Harvard \& Smithsonian, 60 Garden Street, Cambridge, MA 02138, USA.}

\author[0000-0001-5783-6544]{N. P. H. Nesvadba}
\affiliation{Universit\'e de la C\^{o}te d'Azur, Observatoire de la C\^{o}te d'Azur, CNRS, Laboratoire Lagrange, Bd de l'Observatoire, CS 34229,06304 Nice cedex 4, France.} 

\author[0000-0003-1562-5188]{I. Andruchow}
\affiliation{Instituto Argentino de Radioastronom\'{\i}a, CONICET-CICPBA-UNLP, CC5 (1897) Villa Elisa, Prov. de Buenos Aires, Argentina.}
\affiliation{Facultad de Ciencias Astron\'omicas y Geof\'isicas, Universidad Nacional de La Plata, Paseo del Bosque, B1900FWA La Plata, Argentina.} 

\author[0000-0003-0032-9538]{H. A. Pe\~na-Herazo}
\affiliation{East Asian Observatory, 660 N. A'oh$\bar{o}$k$\bar{u}$ Place, Hilo, HI 96720, USA.}

\author[0000-0002-3140-4070]{E. Sani}
\affiliation{European Southern Observatory, Alonso de C\'ordova 3107, Vitacura, Regi\'on Metropolitana, Chile.}

\author[0000-0003-3471-7459]{R. Grossov\'a}
\affiliation{Department of Theoretical Physics and Astrophysics, Faculty of Science, Masaryk University, Kotl\'a\u{r}sk\'a 2, Brno, CZ-611 37, Czech Republic.}
\affiliation{Astronomical Institute of the Czech Academy of Sciences, Bocn\'i II 1401, 141 00, Prague, Czech Republic.}

\author[0000-0002-6472-6711]{V. Reynaldi}
\affiliation{Facultad de Ciencias Astron\'omicas y Geof\'isicas, Universidad Nacional de La Plata, Paseo del Bosque, B1900FWA La Plata, Argentina.} 
\affiliation{Instituto de Astrof\'isica de La Plata (CCT La Plata-CONICET-UNLP), La Plata, Argentina.} 

\author[0000-0002-0765-0511]{R. P. Kraft}
\affiliation{Center for Astrophysics $\mid$ Harvard \& Smithsonian, 60 Garden Street, Cambridge, MA 02138, USA.}

\author[0000-0002-0690-0638]{B. Balmaverde}
\affiliation{Istituto Nazionale di Astrofisica - Osservatorio Astrofisico di Torino, via Osservatorio 20, 10025 Pino Torinese, Italy.}

\author[0000-0002-3866-2726]{S. Cellone}
\affiliation{Facultad de Ciencias Astron\'omicas y Geof\'isicas, Universidad Nacional de La Plata, Paseo del Bosque, B1900FWA La Plata, Argentina.} 
\affiliation{Complejo Astron\'omico El Leoncito (CASLEO), CONICET-UNLP-UNC-UNSJ, Av.\ Espa\~na 1512 (sur), J5402DSP San Juan, Argentina.}

\begin{abstract} 
We recently {constructed} the \cs, a catalog of extragalactic radio sources based on the GLEAM 4-Jy (G4Jy) sample, with the aim of increasing the number of powerful radio galaxies and quasars with {similar selection criteria to those of the} revised release of the Third Cambridge catalog (3CR). The \cs\ consists of a total of 264 radio sources mainly visible from the Southern Hemisphere. Here, we present an initial X-ray analysis of 89 \cs\ radio sources with archival X-ray observations from the Neil Gehrels \swf\ Observatory. {We reduced} a total of 615 \swf\ observations, for about 0.89\,Msec of integrated exposure time, we found X-ray counterparts for 61 radio sources belonging to the \cs, 11 of them showing extended X-ray emission. The remaining 28 sources do not show any X-ray emission associated with their radio cores. Our analysis demonstrates that X-ray snapshot observations, even if lacking uniform exposure times, as those carried out with \swf, allow us {to (i) verify and/or refine the host galaxy identification; (ii) discover the extended X-ray emission around radio galaxies of the intracluster medium when harbored in galaxy clusters, as the case of G4Jy\,1518 and G4Jy\,1664, and (iii) detect X-ray radiation arising from their radio lobes, as for G4Jy\,1863.}
\end{abstract}


\section{Introduction}
\label{sec:intro} 
{The Third Cambridge catalog \citep[3C;][]{edge59} and its revised versions \citep[3CR and 3CRR;][respectively]{bennett62,laing83} is widely considered the gold standard amongst catalogs of powerful radio sources. Since their releases, they have enabled core investigations into the nature of radio-loud active galactic nuclei \citep[i.e., radio galaxies and quasars][]{begelman84,urry95,harvaneck01,hardcastle20}, their environments at all scales and feedback processes occurring therein \citep{mcnamara07,mcnamara12,fabian12,morganti17}.}

The 3C benefits from an extensive list of surveys, all carried out over six decades, searching for counterparts at all wavelengths from radio \citep{willis66,law95,hardcastle00,giovannini05}, to infrared \citep{simpson99,baldi10,werner12,dicken14}, optical \citep{djorgovski88,hiltner91,longair95,mccarthy95,dekoff96,mccarthy96a,mccarthy97,martel99,lehnert99,chiaberge00,buttiglione09,buttiglione10,buttiglione11} and X-ray energies \citep{evans06,hardcastle06,balmaverde12,wilkes13,kuraszkiewicz21}. Many results were also achieved {via} follow up observations performed at all frequencies as those obtained {by}: (i) the Hubble Space Telescope (HST) Snapshot Survey of 3CR Radio Source Counterparts\footnote{https://archive.stsci.edu/prepds/3cr/} \citep{madrid06,privon08,chiaberge15,hilbert16}; (ii) the 3CR \chn\ snapshot survey \citep{massaro10,massaro12a,massaro13a,massaro18,stuardi18,jimenez20} and, more recently, (iii) the MUse RAdio Loud Emission line Snapshot survey \citep[MURALES;][]{balmaverde18a,balmaverde18b,balmaverde19,balmaverde21,balmaverde22}. 

However, despite its success, the 3C suffers from an artificial limitation since it is restricted to the Northern Hemisphere. About 2/3 of 3C sources are not visible from the Southern Hemisphere thus having limited access to modern astronomical facilities and instruments, as the Multi Unit Spectroscopic Explorer \citep[MUSE;][]{bacon10} and the Enhanced Resolution Imager and Spectrograph \citep[ERIS;][]{kenworthy18} mounted at the Very Large Telescope (VLT), the Atacama Large Millimeter/submillimeter Array (ALMA), {the MeerKAT radio telescope \citep[see e.g.,][]{sejake23} and the High Energy Stereoscopic System (HESS) at very high energies. In the near future, new facilities located in the Southern Hemisphere will include the Square Kilometre Array \citep[SKA;][]{mcmullin20}, the Vera Rubin Observatory \citep[a.k.a., Large Synoptic Survey Telescope - LSST;][]{ivezic19}, the Extremely Large Telescope\footnote{elt.eso.org} (ELT) and the Cherenkov Telescope Array (CTA).}

Several attempts were made to create the southern equivalent of the 3C. For instance, Best et al. (1999) {used} the Molonglo Reference Catalogue \citep[MRC;][]{large81} {to select} an equatorial sample of 178 radio sources with flux density above 5\,Jy at 408\,MHz, in the range of declination between -30$^\circ$ and 10$^\circ$ and having Galactic latitudes $|b|\geq10^\circ$. This sample is characterized by a high spectroscopic completeness and its footprint mitigates the 3C observability limitation. Burgess \& Hunstead (2006a,2006b) created the Molonglo Southern 4 Jy sample (MS4), a sample of southern radio sources at 408\,MHz with integrated flux densities  $S_{408}>4.0$ Jy. However it was only until the Murchison Widefield Array \citep[MWA][]{tingay13} became operational in Western Australia that the southern counterpart of the 3C at $\sim$178 MHz was finally created. {We recently built a sample of 264 extragalactic radio sources extracted from the GaLactic and Extragalactic All-sky MWA 4-Jy (G4Jy) catalog \citep{white20a,white20b}, namely: the \cs\ \citep[][hereinafter paper I]{massaro22}, that is the southern equivalent to the 3CR in terms of its nominal flux limit threshold of 9\,Jy \citep{bennett62,spinrad85}.} 

In paper I, we carried out the comparison between archival radio maps and optical images to search for \cs\ host galaxies. This analysis was combined with an extensive literature search performed to identify those radio sources{with} a redshift, $z$, measurement. We found that 79\% of the \cs\ sources (i.e., 208 out of 264) have a clear optical counterpart of their radio cores. For 181 radio sources, the optical counterpart is also coincident with the mid-IR one reported in the G4Jy catalog \citep{white20a,white20b}. Using both NASA Extragalactic Database (NED)\footnote{http://ned.ipac.caltech.edu} and the SIMBAD Astronomical Database\footnote{http://simbad.u-strasbg.fr/simbad/}, we found spectroscopic $z$ measurements for a total of 145 sources (out of 264), corresponding to $\sim$55\% of the \cs\ catalog. To be {considered} for this work, redshifts must have: (i) a published {figure} of the optical spectrum, or (ii) a description of {such spectrum}  with emission and/or absorption lines clearly reported in a table and/or in the publication. These conservative criteria were already adopted in previous analyses and spectroscopic campaigns \citep[see e.g.,][]{massaro16,herazo20,herazo22,kosiba22}. 

In this second paper of the series, we present a first X-ray perspective of the \cs\ catalog based on targeted X-ray observations. By searching the archive of the X-Ray Telescope \citep[XRT;][]{burrows00,burrows05} on board the Neil Gehrels \swf\ Observatory \citep{gehrels04}, we found that a total of 90 \cs\ radio sources, out of 264, were already observed, with at least one observation having nominal exposure time, $T_{exp}$, above 250\,s. The main aims of the current X-ray analysis are to: (i) use the position of X-ray counterparts, when detected, to verify and eventually refine results of the previous optical analysis (see paper I for more details); (ii) test which sources show extended X-ray emission that could be a signature of emission from the intracluster medium (ICM) for those harbored in galaxy clusters and groups and (iii) obtain measurements of their X-ray count rate, useful to plan X-ray follow up observations.

We also compared \swf\ X-ray images with radio maps at different frequencies. Similarly to paper I, we mainly used those radio maps available in the databases of the Very Large Array (VLA) Low-Frequency Sky Survey Redux\footnote{http://cade.irap.omp.eu/dokuwiki/doku.php?id=vlssr} \citep[VLSSr;][]{cohen07}, the Tata Institute of Fundamental Research (TIFR) Giant Metrewave Radio Telescope (GMRT) Sky Survey \citep[TGSS;][]{intema17}, the Sydney University Molonglo Sky Survey \citep[SUMSS;][]{mauch03}, the National Radio Astronomy Observatory (NRAO) VLA Sky Survey \citep[NVSS;][]{condon98}, the Very Large Array (VLA) Sky Survey \citep[VLASS;][]{lacy20}, corresponding to nominal {frequencies} of 74\,MHz, 150\,MHz, 843\,MHz, 1.4\,GHz and 3\,GHz, respectively, and the NRAO VLA Archive Survey (NVAS)\footnote{http://www.vla.nrao.edu/astro/nvas/} databases. This analysis is only devoted to X-ray observations and it does not include any optical and ultraviolet investigation feasible thanks to the observations collected with the Ultra-violet Optical Telescope (UVOT) instrument \citep{roming05} on board \swf. A dedicated paper on broadband photometry for the \cs\ sample, that includes also UVOT data, is currently in preparation \citep{garcia22}. 

This paper is structured as follows: \S~\ref{sec:sample} is dedicated to a brief overview of the \cs\ catalog and the selection criteria underlying the \swf-XRT archival search. \S~\ref{sec:reduc} is devoted to a description of data reduction and analysis procedures adopted here while \S~\ref{sec:results} is dedicated to the results we achieved. A comparison {with previous X-ray analyses is also presented in \S~\ref{sec:xcomp} while \S~\ref{sec:optical} illustrates the search for mid-infrared and optical counterparts using X-ray images.} Summary, conclusions and future perspectives are given in \S~\ref{sec:summary}. Finally, Appendix A reports all \swf\ X-ray images with radio contours overlaid, while Appendix B is devoted to a comparison between \chn\ and \swf\ X-ray observations for four sources having unpublished \chn\ datasets.

As in paper I, we adopt cgs units for numerical results and we assume a flat cosmology with $H_0=69.6$ km s$^{-1}$ Mpc$^{-1}$, $\Omega_\mathrm{M}=0.286$ and $\Omega_\mathrm{\Lambda}=0.714$ \citep{bennett14}. Spectral indices, $\alpha$, are defined by flux density, S$_{\nu}\propto\nu^{-\alpha}$. {For optical images, we used those available in the archives of the Panoramic Survey Telescope \& Rapid Response System \citep[Pan-STARRS;][]{flewelling20} and the Dark Energy Survey \citep[DES;][]{abbott18}. These optical data were complemented by images available in  the red filter of the Digital Sky Survey\footnote{https://archive.eso.org/dss/dss} (DSS), for those sources outside the Pan-STARRS and DES footprints. Pan-STARRS magnitudes are reported in the AB system \citep{oke74,oke83} while DES magnitudes are given in optical filters similar to those of Pan-STARRS and the Sloan Digital Sky Survey \citep[SDSS;][]{ahn12}.}

\section{Sample selection}
\label{sec:sample} 
The \cs\ sample contains a total of 264 extragalactic radio sources {(paper I). The parent sample of the G4Jy-3CRE catalog was the larger} complete sample of extragalactic radio sources with flux density above the threshold of 4\,Jy at 151\,MHz, namely the G4Jy catalog \citep{white20a,white20b} based on the recent GaLactic and Extragalactic All-sky MWA survey \footnote{https://www.mwatelescope.org/gleam} \citep[GLEAM][]{wayth15,hurley17}. It is worth noting that MWA is considered {as} the SKA precursor at low radio frequencies \citep{tingay13}. 

All radio sources included in the \cs\ catalog lie at $Dec.<-5^\circ$, also having Galactic latitudes $|b|>$10$^\circ$ and {are located outside the original footprint of the 3CR catalog. \cs\ sources} have flux {density measurements} at 174 MHz $S_{174}$ and at 181 MHz $S_{181}$, integrated in the GLEAM sub-bands, above 8.13 Jy and 7.85 Jy, respectively, where these thresholds correspond to the 9\,Jy limiting sensitivity, at $\sim$178 MHz, that is the nominal value adopted to create the 3CR. The flux density thresholds were computed assuming a power-law radio spectrum and adopting the spectral index reported in the G4Jy catalog. Thus the \cs\ is  equivalent, in terms of radio flux density selection, to the northern 3CR extragalactic sample \citep{bennett62,spinrad85}. We highlight that relatively bright sources at $Dec.<+30^\circ$ and Galactic latitudes $|b|>10^\circ$, including the Orion Nebula, were all masked in  the GLEAM extragalactic catalog and are not listed in the G4Jy sample \citep[see][and references therein for a list of masked sources and additional details]{white20a,white20b}.

Here, we crossmatched the \cs\ catalog with the \swf\ archive, adopting a search radius of 10\arcmin, and then we selected only sources having, at least, {one} XRT observation with nominal $T_{exp}$ longer than 250\,s. Our search was restricted only to X-ray observations performed in \textsc{Photon Counting} mode \citep{hill04} to ensure that an X-ray image can be obtained and compared with those at other frequencies. Adopting these criteria, we found a total of 90 radio sources with \swf-XRT data available in the archive. We excluded G4Jy\,1038 (a.k.a. 3C\,279), one of the most famous blazars \citep[see e.g.,][]{lynds65,stocke98} because its \swf-XRT observations are already, extensively, discussed in the literature \citep[see e.g.][]{collmar10,hayashida15,larionov20}. 

We processed a total of 615 {individual} observations for 89 radio sources belonging to the \cs\ catalog. All these observations were acquired between May 2005 and November 2022. {Given that G4Jy\,1748 and G4Jy\,1749 lie in the same field by the Swift-XRT, within $\sim$4\arcmin\ angular separation, we processed only the 9 archival observations centered on the former radio source.} 

\begin{figure}[!th]
\begin{center}
\includegraphics[width=5.8cm,height=8.cm,angle=-90]{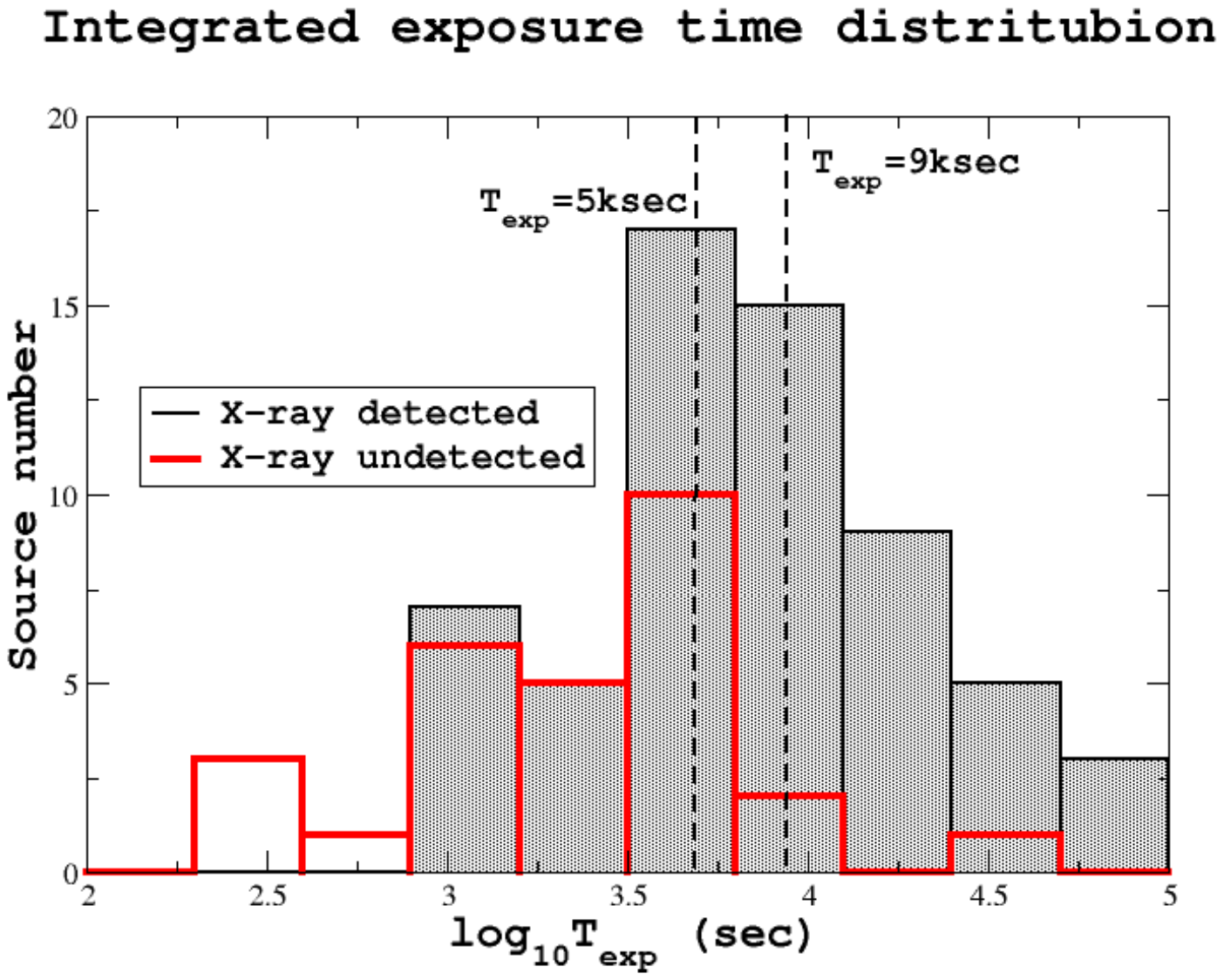}
\end{center}
\caption{Integrated exposure times $T_{exp}$ for all merged event files reduced in our analysis. The black histogram refers to radio sources having an X-ray detected counterpart while the red one represents those for which no X-ray source is spatially coincident with the radio core. A total of 627 observations were retrieved from the \swf-XRT archive and analyzed to investigate the X-ray emission of 89 \cs\ radio sources. The two dashed vertical lines mark the threshold of $5\,\mathrm{ksec}$ and $12\,\mathrm{ksec}$, respectively. We found an X-ray counterpart for all radio sources with integrated $T_{exp}$ {larger} than $\sim12\,\mathrm{ksec}$ and for 85\% of the sample (i.e., 35 out of 41) observed for more than $5\,\mathrm{ksec}$.}
\label{fig:exptime}
\end{figure}

The distribution of the integrated $T_{exp}$ {for the 89 radio sources} is shown in Figure~\ref{fig:exptime}. We remark that even when the integrated $T_{exp}$ appears relatively large (i.e., above $\sim10\,\mathrm{ksec}$), it is the result of merging several, relatively short, observations. Since the integrated $T_{exp}$, for a large fraction of the sources (i.e., $\sim$90\%), is smaller than $\sim25\,\mathrm{ksec}$, this prevents us from performing a uniform spectral analysis. Thus, we focused on searching for X-ray counterparts and refining the mid-IR and optical analysis. 

During our archival search, we also found 4 radio sources, namely, G4Jy\,171, G4Jy\, 260, G4Jy\,411 and G4Jy\,1613 {with} unpublished \chn\ X-ray observations. In these cases, we also reduced these \chn\ data and we compared them with the results obtained from the \swf-XRT images. A brief overview of the \chn\ data reduction procedure adopted here, together with the comparison between X-ray images collected by the two satellites, is reported in Appendix B.

\section{\swf\ X-ray data reduction and analysis}
\label{sec:reduc} 

\subsection{X-ray source detection}
We adopted the same data reduction procedures described in previous \swf-XRT analyses \citep[see e.g.][]{massaro08a,massaro08b,paggi13,marchesini19,marchesini20}, providing results in agreement with those of the \swf-XRT X-Ray point source catalogs \citep{delia13,evans14,evans20}. Therefore, we report here only basic details about the data reduction while additional information can be obtained from the references previously listed. {The entire X-ray analysis performed here and all X-ray images shown in the following sections are restricted to the 0.5-10 keV energy range, unless stated otherwise.}

Raw \swf-XRT data were downloaded from the archive\footnote{https://heasarc.gsfc.nasa.gov/docs/archive.html} and reduced to obtain clean event files using the \textsc{xrtpipeline} task, which is part of the {\it Swift} X-Ray Telescope Data Analysis Software \citep[XRTDAS][]{capalbi05}, distributed within the HEAsoft package (version 6.30.1). Event files were calibrated and cleaned with standard filtering criteria using the \textsc{xrtpipeline} task, combined with the latest calibration files available in the High Energy Astrophysics Science Archive Research Center (HEASARC) calibration database (CALDB) version (v. 20220907)\footnote{https://heasarc.gsfc.nasa.gov/docs/heasarc/caldb/caldb\_supported\_missions.html}. In particular, using \textsc{xselect}, we excluded time intervals (i) with count rates higher than 40 photons per second and (ii) having the CCD temperature exceeding -50$^\circ$C in regions located at the edges of the XRT camera \citep{delia13}. Clean event files, {for the same sources,} were merged using the \textsc{xselect} task, while corresponding exposure maps were merged with the XIMAGE software\footnote{https://heasarc.gsfc.nasa.gov/xanadu/ximage/ximage.html}. 

\begin{figure}[!t]
\begin{center}
\includegraphics[width=3.8cm,height=5.8cm,angle=-90]{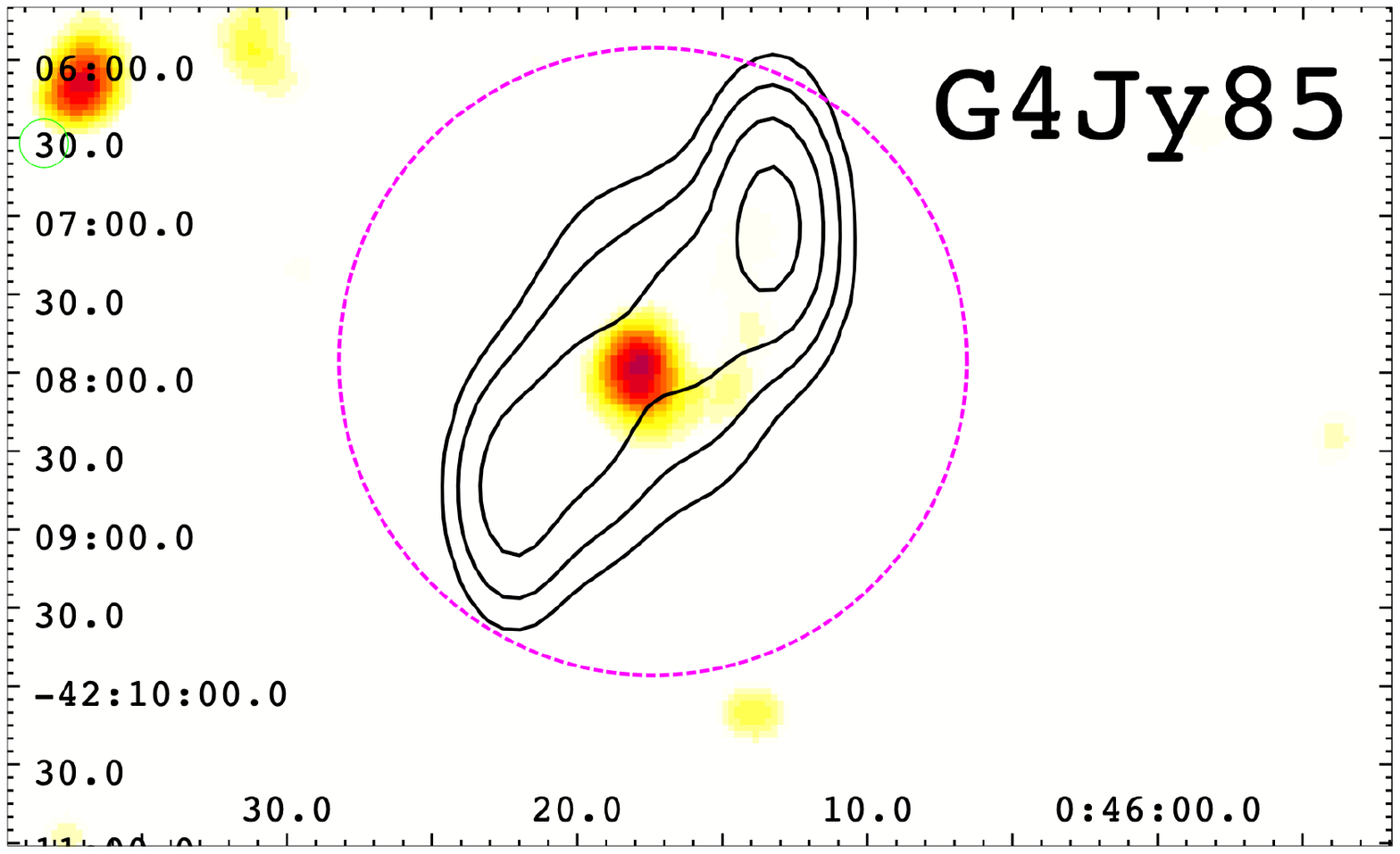}
\includegraphics[width=3.8cm,height=5.8cm,angle=-90]{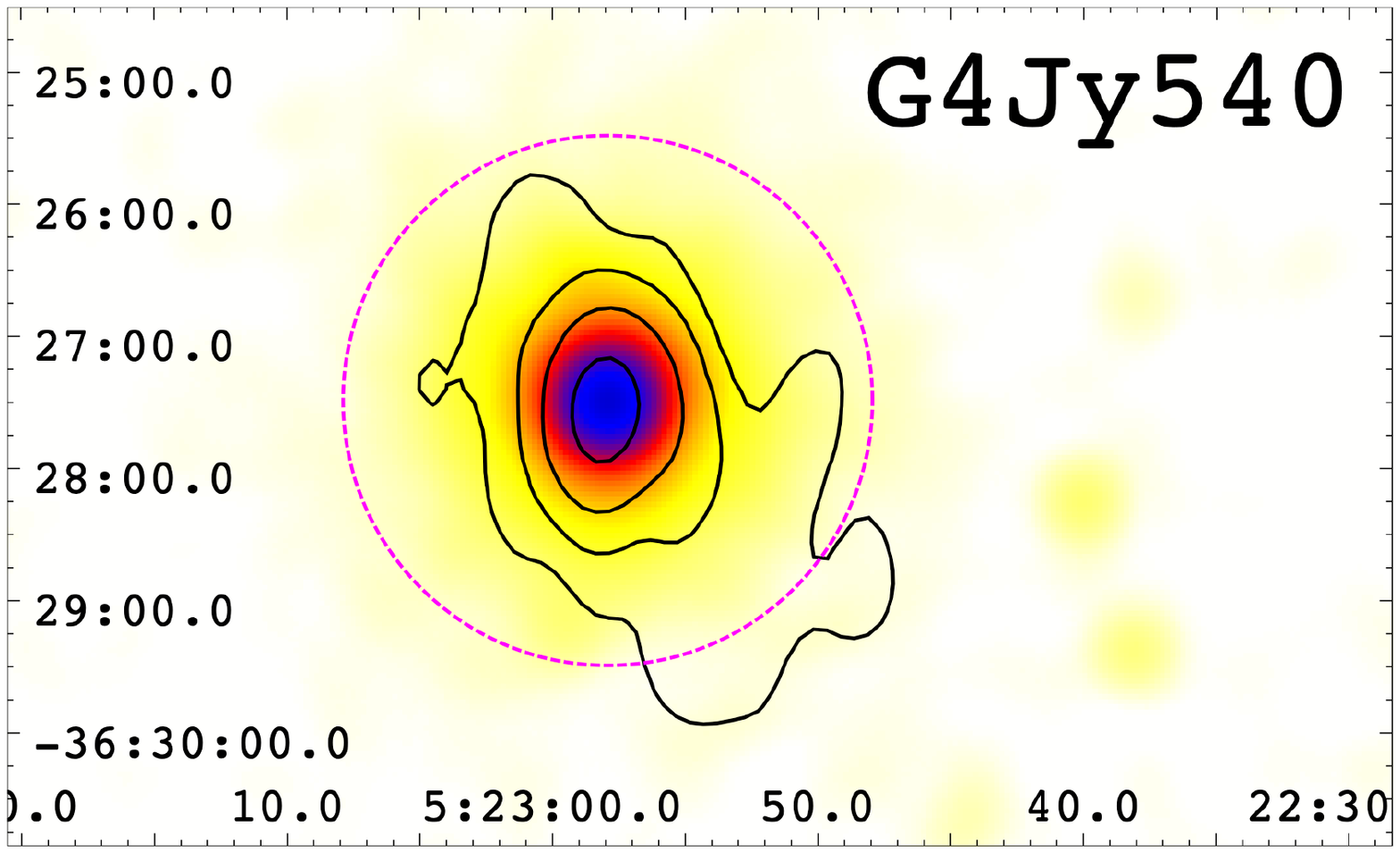}
\includegraphics[width=3.8cm,height=5.8cm,angle=-90]{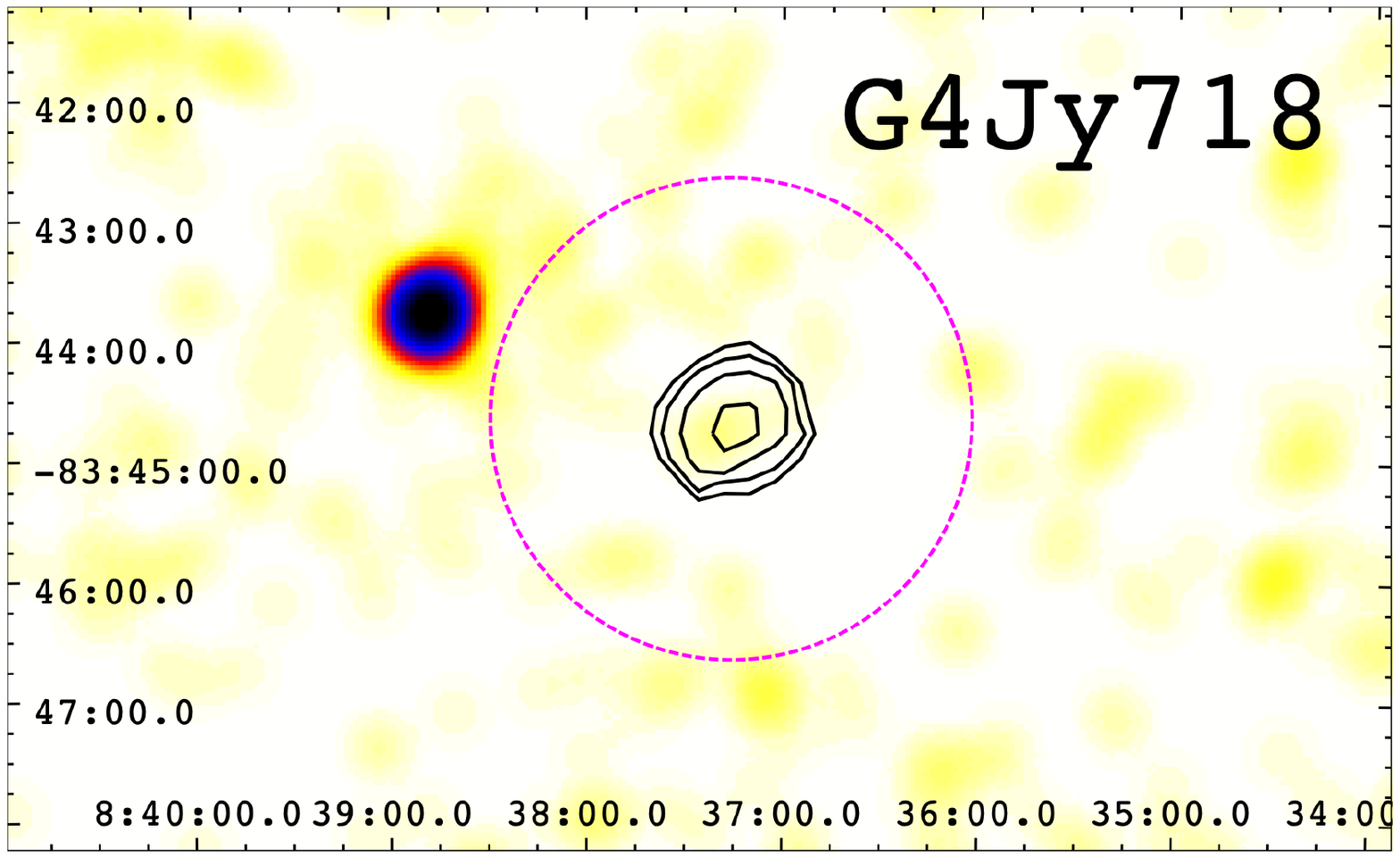}
\end{center}
\caption{{\it Left panel)} The X-ray image, obtained from the merged event file of G4Jy\,85, smoothed with a Gaussian kernel of 10 pixels (i.e., 23.57\arcsec). The magenta circle is centered on the position of the X-ray counterpart associated with this radio sources and it has a radius of 120.207\arcsec\ (i.e., 51 pixels) enclosing 90\% of the \swf-XRT PSF. {Radio contours are drawn using the radio map at 150\,MHz starting at level of 0.3\,Jy/beam and increasing by a factor of 3.} This is a clear example of a radio source for which the X-ray counterpart of its radio core is detected {at the level of the $SNR$} reported in the figure. {\it Central panel)} {same image as the left panel} but for G4Jy\,540, an example of detected extended X-ray emission. {Here radio contours were computed using the archival radio map at 150\,MHz starting at level of 0.1\,Jy/beam and increasing by a factor of 6.} {\it Right panel)} same image as the left panel but for G4Jy\,718, an example of a radio source lacking an X-ray counterpart. In this case the magenta circle is centered on the radio position reported in the G4Jy catalog. {Radio contours overlaid to the X-ray image are drawn from the radio map at 843\,MHz starting at level of 0.2\,Jy/beam and increasing by a factor of 2.}}
\label{fig:example}
\end{figure}

We created images for all merged event files and a detection run was then performed using the sliding cell DETECT (i.e., \textsc{det}) algorithm in XIMAGE \citep{giommi92}. We chose a threshold of the signal-to-noise ratio ($SNR$) equal to 3 to claim an X-ray detection and we adopted the following X-ray detection flags (XDF) to characterize the detection and the identification:
\begin{itemize}
\item {\it x} (X-ray counterpart): all \cs\ sources having a unique X-ray source located at the position of the radio source, {within the X-ray positional uncertainty region}, {e.g.} G4Jy\,85 shown {in left panel} of Figure~\ref{fig:example};
\item {\it e} (extended X-ray source): radio sources for which the X-ray counterpart, detected in the \swf-XRT image, shows diffuse/extended X-ray emission, {e.g.} G4Jy\,540 in {the central panel of}  Figure~\ref{fig:example};
\item {\it u} (undetected X-ray counterpart): those merged event files for which the \textsc{det} algorithm did not report an X-ray detection, above the chosen threshold, at the location of the radio source (see e.g. G4Jy\,718 in {the right panel of} Figure~\ref{fig:example}). We also used this flag for 4 sources for which their X-ray emission is contaminated by a nearby extended objects, thus making it challenging to claim a detection (see following sections for more details).
\end{itemize}

{For all sources that are undetected in X-rays with the \textsc{det} algorithm (labeled \textit{u} in Table~\ref{tab:main}), we carried out a second detection run using the SOurce STAtistic (sosta) tool} available in XIMAGE. {The} \textsc{sosta} algorithm uses a local background to determine the significance of a source detection and to estimate its count rate while detections performed using the \textsc{det} procedure {are} based on the global background. Thus, the former algorithm could be more accurate for observations with relatively low integrated $T_{exp}$. To perform this detection run we were prompted to select the radio source position to compute background, source counts, count rate, and detection significance. Radio sources having a candidate X-ray counterpart detected using \textsc{sosta} have all $SNR$ below 3, {computed using the \textsc{det} task,} and {all the number of photons are reported without uncertainties in Table in Table~\ref{tab:main} } to indicate that {they} correspond to a 3$\sigma$ upper limit. The \textsc{sosta} algorithm does not provide the X-ray source position and thus it is not reported in Table~\ref{tab:main}.

{In} all merged event files, {where} we detected an X-ray counterpart of \cs\ radio sources, we measured: (i) the coordinates of the X-ray centroid (J2000 equinox) using the \textsc{xrtcentroid} task, when detected using the \textsc{det} algorithm; (ii) the number of photons $n_{90}$, within a circular region of radius equal to 120.207\arcsec\ (i.e., 51 pixels) centered on the X-ray coordinates enclosing 90\% of the \swf-XRT PSF; (iii) the expected number of background photons for the same area $n_{b,90}$; (iii) the count rate, in photons/sec and the $SNR$ of the X-ray detection when obtained with the \textsc{det} algorithm. If no statistical uncertainty is reported on the measured count rate, this corresponds to a 3$\sigma$ upper limit obtained with the \textsc{sosta} algorithm. All these parameters, together with the total number of X-ray observations processed per source, the integrated $T_{exp}$ and the value of the XDF are reported in Table~\ref{tab:main}. 

Finally, in Appendix A, we show X-ray images for all merged event files with radio contours overlaid and a magenta dashed circle of 120.207\arcsec\ superimposed, as in Figure~\ref{fig:example}. The circular region is centered on either the location of the X-ray counterpart of each \cs\ radio source, if detected in the \swf-XRT, or on its radio coordinates, as reported in the G4Jy catalog, if not. {The frequencies of the radio maps used to draw the radio contours and the contour parameters are reported in Appendix A (see Table~\ref{tab:radcon} for more details).}

\subsection{Extended X-ray emission}
To determine whether an X-ray source is extended we adopted the following criterion. We assumed a Poissonian distribution for all measured numbers of photons and their uncertainties. We counted the difference $\delta\,n$ between the number of photons ($n_{10}$) {within an annulus of inner radius equal to} 40.069\arcsec\ (i.e., 17 pixels) and {outer} radius of 120.207\arcsec\ (i.e., 51 pixels) and that of background photons $n_{b,10}$ expected in a region having the same area: $\delta\,n=n_{10}-n_{b,10}$. According to the latest model release of the Point Spread function (PSF) for \swf-XRT\footnote{https://heasarc.gsfc.nasa.gov/docs/heasarc/caldb/swift/docs/xrt/SWIFT-XRT-CALDB-10\_v01.pdf} a circular region of 40.069\arcsec\ and 120.207\arcsec\ encloses 80\% and 90\% of the PSF, respectively. Thus, we expect that the measured number of photons $\delta\,n$, for a point-like source, corresponds to $\sim$10\% of the total {flux}. 

Consequently, we computed the 10\% of the total number of expected photons $n_{ex}$ rescaling those (i.e., $n_{90}$) measured within a circular region of 120.207\arcsec\ and subtracting the average background photons ($n_{b,90}$) expected for the same area: $n_{ex} = \frac{1}{9}(n_{90} - n_{b,90})$. We considered {as} extended those X-ray sources for which the difference $(\delta\,n - n_{ex})$ is {larger} than zero within a level of confidence of 3$\sigma$. The number of background photons was obtained by measuring counts within a circular region of 102 pixels in radius, masking other X-ray sources detected using the \textsc{det} algorithm with $SNR>3$, {in all merged event files}, and then rescaled for the ratio of the areas. For extended X-ray sources, if the \textsc{det} algorithm detected more than a single source close to the X-ray intensity peak, we only reported in Table~\ref{tab:main} the one with the highest $SNR$.

\startlongtable
\begin{deluxetable}{rllllrrrrrcrr}
\tabletypesize{\scriptsize}
\rotate
\tablecaption{Results for all \cs\ sources with archival \swf-XRT observations}
\label{tab:main}
\small
\tablehead{\colhead{G4Jy} & \colhead{R.A.$^{(r)}$} & \colhead{Dec.$^{(r)}$} & \colhead{R.A.$^{(x)}$} & \colhead{Dec.$^{(x)}$} & \colhead{$n_{90}(n_{b,90})$} & \colhead{ctr} & \colhead{$SNR$} & \colhead{$T_{exp}$} & \colhead{$N_{obs}$} & \colhead{XDF} & \colhead{$z$} & \colhead{$N_{H,Gal}$} \\ 
\colhead{name} & \colhead{(J2000)} & \colhead{(J2000)} & \colhead{(J2000)} & \colhead{(J2000)} & \colhead{(ph)} & \colhead{(10$^{-2}$ph/s)} & \colhead{} & \colhead{(sec)} & \colhead{} & \colhead{} & \colhead{} & \colhead{(1e20 cm$^{-2}$)} } 
\startdata
20    &  00:10:30.55   &  -44:22:57.00  &  --           &  --           &  24(20)      &  0.2(--)     &  --      &  5932   &  6   &  u    &  --        &  1.12 \\
27    &  00:16:02.68   &  -63:10:07.21  &  --           &  --           &  25(36)      &  0.1(0.05)   &  1.8     &  4072   &  6   &  u    &  --        &  1.52 \\
77    &  00:41:29.21   &  -09:22:40.01  &  --           &  --           &  0(109)      &  --          &  --      &  6722   &  6   &  u    &  --        &  2.69 \\
78    &  00:42:09.19   &  -44:14:01.68  &  00:42:08.5   &  -44:13:57.2  &  47(58)      &  0.48(0.11)  &  4.2     &  6038   &  6   &  x    &  0.346$^{(a)}$  &  2.61 \\
85    &  00:46:16.85   &  -42:07:42.89  &  00:46:17.4   &  -42:07:55.7  &  103(26)     &  1.04(0.14)  &  7.4     &  7128   &  5   &  x    &  0.0526$^{(b)}$  &  1.13 \\
86    &  00:47:33.66   &  -25:17:07.37  &  00:47:33.2   &  -25:17:32.8  &  4474(118)   &  13.5(0.23)  &  59.6    &  32008  &  13  &  e    &  0.00081$^{(c)}$  &  2.79 \\
93    &  00:52:14.88   &  -43:06:29.18  &  00:52:14.6   &  -43:06:27.3  &  44(30)      &  0.45(0.11)  &  4.1     &  5273   &  5   &  x    &  --        &  1.17 \\
120   &  01:05:20.91   &  -45:05:28.00  &  --           &  --           &  14(10)      &  0.51(--)    &  --      &  2267   &  3   &  u    &  --        &  1.64 \\
162   &  01:30:27.96   &  -26:09:58.86  &  --           &  --           &  2(5)        &  1.54(--)    &  --      &  995    &  1   &  u    &  2.34665$^{(d)}$  &  1.31 \\
168   &  01:31:36.53   &  -07:03:57.20  &  --           &  --           &  2(4)        &  2.46(--)    &  --      &  1033   &  1   &  u    &  --        &  3.5  \\
171   &  01:34:00.69   &  -36:29:01.79  &  01:33:57.5   &  -36:29:37.1  &  165(39)     &  0.86(0.1)   &  8.7     &  11744  &  2   &  x    &  0.02977$^{(e)}$  &  1.72 \\
213   &  02:00:12.15   &  -30:53:26.48  &  02:00:12.3   &  -30:53:28.7  &  31(4)       &  2.93(0.57)  &  5.1     &  1221   &  1   &  x    &  0.6768$^{(f)}$  &  1.54 \\
217   &  02:02:14.04   &  -76:20:06.29  &  02:02:12.8   &  -76:19:56.7  &  124(12)     &  7.53(0.74)  &  10.1    &  1750   &  3   &  x    &  0.38939$^{(g)}$  &  5.51 \\
247   &  02:19:02.81   &  -36:26:12.23  &  02:19:03.0   &  -36:26:05.7  &  50(16)      &  5.4(0.9)    &  6.0     &  930    &  1   &  x    &  0.488815 $^{(h)}$  &  1.81 \\
249   &  02:20:08.18   &  -70:22:27.80  &  02:20:08.2   &  -70:22:29.9  &  111(33)     &  2.56(0.28)  &  9.1     &  4606   &  8   &  x    &  --        &  6.06 \\
257   &  02:23:57.55   &  -70:59:46.55  &  --           &  --           &  12(72)      &  0.4(--)     &  --      &  3603   &  5   &  u    &  --        &  6.17 \\
260   &  02:25:02.61   &  -23:12:47.59  &  02:25:02.8   &  -23:12:47.7  &  111(77)     &  13.4(0.31)  &  43.7    &  20139  &  7   &  x    &  0.232241$^{(d)}$  &  1.59 \\
290   &  02:43:44.50   &  -51:12:33.59  &  --           &  --           &  1(1)        &  0.57(--)    &  --      &  311    &  1   &  u    &  --        &  2.0  \\
293   &  02:46:56.26   &  -55:41:22.74  &  --           &  --           &  17(39)      &  0.28(--)    &  --      &  4315   &  7   &  u    &  --        &  2.59 \\
304   &  02:52:46.31   &  -71:04:36.19  &  --           &  --           &  17(19)      &  0.1(0.06)   &  1.6     &  2888   &  2   &  u    &  0.56288$^{(a)}$  &  4.24 \\
373   &  03:38:46.45   &  -35:22:46.26  &  --           &  --           &  0(202)      &  0.0(--)     &  --      &  34042  &  15  &  u    &  0.1126$^{(i)}$  &  1.42 \\
392   &  03:52:31.88   &  -07:11:04.03  &  03:52:30.6   &  -07:11:05.2  &  93(43)      &  0.95(0.14)  &  6.9     &  6599   &  5   &  x    &  0.96354$^{(j)}$  &  4.87 \\
404   &  04:00:16.51   &  -16:10:13.58  &  --           &  --           &  12(3)       &  2.34(--)    &  1.4     &  1033   &  1   &  u    &  0.584$^{(d)}$  &  1.97 \\
411   &  04:05:33.94   &  -13:08:14.39  &  04:05:34.1   &  -13:08:12.3  &  3510(134)   &  9.68(0.19)  &  50.0    &  37306  &  15  &  x    &  0.57055$^{(j)}$  &  4.11 \\
415   &  04:07:48.44   &  -12:11:33.40  &  04:07:48.4   &  -12:11:34.0  &  1375(23)    &  34.4(1.0)   &  33.4    &  4398   &  4   &  x    &  0.5731$^{(k)}$  &  3.54 \\
416   &  04:08:20.25   &  -65:45:10.66  &  --           &  --           &  35(11)      &  0.68(--)    &  --      &  1996   &  1   &  u    &  0.962$^{(l)}$  &  3.68 \\
427   &  04:12:48.02   &  -56:00:48.82  &  --           &  --           &  33(37)      &  0.29(--)    &  --      &  4448   &  3   &  u    &  --        &  0.9  \\
446   &  04:20:56.43   &  -62:23:36.94  &  --           &  --           &  15(49)      &  0.13(0.08)  &  1.7     &  4002   &  5   &  u    &  --        &  1.82 \\
464   &  04:29:40.22   &  -36:30:56.02  &  04:29:40.0   &  -36:30:50.7  &  58(16)      &  0.99(0.21)  &  4.8     &  3678   &  2   &  x    &  --        &  1.34 \\
492   &  04:44:37.45   &  -28:09:48.33  &  04:44:37.6   &  -28:09:53.7  &  1661(70)    &  11.5(0.3)   &  38.9    &  15533  &  3   &  x    &  0.147$^{(a)}$  &  2.46 \\
506   &  04:55:14.31   &  -30:06:47.20  &  04:55:13.9   &  -30:06:50.4  &  38(23)      &  0.41(0.12)  &  3.4     &  4533   &  2   &  x    &  --        &  1.06 \\
510   &  04:56:09.84   &  -21:59:15.98  &  04:56:08.9   &  -21:59:11.5  &  489(18)     &  10.8(0.51)  &  21.0    &  4982   &  1   &  x    &  0.533$^{(m)}$  &  2.77 \\
518   &  05:06:43.75   &  -61:09:41.51  &  05:06:43.8   &  -61:09:41.1  &  4586(771)   &  4.97(0.09)  &  53.6    &  96348  &  82  &  x    &  1.093$^{(n)}$  &  1.95 \\
540   &  05:22:57.90   &  -36:27:29.67  &  05:22:57.9   &  -36:27:29.3  &  44910(376)  &  56.6(0.27)  &  210.9   &  88761  &  50  &  e    &  0.056546$^{(o)}$  &  3.24 \\
563   &  05:36:13.67   &  -49:44:23.10  &  05:36:13.9   &  -49:44:20.3  &  35(30)      &  0.48(0.11)  &  4.2     &  5120   &  6   &  x    &  --        &  3.01 \\
580   &  05:49:23.49   &  -40:51:13.04  &  --           &  --           &  43(47)      &  0.11(0.06)  &  2.0     &  7479   &  7   &  u    &  --        &  3.57 \\
613   &  06:26:47.21   &  -54:32:46.20  &  06:26:49.8   &  -54:32:35.9  &  447(41)     &  0.54(0.11)  &  4.9     &  6629   &  2   &  e    &  0.051976$^{(p)}$  &  7.83 \\
614   &  06:27:06.94   &  -35:29:12.93  &  06:27:06.7   &  -35:29:16.2  &  7613(578)   &  24.7(0.33)  &  73.9    &  33736  &  14  &  x    &  0.054855$^{(q)}$  &  6.32 \\
618   &  06:35:45.05   &  -75:16:15.31  &  06:35:46.3   &  -75:16:15.8  &  9052(282)   &  16.7(0.21)  &  80.5    &  62843  &  67  &  x    &  0.653$^{(r)}$  &  7.23 \\
619   &  06:36:31.30   &  -20:34:38.00  &  06:36:32.2   &  -20:34:53.5  &  194(193)    &  0.34(0.05)  &  7.5     &  25274  &  10  &  x    &  0.055198$^{(s)}$  &  20.6 \\
644   &  07:09:21.90   &  -36:02:10.26  &  07:09:14.2   &  -36:01:22.6  &  1260(53)    &  6.83(0.23)  &  29.5    &  20319  &  5   &  x    &  0.11$^{(t)}$  &  16.3 \\
651   &  07:17:06.38   &  -36:21:52.95  &  07:17:08.9   &  -36:22:08.0  &  156(21)     &  0.98(0.26)  &  3.8     &  2061   &  2   &  e    &  0.031358$^{(h)}$  &  19.1 \\
672   &  07:43:32.63   &  -67:26:28.61  &  07:43:31.8   &  -67:26:24.4  &  199(9)      &  8.61(0.68)  &  12.7    &  2868   &  2   &  x    &  1.512$^{(u)}$  &  8.95 \\
718   &  08:37:14.08   &  -83:44:40.49  &  --           &  --           &  42(25)      &  0.33(--)    &  --      &  4473   &  4   &  u    &  --        &  10.8 \\
721   &  08:39:50.80   &  -12:14:26.56  &  08:39:50.5   &  -12:14:34.4  &  8752(69)    &  33.4(0.42)  &  79.3    &  27184  &  7   &  x    &  0.19787$^{(v)}$  &  5.43 \\
723   &  08:41:26.19   &  -75:40:31.12  &  08:41:26.6   &  -75:40:25.6  &  101(17)     &  7.18(0.77)  &  9.4     &  1414   &  1   &  x    &  0.521$^{(w)}$  &  7.52 \\
835   &  10:18:09.23   &  -31:44:15.50  &  10:18:08.9   &  -31:44:12.8  &  70(22)      &  1.68(0.24)  &  7.0     &  3786   &  3   &  x    &  1.346$^{(x)}$  &  5.1  \\
836   &  10:19:43.84   &  -42:24:50.34  &  --           &  --           &  26(37)      &  0.37(--)    &  1.2     &  5403   &  7   &  u    &  --        &  7.91 \\
854   &  10:33:13.07   &  -34:18:43.99  &  --           &  --           &  20(14)      &  0.67(--)    &  1.6     &  3360   &  5   &  u    &  --        &  5.5  \\
876   &  10:51:30.35   &  -09:18:13.19  &  10:51:29.7   &  -09:18:09.3  &  1028(53)    &  11.5(0.38)  &  30.5    &  9523   &  10  &  x    &  0.345296$^{(k)}$  &  2.96 \\
917   &  11:25:54.63   &  -35:23:22.66  &  --           &  --           &  11(4)       &  4.04(--)    &  0.8     &  451    &  1   &  u    &  0.033753$^{(y)}$  &  5.97 \\
927   &  11:34:23.58   &  -17:27:51.70  &  11:34:23.8   &  -17:27:49.4  &  137(73)     &  0.59(0.08)  &  7.2     &  14221  &  14  &  x    &  1.618$^{(d)}$  &  2.82 \\
933   &  11:39:10.69   &  -13:50:42.90  &  11:39:10.5   &  -13:50:40.9  &  530(34)     &  7.11(0.36)  &  20.0    &  8066   &  1   &  x    &  0.556458$^{(a)}$  &  3.27 \\
950   &  11:45:30.94   &  -48:36:10.40  &  11:45:30.9   &  -48:36:10.9  &  161(46)     &  2.9(0.3)    &  9.8     &  5203   &  7   &  x    &  --        &  10.4 \\
1034  &  12:54:37.02   &  -12:33:29.47  &  12:54:35.4   &  -12:34:00.6  &  24(13)      &  1.21(0.4)   &  3.0     &  990    &  3   &  x    &  0.015464$^{(z)}$  &  3.07 \\
1071  &  13:30:07.18   &  -21:42:04.50  &  13:30:07.0   &  -21:41:56.9  &  63(4)       &  5.58(0.77)  &  7.2     &  1279   &  3   &  x    &  0.528$^{(\alpha)}$  &  7.1  \\
1079  &  13:34:16.65   &  -10:09:23.54  &  13:34:18.6   &  -10:09:26.0  &  75(36)      &  0.26(0.08)  &  3.4     &  10533  &  1   &  x    &  --        &  2.34 \\
1080  &  13:36:38.91   &  -33:58:03.45  &  13:36:38.8   &  -33:57:57.6  &  294(78)     &  2.75(0.21)  &  12.8    &  7562   &  10  &  x    &  0.0125$^{(\beta)}$  &  3.96 \\
1135  &  14:16:33.65   &  -36:40:47.02  &  14:16:33.1   &  -36:40:51.4  &  167(39)     &  2.05(0.19)  &  10.8    &  7108   &  5   &  x    &  0.0747$^{(\gamma)}$  &  4.0  \\
1136  &  14:16:41.12   &  -21:46:18.01  &  14:16:40.8   &  -21:46:12.1  &  165(73)     &  0.37(0.05)  &  7.0     &  20186  &  10  &  x    &  1.116$^{(\delta)}$  &  7.74 \\
1145  &  14:19:49.70   &  -19:28:32.22  &  14:19:49.6   &  -19:28:22.9  &  1500(40)    &  17.4(0.52)  &  33.8    &  8934   &  3   &  x    &  0.1195$^{(\epsilon)}$  &  7.19 \\
1148  &  14:20:04.09   &  -49:35:47.49  &  14:20:03.0   &  -49:35:40.3  &  360(31)     &  1.91(0.26)  &  7.5     &  4285   &  3   &  e    &  0.09142$^{(\gamma)}$  &  13.9 \\
1158  &  14:24:31.90   &  -49:13:48.13  &  14:24:31.9   &  -49:13:54.5  &  65(35)      &  0.77(0.16)  &  4.8     &  4977   &  8   &  x    &  0.662$^{(\zeta)}$  &  14.5 \\
1192  &  14:48:29.05   &  -47:01:39.90  &  14:48:28.3   &  -47:01:45.3  &  33(18)      &  0.41(0.14)  &  3.1     &  3600   &  4   &  x    &  --        &  9.33 \\
1203  &  14:54:28.53   &  -36:39:57.32  &  --           &  --           &  86(74)      &  0.3(--)     &  1.7     &  8808   &  9   &  u    &  --        &  5.31 \\
1225  &  15:10:53.55   &  -05:43:07.10  &  15:10:53.4   &  -05:43:07.4  &  314(108)    &  1.87(0.13)  &  14.4    &  16150  &  7   &  x    &  1.191$^{(\eta)}$  &  5.96 \\
1279  &  15:48:59.13   &  -32:17:04.14  &  --           &  --           &  51(29)      &  0.51(--)    &  0.5     &  2896   &  1   &  u    &  0.1082$^{(\gamma)}$  &  6.9  \\
1411  &  17:25:26.13   &  -80:04:46.04  &  --           &  --           &  3.0(3)      &  1.66(--)    &  --      &  817    &  2   &  u    &  --        &  6.25 \\
1423  &  17:37:37.14   &  -56:33:50.71  &  17:37:35.7   &  -56:34:04.3  &  80(64)      &  2.82(0.38)  &  7.5     &  2482   &  4   &  x    &  0.09846$^{(a)}$  &  6.64 \\
1432  &  17:42:01.59   &  -60:55:22.39  &  17:42:01.2   &  -60:55:12.6  &  1707(96)    &  18.1(0.52)  &  35.0    &  10350  &  5   &  x    &  --        &  6.2  \\
1472  &  18:19:34.76   &  -63:45:48.27  &  18:19:34.7   &  -63:45:47.7  &  312(34)     &  4.99(0.33)  &  15.3    &  6070   &  7   &  x    &  0.06412$^{(\theta)}$  &  6.06 \\
1477  &  18:22:16.23   &  -63:59:16.76  &  18:22:16.3   &  -63:59:17.1  &  27(16)      &  1.58(0.37)  &  4.2     &  1542   &  2   &  x    &  --        &  6.79 \\
1498  &  18:37:42.05   &  -43:35:32.77  &  --           &  --           &  5.0(2)      &  5.55(--)    &  --      &  263    &  1   &  u    &  --        &  6.5  \\
1518  &  19:15:47.88   &  -26:52:57.84  &  19:15:48.8   &  -26:52:51.9  &  189(37)     &  1.3(0.21)   &  6.0     &  4017   &  6   &  e    &  0.226$^{(d)}$  &  7.66 \\
1605  &  20:10:28.97   &  -56:26:28.75  &  --           &  --           &  0.0(11)     &  --          &  --      &  837    &  2   &  u    &  --        &  4.48 \\
1613  &  20:18:05.16   &  -55:40:28.39  &  20:18:01.0   &  -55:39:29.0  &  392(64)     &  1.85(0.12)  &  15.5    &  17280  &  4   &  x    &  0.060629$^{(y)}$  &  4.77 \\
1635  &  20:33:16.60   &  -22:53:20.90  &  20:33:16.6   &  -22:53:17.0  &  594(39)     &  7.43(0.35)  &  21.0    &  8939   &  4   &  x    &  0.131491$^{(r)}$  &  4.01 \\
1640  &  20:35:47.60   &  -34:54:02.67  &  --           &  --           &  6.0(25)     &  1.46(--)    &  --      &  827    &  2   &  u    &  --        &  2.58 \\
1664  &  20:56:04.34   &  -19:56:35.41  &  20:56:04.2   &  -19:56:30.6  &  88(19)      &  1.39(0.31)  &  4.5     &  1968   &  8   &  e    &  0.15662$^{(\lambda)}$  &  5.09 \\
1671  &  21:01:38.97   &  -28:01:45.58  &  --           &  --           &  33(13)      &  0.52(0.29)  &  1.8     &  1346   &  2   &  u    &  0.039444$^{(y)}$  &  7.31 \\
1708  &  21:37:44.45   &  -14:32:54.22  &  21:37:45.3   &  -14:32:53.9  &  2251(98)    &  20.7(0.51)  &  40.9    &  12461  &  17  &  x    &  0.20047$^{(\mu)}$  &  4.15 \\
1717  &  21:43:33.18   &  -43:12:47.79  &  21:43:33.2   &  -43:12:46.1  &  77(26)      &  1.03(0.19)  &  5.5     &  4363   &  8   &  x    &  --        &  1.48 \\
1723  &  21:47:23.97   &  -81:32:07.79  &  --           &  --           &  2.0(4)      &  3.46(--)    &  --      &  333    &  1   &  u    &  --        &  8.3  \\
1748  &  21:57:06.78   &  -69:41:21.66  &  21:57:05.8   &  -69:41:22.0  &  5176(99)    &  32.4(0.46)  &  70.4    &  18150  &  9   &  e    &  0.0281$^{(\nu)}$  &  2.49 \\
1749  &  21:57:47.60   &  -69:41:53.59  &  21:57:48.3   &  -69:41:53.3  &  264(99)     &  0.41(0.06)  &  6.9     &  18150  &  9   &  x    &  --        &  2.5  \\
1757  &  22:06:10.33   &  -18:35:39.01  &  --           &  --           &  23(18)      &  0.67(--)    &  1.3     &  3641   &  4   &  u    &  0.6185$^{(\xi)}$  &  2.3  \\
1822  &  23:21:02.01   &  -16:23:05.21  &  23:21:01.8   &  -16:23:03.6  &  33(21)      &  0.61(0.15)  &  4.0     &  3337   &  2   &  x    &  1.414$^{(d)}$  &  1.69 \\
1840  &  23:34:26.63   &  -41:25:26.02  &  --           &  --           &  8.0(12)     &  1.26(--)    &  1.3     &  1700   &  4   &  u    &  0.907$^{(x)}$  &  1.5  \\
1863  &  23:59:03.71   &  -60:55:13.55  &  23:59:04.1   &  -60:54:59.8  &  236(31)     &  2.39(0.22)  &  11.0    &  7118   &  1   &  e    &  0.0959$^{(y)}$  &  1.4   \\
\enddata
\tablecomments{
Col. (1) the G4Jy name of the radio source, also adopted in the \cs\ catalog; col. (2, 3) the Right Ascension and the Declination (Equinox J2000) of the brightness-weighted radio centroid as reported in the G4Jy catalog; col. (4, 5) the Right Ascension and the Declination (Equinox J2000) measured from the distribution of the X-ray photons for radio sources with XDF=x or XDF=e; col. (6) the number of photons measured in the 0.5--10 keV energy range within a circular region enclosing 90\% of the \swf-XRT PSF, together with the average number of photons expected in the background for the same area, as reported in parenthesis; col. (7) the X-ray count rate measured in photons/sec with the 1$\sigma$ uncertainty in parenthesis as obtained using \textsc{det} algorithm. For those sources having not statistical uncertainty reported, thus indicated with a dashed line, the count rate corresponds to a 3$\sigma$ upper limit obtained using the \textsc{sosta} algorithm; col. (8) the value of the $SNR$ for all radio sources with an X-ray counterpart detected using the \textsc{det} algorithm. X-ray counterparts, at $SNR<3$ correspond to those detected running \textsc{sosta} algorithm and in these cases no X-ray coordinates were computed; col. (9) the integrated $T_{exp}$; col. (10) the total number of sources reduced and analyzed; col. (11) the XDF assigned in our analysis (``x '' labels radio sources with a detected X-ray point-like counterpart; ``e'' stands for radio sources with extended X-ray emission around their radio cores and ``u'' for \cs\ with radio cores undetected in the X-ray band); col. (12) the redshift $z$ obtained from the literature analysis presented in paper I, question mark indicated sources with uncertain $z$ estimates; col. (13) the Galactic column density $N_{H,Gal}$ \citep{hi4pi}.
}
\tablerefs{
(a) Tadhunter et al. (1993); (b) Whiteoak (1972); (c) Springob et al. (2005); (d) Best et al. (1999); (e) Burbidge \& Burbidge (1972); (f) Jones et al. (2004); (g) Jauncey et al. (1978); (h) Jones et al. (2009); (i) Carter et al. (1983); (j) Lynds (1967); (k) Kinman \& Burbidge (1967); (l) Labiano et al. (2007); (m) Wright et al. (1979); (n) Wright et al. (1977); (o) Sbarufatti et al. (2006); (p) Cava et al. (2009); (q) Quintana \& Ramirez (1995); (r) Hunstead et al. (1978); (s) Storchi-Bergmann et al. (1996); (t) Parisi et al. (2014); (u) Bergeron \& Kunth (1984); (v) Ho \& Minjin (2009); (w) Browne \& Savage 1977); (x) di Serego et al. (1994); (y) Tritton (1972); (z) Borne \& Hoessel (1984); ($\alpha$) Burbidge \& Kinman (1966); ($\beta$) Younis et al. (1985); ($\gamma$) Simpson et al. (1993); ($\delta$) Best et al. (2000); ($\epsilon$) Burbidge (1967); ($\zeta$) Marshall et al. (2011); ($\eta$) Peterson \& Bolton (1972); ($\theta$) Thompson et al. (1990); ($\lambda$) Stickel \& Kuehr (1994); ($\mu$) Baldwin (1975); ($\nu$) Marenbach \& Appenzeller (1982); ($\xi$) Morton \& Tritton (1982), (see also paper I for more details).
}
\end{deluxetable}

\section{Results}
\label{sec:results} 

\subsection{An X-ray overview of the whole sample}
\label{sec:xrays} 
We found that all 25 \cs\ sources with an integrated exposure time above $T_{exp}\simeq9\,\mathrm{ksec}$ have an X-ray counterpart. {The only exception is} G4Jy\,373, an FR\,II radio galaxy at $z=0.1126$ \citep{carter83} lying behind the Fornax Cluster at an angular separation of $\sim$5\,\arcmin\ from NGC\,1399 \citep[see e.g.,][]{killeen88a,hilker99}, for which the extended X-ray emission of the nearby galaxy cluster prevents us from detecting its X-ray counterpart. When selecting a threshold of integrated $T_{exp}$ above $5\,\mathrm{ksec}$, 88\% of sources (i.e., 37 out of 41) have an X-ray counterpart, as shown in Figure~\ref{fig:exptime}.

There are 28 radio sources with no X-ray emission associated with their radio cores, flagged as undetected. This list also includes G4Jy\,77 and G4Jy\,1605 that are two extended radio sources, the former is a radio phoenix of the galaxy cluster Abell\,85 \citep[see e.g.,][]{bagchi98,kempner04,ichinohe15} while the latter is the radio relic of Abell\,3667 \citep[see e.g.,][]{johnston08,owers09}. These sources are not expected to be associated with an X-ray point-like counterpart. 

We also found 11 radio sources with extended X-ray emission around them. However, {for three out of 11 sources}, namely: G4Jy\,416, G4Jy\,651, G4Jy\,1749, such diffuse emission is contaminated {by that of a nearby X-ray object}. {These sources} are all marked as undetected. Signatures of extended X-ray emission are also evident in G4Jy\,1863, the only source for which we claim the X-ray detection of its radio lobes, as described {below}. The remaining 50 radio sources have a detected X-ray point-like counterpart associated with their radio core. 

{Five out of the 11 sources with extended X-ray emission are well known} since their \chn\ and \xmm\ observations have been extensively discussed in the literature. These radio sources are: (i) G4Jy\,86 \citep[a.k.a. Sculptor Galaxy; see e.g.,][]{hoopes96} a nearby star forming galaxy at $z=0.00081$ \citep{springob05}; (ii) G4Jy\,540 (a.k.a. PKS\,0521-36) a $\gamma$-ray emitting blazar \citep{abdollahi20} with an X-ray jet detected by \chn\ \citep[see e.g.,][]{birkinshaw02,massaro11a}; (iii) G4Jy\,613 known to be associated with the galaxy cluster Abell\,3395 \citep{abell58,abell89,quintana95a,ebeling96,sun09};  (iv) G4Jy\,1148 (a.k.a. PKS\,1416-49) harbored in a non cool-core galaxy cluster detected by \chn\ \citep{worrall17} and (v) G4Jy\,1748 (a.k.a PKS\,2152-69) a lobe dominated quasar hosted in a galaxy cluster with X-ray cavities inflated {by} radio plasma \citep{young05} and hotspots detected in the X-rays \citep[see also][]{ly05,massaro11a}. 

Finally, we remark that adopting the same criteria described in \S~\ref{sec:reduc}, to determine the presence of diffuse X-ray emission, but using an earlier PSF model \citep{moretti04}, extensively adopted in the literature to search for galaxy clusters in \swf-XRT deep observations \citep[see e.g.,][]{tundo12,liu13,tozzi14,liu15,dai15}, the following sources would be also identified as extended: G4Jy\,260, G4Jy\,411, G4Jy\,415, G4Jy\,492, G4Jy\,614, G4Jy\,618, G4Jy721, G4Jy\,876 and G4Jy\,1080. In particular, for some of these radio sources we found in the literature that: G4Jy\,415 is a quasar at $z=0.5731$ that resides in a galaxy group \citep{johnson18}, and diffuse X-ray emission has been previously detected around G4Jy\,492 \citep{mingo17}, G4Jy\,721, an FR\,II radio galaxy harbored in a galaxy cluster \citep{yee83,ellingson87,ellingson89,yates89} and G4Jy\,876 that also lie in a galaxy-rich large-scale environment \citep{hintzen83,ellingson88,ellingson91,hutchings96}. G4Jy\,1080 is the brightest galaxy \citep[a.k.a. IC\,4296; see e.g.,][and references therein]{killeen86,wegner03,grossova19,condon21,grossova22} of the Abell\,3565 galaxy cluster \citep{abell58,abell89} at $z$=0.0125 \citep[see e.g.,][]{sandage78,efstathiou80} while G4Jy\,614, the BL Lac detected at TeV energies (a.k.a. PKS 0625-35) also belongs to the galaxy cluster Abell\,3392 \citep{abell58,abell89,quintana95a,ebeling96,sun09}.

\subsection{Extended X-ray emission around G4Jy\,1518 and G4Jy\,1664}
\label{sec:G4Jy1518-1664} 
{We found clear signatures of extended X-ray emission around two radio sources {in the \swf-XRT images}: G4Jy\,1518 and G4Jy\,1664. 

G4Jy\,1518 belongs to the equatorial sample of Best et al. (1999) and it is classified as a radio galaxy lying at $z$=0.226. It is also listed in the Molonglo Reference Catalogue of radio sources \citep[MRC\,1912-269][]{large81}. From a radio perspective, G4Jy\,1518 appears to be a ``restarted'' radio galaxy \citep{}, as shown in the archival radio maps of Figure~\ref{fig:ext}. The host-galaxy identification is consistent with detection of the radio core, as recently shown using MeerKAT observations \citep{sejake23}.
\begin{figure}[!th]
\begin{center}
\includegraphics[width=10.cm,height=16.cm,angle=-90]{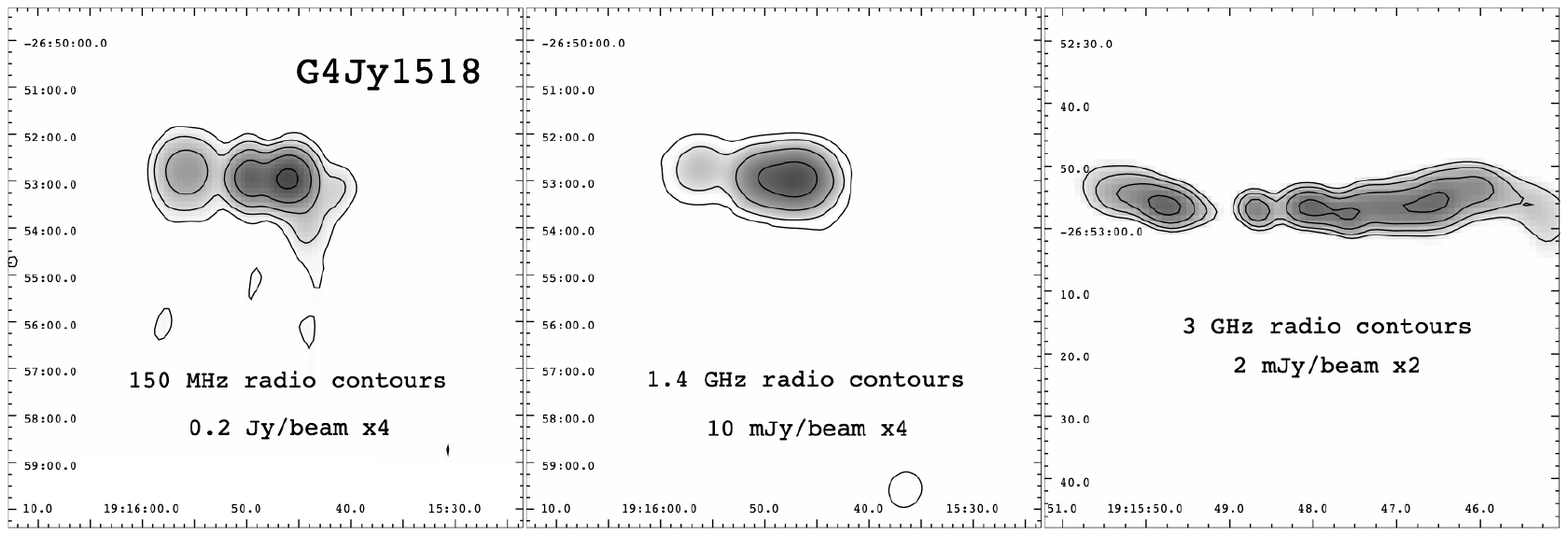}
\end{center}
\caption{Radio maps for G4Jy\,1518 retrieved from the TGSS, NVSS and VLASS archives (left to right panels), at 150\,MHz, 1.4\,GHz and 3\,GHz, respectively. The frequency of the radio map from which contours were drawn is reported together with the intensity of the first level. All radio contours increase in level by a binning factor as indicated. The radio morphology of G4Jy\,1518 shows a classical double-double radio structure similar to ``restarted'' radio galaxies.}
\label{fig:ext}
\end{figure}

G4Jy\,1664 is a radio galaxy at $z$=0.15662 \citep[a.k.a. 6dF\,J2056043-195635][]{stickel94,jones09} with a classical FR\,II radio morphology, as shown in Figure~\ref{fig:ext2}. The redshift estimate was originally derived from stellar absorption features, clearly present in its optical spectrum. The radio core emission appears to be similar to Giga-Peaked radio Sources \citep{odea91}.} 
\begin{figure}[!th]
\begin{center}
\includegraphics[width=10.cm,height=16.cm,angle=-90]{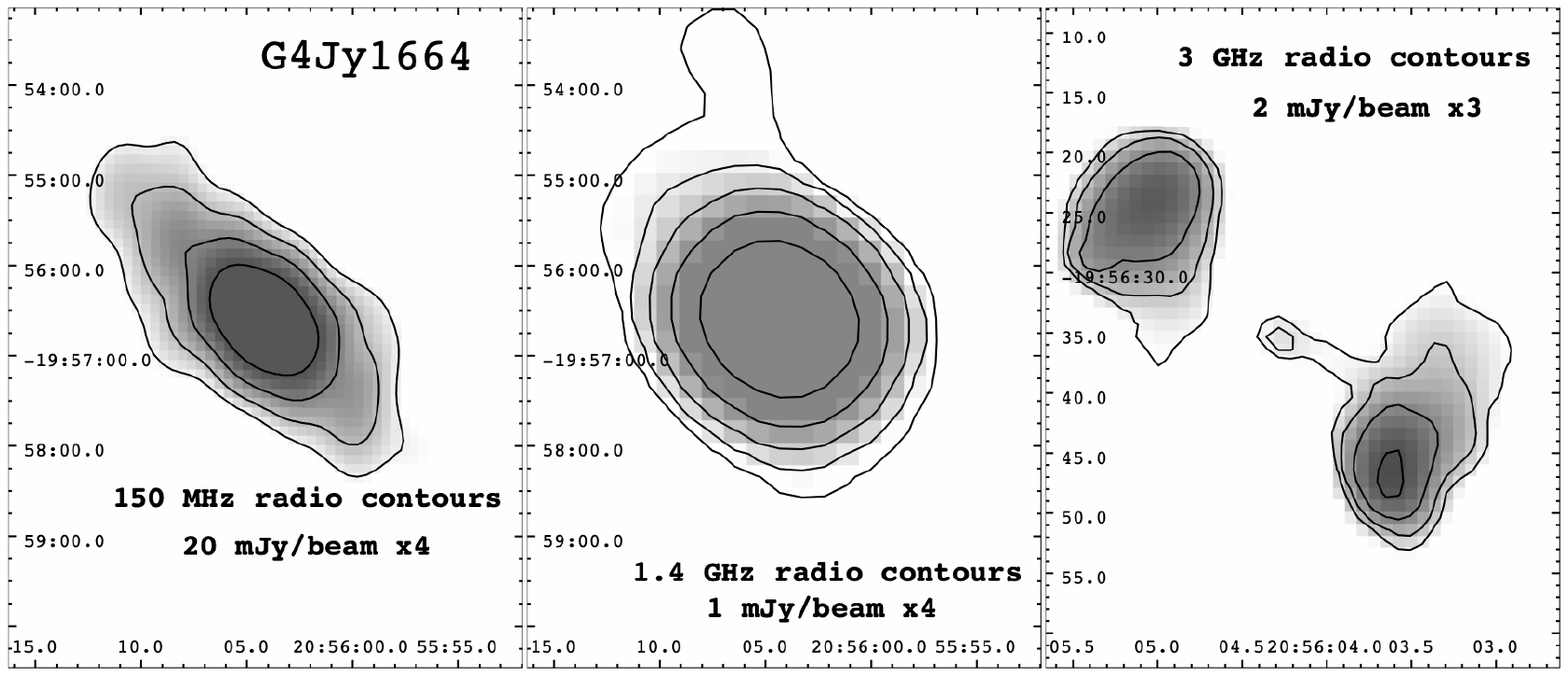}
\end{center}
\caption{Same as Figure~\ref{fig:ext} for G4Jy\,1664. The radio morphology of G4Jy\,1664 is typical of FR\,II radio sources at 3\,GHz while it appears a bit extended at lower frequencies where the double radio structure does not appear.}
\label{fig:ext2}
\end{figure}

\subsection{X-ray counterparts of radio lobes in G4Jy\,1863}
\label{sec:G4Jy1863} 
\begin{figure}[!th]
\begin{center}
\includegraphics[width=10.cm,height=16.cm,angle=-90]{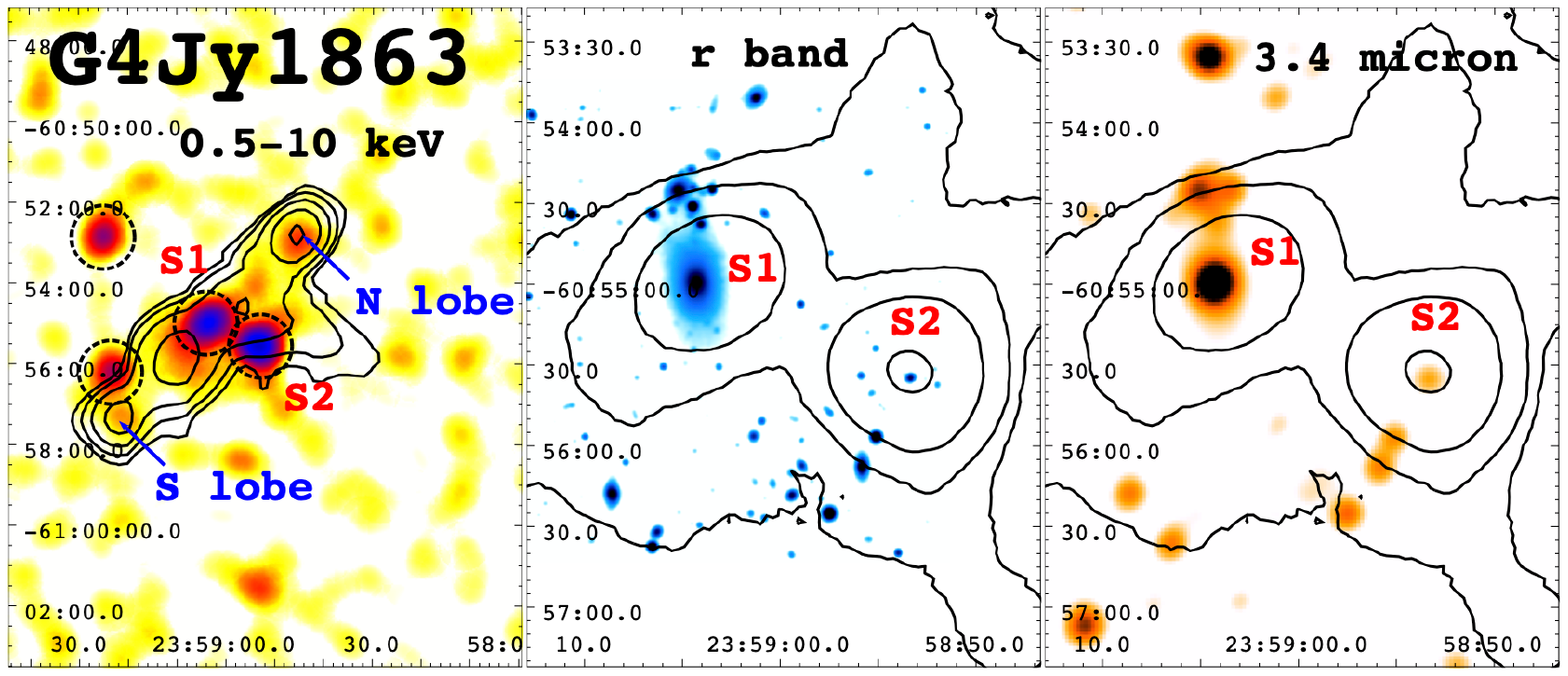}
\end{center}
\caption{The X-ray image (left panel), smoothed with a Gaussian kernel of 42.426\,\arcsec, for the radio source G4Jy\,1863. Radio contours are drawn using a radio map at 843\,MHz starting at level of 90\,mJy/beam and increasing by a factor of 3. Black circles, of 20 pixel radius, mark the location of point-like sources detected at $SNR<3$ while {blue} arrows mark the location of both the northern (N) and the southern (S) radio lobe. The X-ray counterpart of the radio core (S1) and that of the nearby companion galaxy (S2) have a clear counterpart in the optical (r-band) image retrieved from DES $r$ band (central panel) as well as at mid-infrared frequencies at 3.4$\mu$m obtained with the WISE All-sky survey (right panel). X-ray intensity contours, obtained from the \swf-XRT event files and drawn starting at a level of 0.01 photons, are overlaid to both the optical and the mid-infrared image to help identifying the location of the low energy counterparts for S1 and S2. The X-ray emission of both the northern and the southern radio lobes is detected, above 3$\sigma$ level of confidence, comparing the number of photons measured within a circular region, of 20 pixel radius, centered on their radio intensity peak and that expected, on average, on a background region of the same area.}
\label{fig:G4Jy1863ext}
\end{figure}

An intriguing case is G4Jy\,1863, a narrow-line hard X-ray selected giant radio galaxy \citep{cusumano10,oh18,bruni20} at $z$=0.0959 \citep{tritton72,danziger83,ramos11} lying in the direction of the galaxy cluster Abell 4067 \citep{abell58,abell89,teague90} and shown in Figure~\ref{fig:G4Jy1863ext}. G4Jy\,1863 shows several signatures of a past merger in its optical image, {for instance it has} several irregular shells and two faint arcs \citep{ramos11}. 

In Figure~\ref{fig:G4Jy1863ext} the red circles mark the positions of detected X-ray sources close to the radio structure of G4Jy\,1863, and in particular those labelled as S1 and S2. These two point-like sources have a clear optical and mid-infrared counterpart, as shown in the central and in the right panel of the same figure. The proximity of these two X-ray sources prevented us from claiming the presence of extended X-ray emission according to the procedure previously described. 

However, considering the number of photons associated with both the northern and the southern radio lobes, in circular regions of 20 pixel radius, in comparison with those expected in the background for the same area, we claim a detection above 3$\sigma$ level of confidence for the X-ray counterparts of its radio lobes. This could be due to inverse Compton scattering off seed photons arising from the Cosmic Microwave Background \citep[see e.g.,][]{hoyle65,longair70,harris79,hardcastle02,harris06,croston05,worrall09}.

\section{Comparison with previous X-ray analyses}
\label{sec:xcomp} 
Results obtained with our detection analysis are in agreement with those achievable when crossmatching radio positions, as well as mid-IR and optical ones (when available) with X-ray sources listed in the clean sample of the Second \swf-XRT Point Source Catalog\footnote{https://heasarc.gsfc.nasa.gov/W3Browse/swift/swift2sxps.html} \citep[2SXPS][]{evans20}. The only differences we found are {summarized below}: 
\begin{itemize}
\item For four sources, namely: G4Jy\,27, G4Jy\,446, G4Jy\,854 and G4Jy\,1757, all marked with XDF=u, there is a counterpart listed in the 2SXPS catalog found within the X-ray positional uncertainty at 90\% level of confidence and having the X-ray position consistent with their radio core. We also reported their X-ray detection based on the run carried out with the \textsc{sosta} algorithm having relatively low values of $SNR$, below the threshold adopted for the \textsc{det} algorithm. This difference is based on the fact that the 2SXPS catalog used a lower threshold on the $SNR$ to claim a detection. {Moreover the 2SXPS catalog is} built reducing \swf-XRT observations with nominal $T_{exp}$ lower {than the threshold of }250\,sec adopted here. 

\item Two more sources, namely G4Jy\,45 and G4Jy\,1038 (a.k.a. 3C\,279) have an X-ray counterpart listed in the 2SXPS catalog, that we did not analyze here. G4Jy\,45 has XRT observations beyond the threshold of 10\arcmin\ separation set to select sources and thus was automatically excluded when retrieving datasets from the \swf\ archive. G4Jy\,1038, as previously stated, was not selected {as it is} extensively discussed in the literature.
\end{itemize}

We also compared our results with those recently presented by Maselli et al. (2022) that analyzed \swf\ observations of 31 radio sources listed in the SMS4 catalog \citep{burgess06a,burgess06b}, 22 in common with our sample. Our results are in agreement with those presented Maselli et al. (2022). {Ten sources (G4Jy\,93, G4Jy\,249, G4Jy\,506, G4Jy\,563, G4Jy\,672, G4Jy\,723, G4Jy\,950, G4Jy\,1135, G4Jy\,1192 and G4Jy\,1432, we detect the same X-ray counterpart while G4Jy\,257, G4Jy\,293, G4Jy\,416, G4Jy\,718 and G4Jy\,836), all marked as XDF=u according to our analysis, have only upper limits on their X-ray count rate measured by Maselli et al. (2022).} Moreover, as occurred when comparing our results with those of the 2SXPS catalog, for 5 sources, namely: G4Jy\,27, G4Jy\,446, G4Jy\,580, G4Jy\,854 and G4Jy\,1203, we detected the X-ray counterpart only using the \textsc{sosta} algorithm, the same procedure adopted by Maselli et al. (2022). 

Finally, the only difference between our analysis and that of the 2SXPS catalog and/or the literature \citep[i.e.,][]{maselli22} is for G4Jy\,20 that is marginally detected in both analyses but undetected in our investigation. The main reason is due to the threshold on $T_{exp}$ chosen to retrieve \swf-XRT archival observations. Including all datasets with nominal $T_{exp}$ above 50\,sec we also found a marginal detection, adopting the \textsc{sosta} algorithm, with $SNR=1.6$ in agreement with previous analyses.

\section{Mid-infrared and optical counterparts identified with X-ray observations}
\label{sec:optical} 
We also compared {the} results of the X-ray analysis with those obtained from the search for optical counterparts, {and all results of this comparison are summarized as follows.}

\begin{figure}[!th]
\begin{center}
\includegraphics[width=10.cm,height=16.cm,angle=-90]{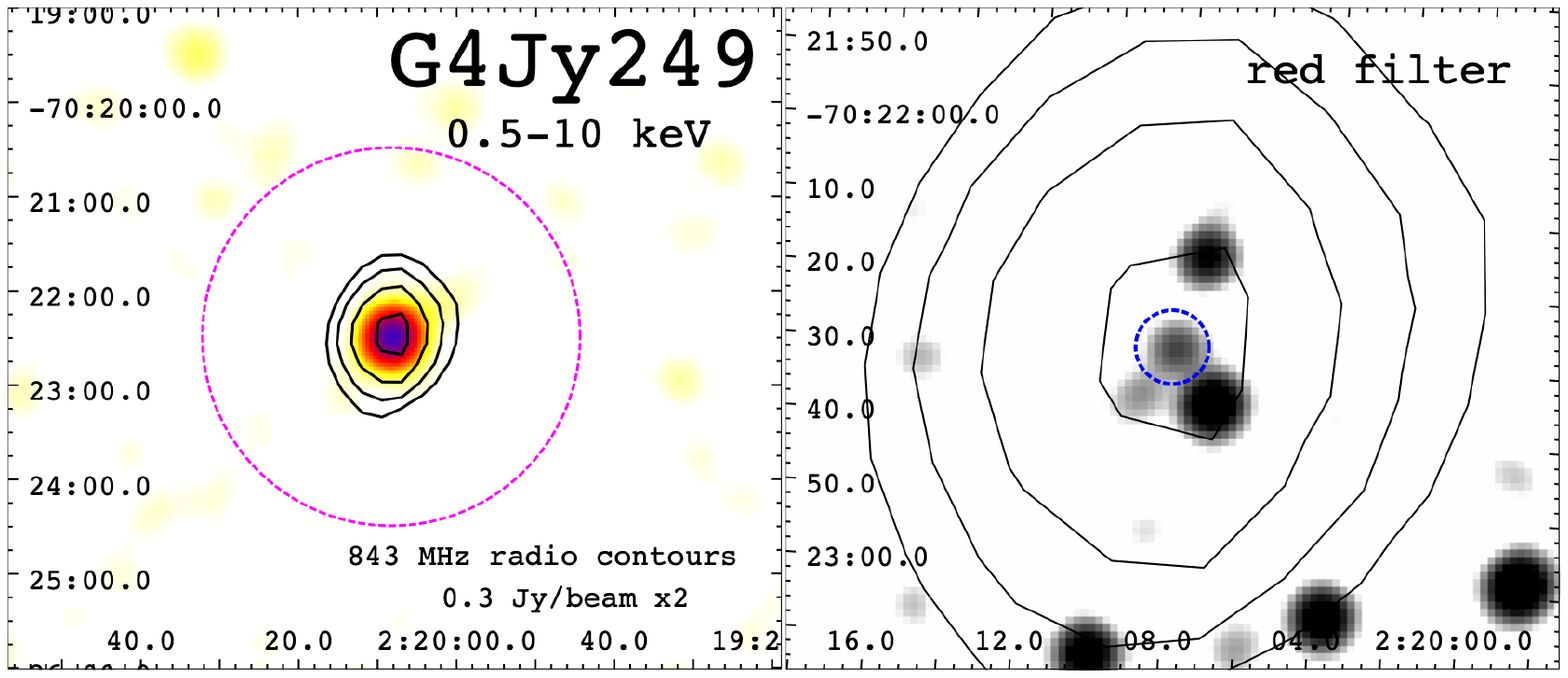}\\
\includegraphics[width=10.cm,height=16.cm,angle=-90]{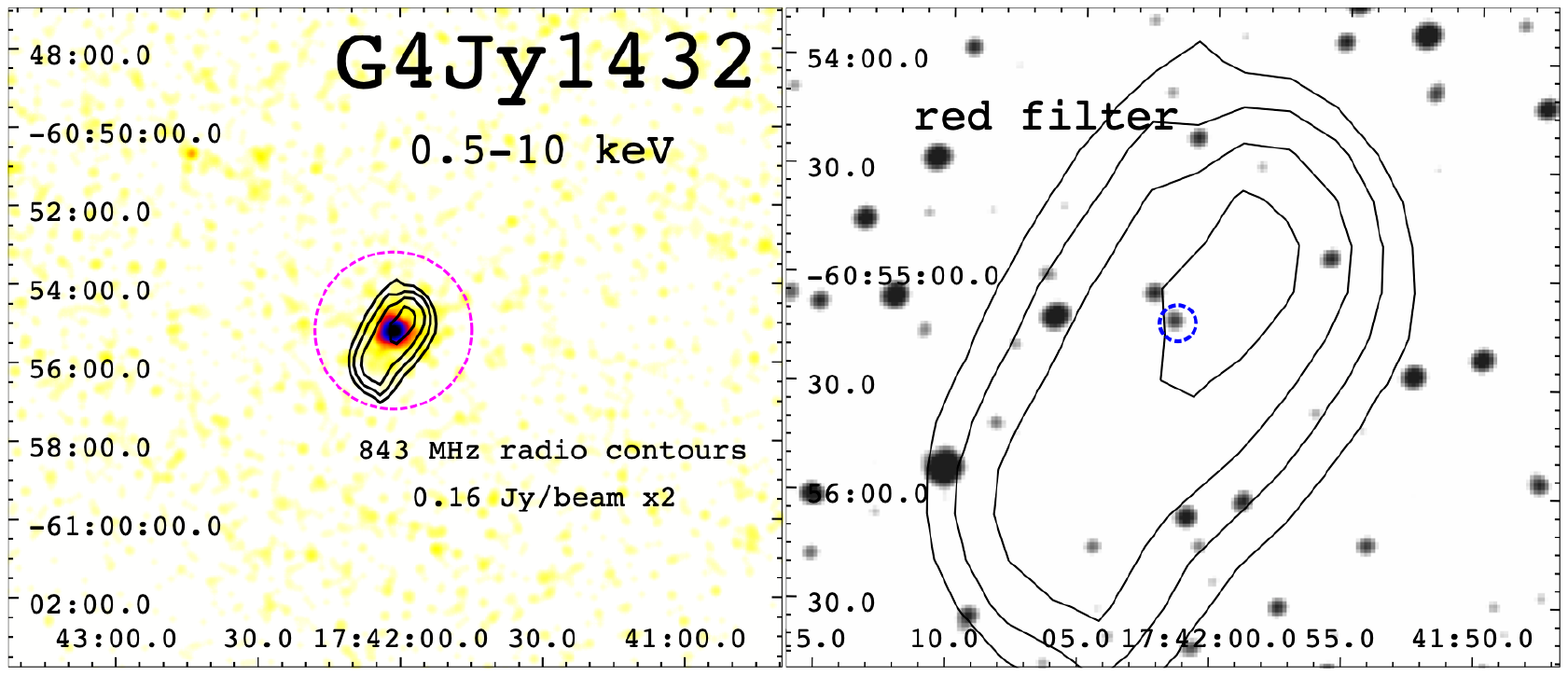}
\end{center}
\caption{{\it (Upper panels)} the left image is obtained from the merged XRT event files, smoothed with a Gaussian kernel of 10 pixels (i.e., 23.57\arcsec), as in Figure~\ref{fig:example} while the right one is the optical image retrieved from the DSS archive in the red filter. The blue circle in the right panel is centered on the location of the X-ray counterpart to mark its position and that of the corresponding host galaxy. Radio contours are overlaid to both images drawn at the frequency indicated (i.e., 843\,MHz obtained from the SUMSS archive) together with the intensity of the first contour level and increasing in level by a binning factor of 2. {\it (Lower panels)} same as upper panels for the radio source G4Jy\,1432. Thanks to this optical-to-X-ray comparison we have been able to distinguish which source is the optical counterpart for these two radio sources having radio maps with insufficient angular resolution to locate precisely the radio core. Cyan cross marks the location of the radio source (i.e., brightness-weighted radio centroid as reported in the G4Jy catalog) while the red cross marks the mid-IR assigned counterpart, associated in the analysis of the G4Jy sample.}
\label{fig:confused}
\end{figure}

\begin{figure}[!th]
\begin{center}
\includegraphics[width=10.cm,height=16.cm,angle=-90]{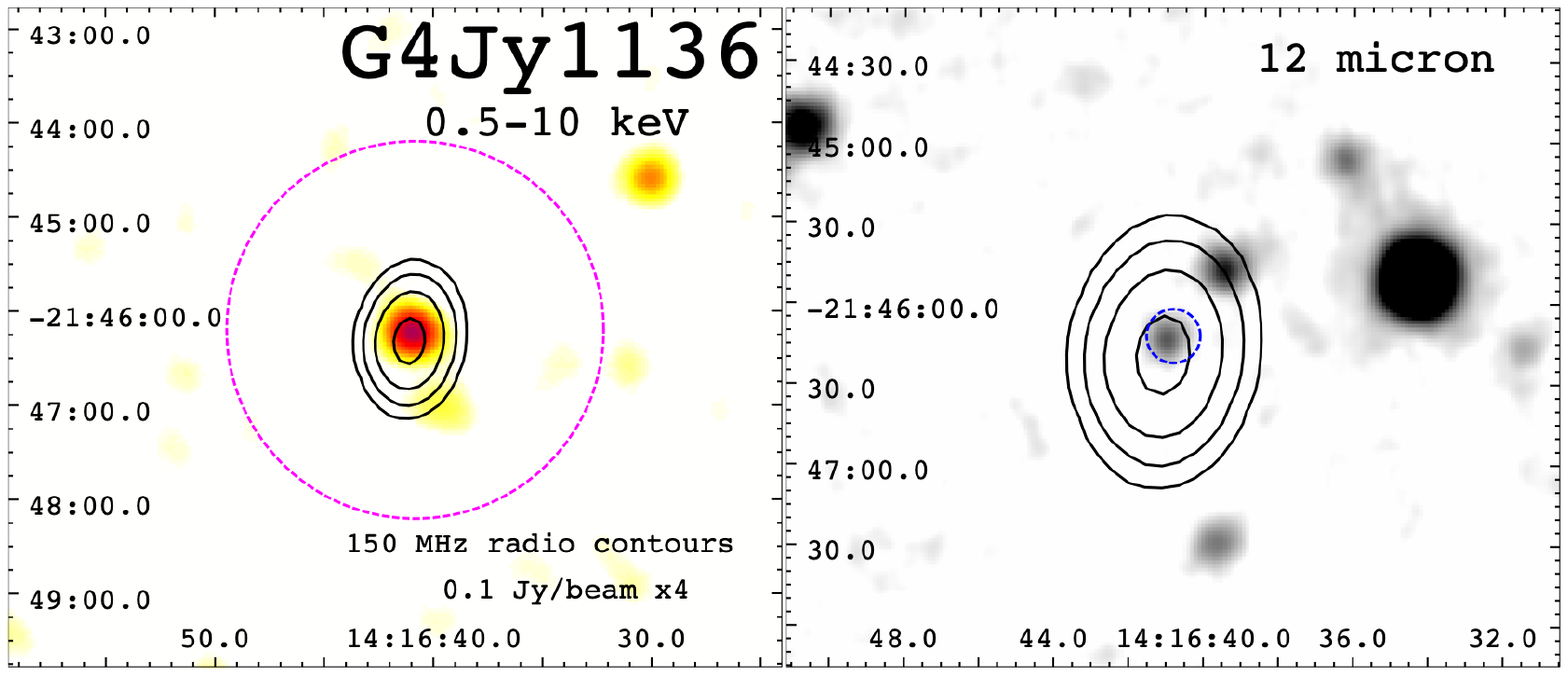} \\
\end{center}
\caption{{\it Left panel)} The X-ray image of G4Jy\,1136 obtained from the merged XRT event file as in Figure~\ref{fig:example}, smoothed with a Gaussian kernel of 10 pixels (i.e., 23.57\arcsec) and with radio contours drawn from the TGSS radio map overlaid. These were computed starting at level of 0.1 Jy/beam and increasing in level by a binning factor of 4. {\it Right panel)} The same field but observed as part of the WISE All-sky survey at 12$\mu$m. The magenta dashed circle in the right panel is centered on the location of the X-ray counterpart to mark its position and that the mid-IR associated source (position indicated by the red cross in the right panel). According to our previous analysis (paper I) this radio source lacks an  optical counterpart but the \swf-XRT analysis revealed that the mid-IR counterpart assigned in the G4Jy catalog is correct as it overlaps with its X-ray counterpart. It is worth highlighting the presence of a relatively bright star located in the north-western direction with respect to the position of G4Jy\,1136.  However it appears fainter in the 12$\mu$m WISE image, shown here, rather than in that at 3.4$\mu$m used in paper I. The cyan cross, if present, marks the location of the radio source (i.e., brightness-weighted radio centroid as reported in the G4Jy catalog) while the red one that of the assigned mid-IR counterpart associated in the analysis of the G4Jy sample.}
\label{fig:noopt1}
\end{figure}

\begin{figure}[!th]
\begin{center}
\includegraphics[width=10.cm,height=16.cm,angle=-90]{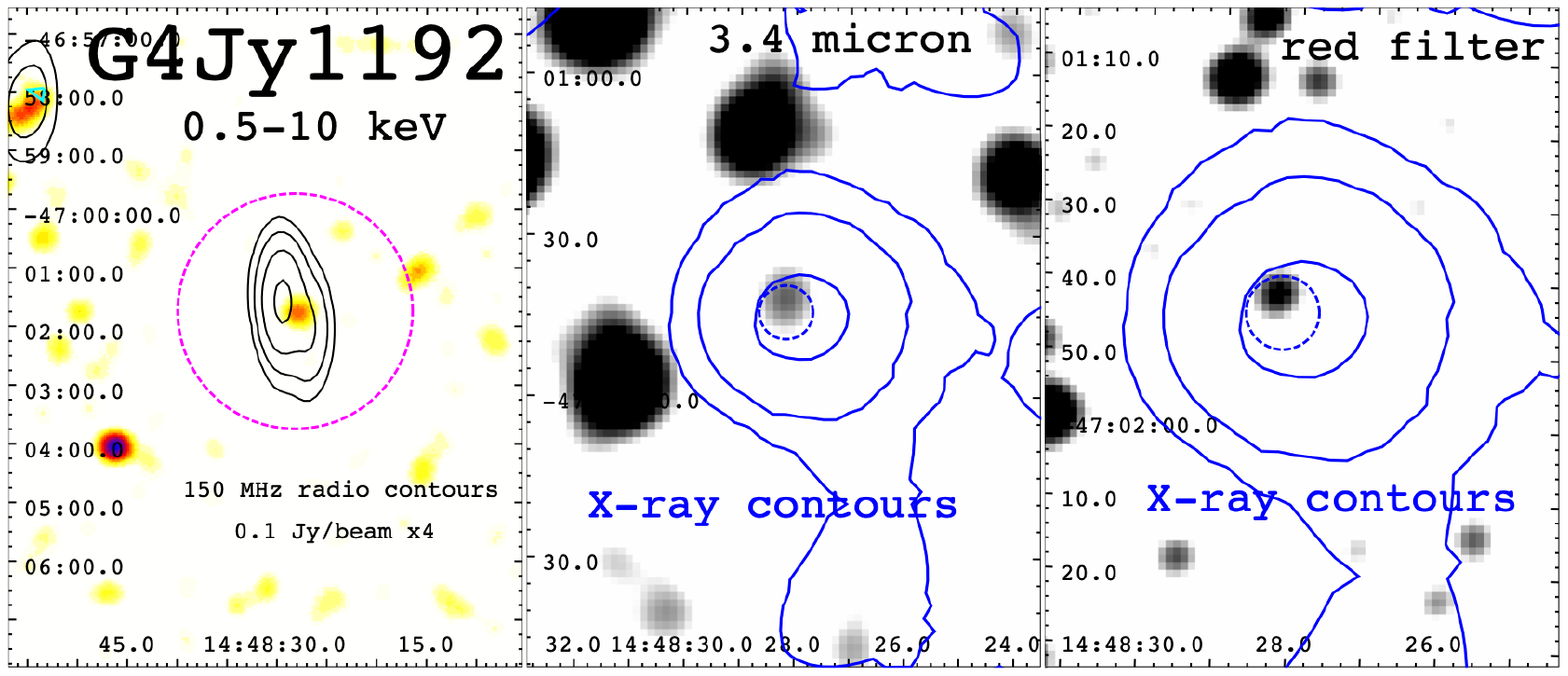}
\end{center}
\caption{{\it Left panel)} The X-ray image of G4Jy\,1192 obtained from the merged XRT event file smoothed with a Gaussian kernel of 10 pixels (i.e., 23.57\arcsec), as in Figure~\ref{fig:example} and in Figure~\ref{fig:noopt1} with radio contours drawn from the TGSS radio map overlaid. These contours were computed starting at level of 0.1 Jy/beam and increase in level by a binning factor of 4. {\it Middle and Right panels)} The same field but observed as part of the WISE All-sky survey at 3.4$\mu$m and in the red filter collected from the DSS archive, respectively. The magenta dashed circle in the left panel is centered on the location of the X-ray counterpart to mark its position, the same is marked by the blue dashed circles in the other two panels, while the cyan cross marks the location of the radio source (i.e., brightness-weighted radio centroid) as reported in the G4Jy catalog. In our previous analysis (paper I) G4Jy\,1192 lacks both mid-IR and optical counterparts. However, thanks to the \swf-XRT analysis, we have been able to identify these counterparts. X-ray contours obtained from the image in the left panel are overlaid to both mid-IR and optical images for comparison.}
\label{fig:noopt2}
\end{figure}

\begin{itemize}
\item For the radio source G4Jy\,672, with an assigned mid-IR counterpart reported in paper I, the X-ray counterpart lies within an angular separation of $\sim$2\arcsec\ from the optical host galaxy, thus confirming its previous association. 

\item There are two more sources, namely, G4Jy\,249 and G4Jy\,1432 (see Figure~\ref{fig:confused}), for which available radio maps lack the angular resolution needed to precisely locate their radio cores and consequently their host galaxies in the optical band. However, using the \swf-XRT images it has been possible to determine the position of their optical counterparts using their X-ray emission. We are also able to confirm that X-ray detected sources is spatially coincident with their mid-IR counterparts associated in the G4Jy catalog.

\item Despite the lack of a potential optical counterpart, the X-ray emission of G4Jy\,1136 is spatially associated with its mid-IR counterpart (see also Figure~\ref{fig:noopt1}) as previously stated in the G4Jy catalog \citep{white20a,white20b}. {On the other hand, the comparison with mid-IR, optical, and X-ray data carried out for} G4Jy\,1192, revealed the location of its optical host galaxy, as shown in the right panel of Figure~\ref{fig:noopt2}, in agreement with the mid-IR association described in paper I.
\end{itemize}

\section{Summary, conclusions and future perspectives}
\label{sec:summary} 
In this second paper of the series, we present a first overview of the X-ray archival observations for \cs\ radio sources, mainly focusing on those observed with the XRT instrument on board the Neil Gherels \swf\ Observatory.

The \cs\ sample was recently extracted from the G4Jy catalog \citep{white20a,white20b} to obtain a list of powerful radio sources, selected at low radio frequencies, equivalent, in terms of flux density, to the 3C catalog but having all sources located in the Southern Hemisphere (paper I). The main advantage underlying the selection of this sample is the opportunity to study powerful radio sources and emission processes occurring in their large scale environments with a catalog observable with modern Southern Hemisphere telescopes and instruments such as MUSE and ERIS at VLT and/or ALMA and, in the near future, also SKA, LSST and ELT.

We retrieved, reduced and analyzed all archival \swf-XRT datasets collected between May 2005 and November 2022, having nominal $T_{exp}$ larger than 250\,sec for a total of 615 observations processed for 89 sources. We found 28 radio sources with no X-ray counterpart detected above the chosen threshold of $SNR=3$. We found 11 radio sources with extended/diffuse X-ray emission, estimated according to the criterion based on the \swf-XRT PSF model. In particular, G4Jy\,1518 and G4Jy\,1664 show clear extended X-ray emission, previously unknown, suggesting that they could be hosted in galaxy clusters, while for G4Jy\,1863 we claim the X-ray counterpart detection of its radio lobes. All remaining 40 radio sources have a detected X-ray point-like counterpart of their radio core. The main reason underlying the non-detection of X-ray counterparts for all radio sources analyzed here is due to their relatively small integrated $T_{exp}$. 

\swf\ X-ray observations can provide first insights into the presence of diffuse/extended X-ray emission, {used as the basis for future studies, and also to} (i) confirm previous mid-IR and optical associations and to (ii) locate precisely the host galaxy position for those sources for which the angular resolution of the radio maps was not sufficient to reveal the position of the radio core. 

We also obtained additional results that refine our previous radio-to-optical comparison (paper I):
\begin{itemize}
\item For G4Jy\,672, having no assigned mid-IR counterpart in the original G4Jy catalog, the X-ray counterpart lies within an angular separation of $\sim$2\arcsec\ from the optical host galaxy, thus confirming its optical association since it is smaller than the typcial \swf-XRT positional uncertainty for that $T_{exp}$ \citep[see e.g.][]{evans20}. There is further confirmation of the host-galaxy position via detection of the radio core in a recent MeerKAT image \citep{sejake23}.
\item Our previous radio-to-optical comparison did not allow us to firmly establish the host galaxies of G4Jy\,249 and G4Jy\,1432 due to the relatively poor angular resolution of the radio maps used to locate their radio cores. However, using merged \swf-XRT images we determined precisely the location of their optical counterpart from their X-ray emission. 
\item We also confirmed that the X-ray emission of G4Jy\,1136 is spatially associated with its assigned mid-IR counterpart as listed in the original G4Jy catalog.
\item The mid-IR, optical and X-ray comparison, carried out for G4Jy\,1192, also revealed the location of its optical host galaxy.
\end{itemize}

\swf-XRT archival datasets, even if not homogeneous in terms of integrated $T_{exp}$ that span a range between 5 to $96\,\mathrm{ksec}$, proved to be a powerful tool to refine the optical search for host galaxies of powerful radio sources and discover the presence of extended X-ray emission. In the case of G4Jy\,1518 and G4Jy\,1664, having snapshot XRT observations of $\sim$4 and $\sim2\,\mathrm{ksec}$ integrated $T_{exp}$, it has been possible to reveal the presence of the surrounding ICM. 

\swf-XRT observations are certainly paving the path to identify those radio sources deserving X-ray follow up observations to perform {detailed} X-ray spectral analyses. This is crucial to complete optical spectroscopic observations necessary to (i) obtain the $z$ measurements for the whole \cs\ catalog and (ii) classify all radio sources listed therein from an optical perspective.  

Additional X-ray datasets have been already requested and approved for the \xmm\ satellite, as ``filler programs'' and as snapshot observations, to achieve a more complete high energy overview of the \cs\ catalog while follow up X-ray observations will be also requested in the near future to deeply investigate the large-scale environment of those radio sources harbored in the galaxy clusters/groups.

\newpage
\acknowledgments
We thank the anonymous referee for useful and valuable comments and suggestions that led to improvements in the paper.
We wish to dedicate this paper to D. E. Harris and R. W. Hunstead, their insight, passion and contributions to radio astronomy are an inspiration for most of us.

F. M. wishes to thank Dr. A. Moretti for their valuable discussions on the \swf-XRT PSF models and for all useful suggestions provided.  

A. J. acknowledges the financial support (MASF\_CONTR\_FIN\_18\_01) from the Italian National Institute of Astrophysics under the agreement with the Instituto de Astrofisica de Canarias for the ``Becas Internacionales para Licenciados y/o Graduados Convocatoria de 2017’’. S.V. W. acknowledges financial assistance of the South African Radio Astronomy Observatory (SARAO)\footnote{https://www.sarao.ac.za}. W.F. and R.K. acknowledge support from the Smithsonian Institution and the \chn\ High Resolution Camera Project through NASA contract NAS8-03060.  W.F. also acknowledges support from NASA Grants 80NSSC19K0116, GO1-22132X, and GO9-20109X. This investigation is supported by the National Aeronautics and Space Administration (NASA) grants GO9-20083X, GO0-21110X and GO1-22087X. I.A., S.A.C. and V.R. are partially supported by grant PIP 1220200102169CO, Argentine Research Council (CONICET), and by the UNLP research project, G178 (2022-2025). A.G.-P. acknowledges support from the CONACyT program for their PhD studies. V.C. acknowledges support from the Fulbright — García Robles scholarship. This work was partially supported by CONACyT (Consejo Nacional de Ciencia y Tecnología) research grant 280789.

This research has made use of the NASA/IPAC Extragalactic Database (NED), which is operated by the Jet Propulsion Laboratory, California Institute of Technology, under contract with the National Aeronautics and Space Administration.
This research has made use of the SIMBAD database, operated at CDS, Strasbourg, France \citep{wenger00}.

This research has made use of the CIRADA cutout service at URL cutouts.cirada.ca, operated by the Canadian Initiative for Radio Astronomy Data Analysis (CIRADA). CIRADA is funded by a grant from the Canada Foundation for Innovation 2017 Innovation Fund (Project 35999), as well as by the Provinces of Ontario, British Columbia, Alberta, Manitoba and Quebec, in collaboration with the National Research Council of Canada, the US National Radio Astronomy Observatory and Australia’s Commonwealth Scientific and Industrial Research Organisation.
The National Radio Astronomy Observatory is a facility of the National Science Foundation operated under cooperative agreement by Associated Universities, Inc.
Part of this work is based on the NVSS (NRAO VLA Sky Survey): The National Radio Astronomy Observatory is operated by Associated Universities, Inc., under contract with the National Science Foundation and on the VLA low-frequency Sky Survey (VLSS). 
We thank the staff of the GMRT that made these observations possible. GMRT is run by the National Centre for Radio Astrophysics of the Tata Institute of Fundamental Research.
The Molonglo Observatory site manager, Duncan Campbell-Wilson, and the staff, Jeff Webb, Michael White and John Barry, are responsible for the smooth operation of Molonglo Observatory Synthesis Telescope (MOST) and the day-to-day observing programme of SUMSS. The SUMSS survey is dedicated to Michael Large whose expertise and vision made the project possible. The MOST is operated by the School of Physics with the support of the Australian Research Council and the Science Foundation for Physics within the University of Sydney. 
This scientific work makes use of the Murchison Radio-astronomy Observatory, operated by CSIRO. We acknowledge the Wajarri Yamatji people as the traditional owners of the Observatory site. Support for the MWA comes from the US National Science Foundation (grants AST-0457585, PHY-0835713, CAREER-0847753, and AST-0908884), the Australian Research Council (LIEF grants LE0775621 and LE0882938), the US Air Force Office of Scientific Research (grant FA9550-0510247), and the Centre for All-sky Astrophysics (an Australian Research Council Centre of Excellence funded by grant CE110001020). Support is also provided by the Smithsonian Astrophysical Observatory, the MIT School of Science, the Raman Research Institute, the Australian National University, and the Victoria University of Wellington (via grant MED-E1799 from the New Zealand Ministry of Economic Development and an IBM Shared University Research Grant). The Australian Federal government provides additional support via the Commonwealth Scientific and Industrial Research Organisation (CSIRO), National Collaborative Research Infrastructure Strategy, Education Investment Fund, and the Australia India Strategic Research Fund, and Astronomy Australia Limited, under contract to Curtin University. This work was supported by resources provided by the Pawsey Supercomputing Centre with funding from the Australian Government and the Government of Western Australia. We acknowledge the iVEC Petabyte Data Store, the Initiative in Innovative Computing, and the CUDA Center for Excellence sponsored by NVIDIA at Harvard University, and the International Centre for Radio Astronomy Research (ICRAR), a Joint Venture of Curtin University, and The University of Western Australia, funded by the Western Australian State government.

This publication makes use of data products from the Wide-field Infrared Survey Explorer, which is a joint project of the University of California, Los Angeles, and the Jet Propulsion Laboratory/California Institute of Technology, funded by the National Aeronautics and Space Administration.

This project used public archival data from the Dark Energy Survey (DES). Funding for the DES Projects has been provided by the U.S. Department of Energy, the U.S. National Science Foundation, the Ministry of Science and Education of Spain, the Science and Technology FacilitiesCouncil of the United Kingdom, the Higher Education Funding Council for England, the National Center for Supercomputing Applications at the University of Illinois at Urbana-Champaign, the Kavli Institute of Cosmological Physics at the University of Chicago, the Center for Cosmology and Astro-Particle Physics at the Ohio State University, the Mitchell Institute for Fundamental Physics and Astronomy at Texas A\&M University, Financiadora de Estudos e Projetos, Funda{\c c}{\~a}o Carlos Chagas Filho de Amparo {\`a} Pesquisa do Estado do Rio de Janeiro, Conselho Nacional de Desenvolvimento Cient{\'i}fico e Tecnol{\'o}gico and the Minist{\'e}rio da Ci{\^e}ncia, Tecnologia e Inova{\c c}{\~a}o, the Deutsche Forschungsgemeinschaft, and the Collaborating Institutions in the Dark Energy Survey.
The Collaborating Institutions are Argonne National Laboratory, the University of California at Santa Cruz, the University of Cambridge, Centro de Investigaciones Energ{\'e}ticas, Medioambientales y Tecnol{\'o}gicas-Madrid, the University of Chicago, University College London, the DES-Brazil Consortium, the University of Edinburgh, the Eidgen{\"o}ssische Technische Hochschule (ETH) Z{\"u}rich,  Fermi National Accelerator Laboratory, the University of Illinois at Urbana-Champaign, the Institut de Ci{\`e}ncies de l'Espai (IEEC/CSIC), the Institut de F{\'i}sica d'Altes Energies, Lawrence Berkeley National Laboratory, the Ludwig-Maximilians Universit{\"a}t M{\"u}nchen and the associated Excellence Cluster Universe, the University of Michigan, the National Optical Astronomy Observatory, the University of Nottingham, The Ohio State University, the OzDES Membership Consortium, the University of Pennsylvania, the University of Portsmouth, SLAC National Accelerator Laboratory, Stanford University, the University of Sussex, and Texas A\&M University.
Based in part on observations at Cerro Tololo Inter-American Observatory, National Optical Astronomy Observatory, which is operated by the Association of Universities for Research in Astronomy (AURA) under a cooperative agreement with the National Science Foundation.
The Pan-STARRS1 Surveys (PS1) have been made possible through contributions of the Institute for Astronomy, the University of Hawaii, the Pan-STARRS Project Office, the Max-Planck Society and its participating institutes, the Max Planck Institute for Astronomy, Heidelberg and the Max Planck Institute for Extraterrestrial Physics, Garching, The Johns Hopkins University, Durham University, the University of Edinburgh, Queen's University Belfast, the Harvard-Smithsonian Center for Astrophysics, the Las Cumbres Observatory Global Telescope Network Incorporated, the National Central University of Taiwan, the Space Telescope Science Institute, the National Aeronautics and Space Administration under Grant No. NNX08AR22G issued through the Planetary Science Division of the NASA Science Mission Directorate, the National Science Foundation under Grant No. AST-1238877, the University of Maryland, and Eotvos Lorand University (ELTE).
Based on photographic data obtained using The UK Schmidt Telescope. The UK Schmidt Telescope was operated by the Royal Observatory Edinburgh, with funding from the UK Science and Engineering Research Council, until 1988 June, and thereafter by the Anglo-Australian Observatory. Original plate material is copyright (c) of the Royal Observatory Edinburgh and the Anglo-Australian Observatory. The plates were processed into the present compressed digital form with their permission. The Digitized Sky Survey was produced at the Space Telescope Science Institute under US Government grant NAG W-2166.
We acknowledge the efforts of the staff of the Anglo-Australian Observatory, who have undertaken the observations and developed the 6dF instrument.
This dataset or service is made available by the Infrared Science Archive (IRSA) at IPAC, which is operated by the California Institute of Technology under contract with the National Aeronautics and Space Administration.

SAOImageDS9 development has been made possible by funding from the \chn\ X-ray Science Center (CXC), the High Energy Astrophysics Science Archive Center (HEASARC) and the JWST Mission office at Space Telescope Science Institute \citep{joye03}.
This research has made use of data obtained from the high-energy Astrophysics Science Archive Research Center (HEASARC) provided by NASA’s Goddard Space Flight Center. We acknowledge the use of NASA's SkyView facility (http://skyview.gsfc.nasa.gov) located at NASA Goddard Space Flight Center.
TOPCAT and STILTS astronomical software \citep{taylor05} were used for the preparation and manipulation of the tabular data and the images.
The analysis is partially based on the OCCAM computing facility hosted by C3S\footnote{http://c3s.unito.it/} at UniTO \citep{aldinucci17}.

\newpage
\appendix

\section{X-ray images}
\label{app:ximages}
{X-ray images for all sources, obtained from merged event files. These images were restricted to the 0.5-10 keV energy range and smoothed with a Gaussian kernel of 10 pixels (i.e., 23.57\arcsec). Radio contours overlaid were drawn using archival radio maps using the parameters reported in Table~\ref{tab:radcon}. Magenta circles, having a radius of 120.207\arcsec\ (i.e., 51 pixels) thus enclosing 90\% of the \swf-XRT PSF, are centered on the position of the X-ray counterpart associated with each radio source, when detected.}

\begin{figure*}[!th]
\begin{center}
\includegraphics[width=3.8cm,height=6.4cm,angle=-90]{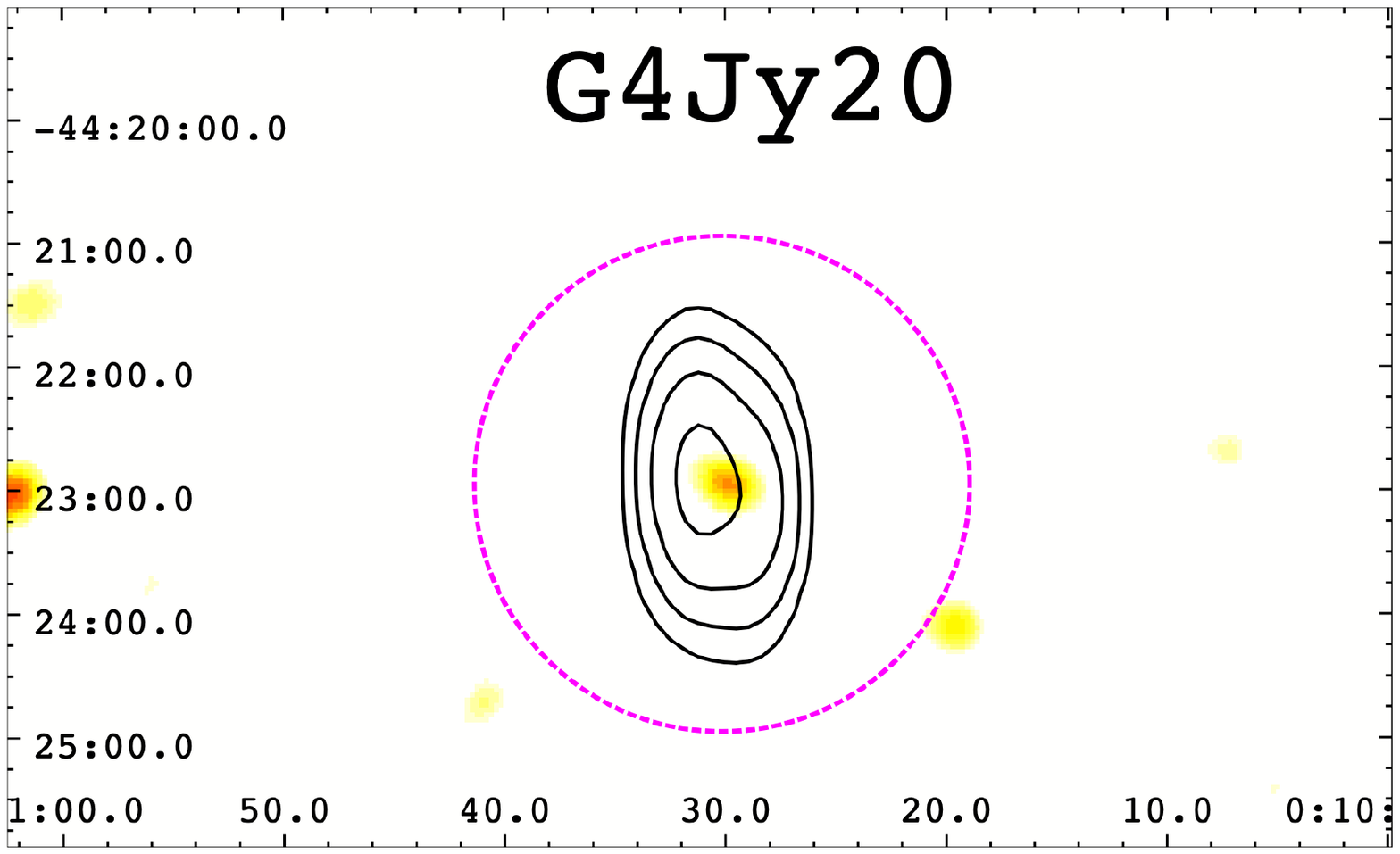}
\includegraphics[width=3.8cm,height=6.4cm,angle=-90]{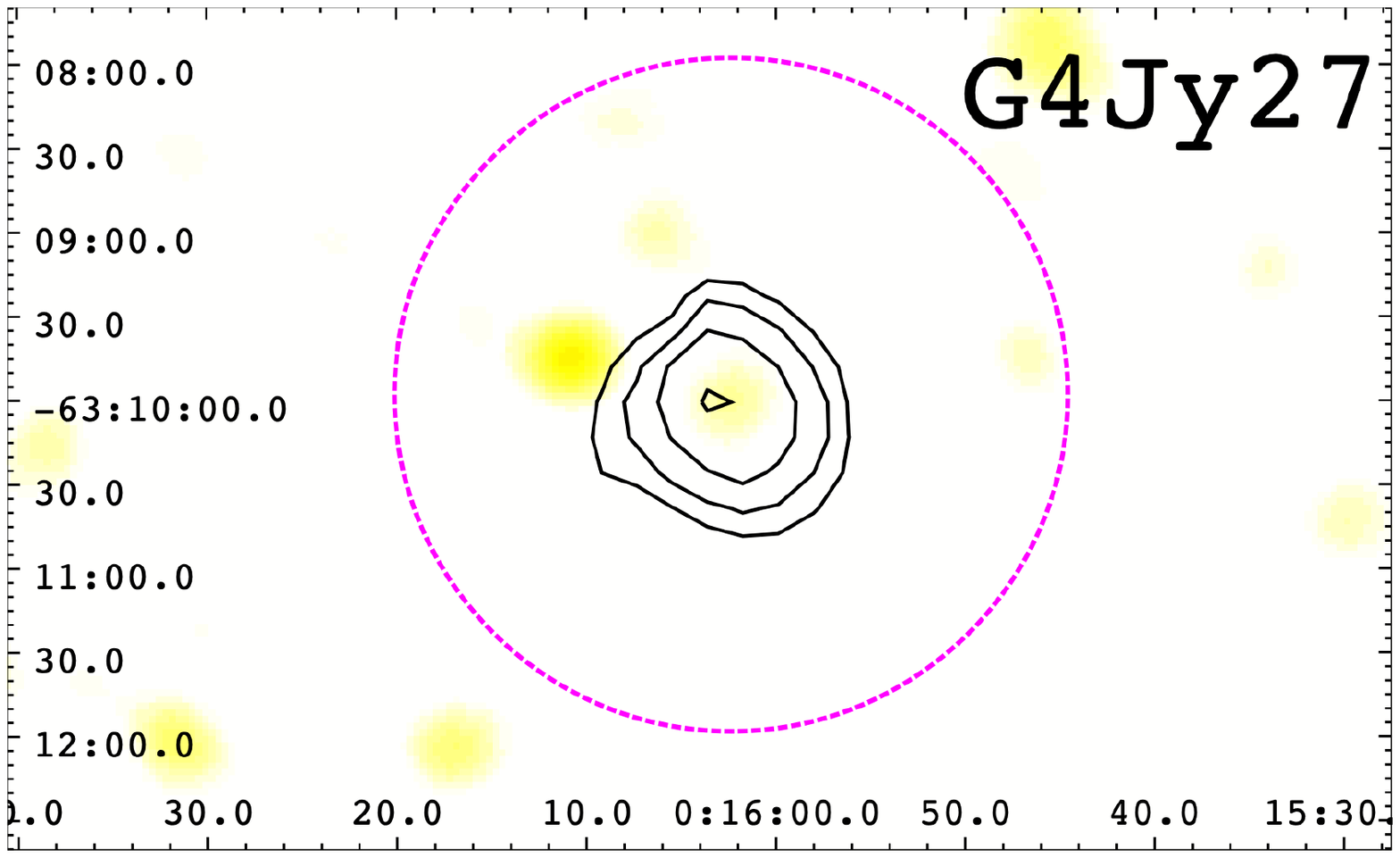}
\includegraphics[width=3.8cm,height=6.4cm,angle=-90]{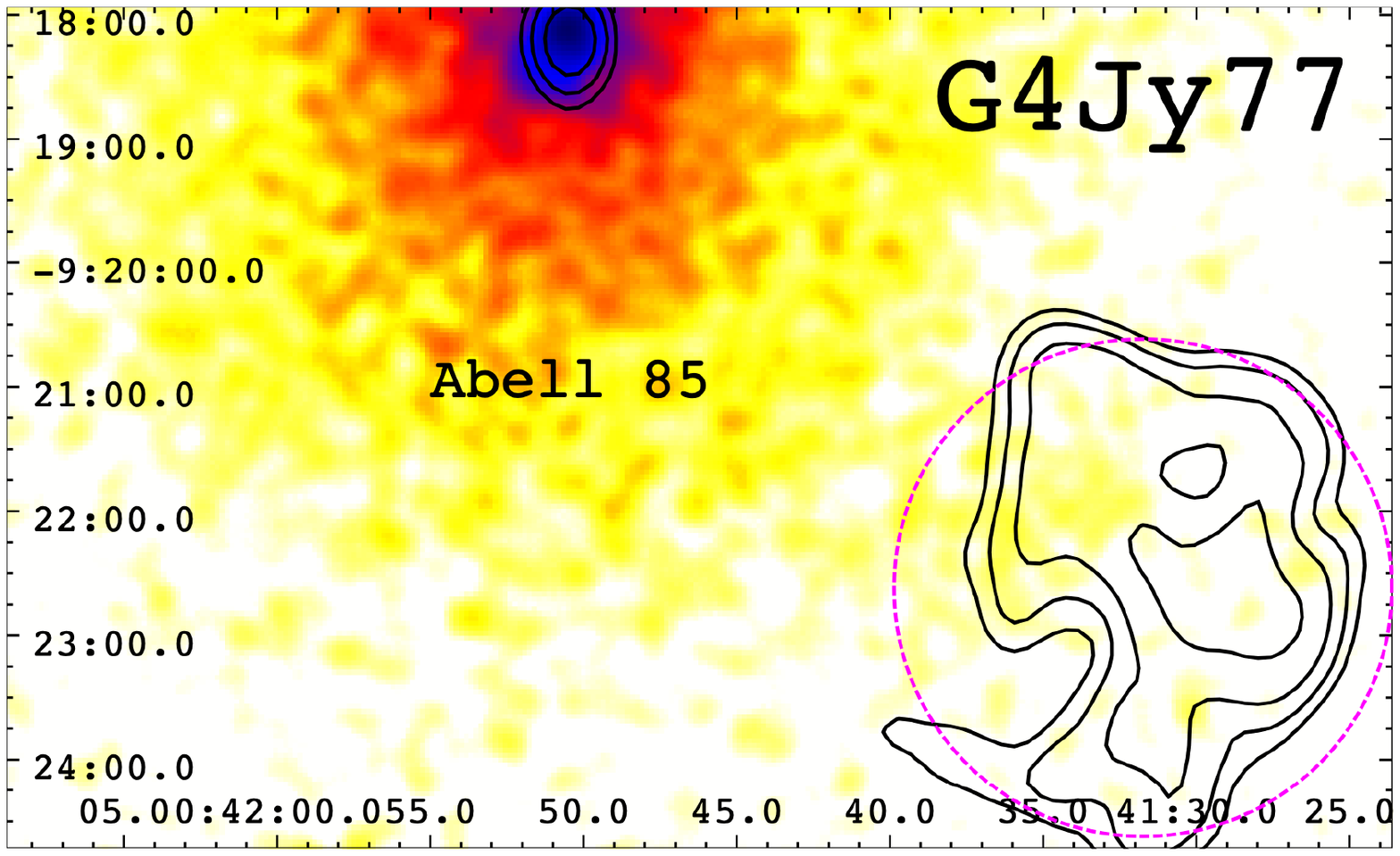}
\includegraphics[width=3.8cm,height=6.4cm,angle=-90]{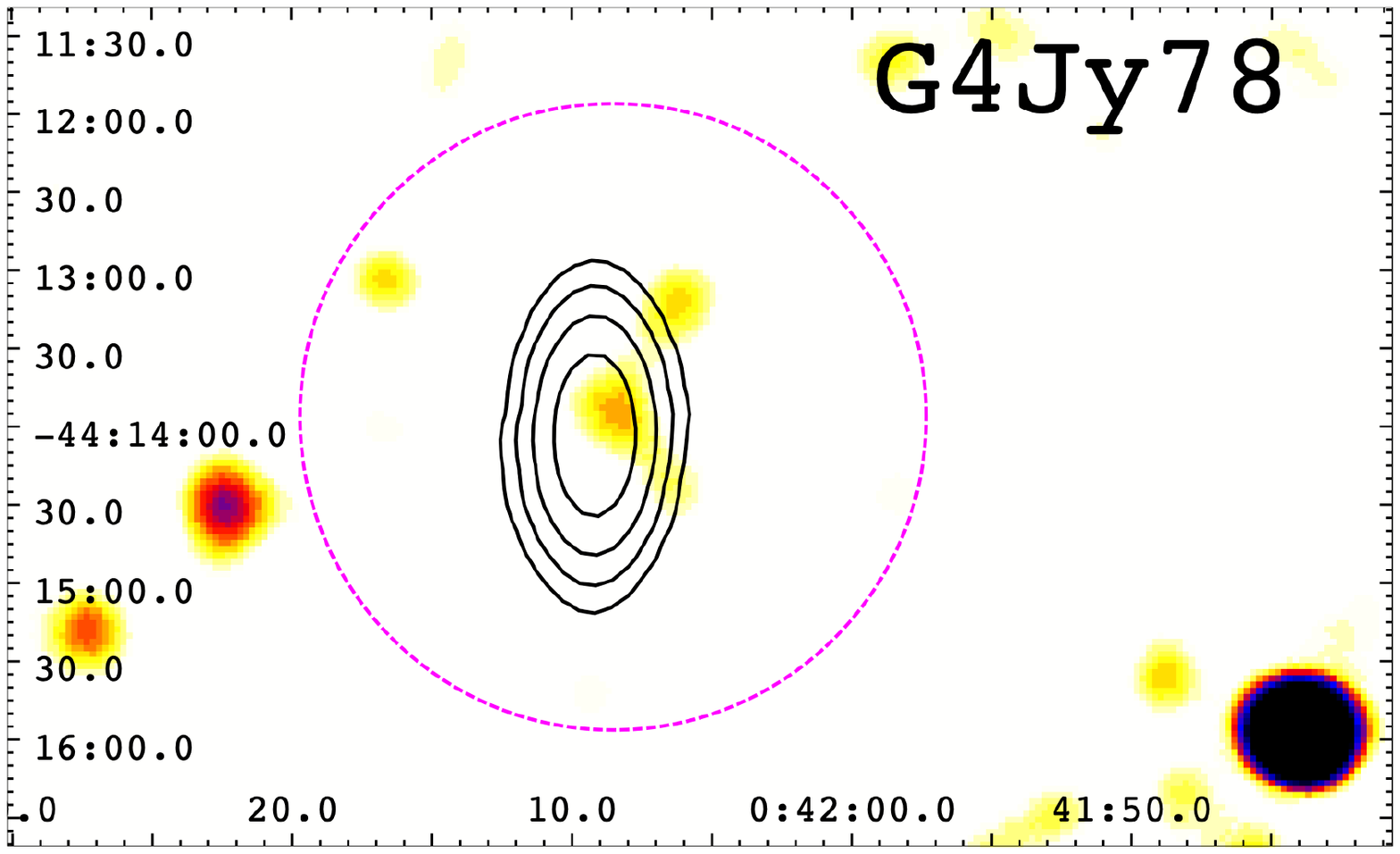}
\includegraphics[width=3.8cm,height=6.4cm,angle=-90]{./G4Jy85xrt.pdf}
\includegraphics[width=3.8cm,height=6.4cm,angle=-90]{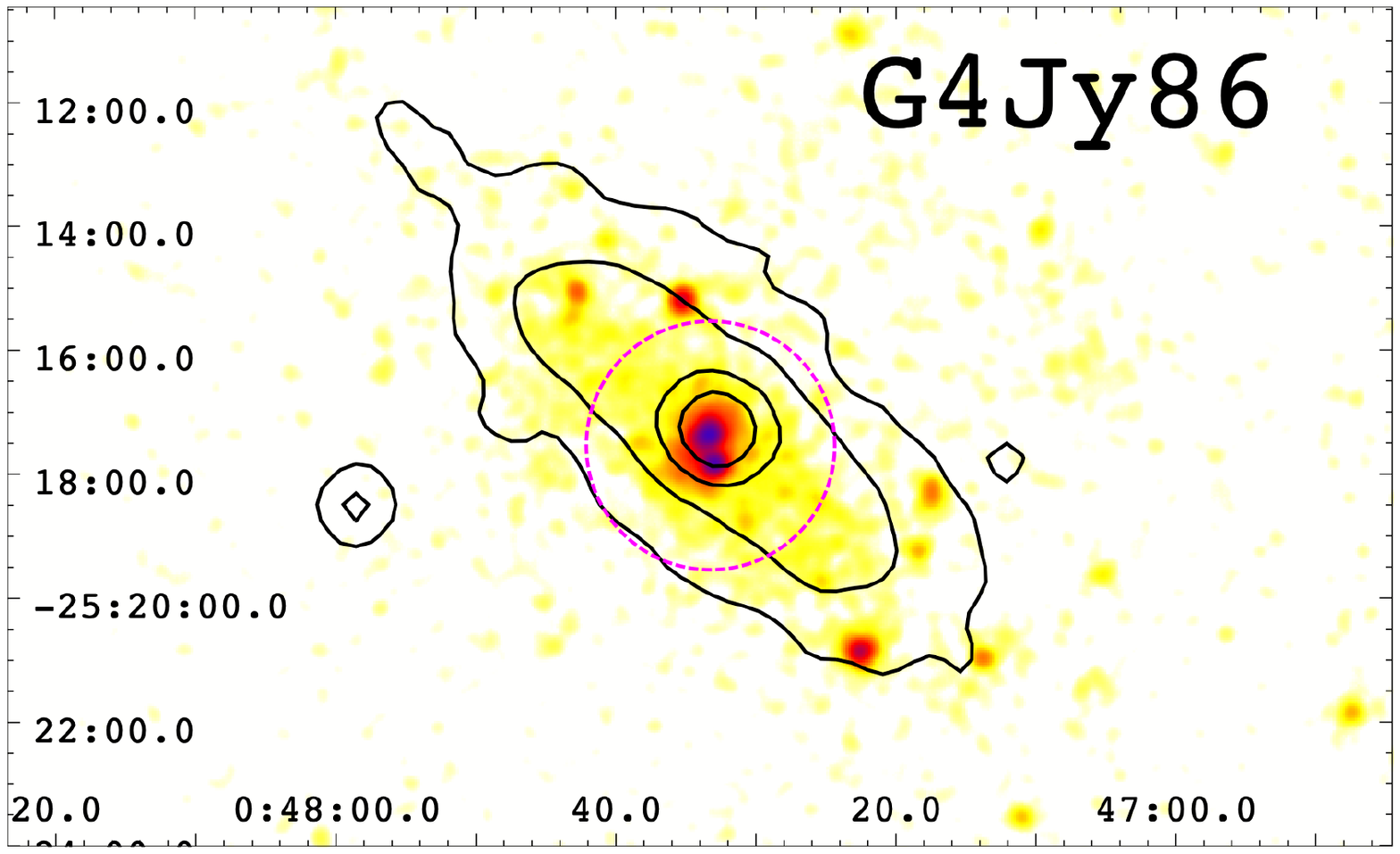}
\includegraphics[width=3.8cm,height=6.4cm,angle=-90]{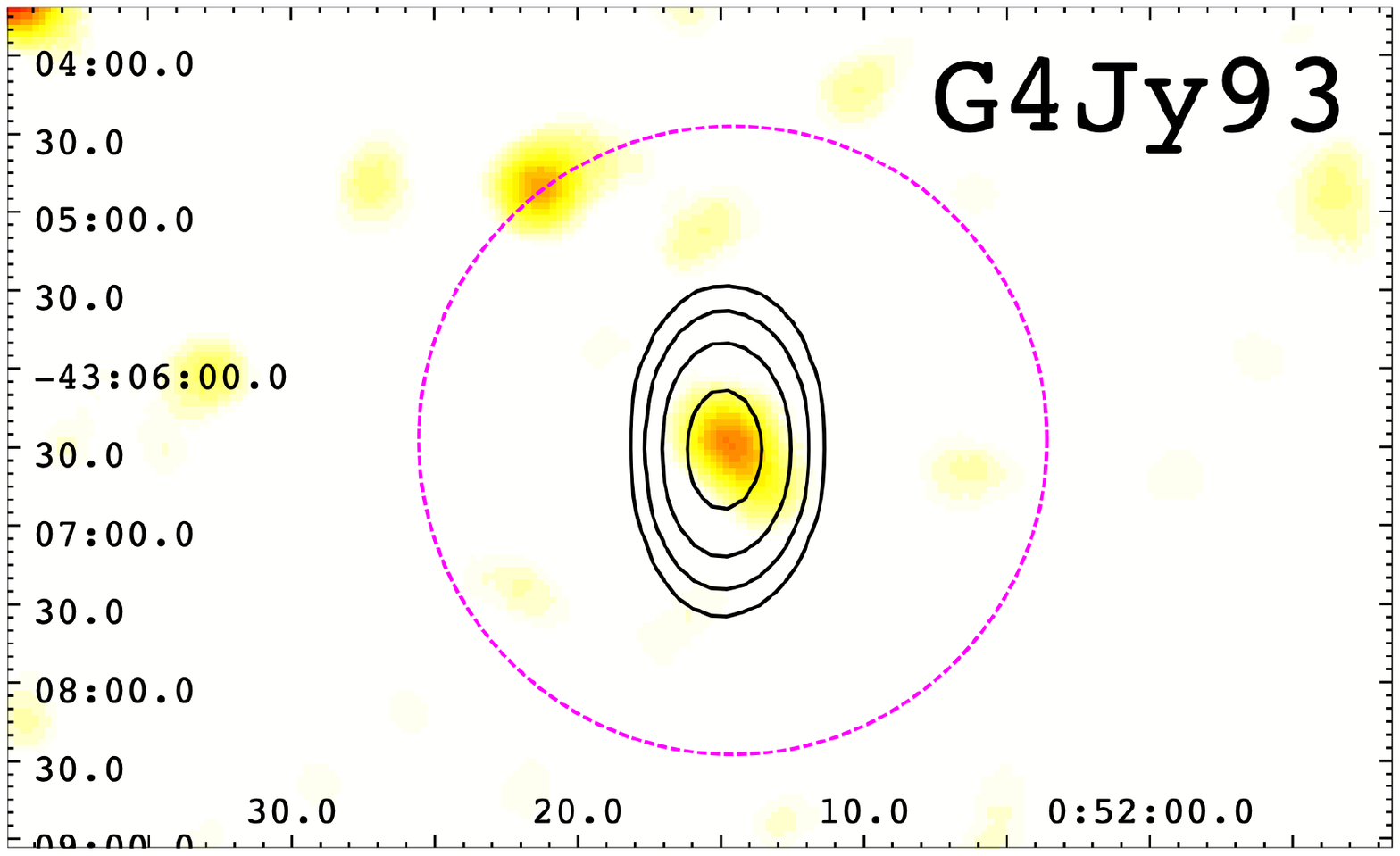}
\includegraphics[width=3.8cm,height=6.4cm,angle=-90]{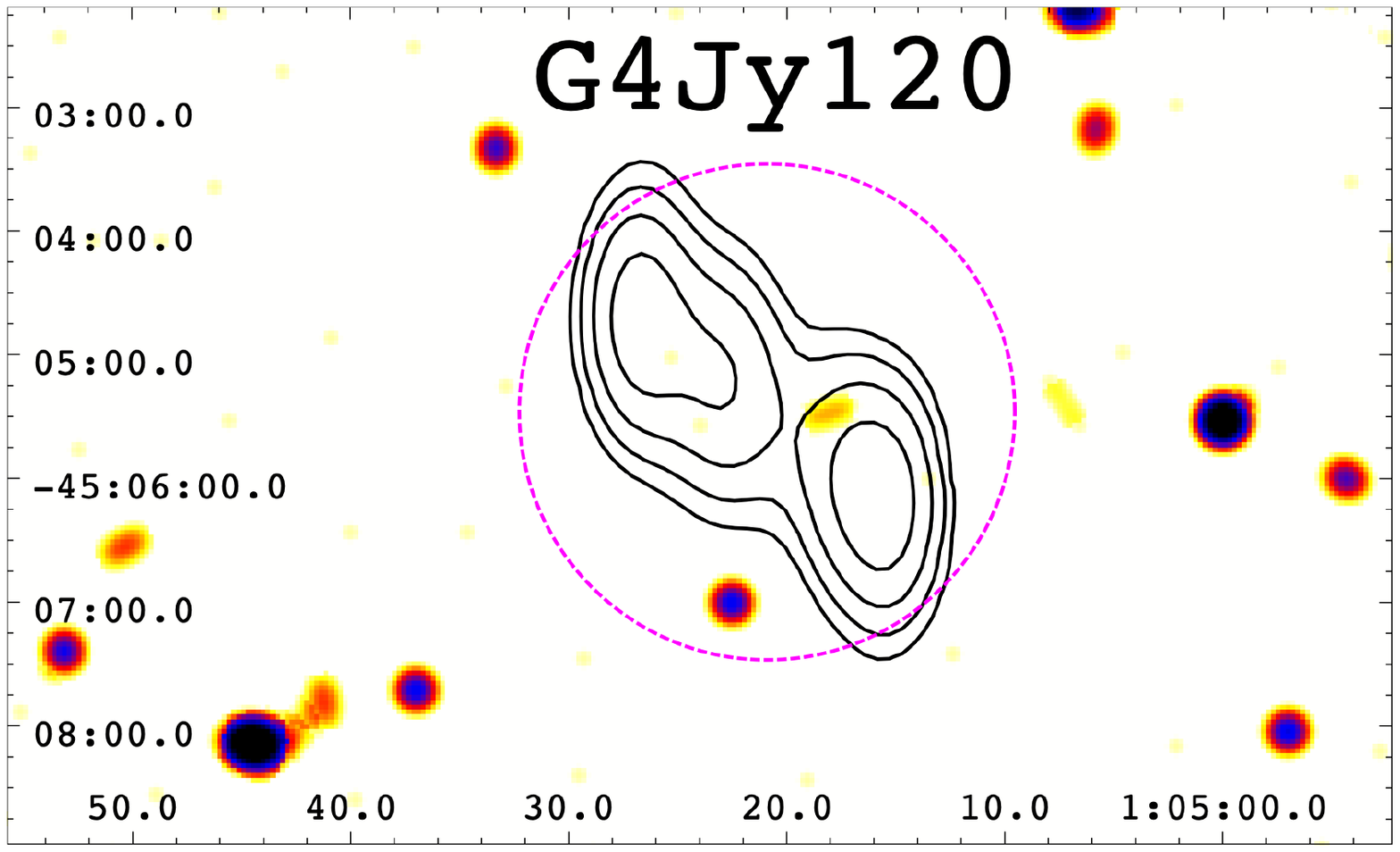}
\includegraphics[width=3.8cm,height=6.4cm,angle=-90]{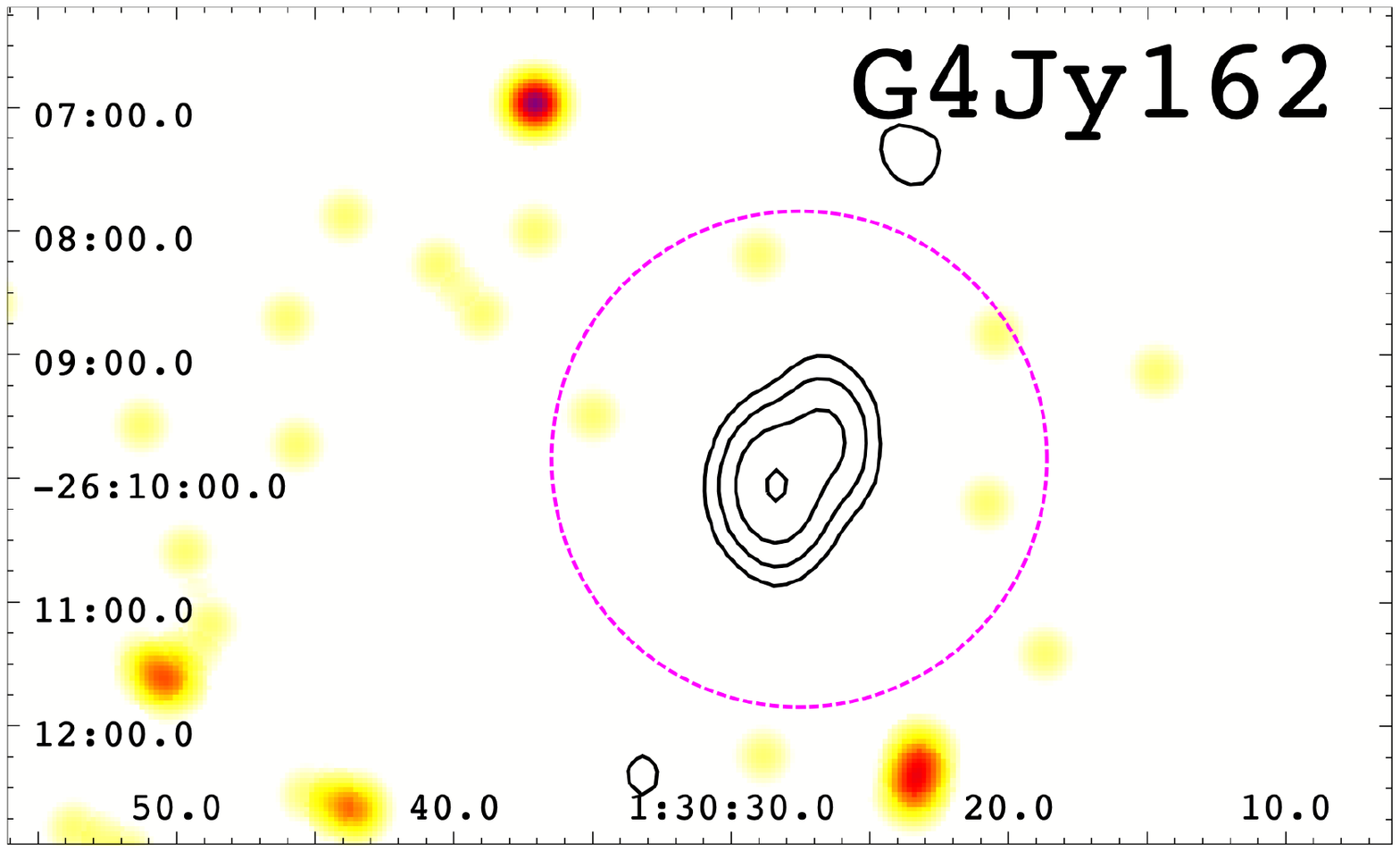}
\includegraphics[width=3.8cm,height=6.4cm,angle=-90]{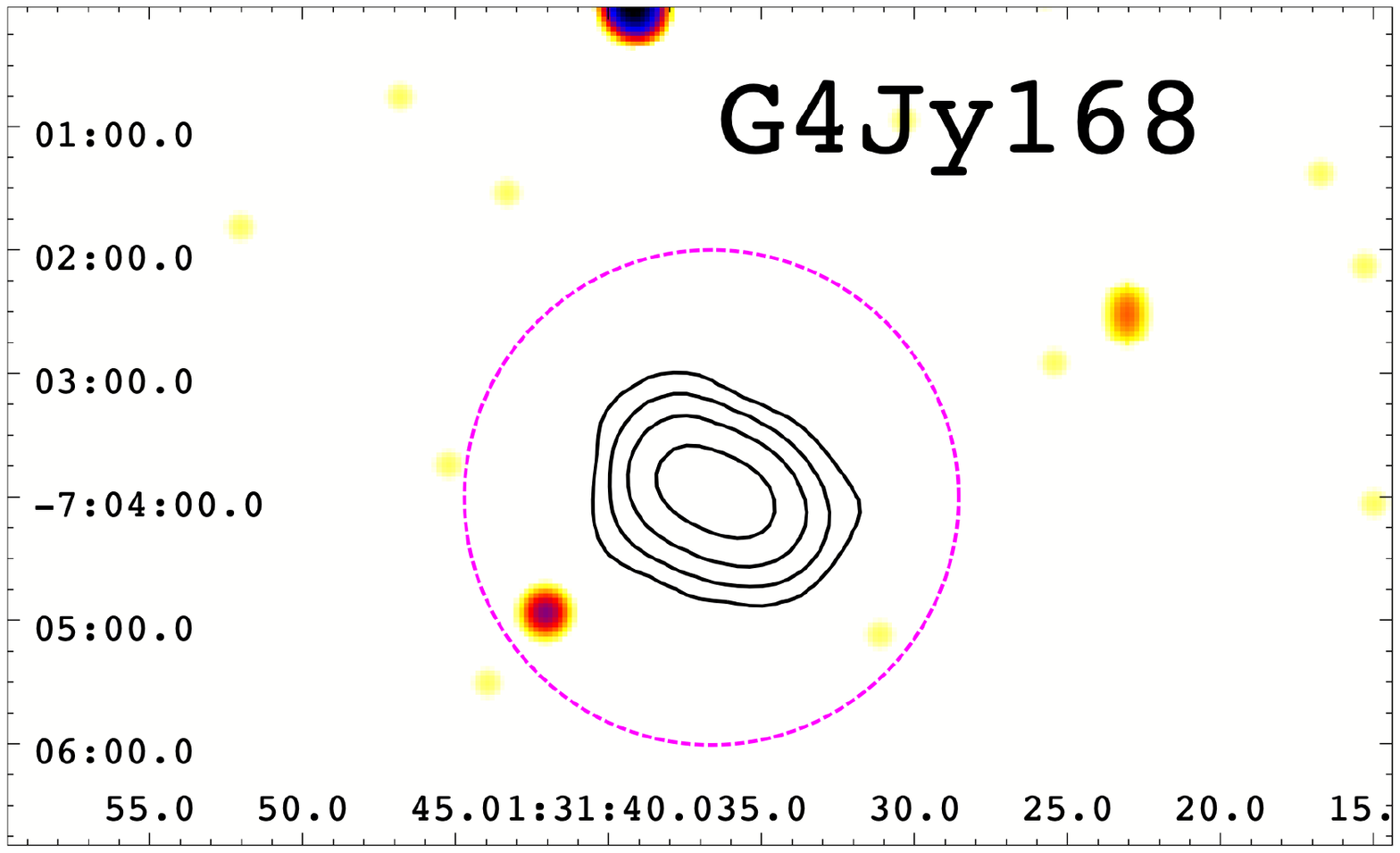}
\includegraphics[width=3.8cm,height=6.4cm,angle=-90]{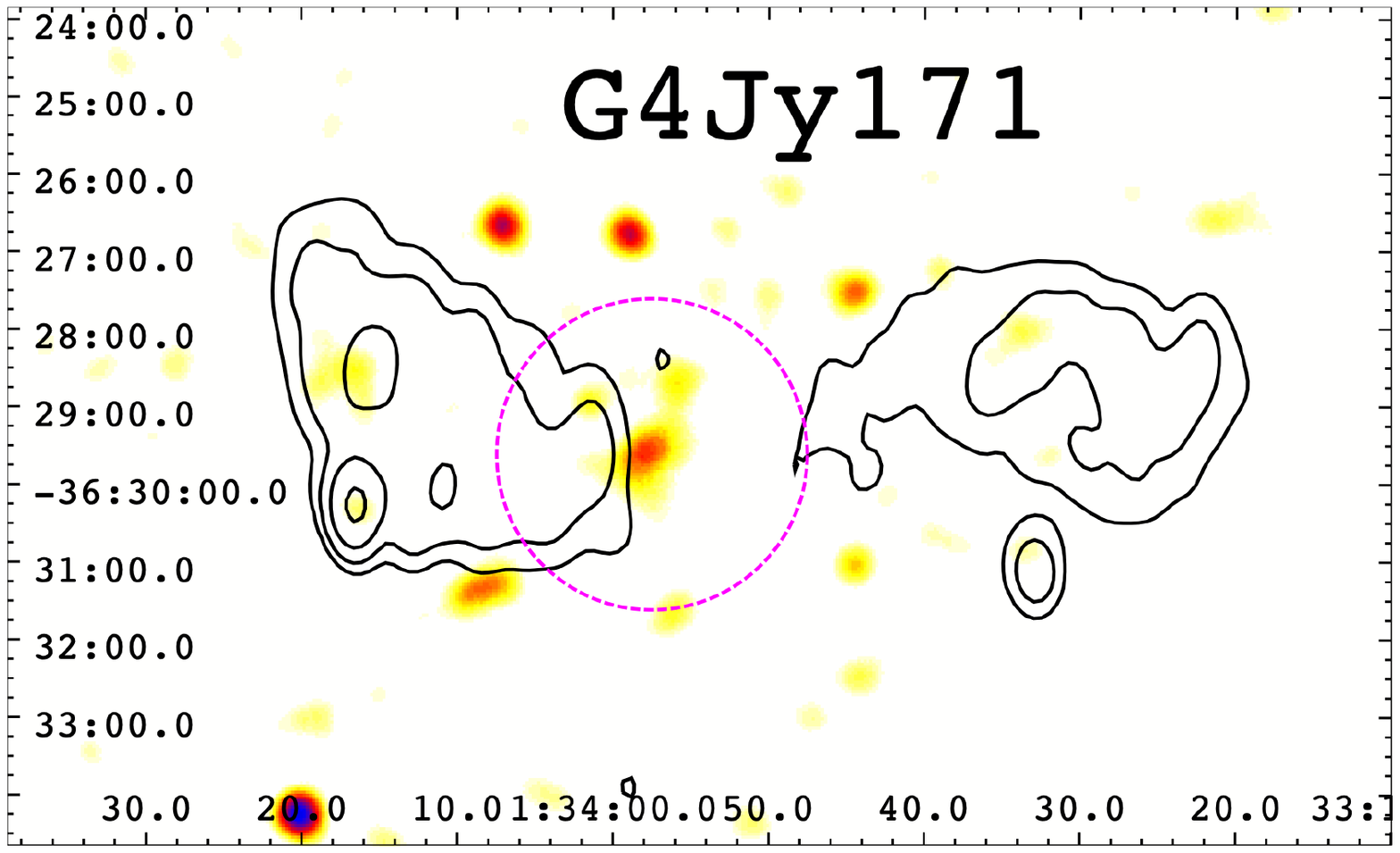}
\includegraphics[width=3.8cm,height=6.4cm,angle=-90]{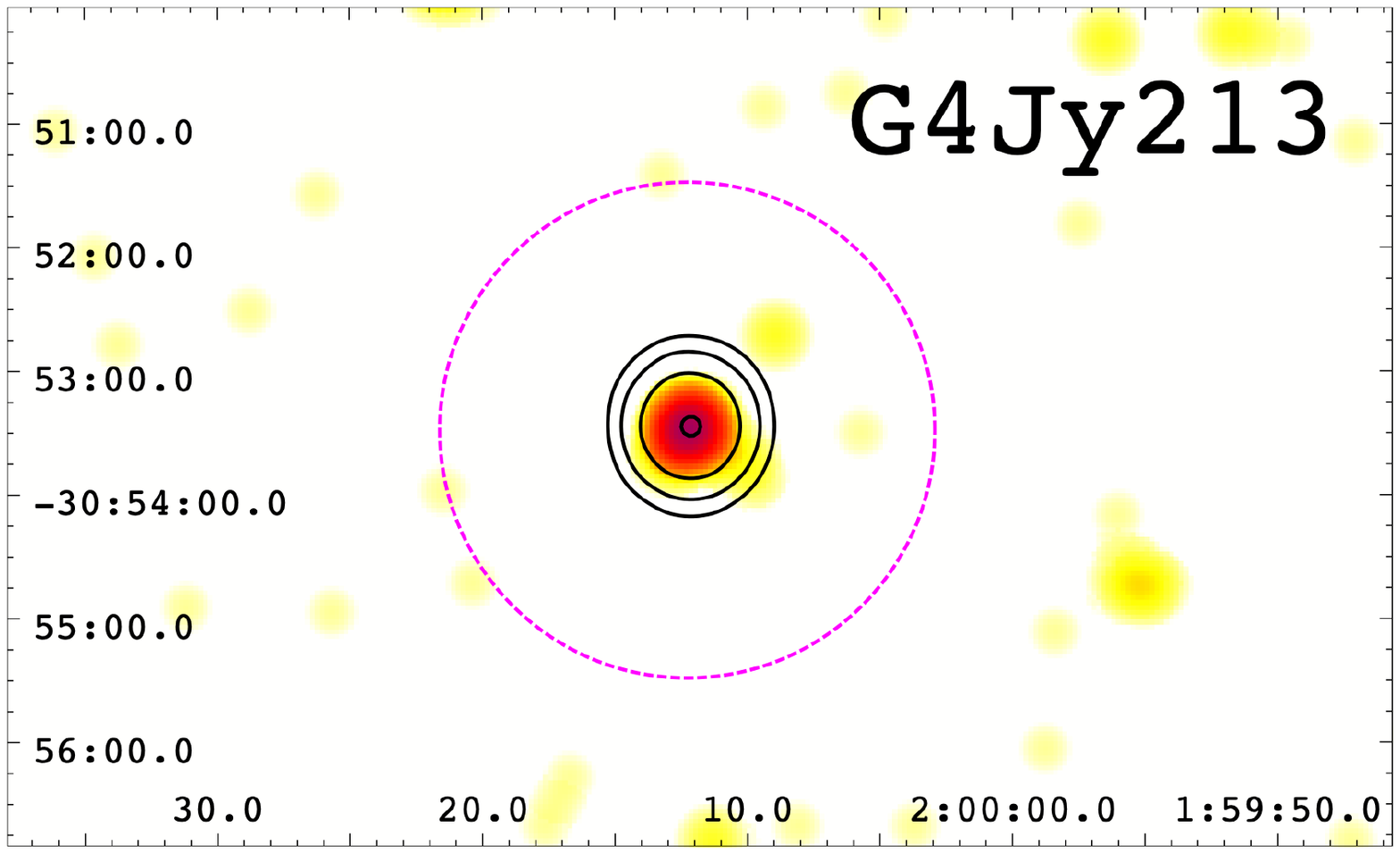}
\caption{Same as Figure~\ref{fig:example} for the following \cs\ radio sources:
G4Jy\,20, G4Jy\,27, G4Jy\,77, G4Jy\,78, G4Jy\,85, G4Jy\,86, G4Jy\,93, G4Jy\,120, G4Jy\,162, G4Jy\,168, G4Jy\,171, G4Jy\,213.}
\end{center}
\end{figure*}

\begin{figure*}[!th]
\begin{center}
\includegraphics[width=3.8cm,height=6.4cm,angle=-90]{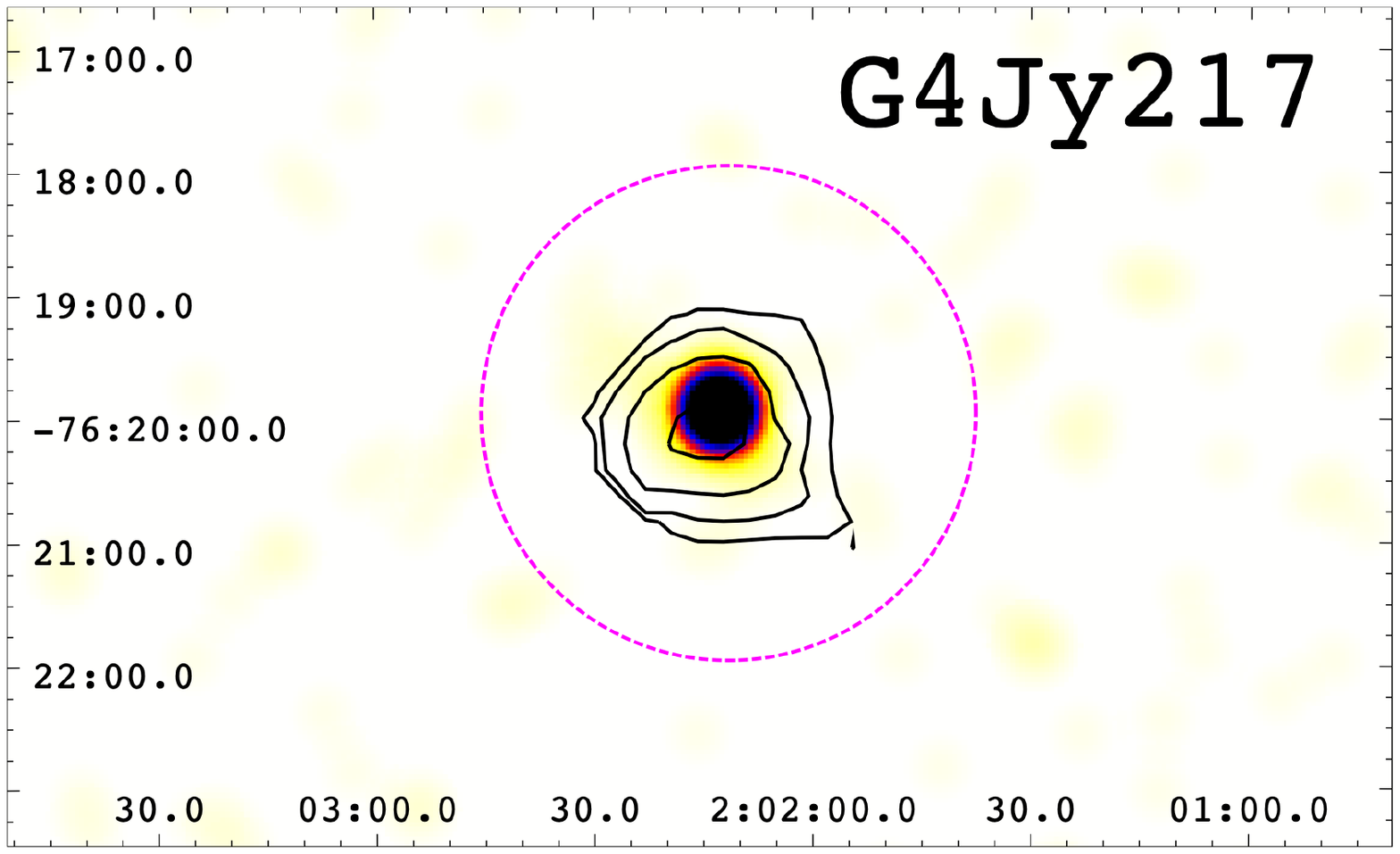}
\includegraphics[width=3.8cm,height=6.4cm,angle=-90]{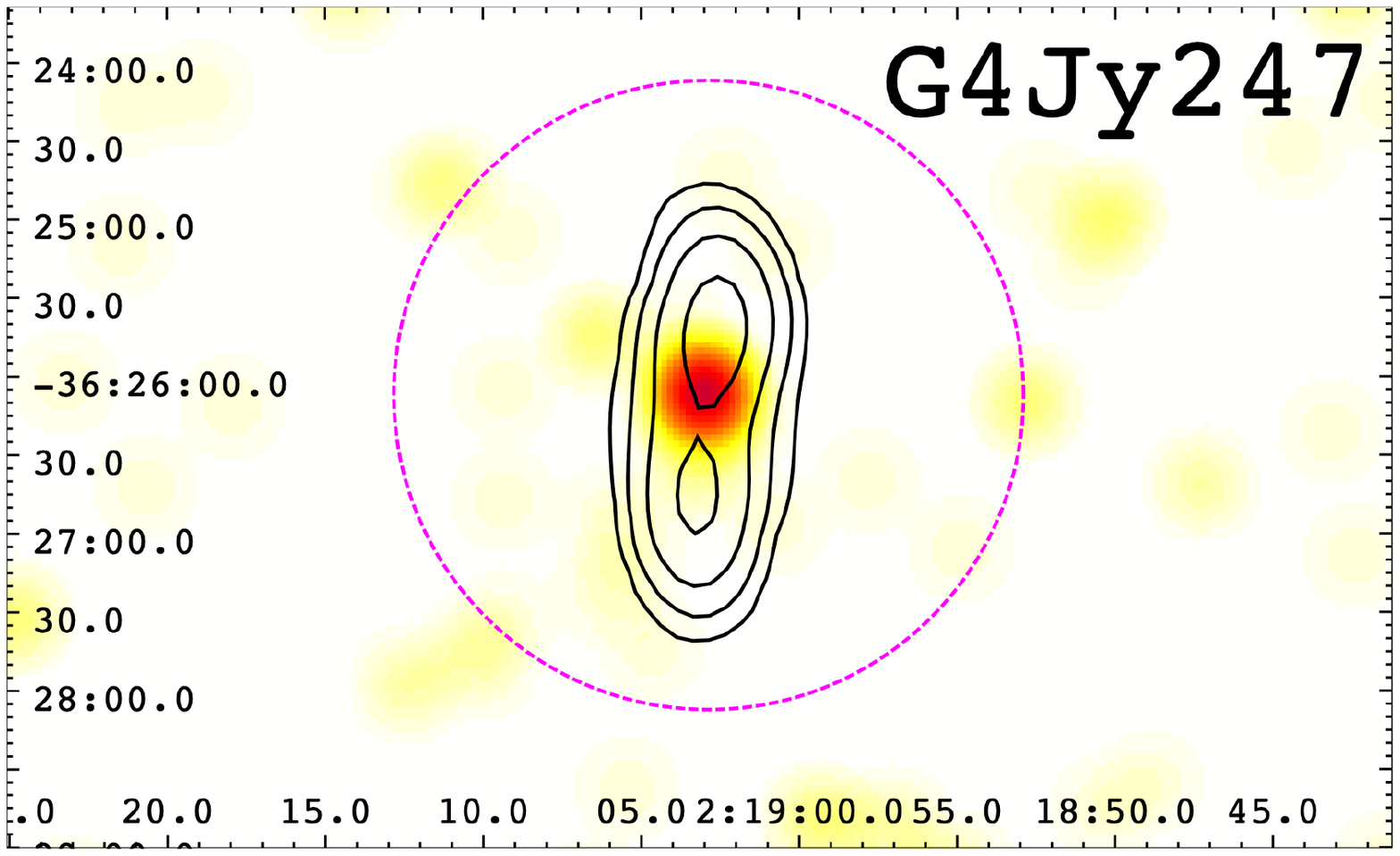}
\includegraphics[width=3.8cm,height=6.4cm,angle=-90]{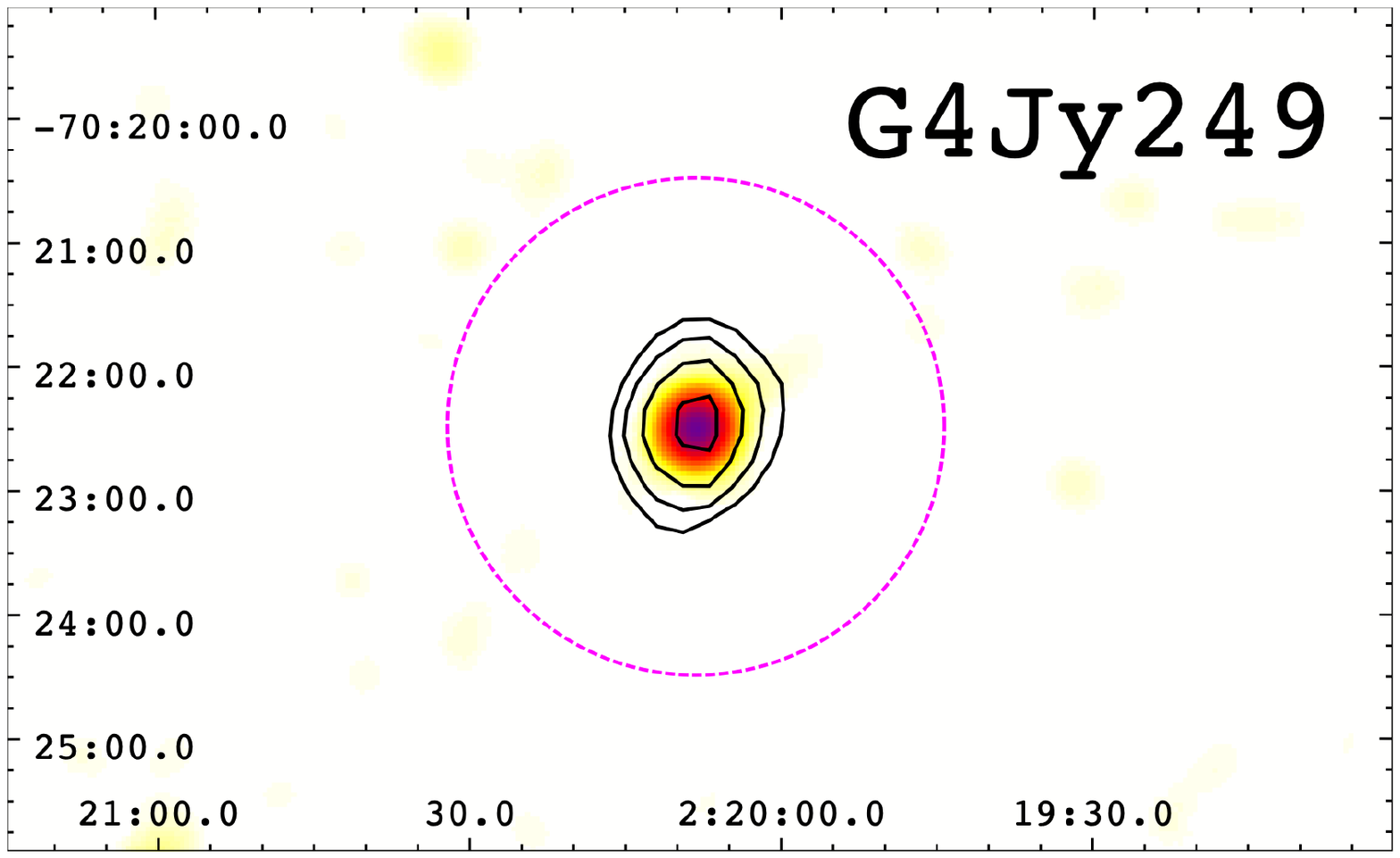}
\includegraphics[width=3.8cm,height=6.4cm,angle=-90]{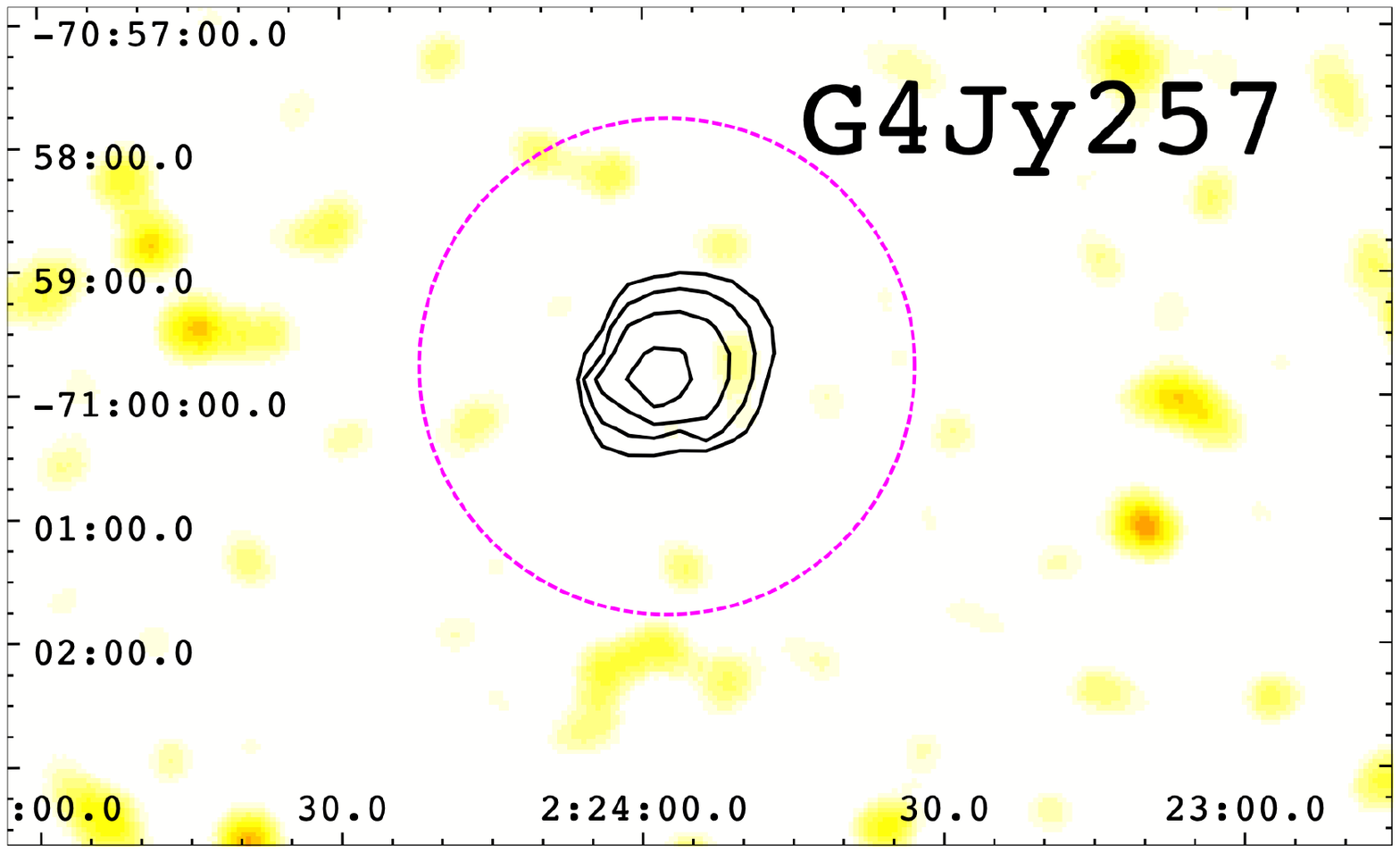}
\includegraphics[width=3.8cm,height=6.4cm,angle=-90]{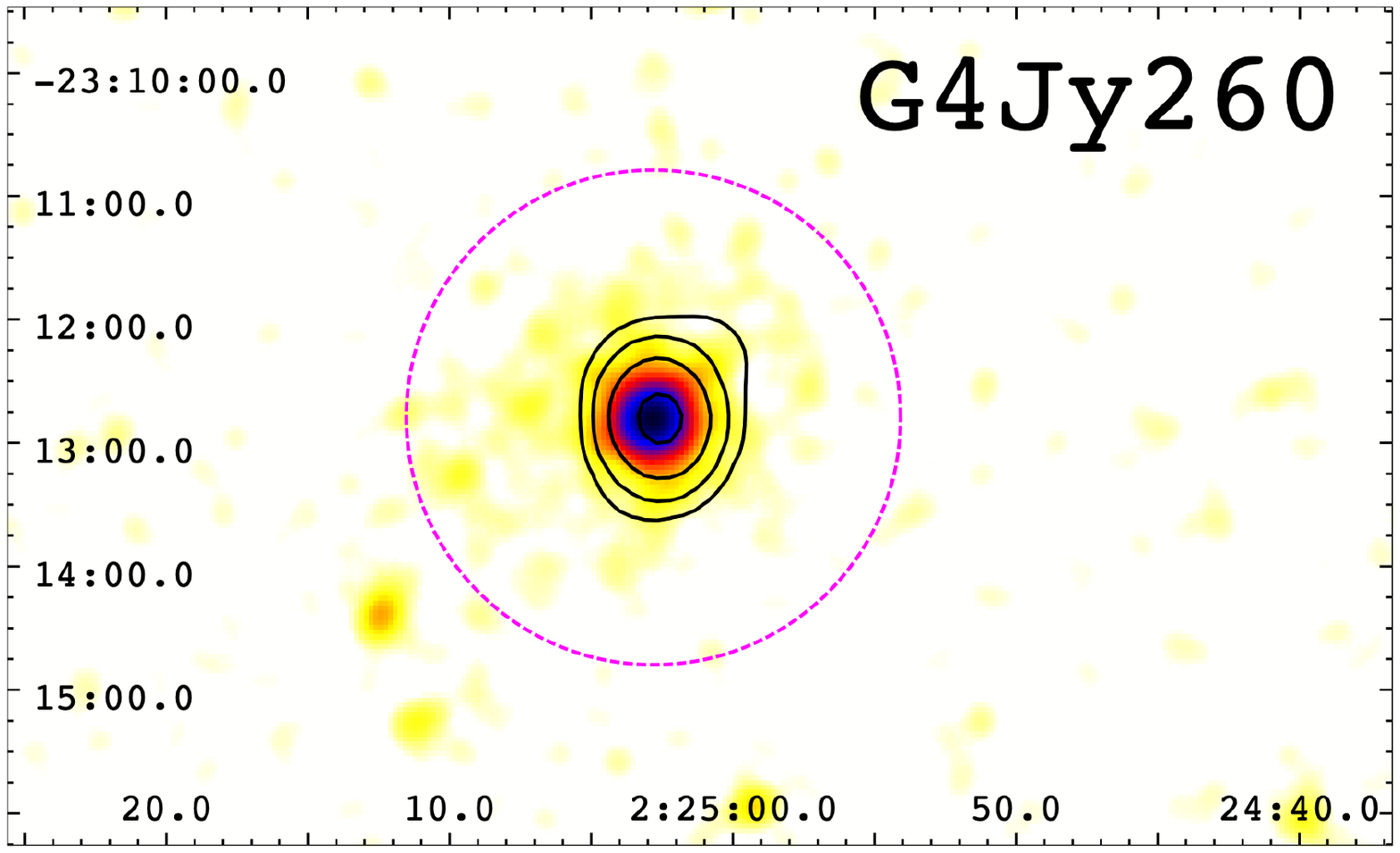}
\includegraphics[width=3.8cm,height=6.4cm,angle=-90]{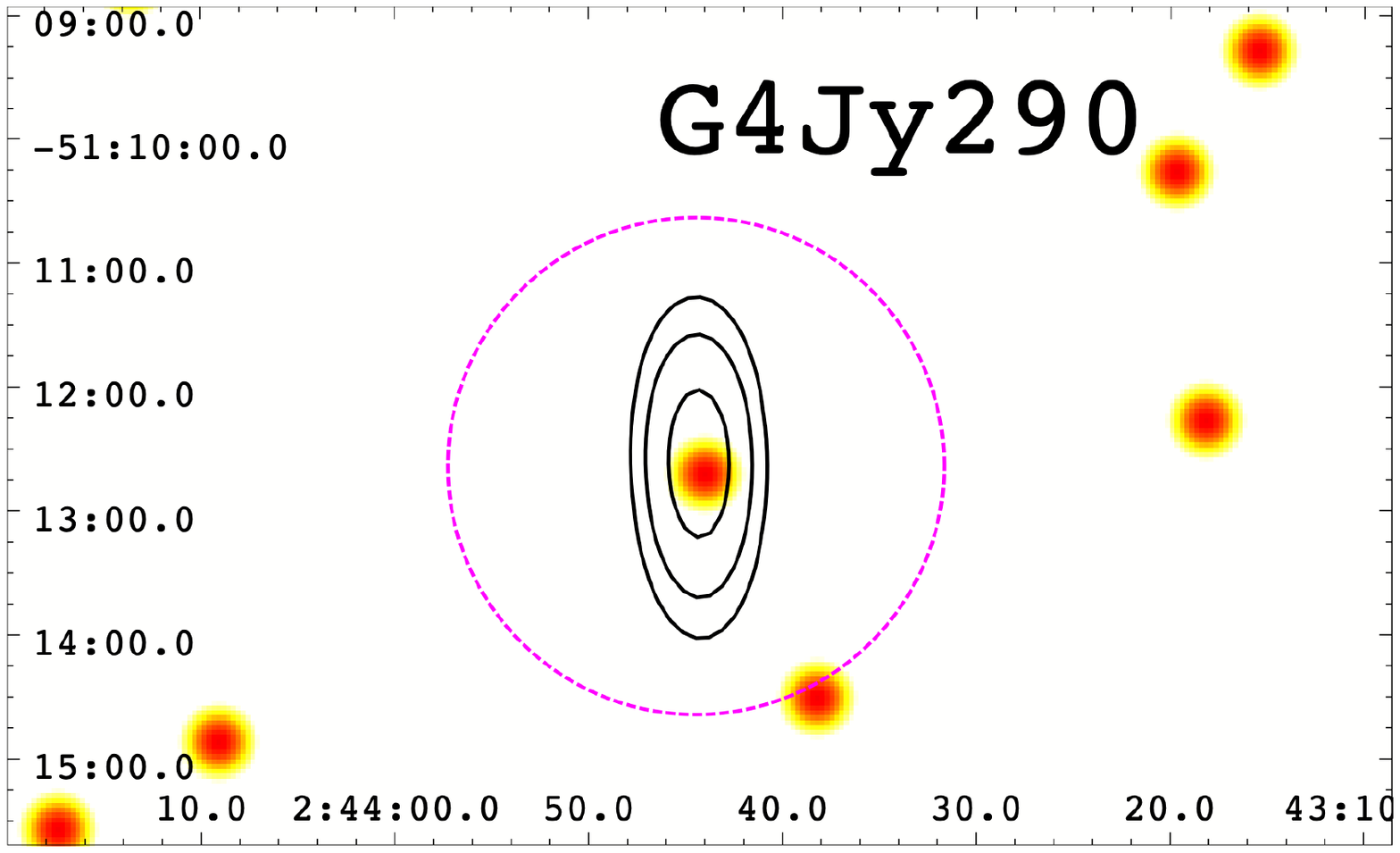}
\includegraphics[width=3.8cm,height=6.4cm,angle=-90]{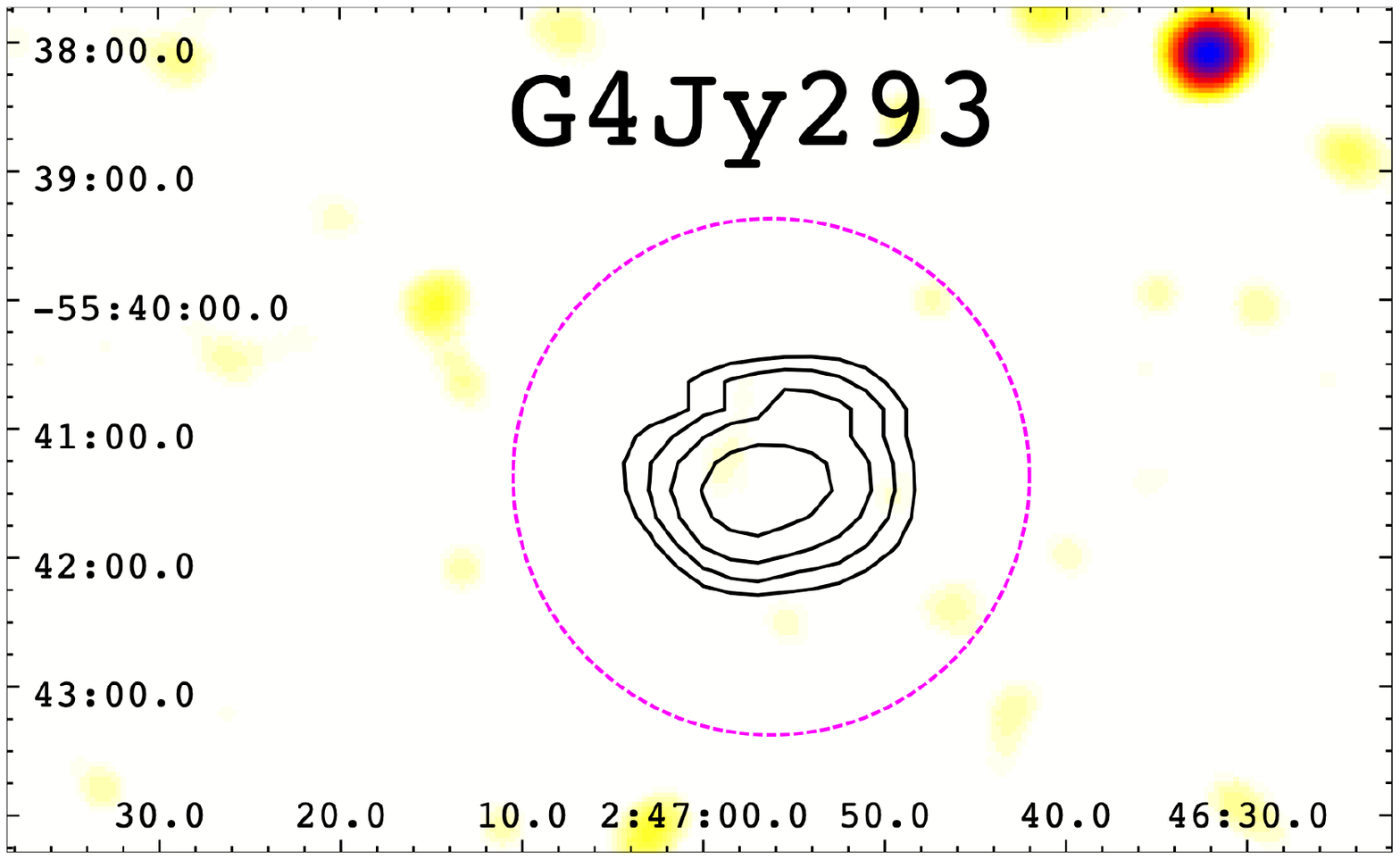}
\includegraphics[width=3.8cm,height=6.4cm,angle=-90]{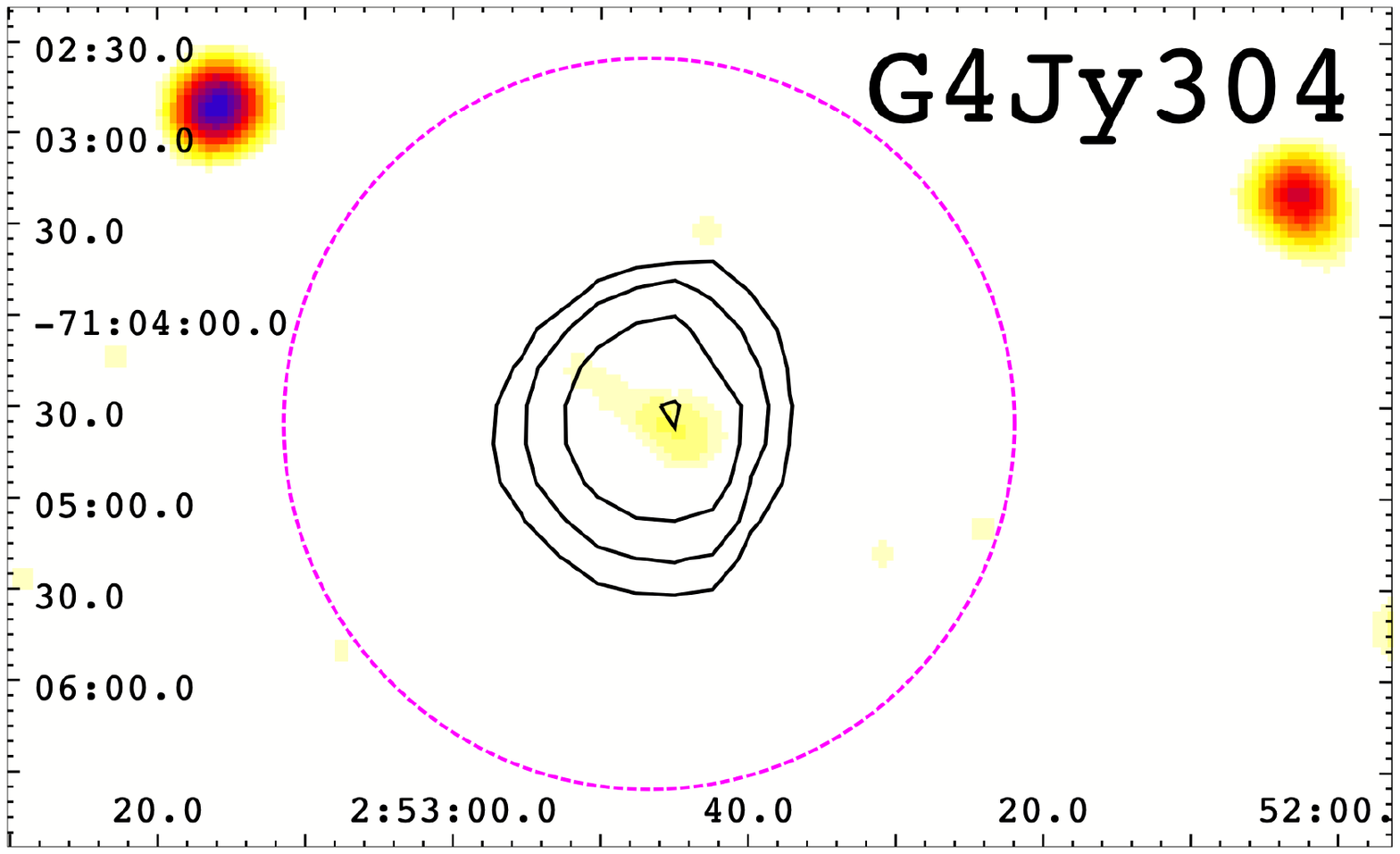}
\includegraphics[width=3.8cm,height=6.4cm,angle=-90]{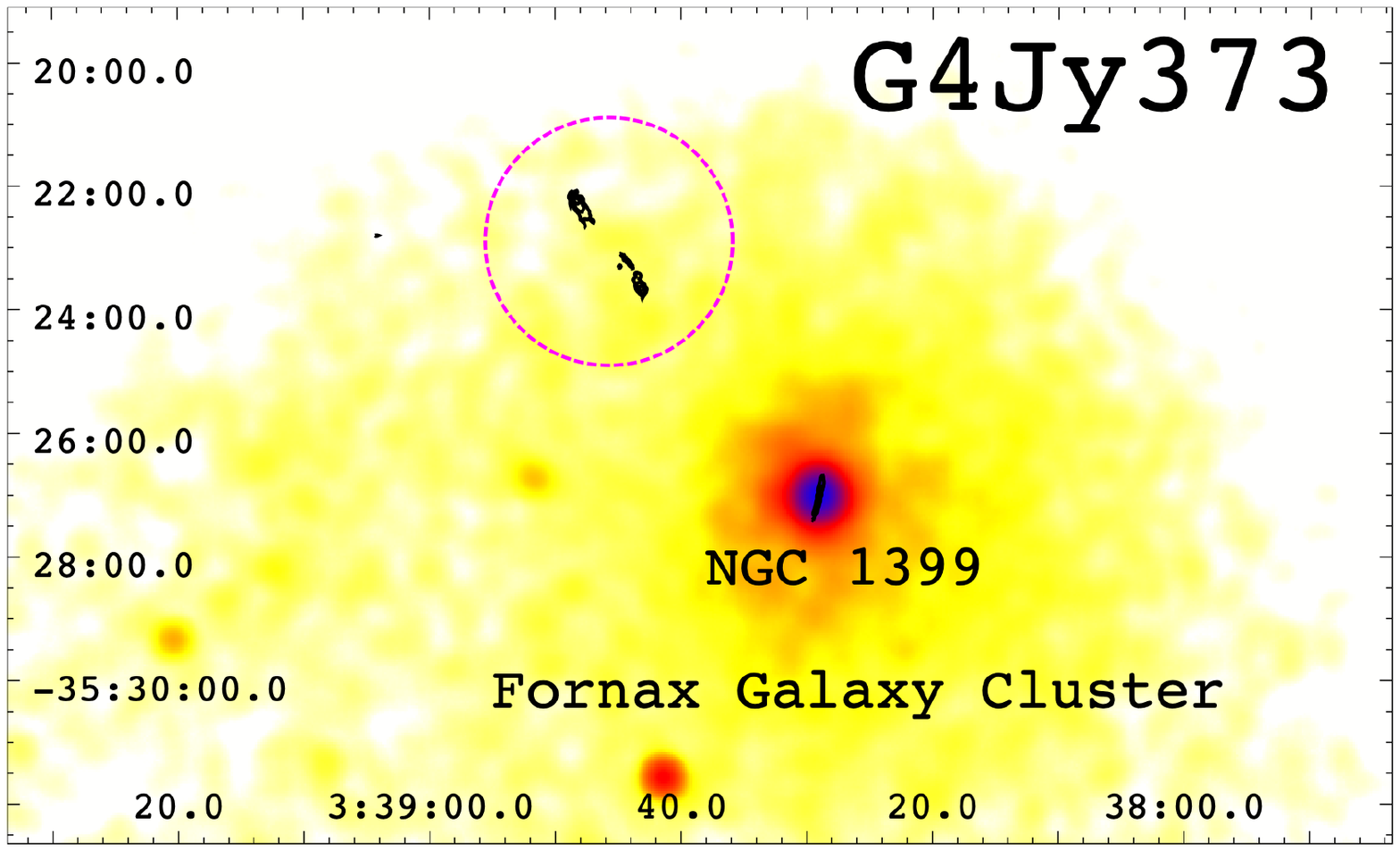}
\includegraphics[width=3.8cm,height=6.4cm,angle=-90]{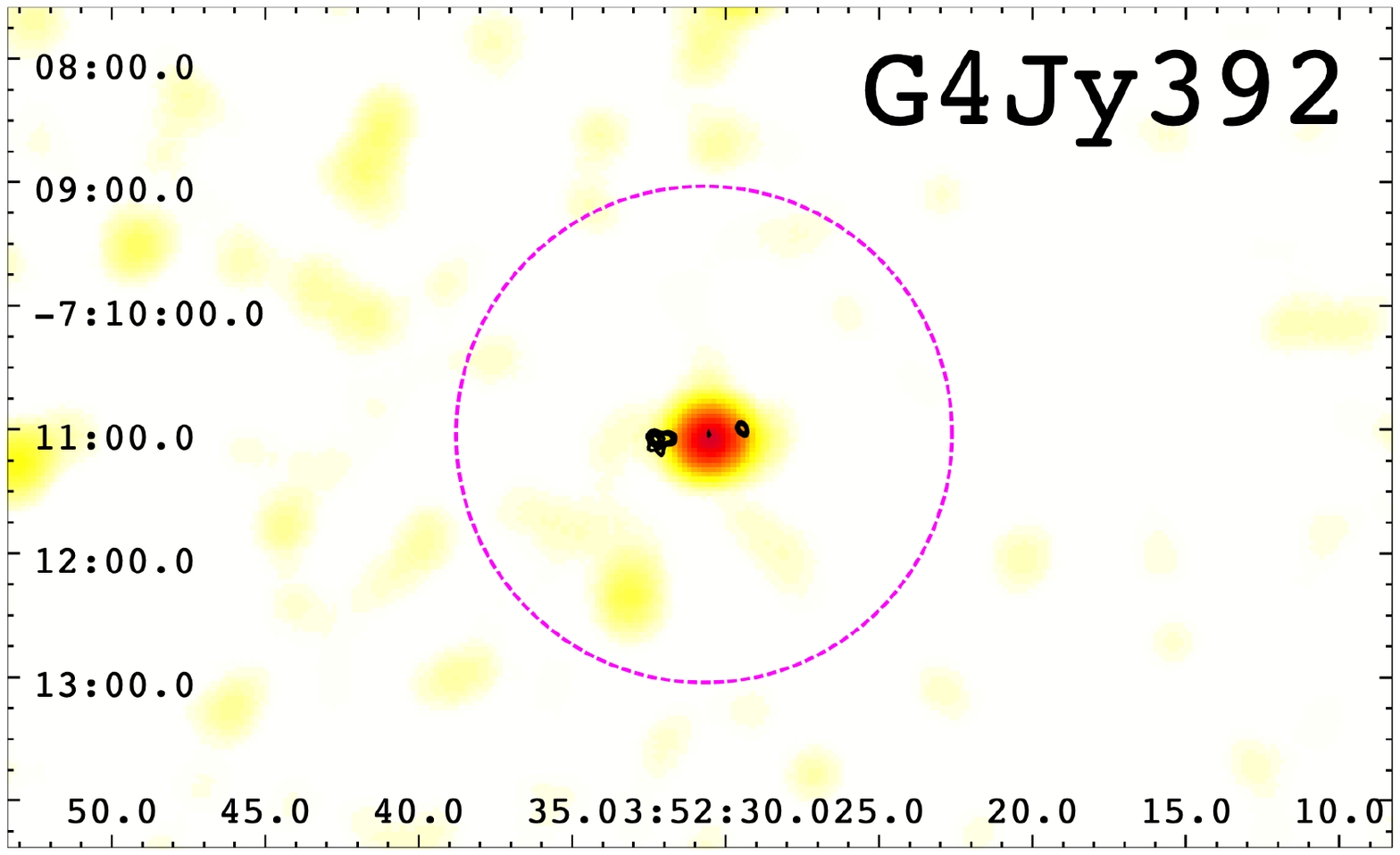}
\includegraphics[width=3.8cm,height=6.4cm,angle=-90]{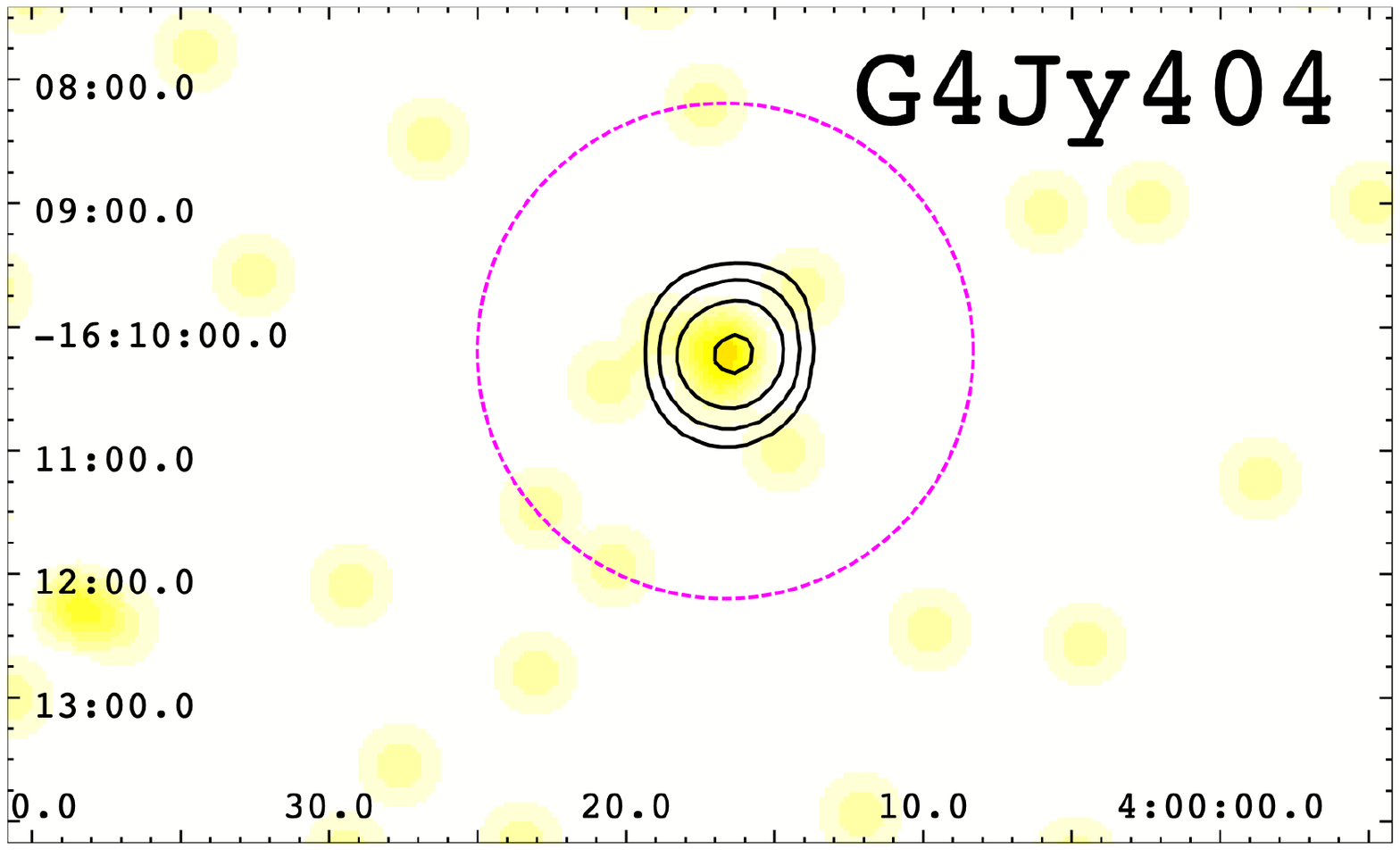}
\includegraphics[width=3.8cm,height=6.4cm,angle=-90]{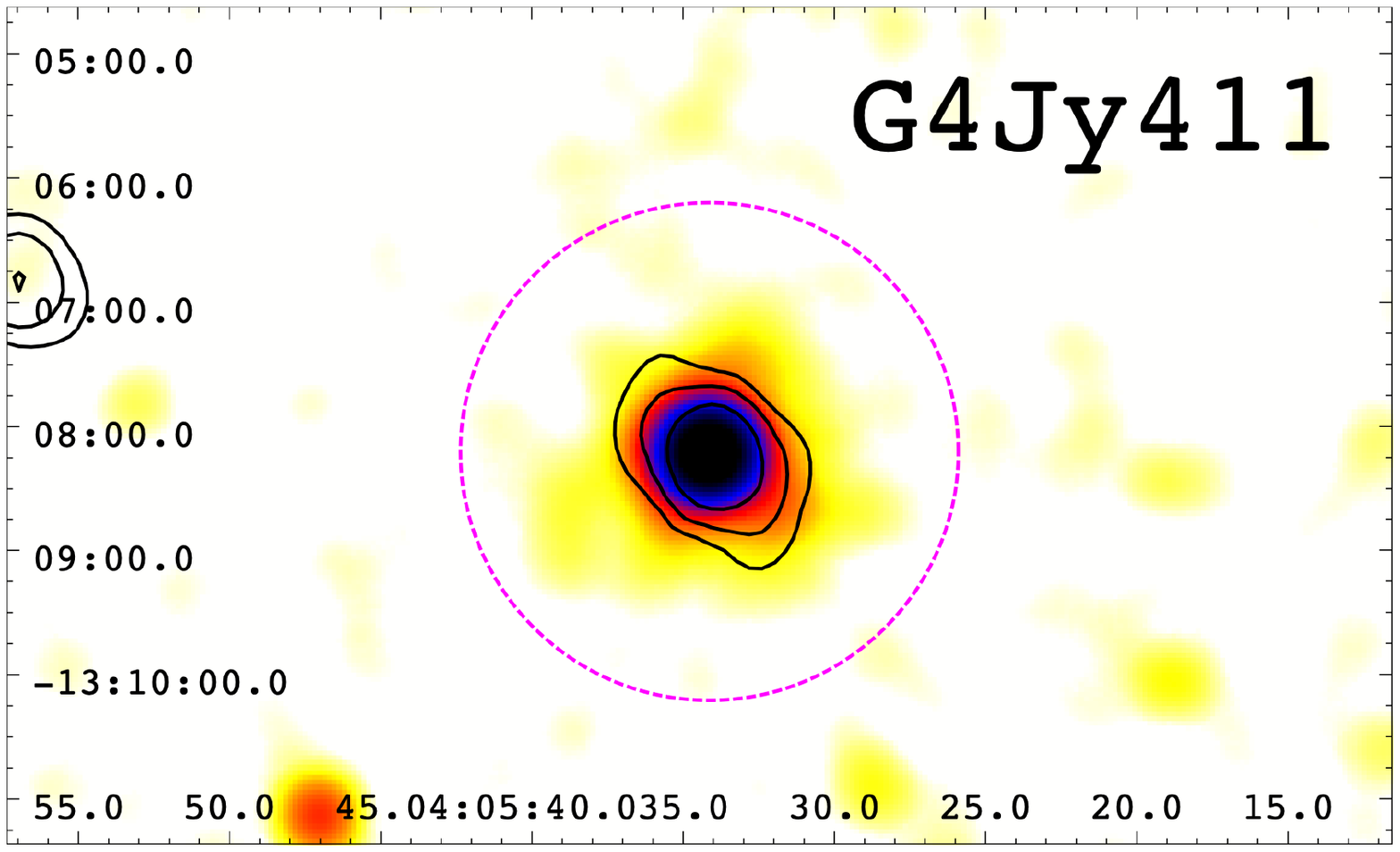}
\caption{Same as Figure~\ref{fig:example} for the following \cs\ radio sources:
G4Jy\,217, G4Jy\,247, G4Jy\,249, G4Jy\,257, G4Jy\,260, G4Jy\,290, G4Jy\,293, G4Jy\,304, G4Jy\,373, G4Jy\,392, G4Jy\,404, G4Jy\,411.}
\end{center}
\end{figure*}

\begin{figure*}[!th]
\begin{center}
\includegraphics[width=3.8cm,height=6.4cm,angle=-90]{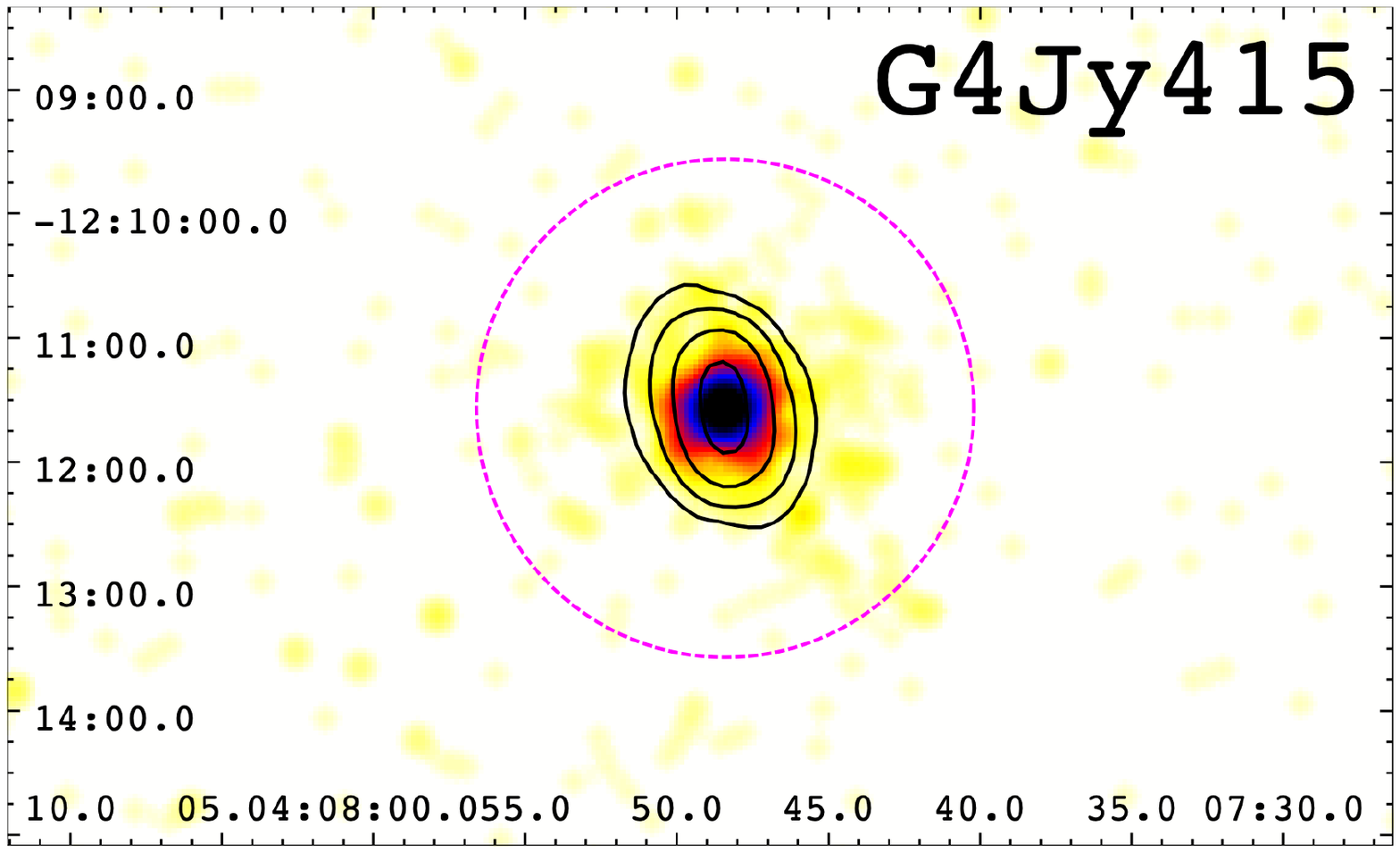}
\includegraphics[width=3.8cm,height=6.4cm,angle=-90]{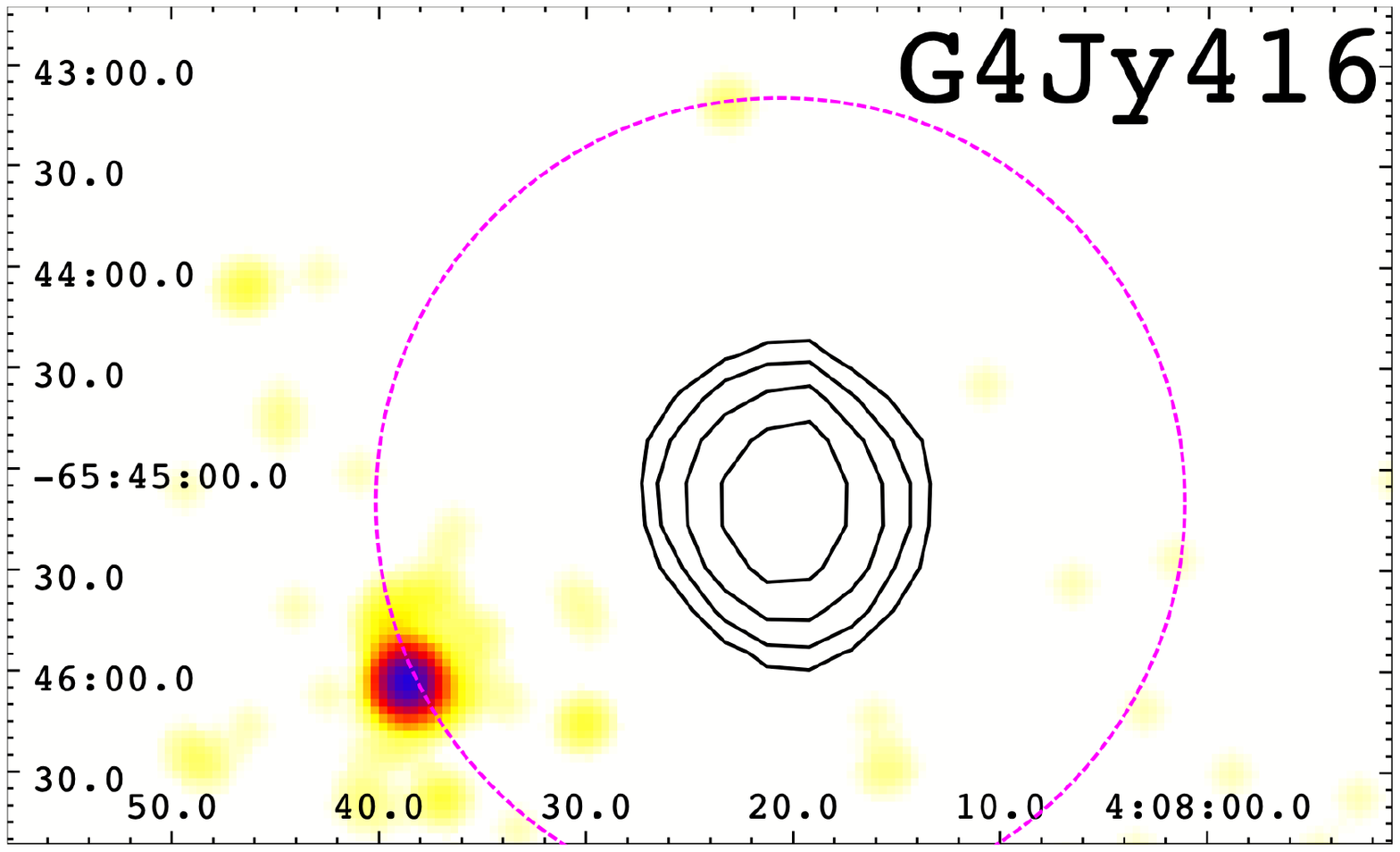}
\includegraphics[width=3.8cm,height=6.4cm,angle=-90]{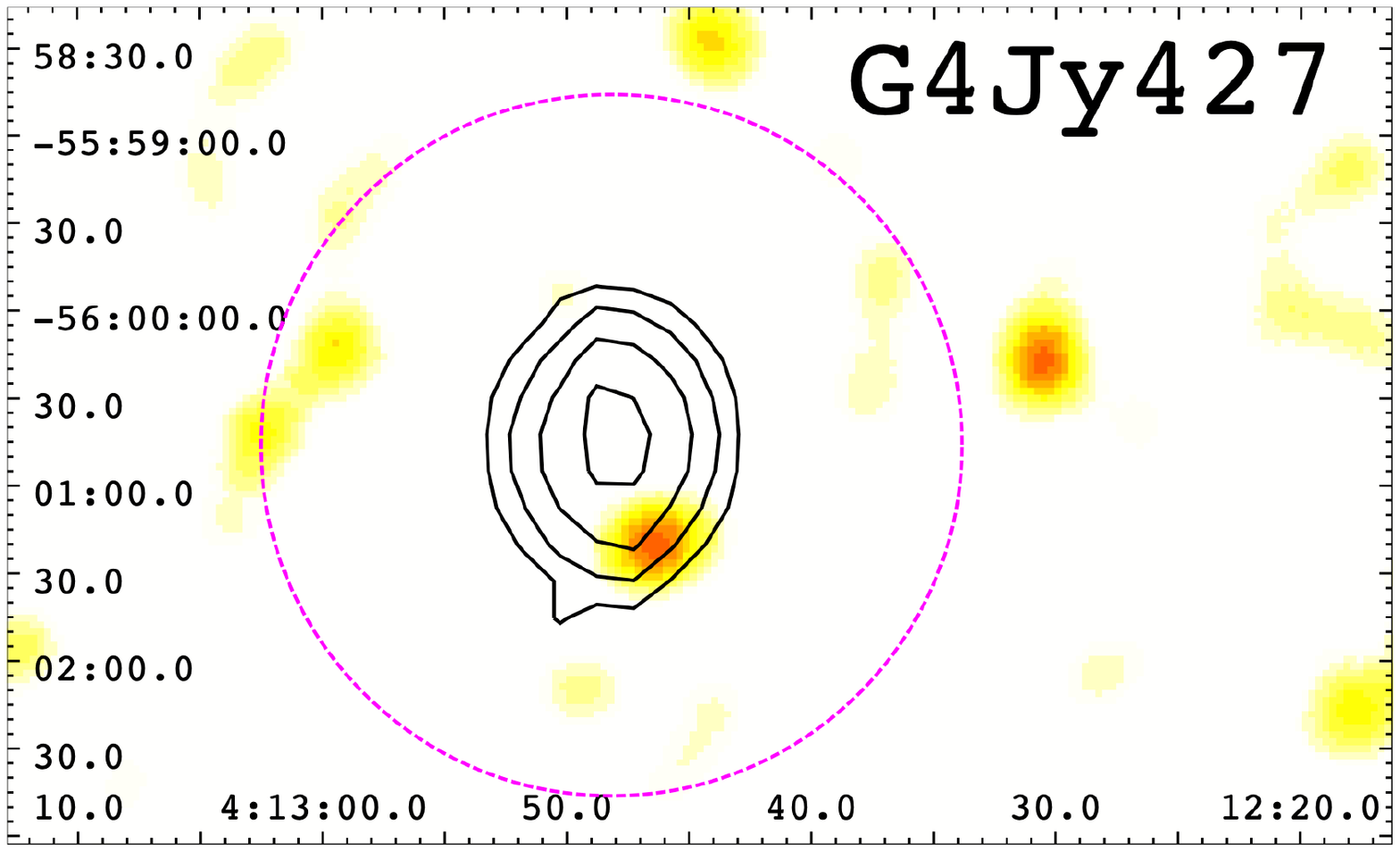}
\includegraphics[width=3.8cm,height=6.4cm,angle=-90]{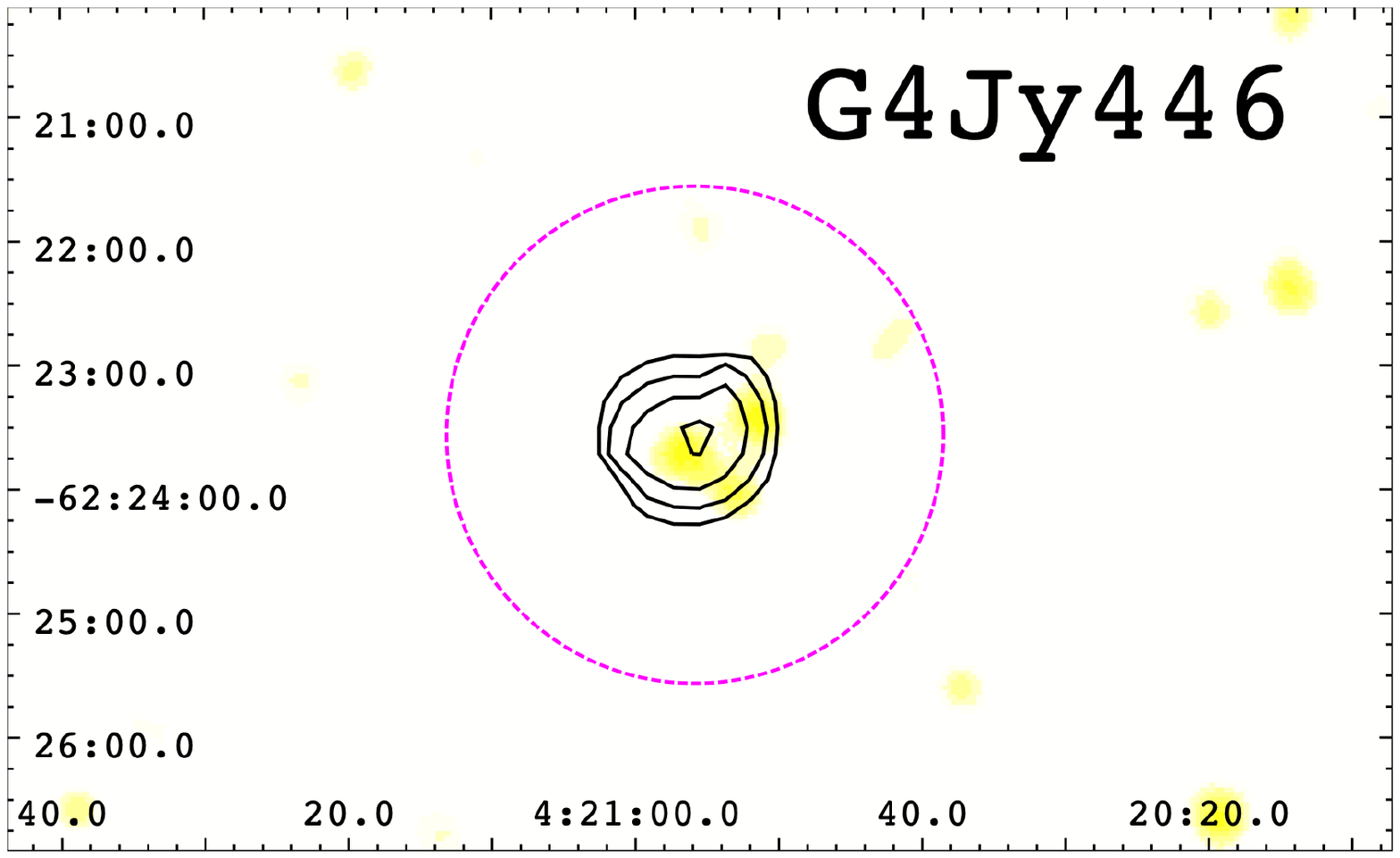}
\includegraphics[width=3.8cm,height=6.4cm,angle=-90]{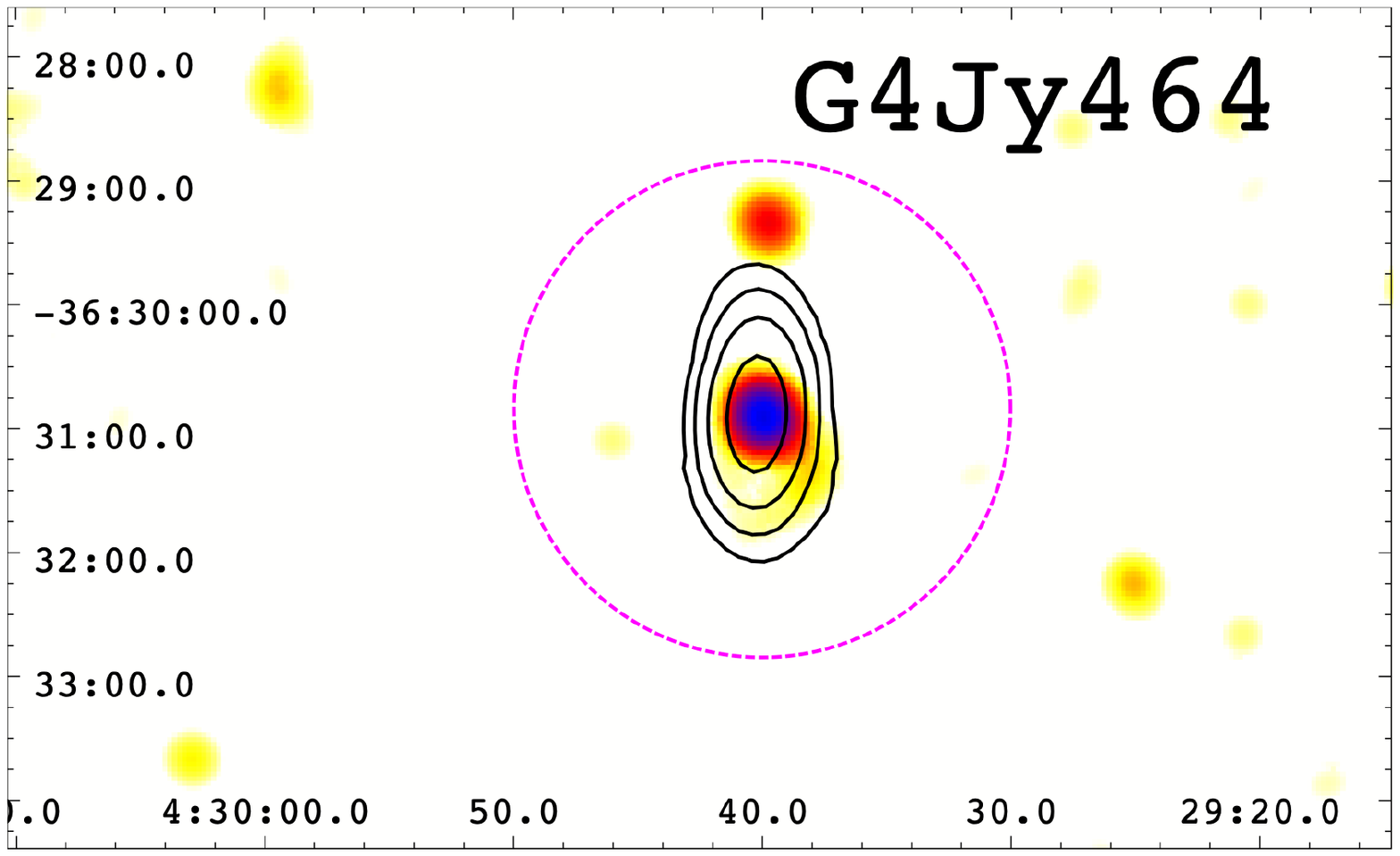}
\includegraphics[width=3.8cm,height=6.4cm,angle=-90]{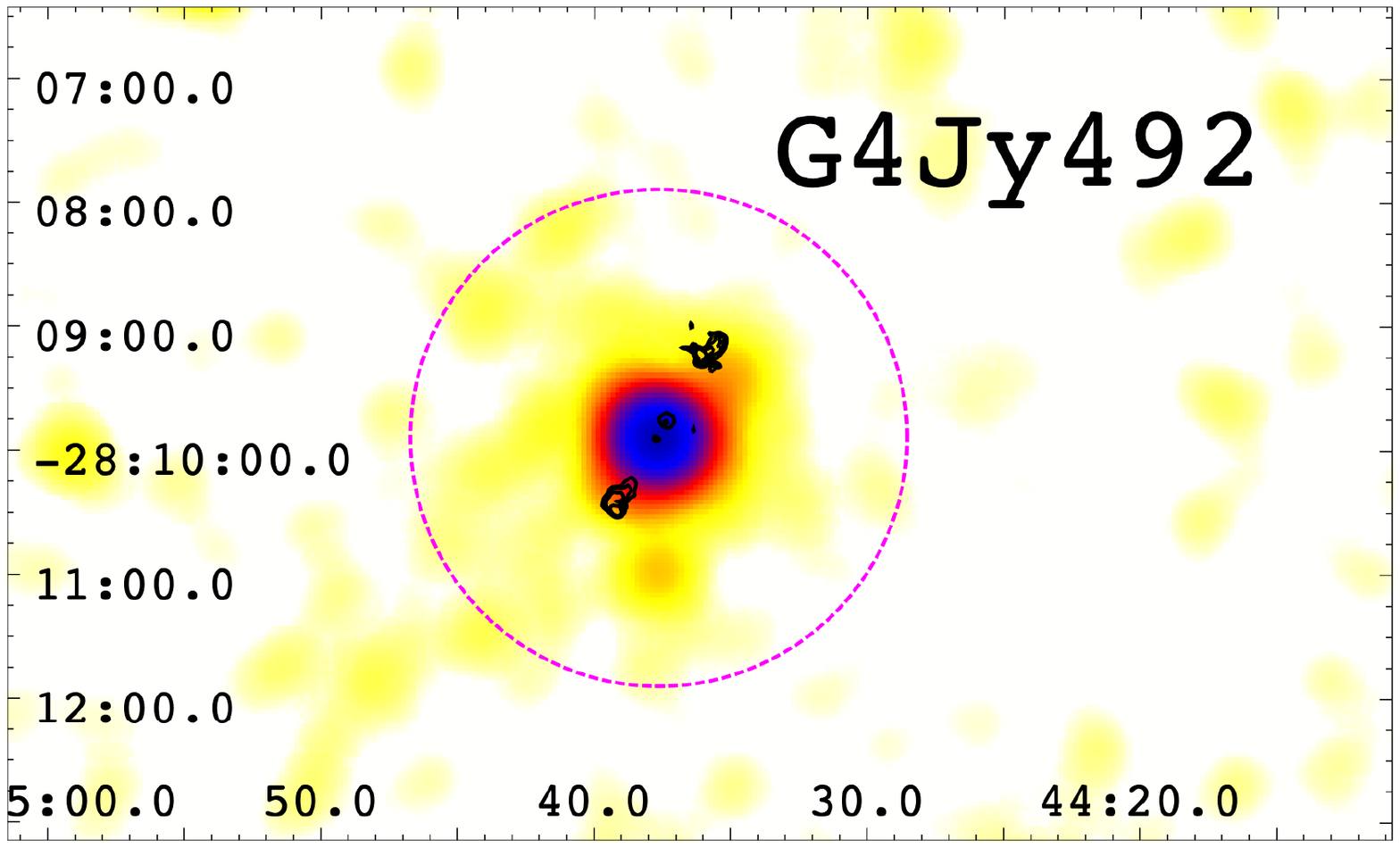}
\includegraphics[width=3.8cm,height=6.4cm,angle=-90]{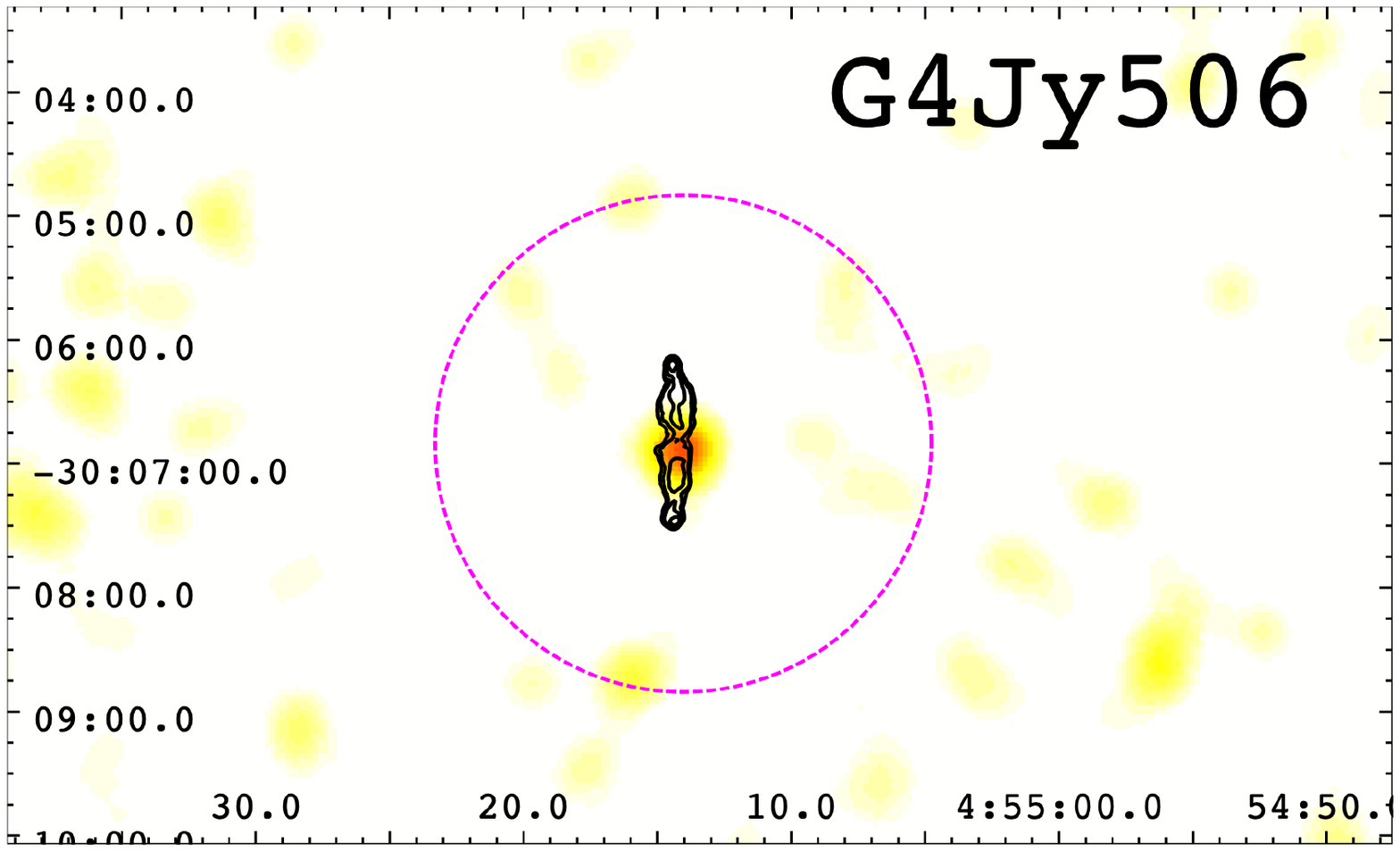}
\includegraphics[width=3.8cm,height=6.4cm,angle=-90]{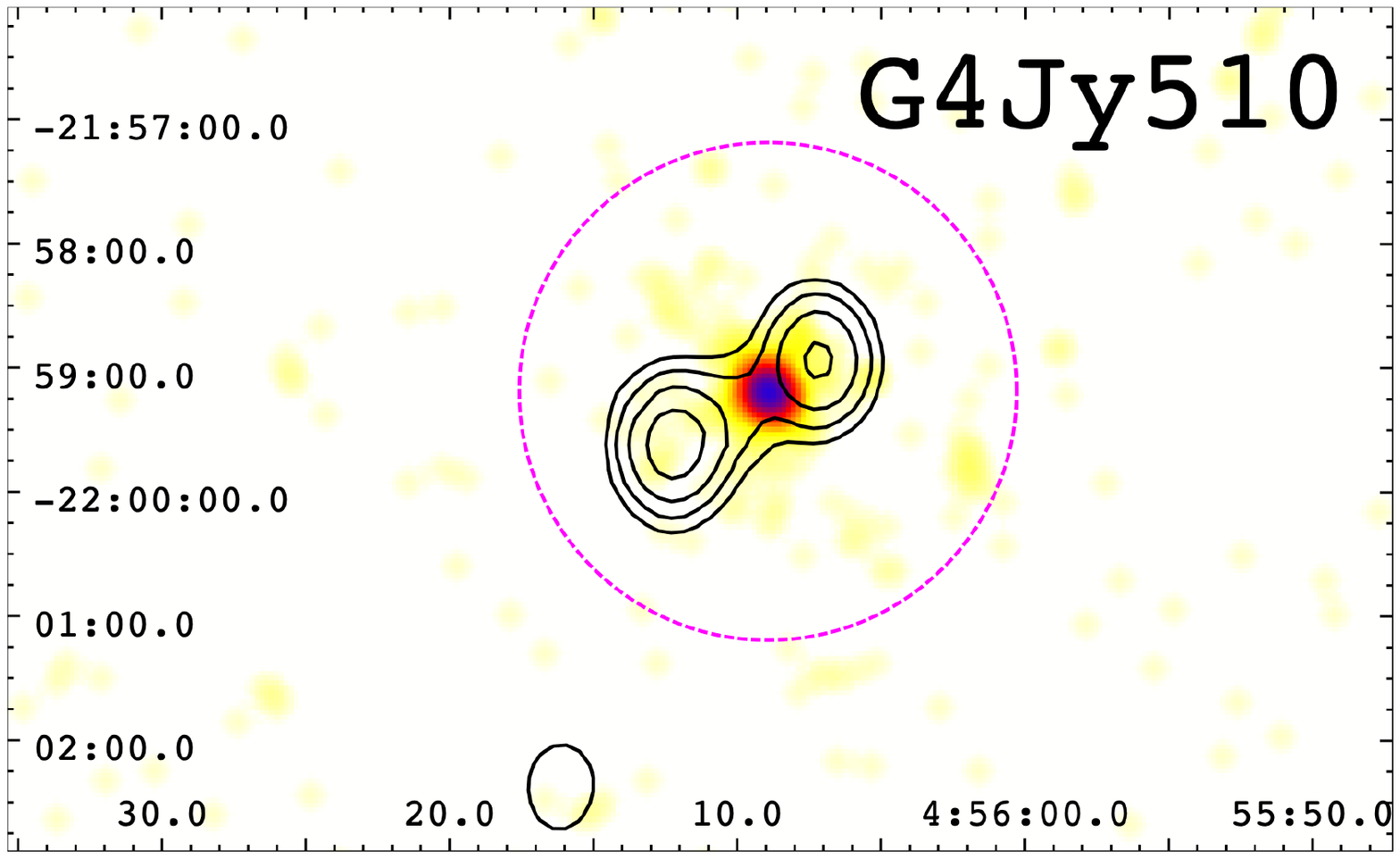}
\includegraphics[width=3.8cm,height=6.4cm,angle=-90]{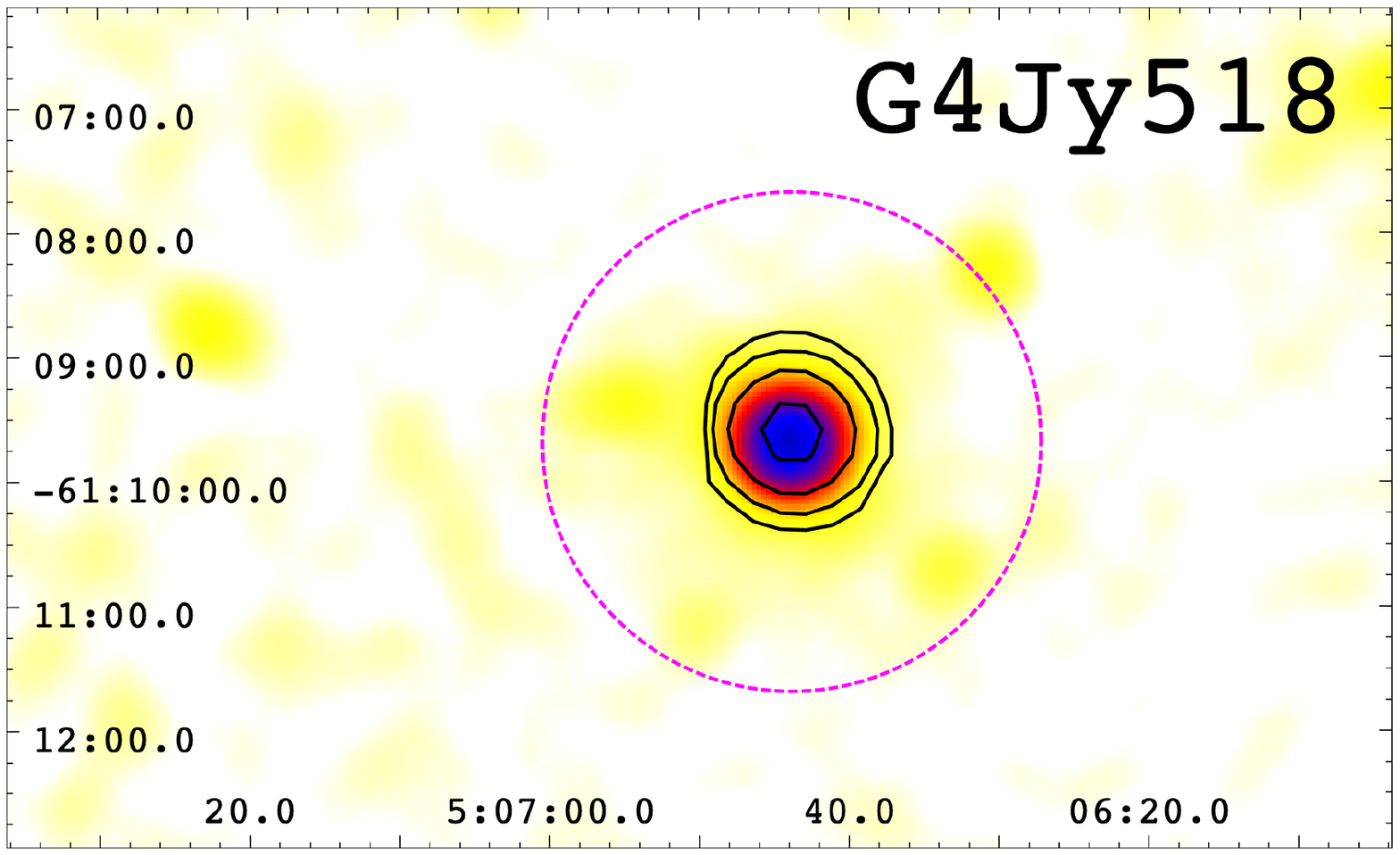}
\includegraphics[width=3.8cm,height=6.4cm,angle=-90]{./G4Jy540xrt.pdf}
\includegraphics[width=3.8cm,height=6.4cm,angle=-90]{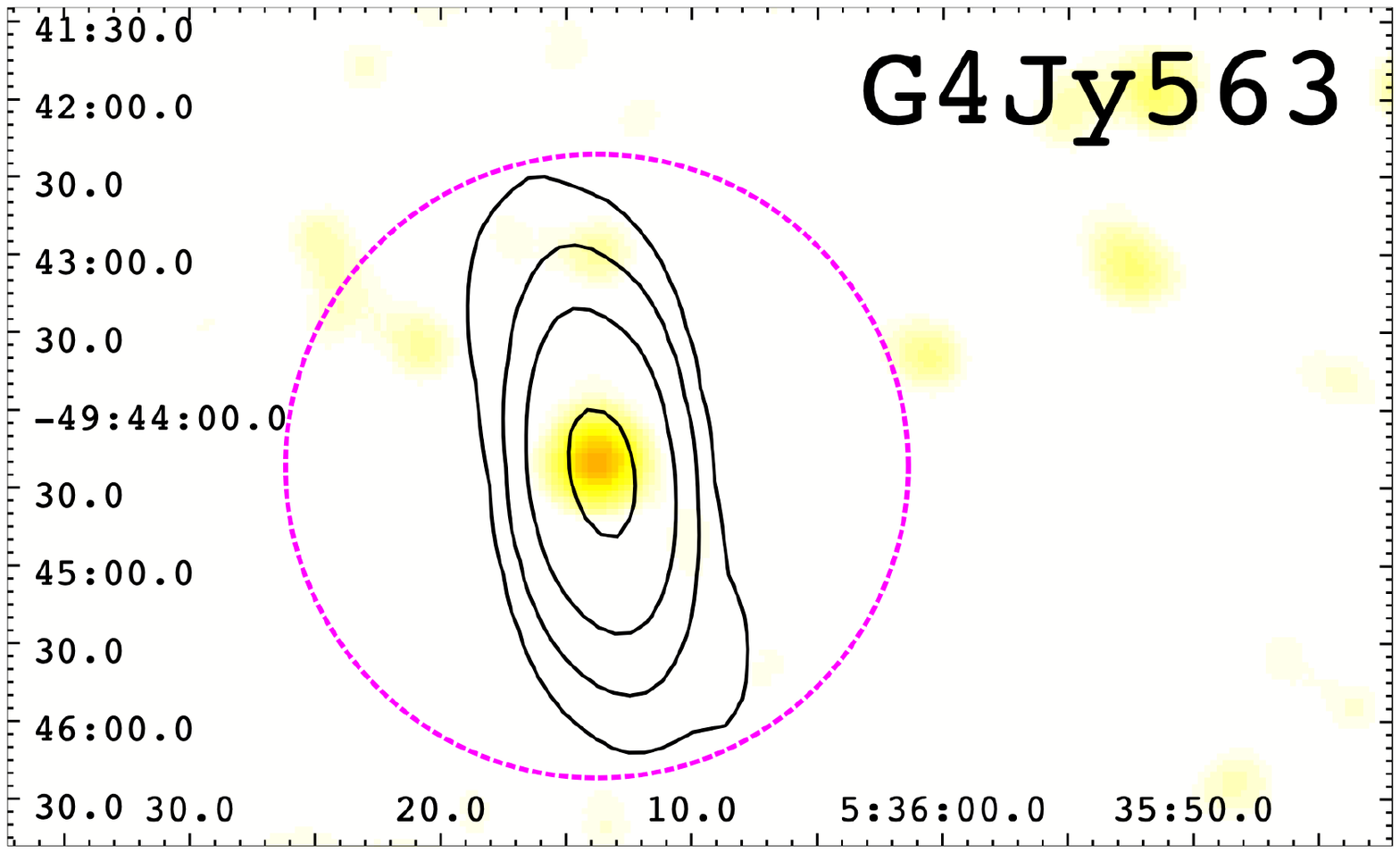}
\includegraphics[width=3.8cm,height=6.4cm,angle=-90]{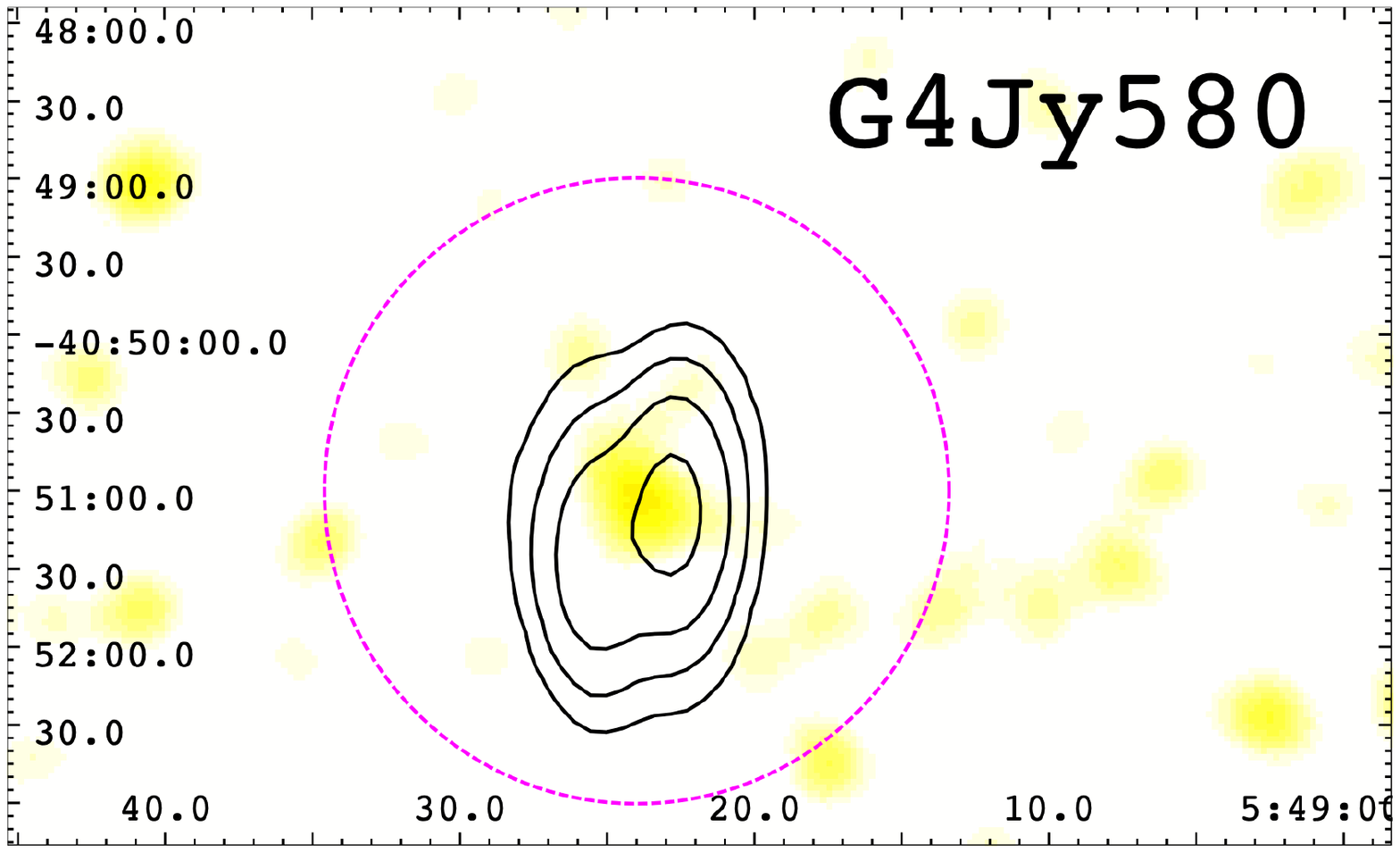}
\caption{Same as Figure~\ref{fig:example} for the following \cs\ radio sources:
G4Jy\,415, G4Jy\,416, G4Jy\,427, G4Jy\,446, G4Jy\,464, G4Jy\,492, G4Jy\,506, G4Jy\,510, G4Jy\,518, G4Jy\,540, G4Jy\,563, G4Jy\,580.}
\end{center}
\end{figure*}

\begin{figure*}[!th]
\begin{center}
\includegraphics[width=3.8cm,height=6.4cm,angle=-90]{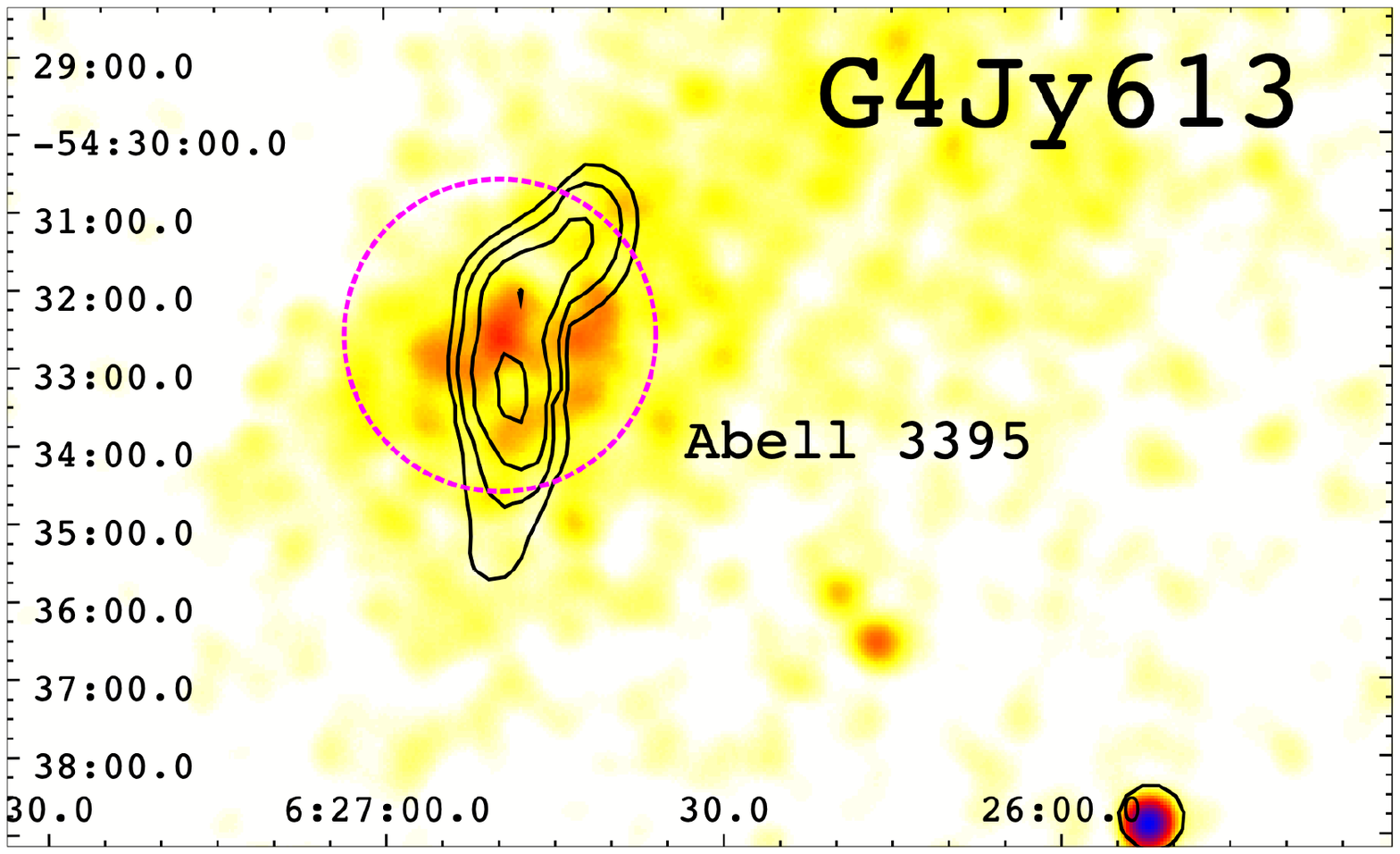}
\includegraphics[width=3.8cm,height=6.4cm,angle=-90]{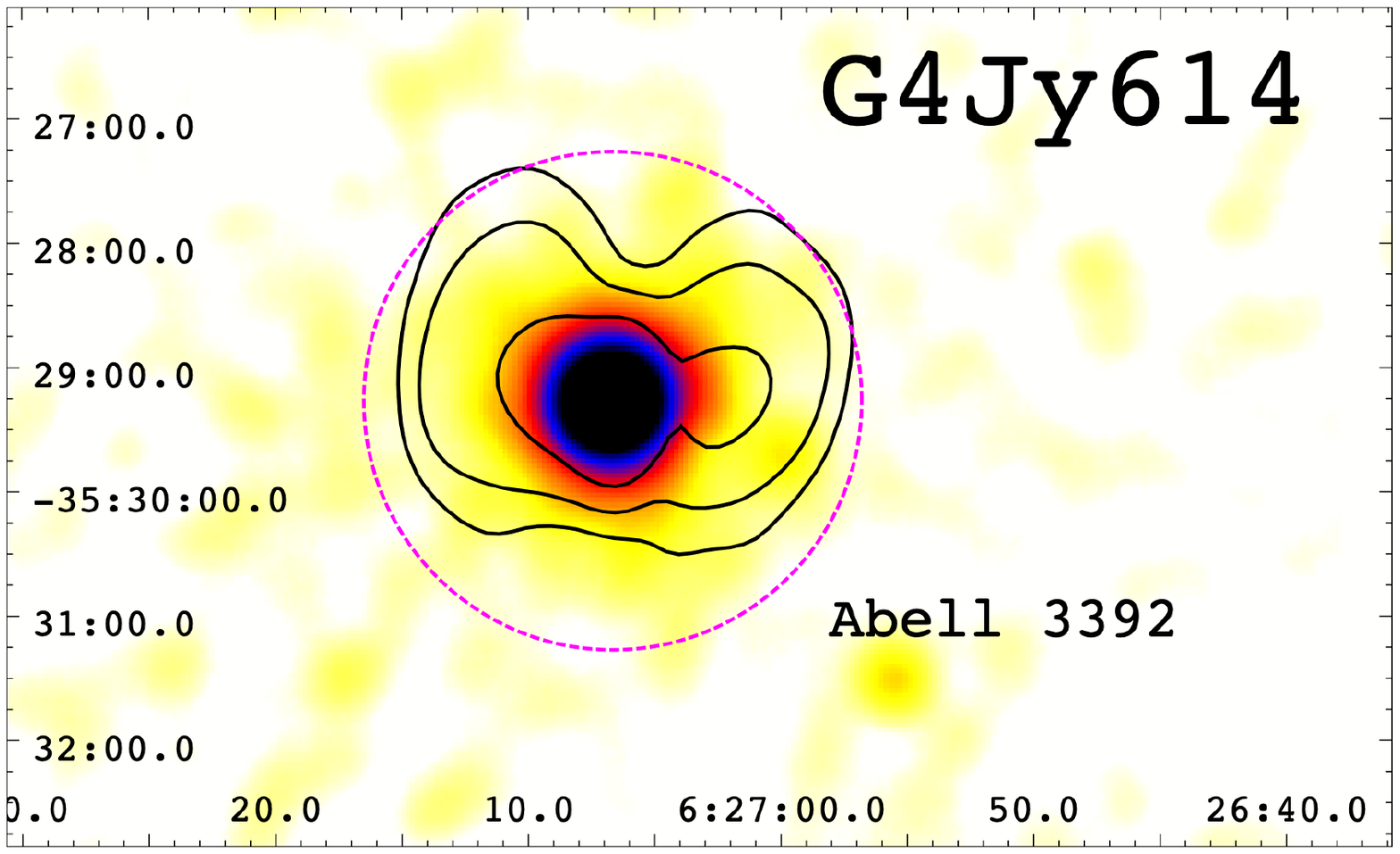}
\includegraphics[width=3.8cm,height=6.4cm,angle=-90]{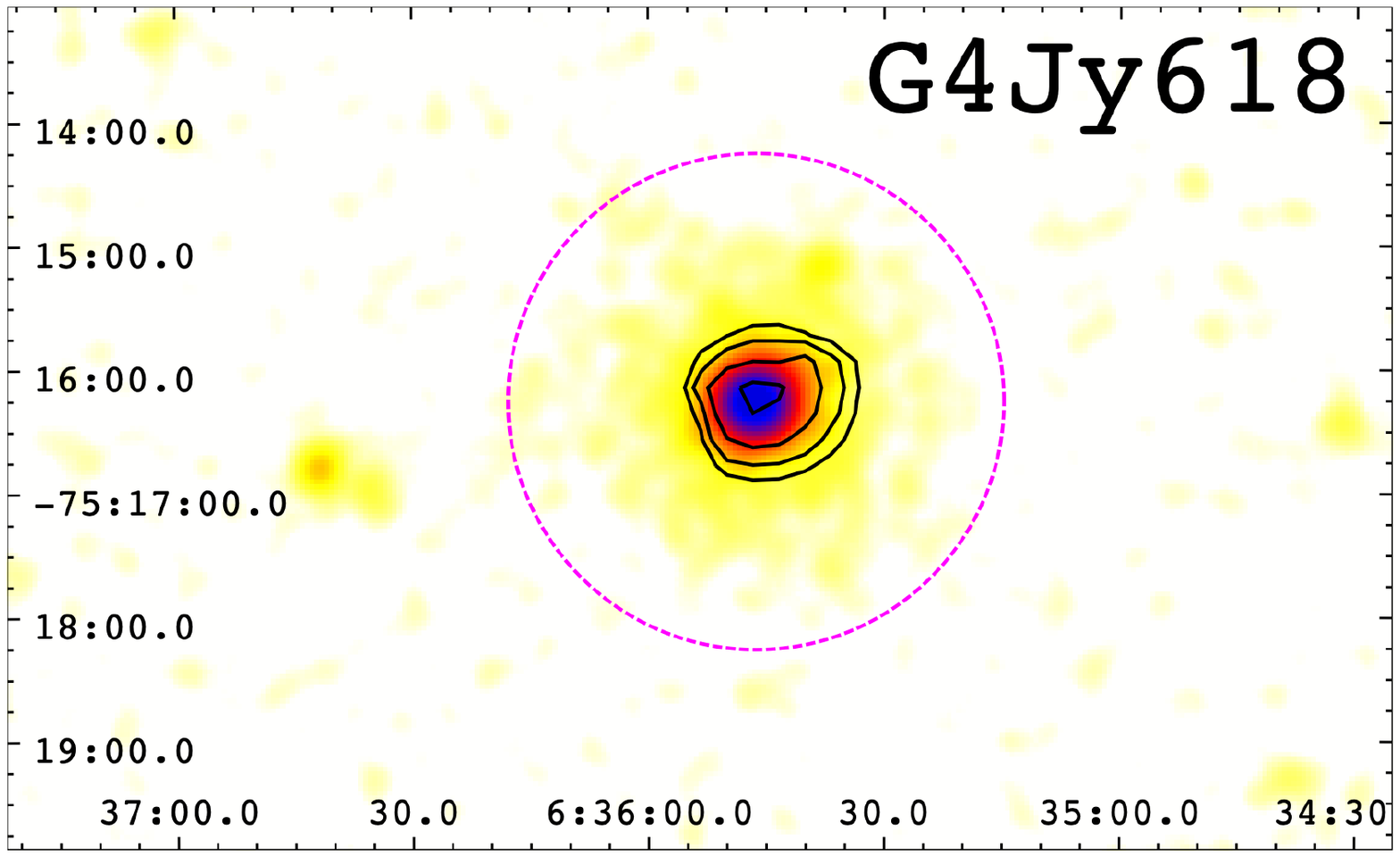}
\includegraphics[width=3.8cm,height=6.4cm,angle=-90]{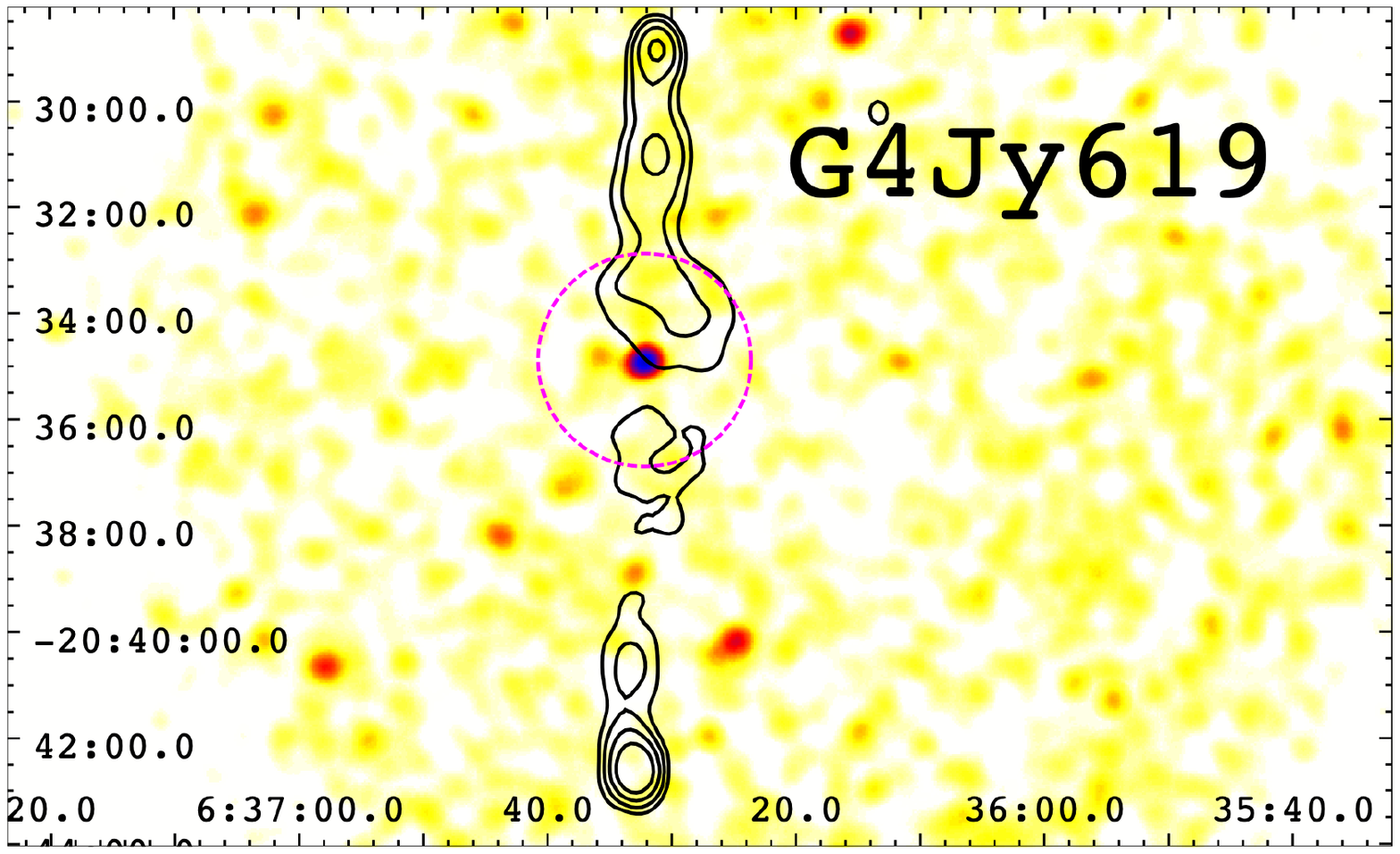}
\includegraphics[width=3.8cm,height=6.4cm,angle=-90]{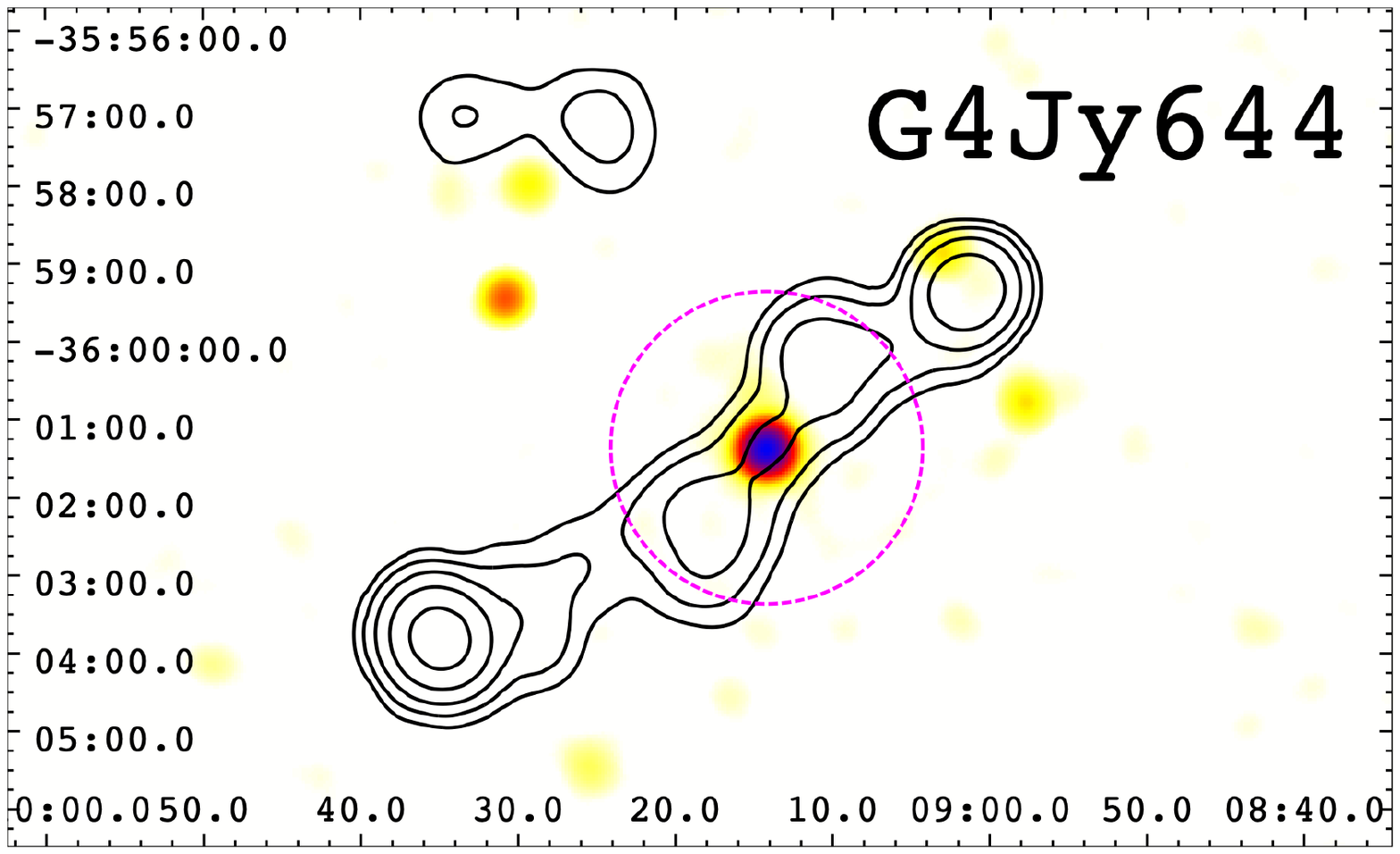}
\includegraphics[width=3.8cm,height=6.4cm,angle=-90]{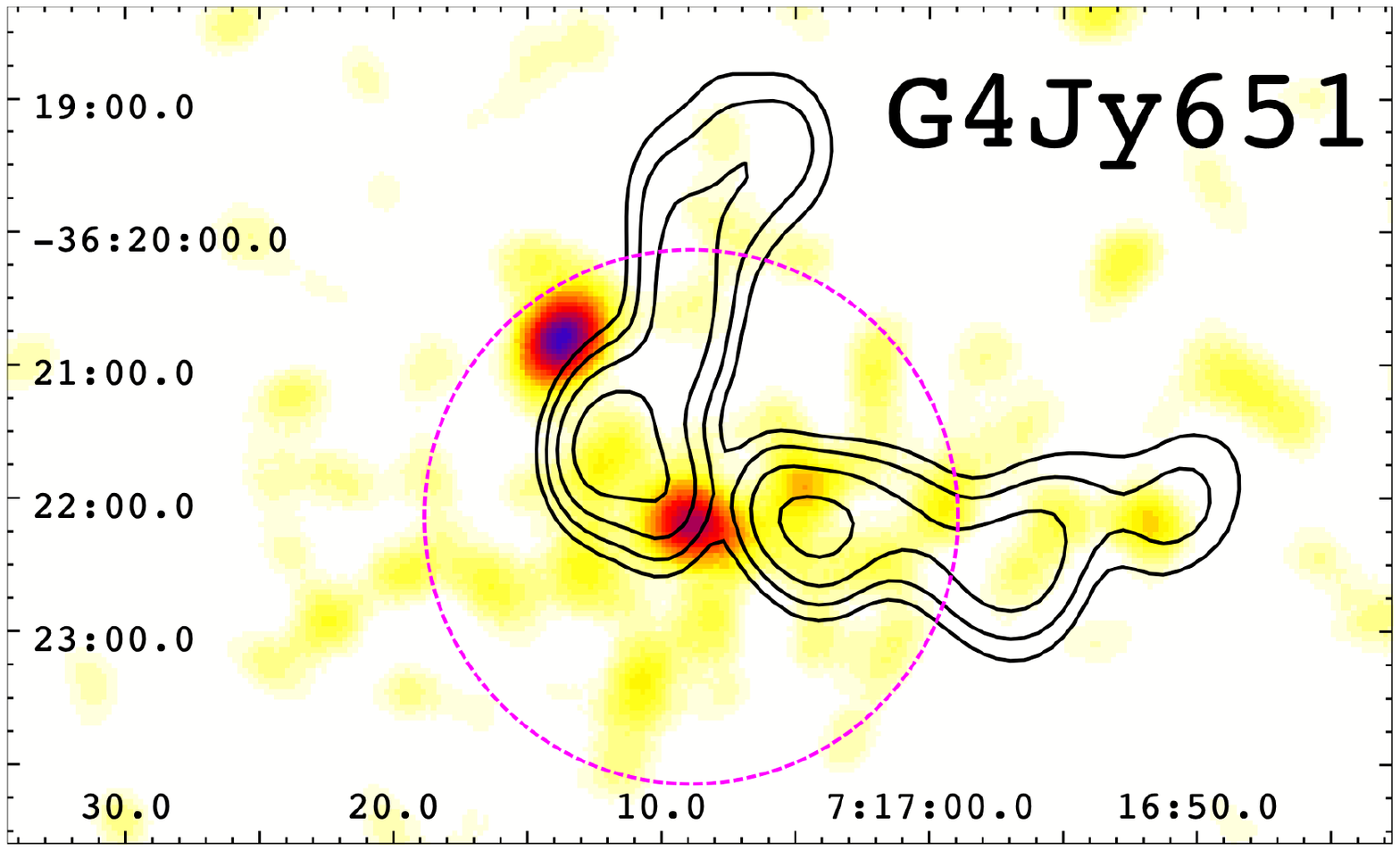}
\includegraphics[width=3.8cm,height=6.4cm,angle=-90]{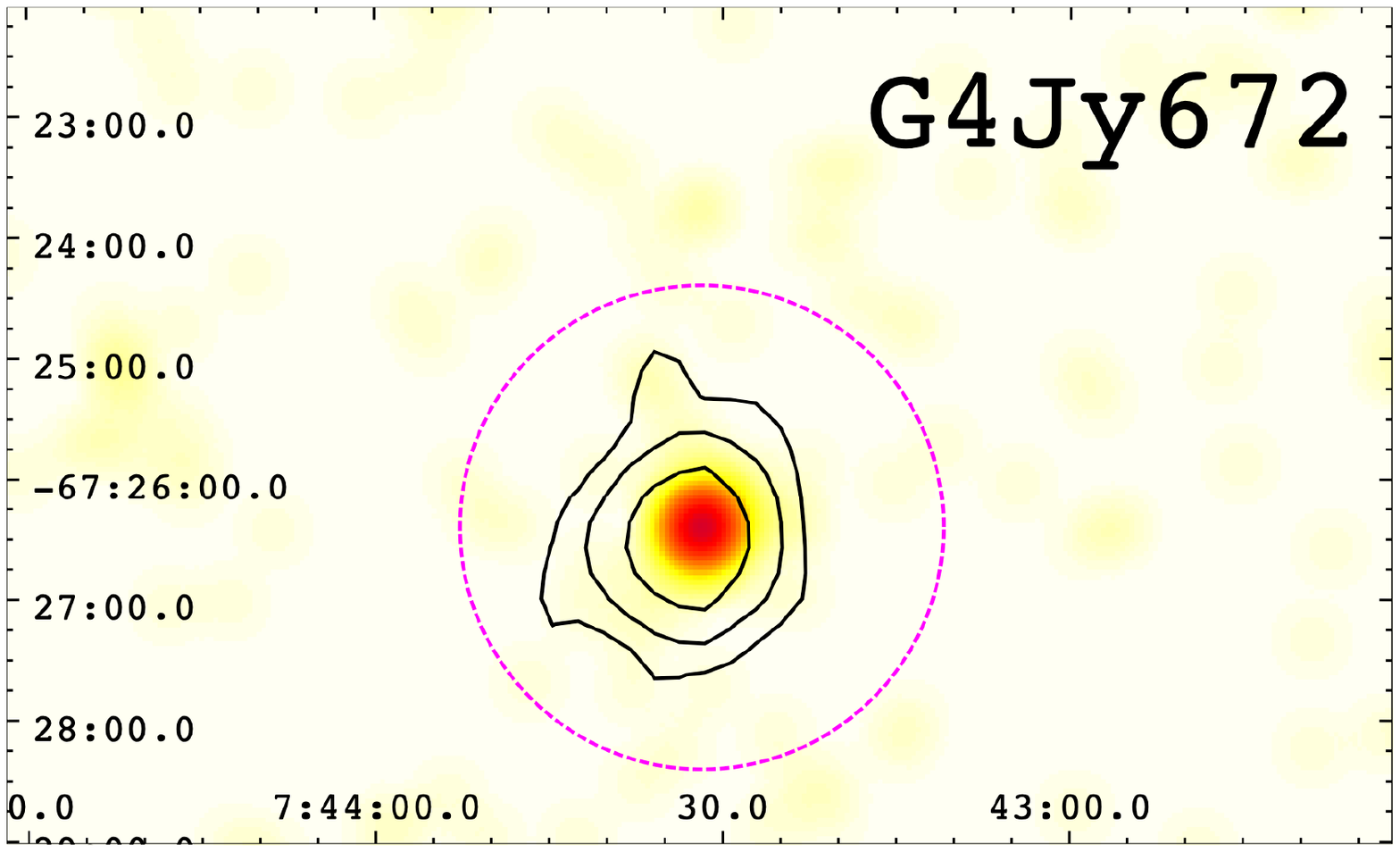}
\includegraphics[width=3.8cm,height=6.4cm,angle=-90]{./G4Jy718xrt.pdf}
\includegraphics[width=3.8cm,height=6.4cm,angle=-90]{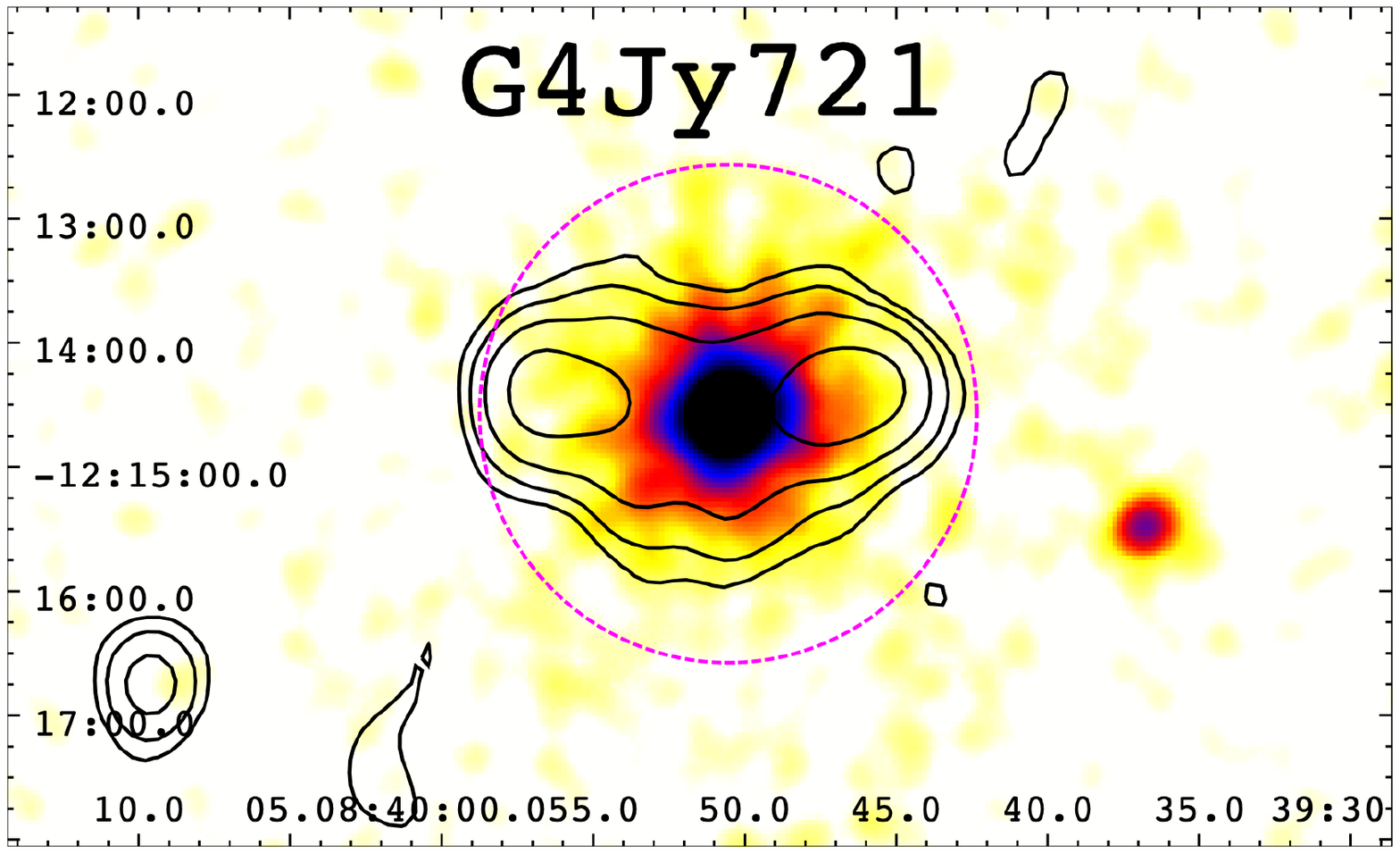}
\includegraphics[width=3.8cm,height=6.4cm,angle=-90]{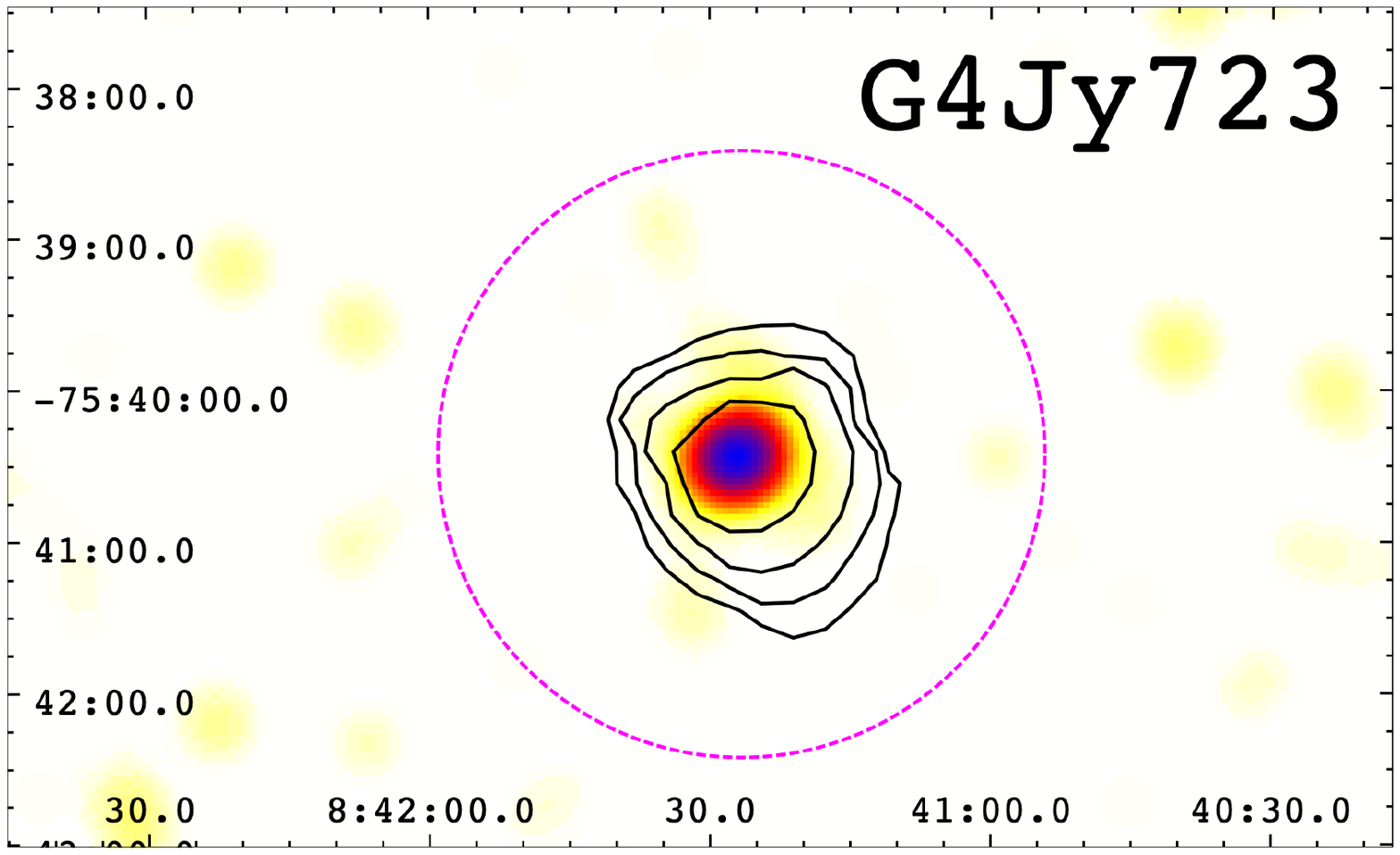}
\includegraphics[width=3.8cm,height=6.4cm,angle=-90]{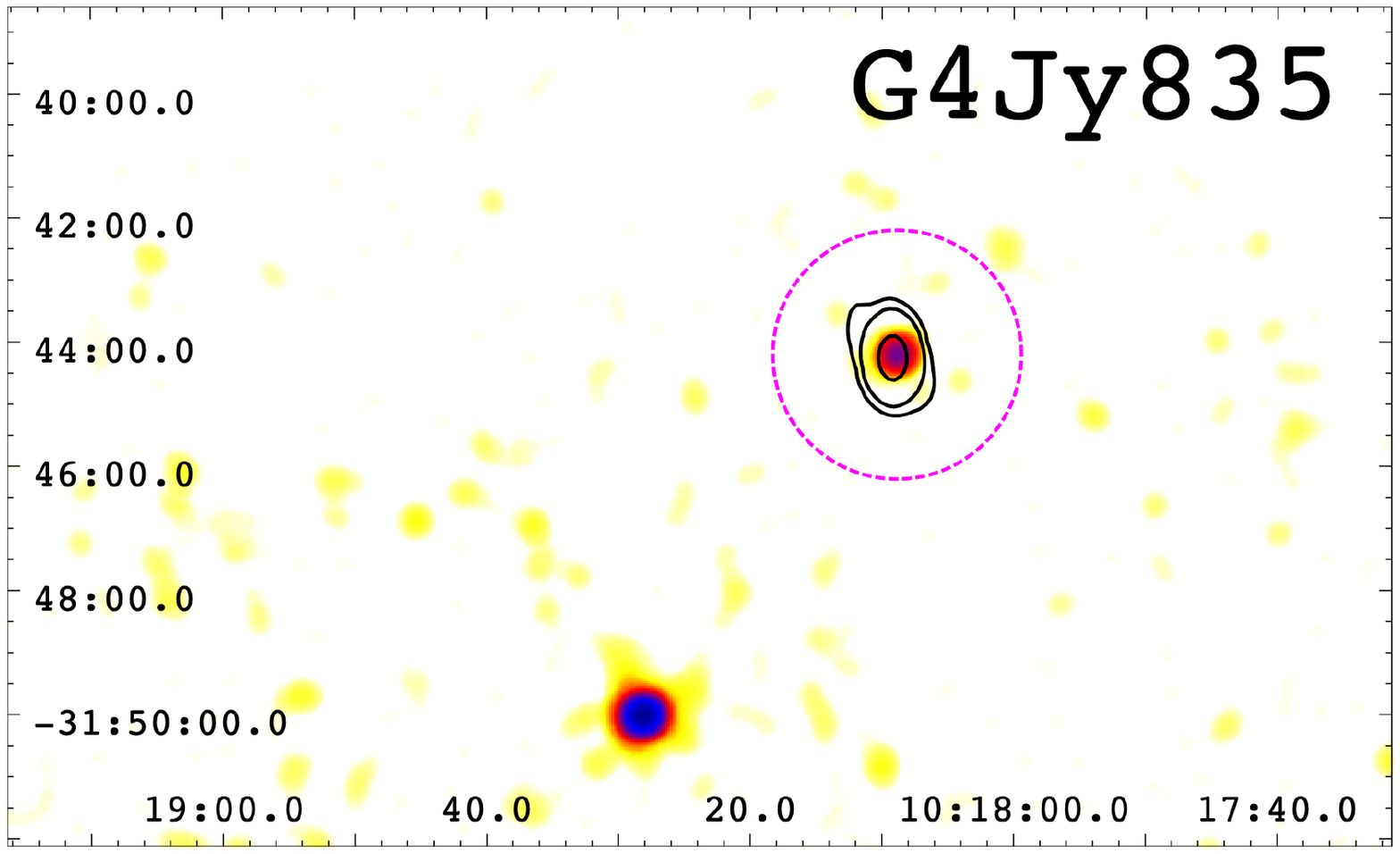}
\includegraphics[width=3.8cm,height=6.4cm,angle=-90]{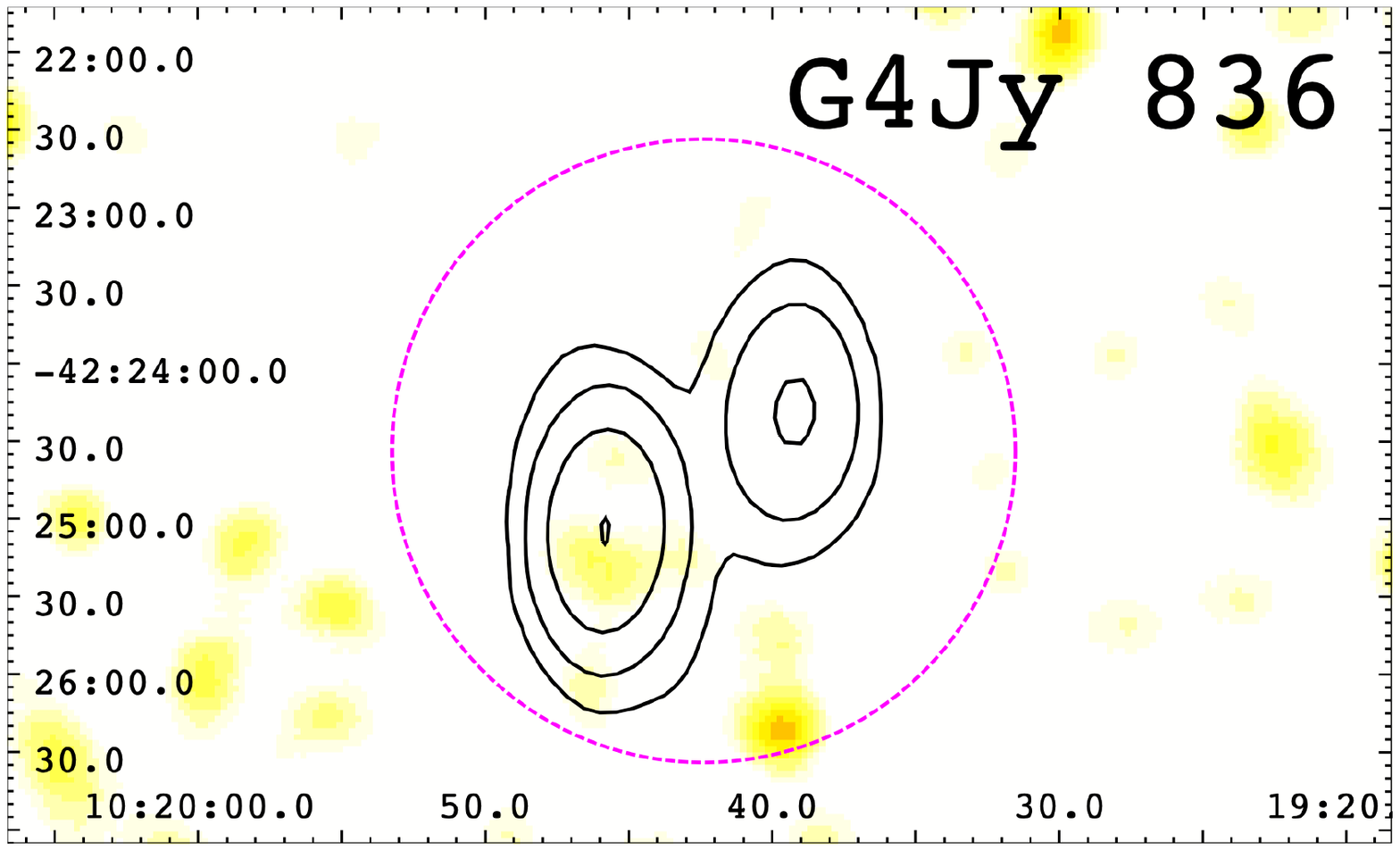}
\caption{Same as Figure~\ref{fig:example} for the following \cs\ radio sources:
G4Jy\,613, G4Jy\,614, G4Jy\,618, G4Jy\,619, G4Jy\,644, G4Jy\,651, G4Jy\,672, G4Jy\,718, G4Jy\,721, G4Jy\,723, G4Jy\,835, G4Jy\,836.}
\end{center}
\end{figure*}

\begin{figure*}[!th]
\begin{center}
\includegraphics[width=3.8cm,height=6.4cm,angle=-90]{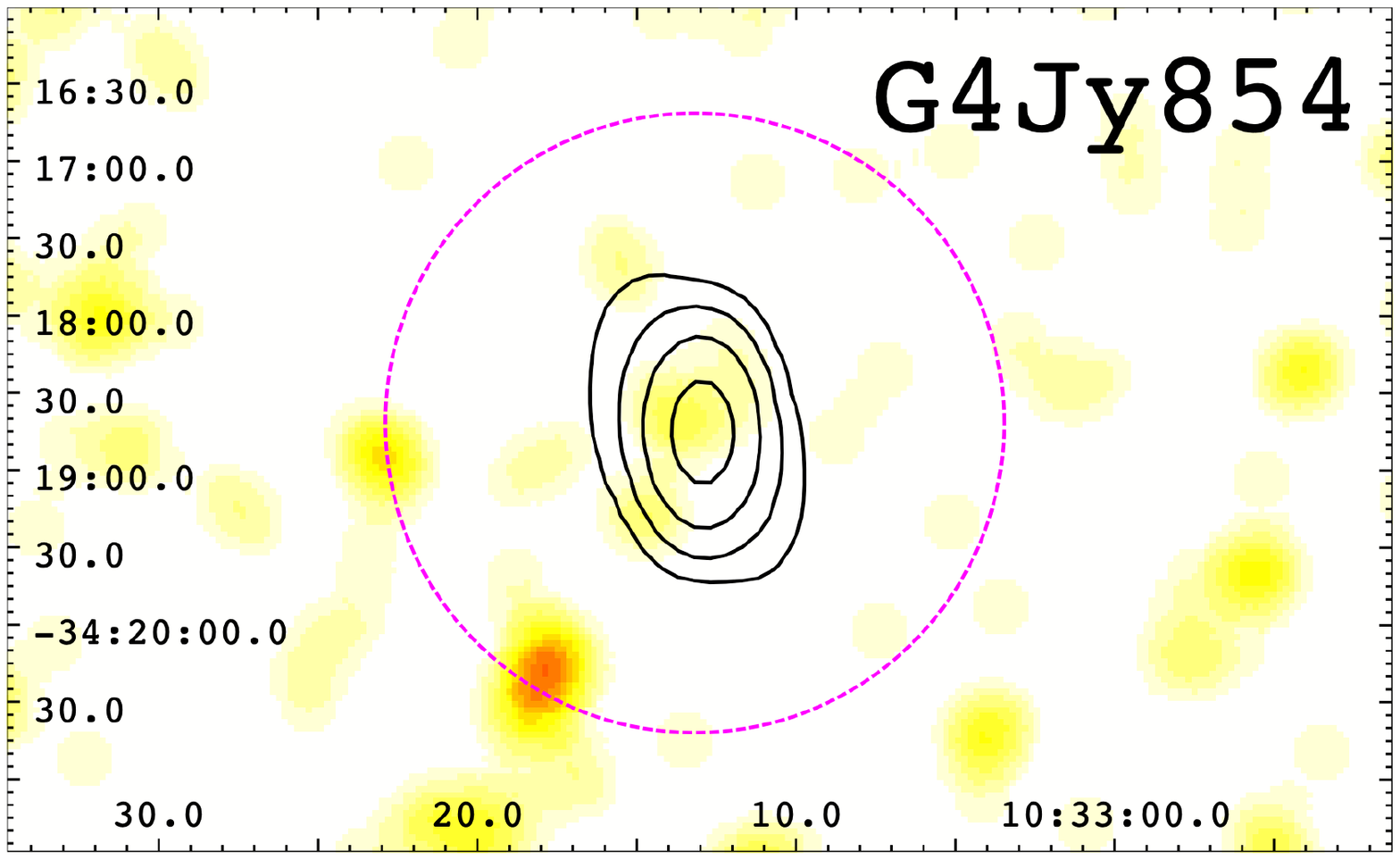}
\includegraphics[width=3.8cm,height=6.4cm,angle=-90]{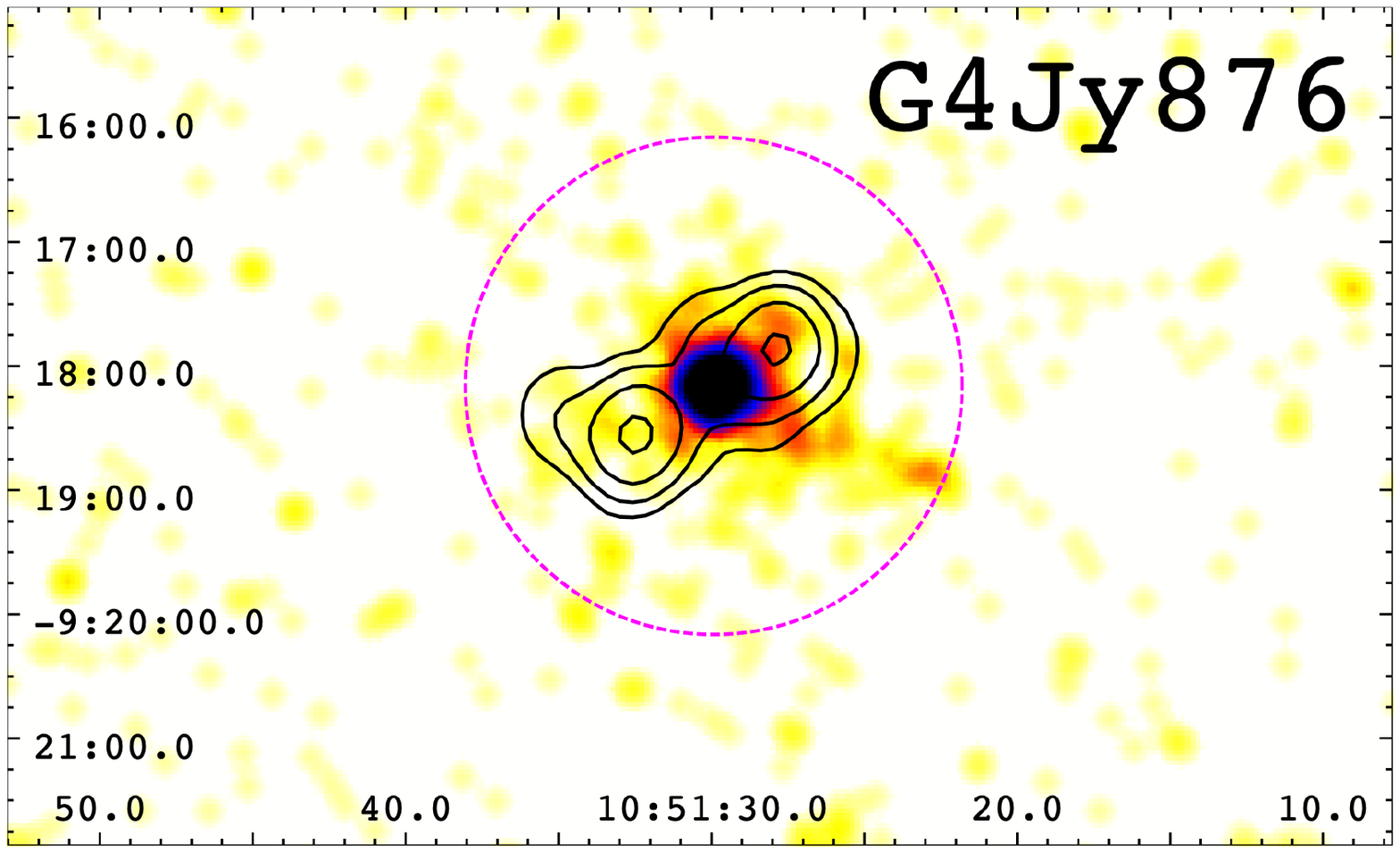}
\includegraphics[width=3.8cm,height=6.4cm,angle=-90]{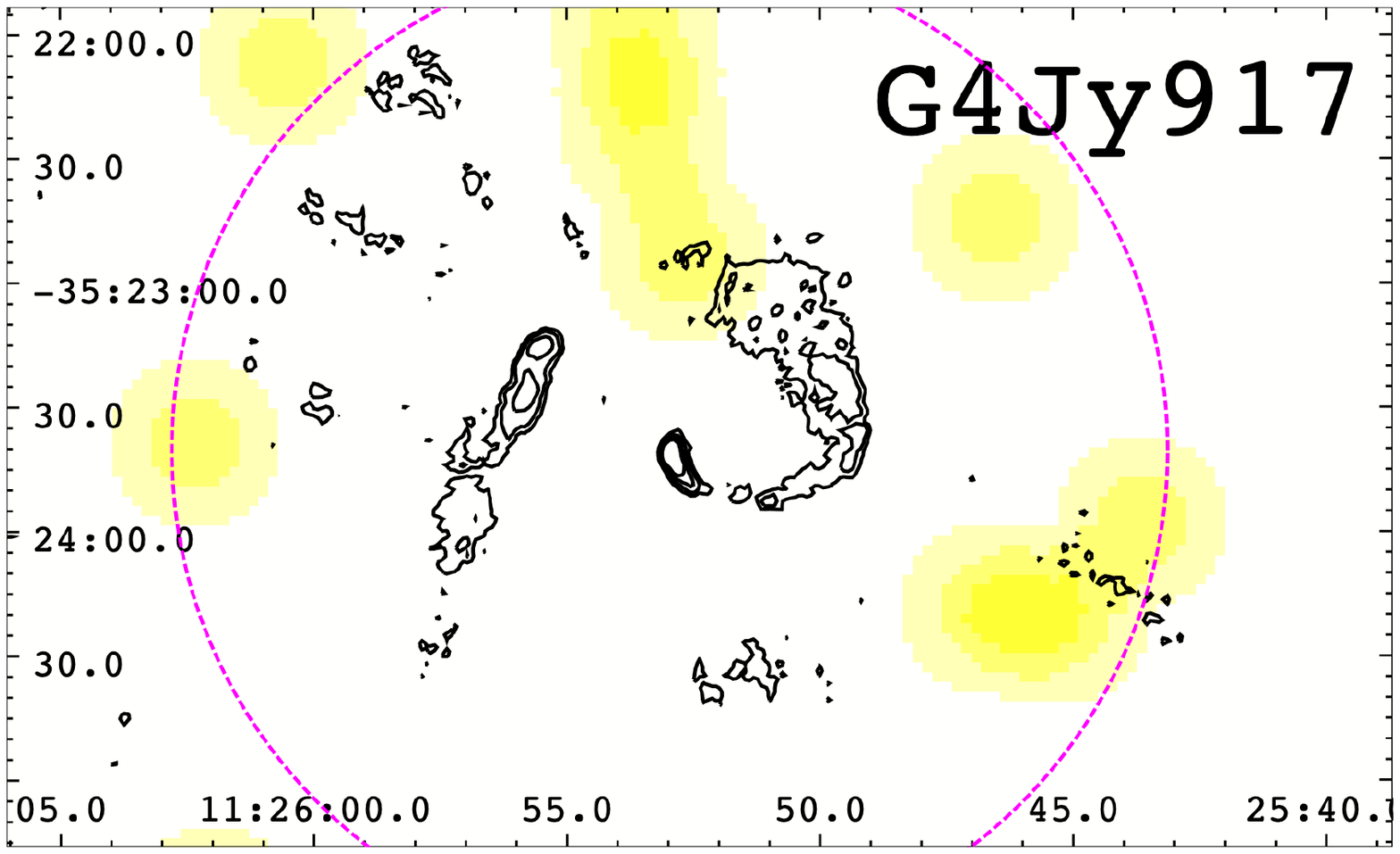}
\includegraphics[width=3.8cm,height=6.4cm,angle=-90]{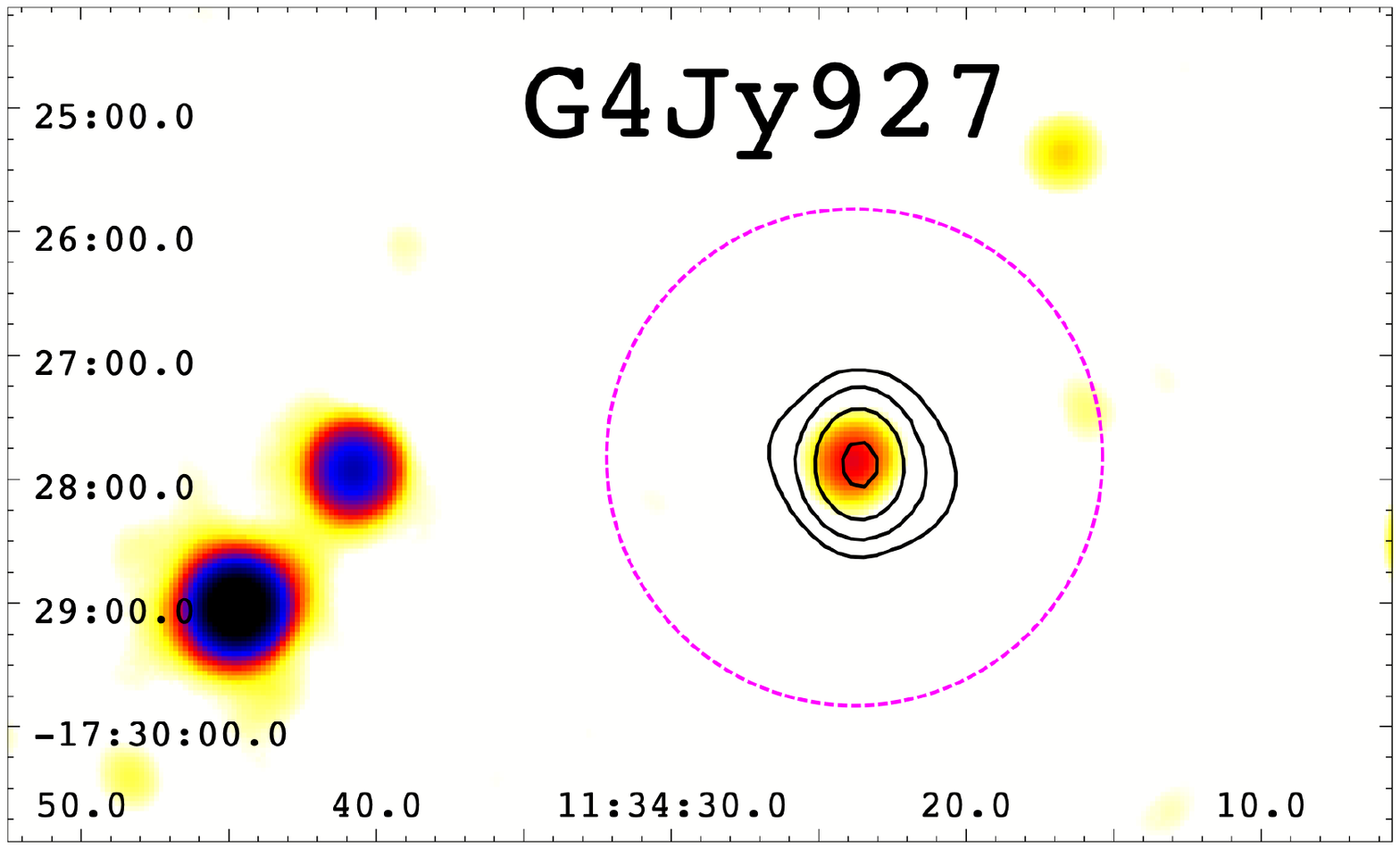}
\includegraphics[width=3.8cm,height=6.4cm,angle=-90]{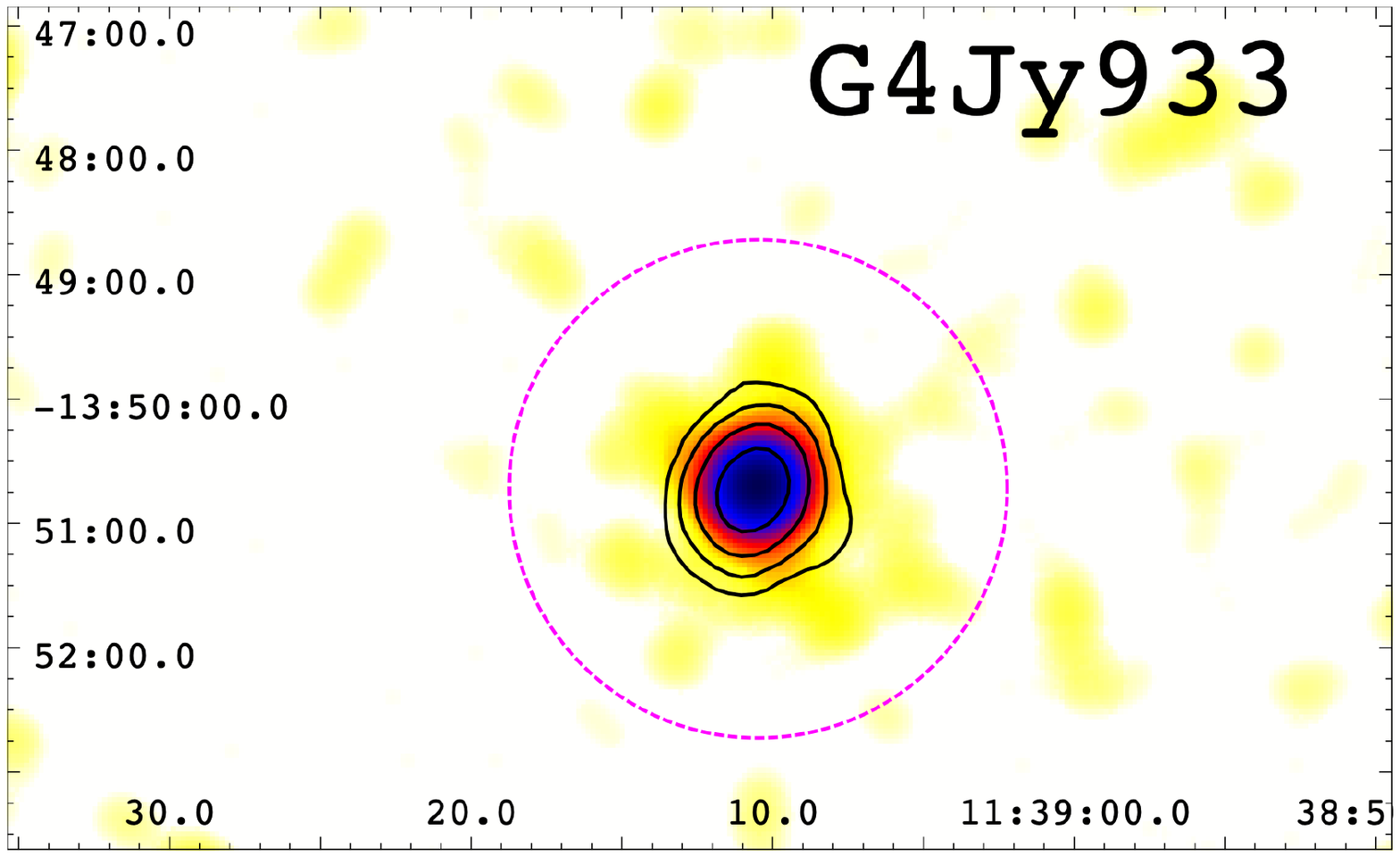}
\includegraphics[width=3.8cm,height=6.4cm,angle=-90]{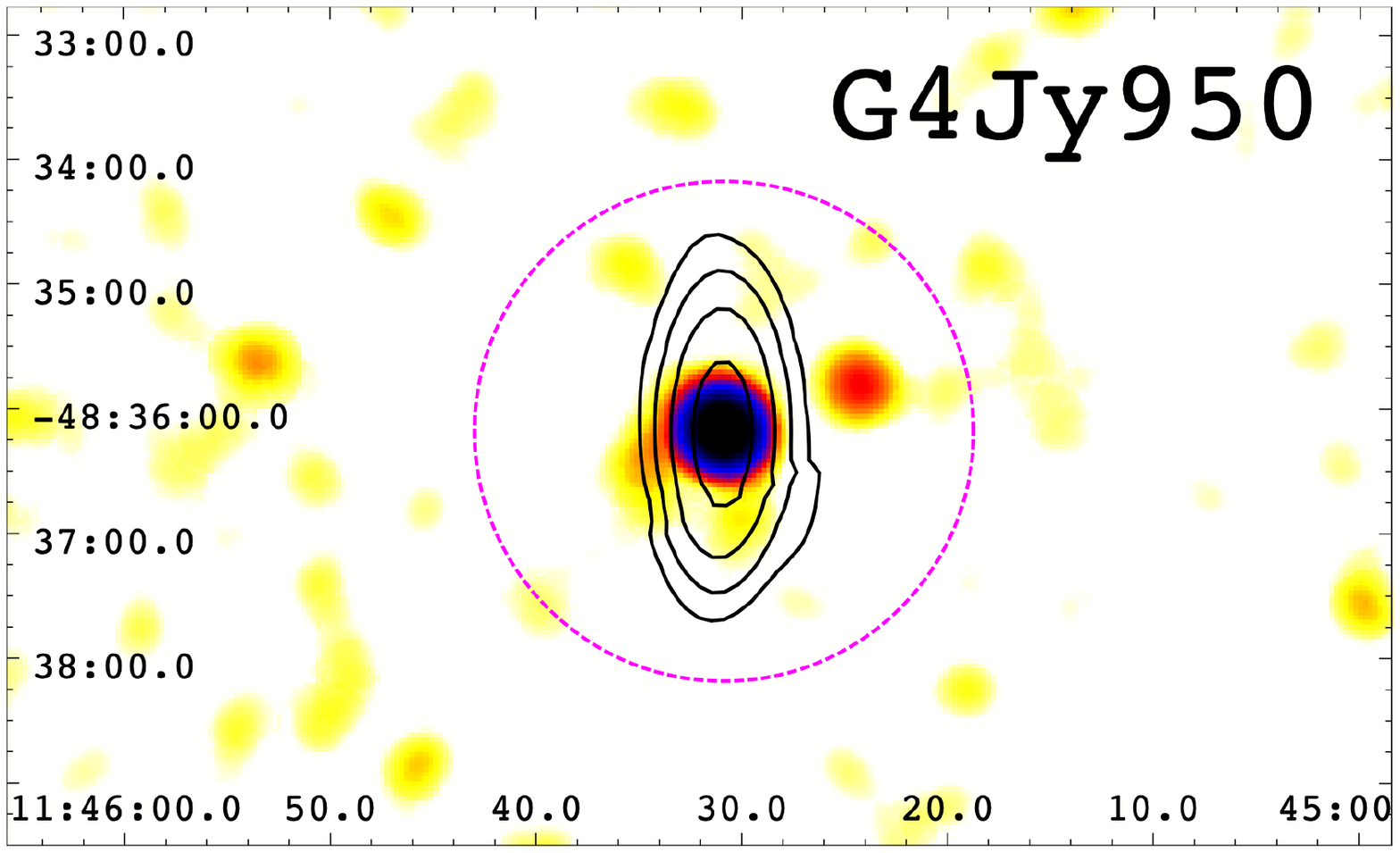}
\includegraphics[width=3.8cm,height=6.4cm,angle=-90]{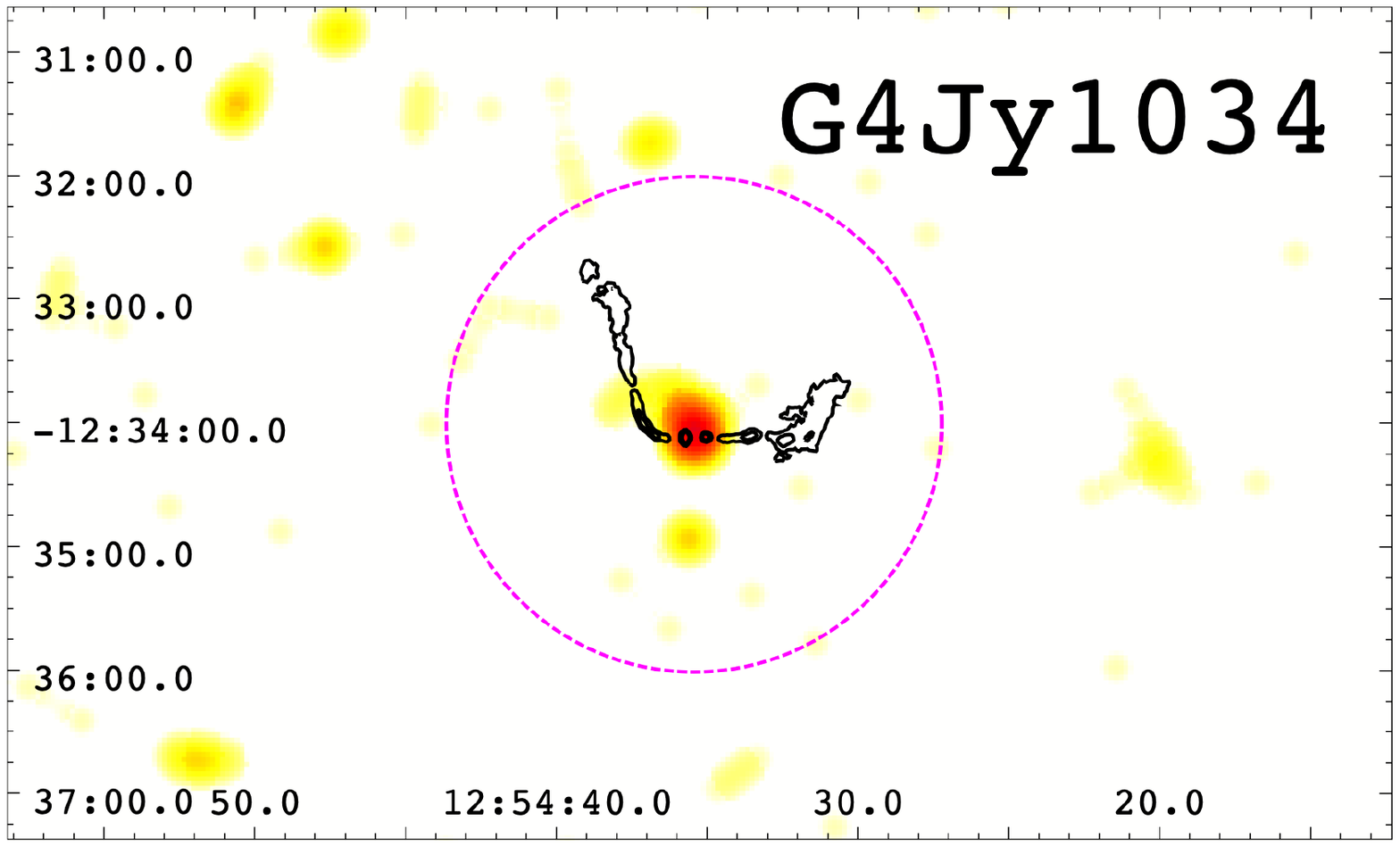}
\includegraphics[width=3.8cm,height=6.4cm,angle=-90]{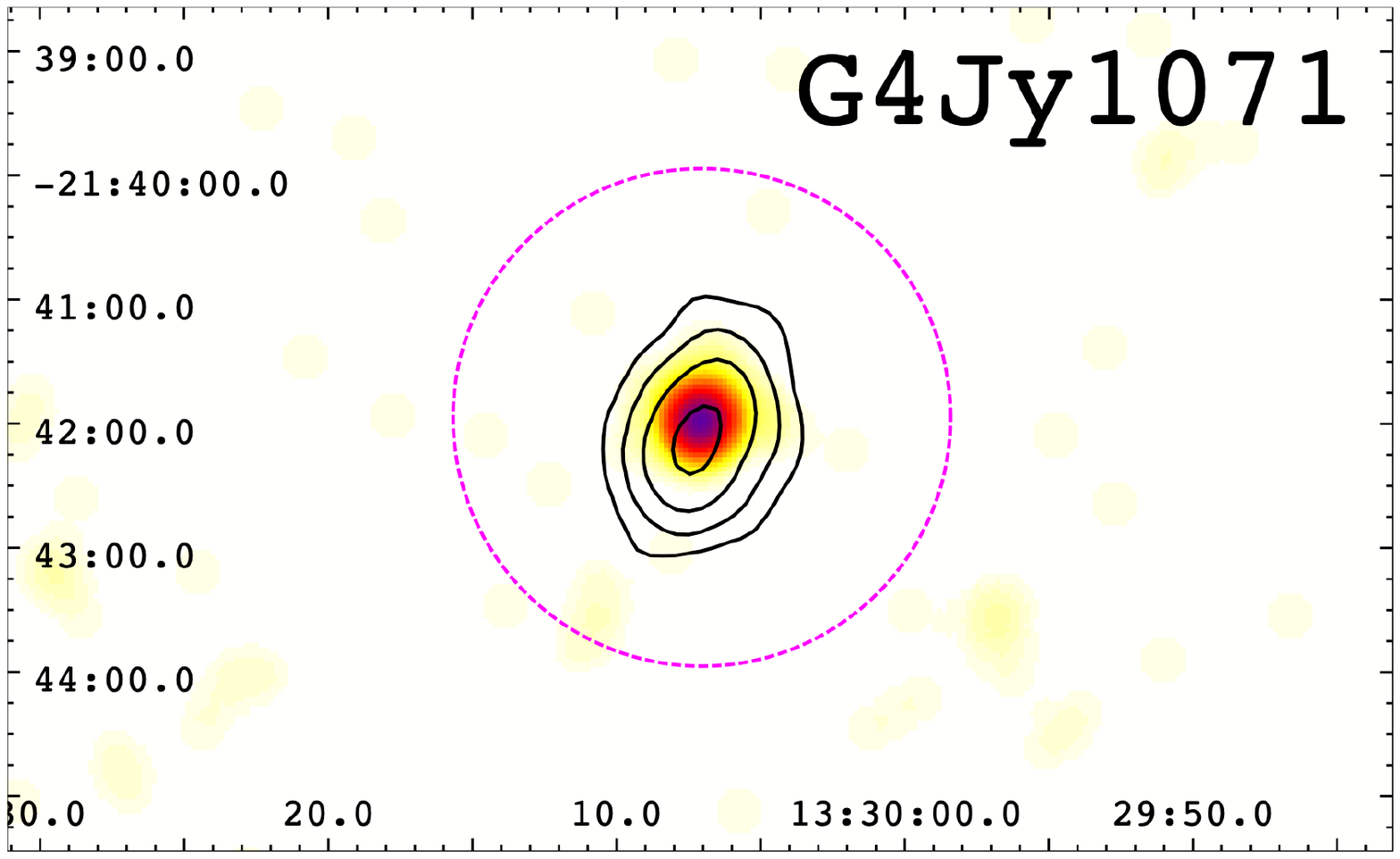}
\includegraphics[width=3.8cm,height=6.4cm,angle=-90]{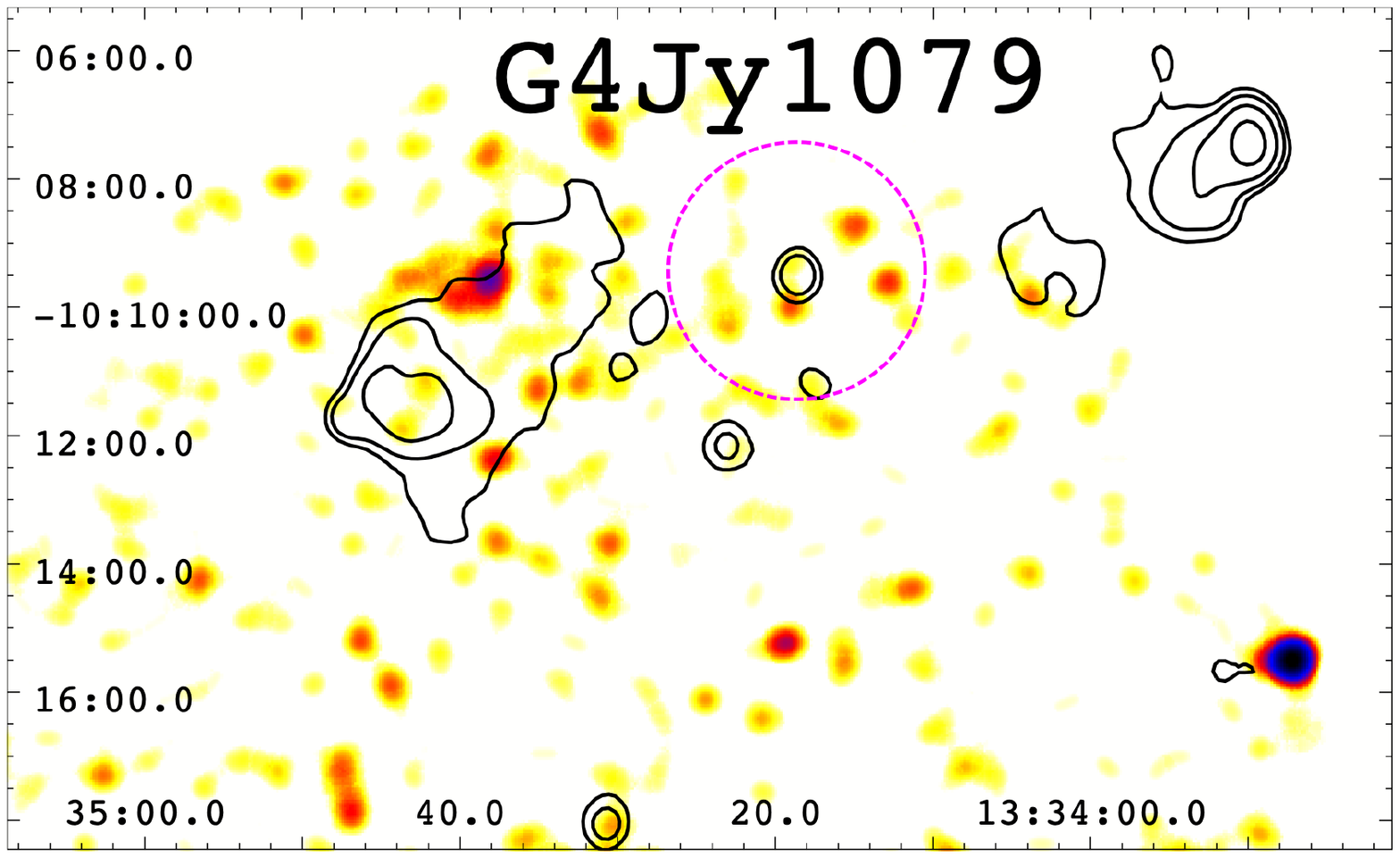}
\includegraphics[width=3.8cm,height=6.4cm,angle=-90]{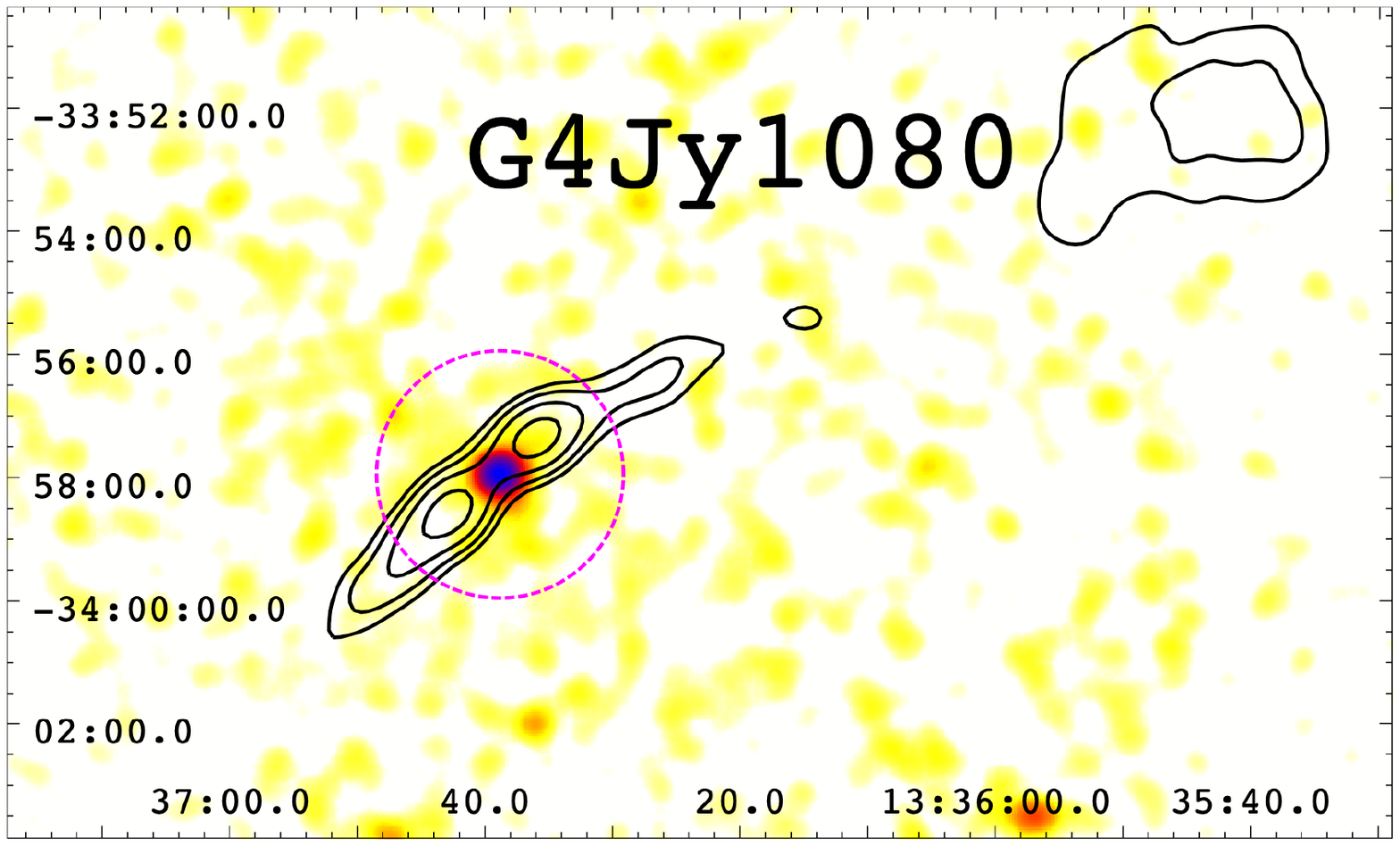}
\includegraphics[width=3.8cm,height=6.4cm,angle=-90]{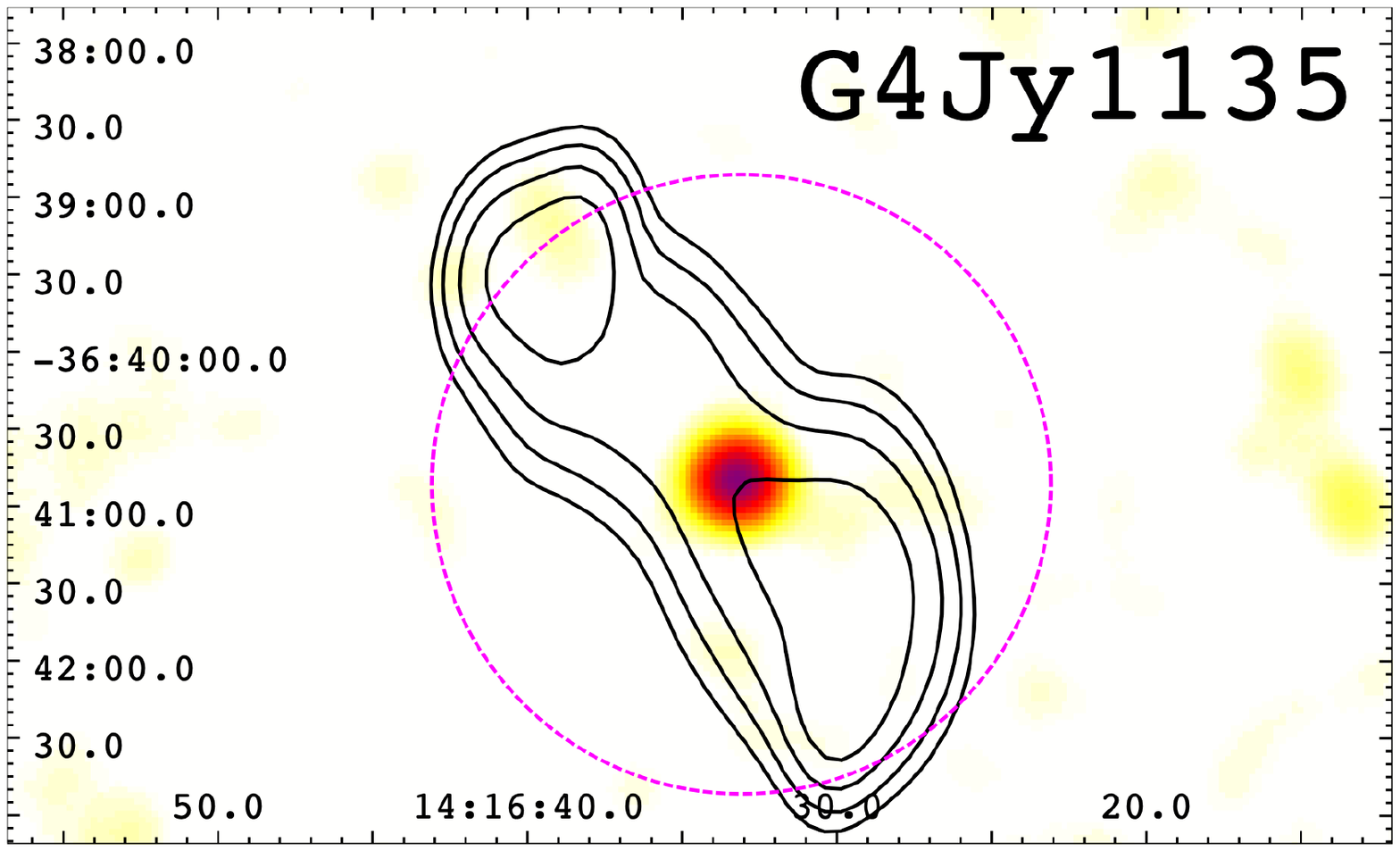}
\includegraphics[width=3.8cm,height=6.4cm,angle=-90]{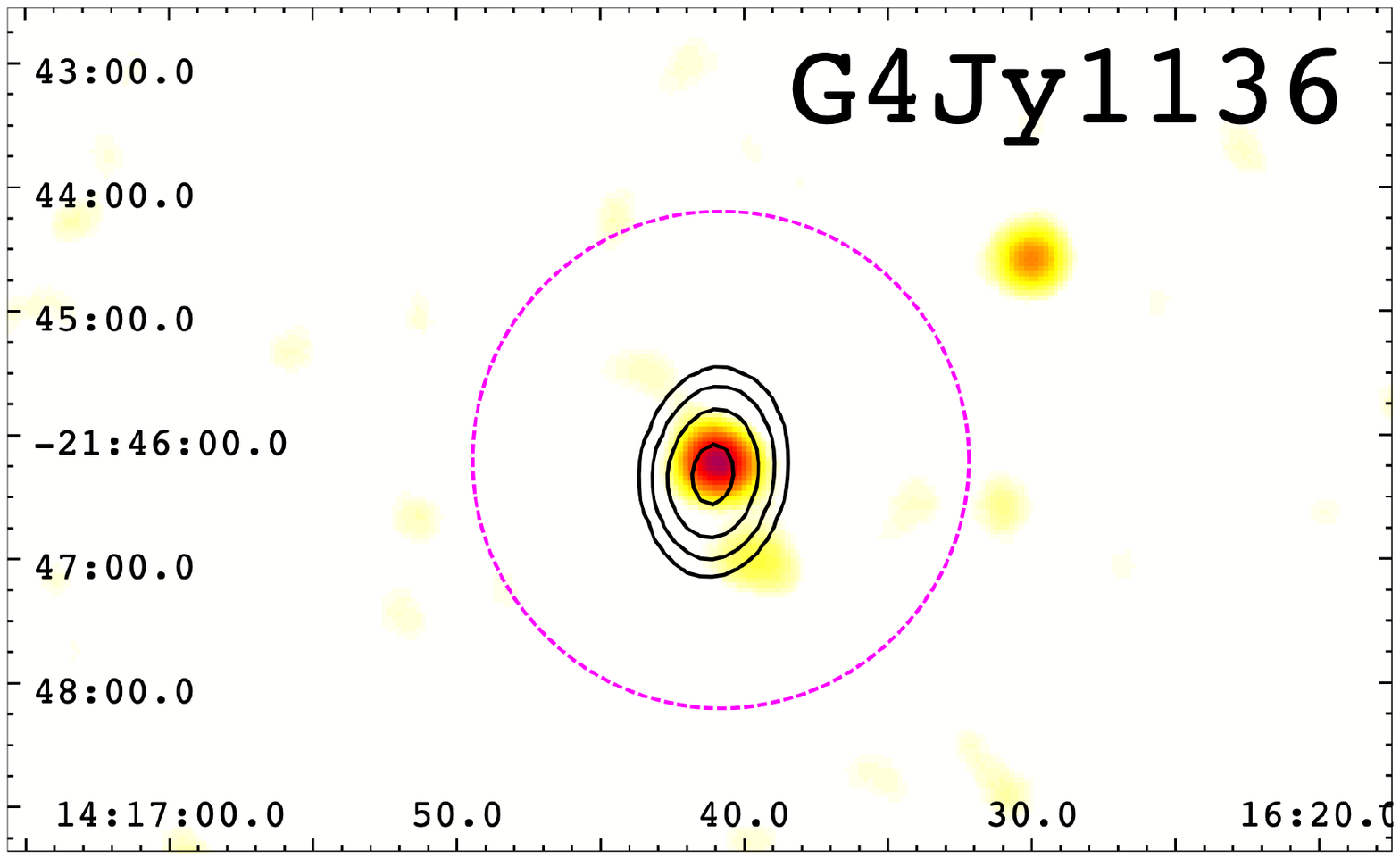}
\caption{Same as Figure~\ref{fig:example} for the following \cs\ radio sources:
G4Jy\,854, G4Jy\,876, G4Jy\,917, G4Jy\,927, G4Jy\,933, G4Jy\,950, G4Jy\,1034, G4Jy\,1071, G4Jy\,1079, G4Jy\,1080, G4Jy\,1135, G4Jy\,1136.}
\end{center}
\end{figure*}

\begin{figure*}[!th]
\begin{center}
\includegraphics[width=3.8cm,height=6.4cm,angle=-90]{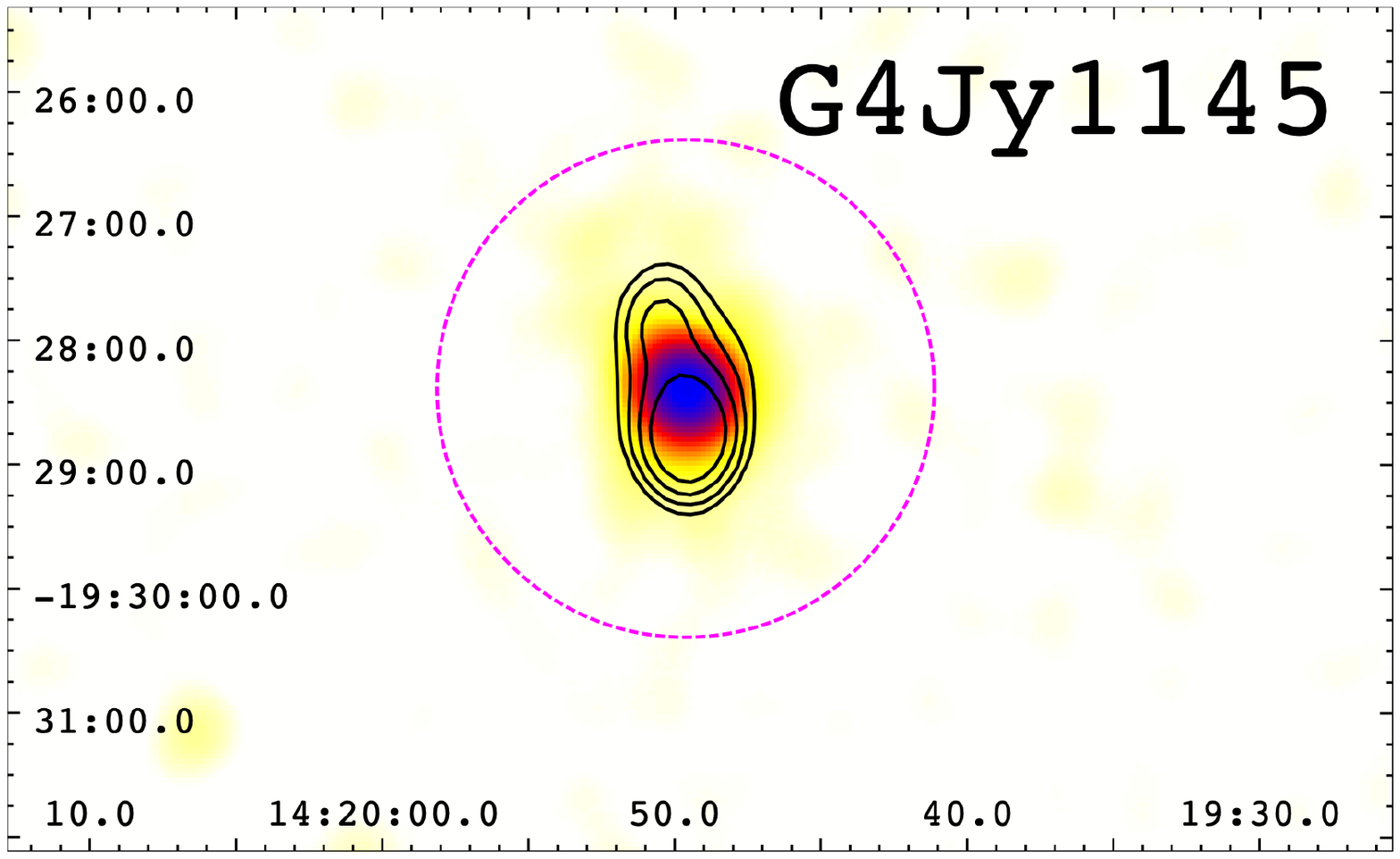}
\includegraphics[width=3.8cm,height=6.4cm,angle=-90]{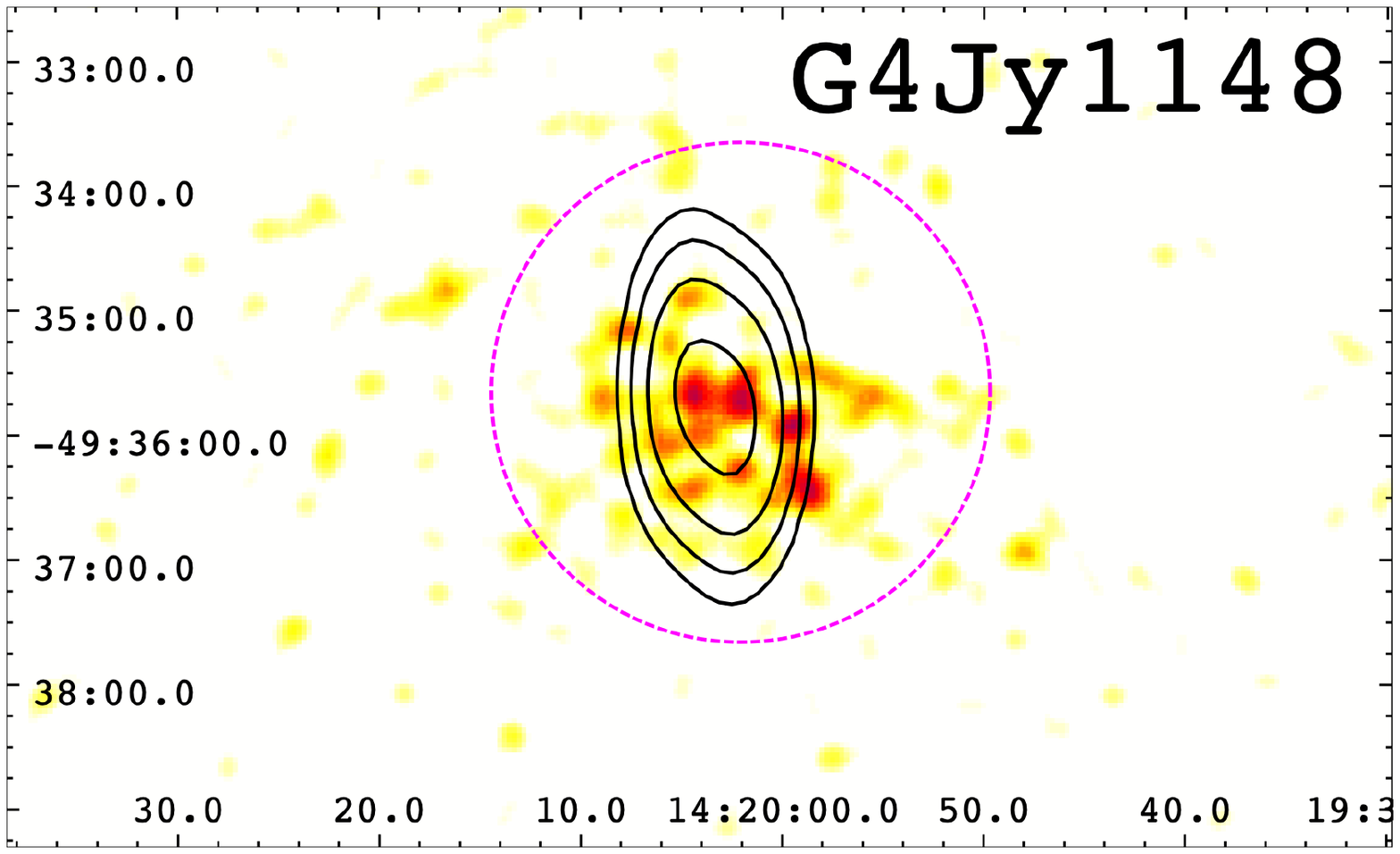}
\includegraphics[width=3.8cm,height=6.4cm,angle=-90]{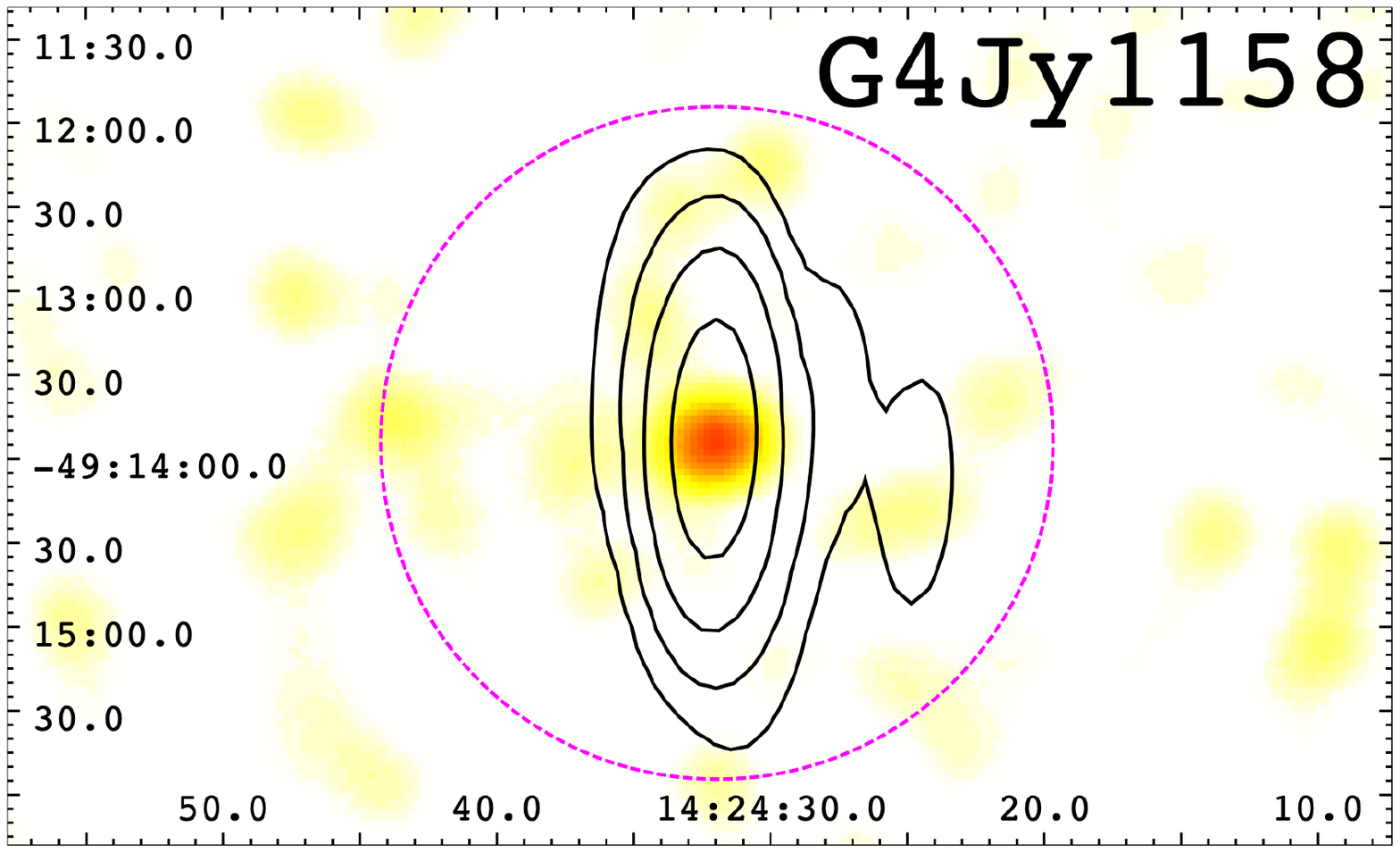}
\includegraphics[width=3.8cm,height=6.4cm,angle=-90]{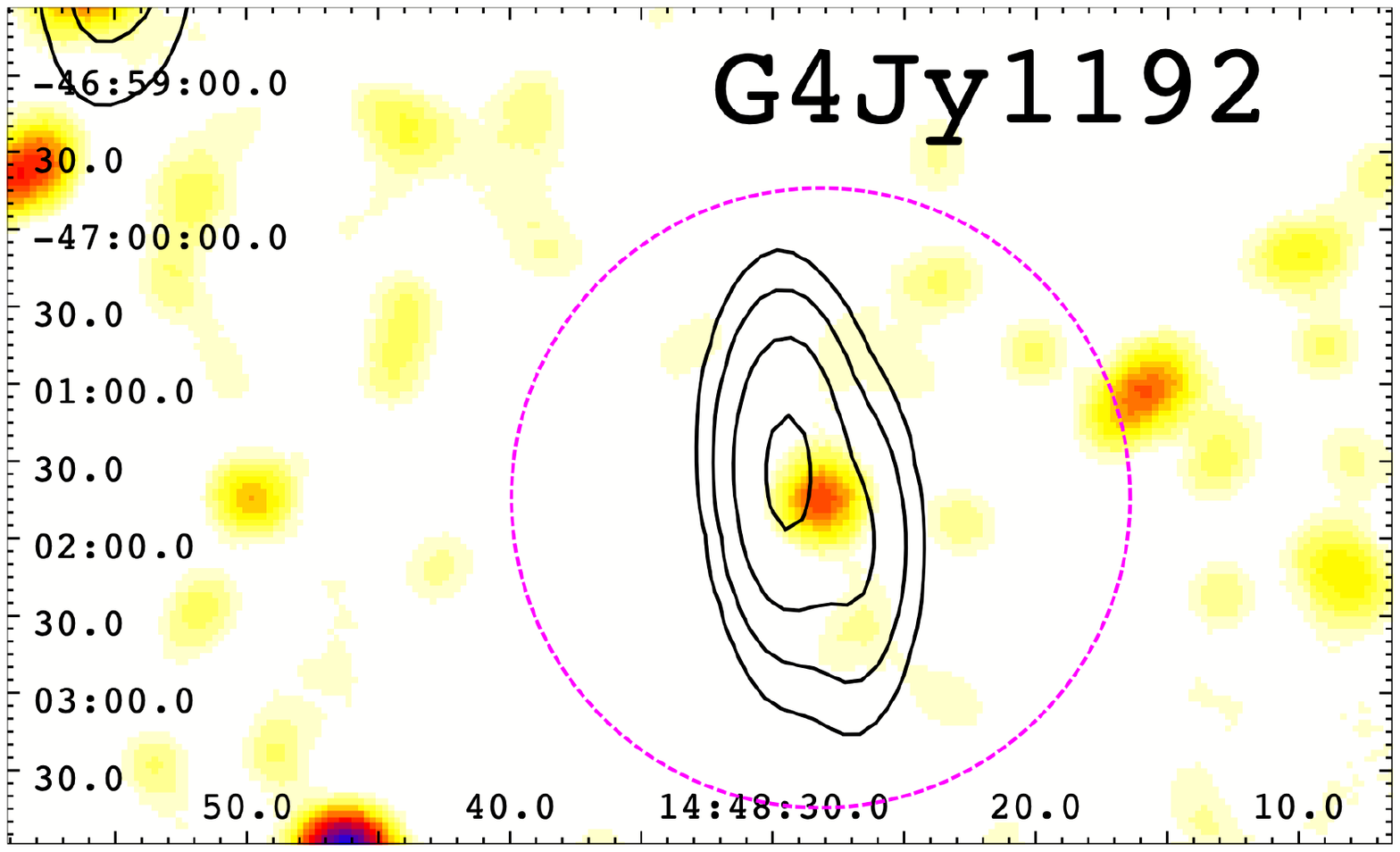}
\includegraphics[width=3.8cm,height=6.4cm,angle=-90]{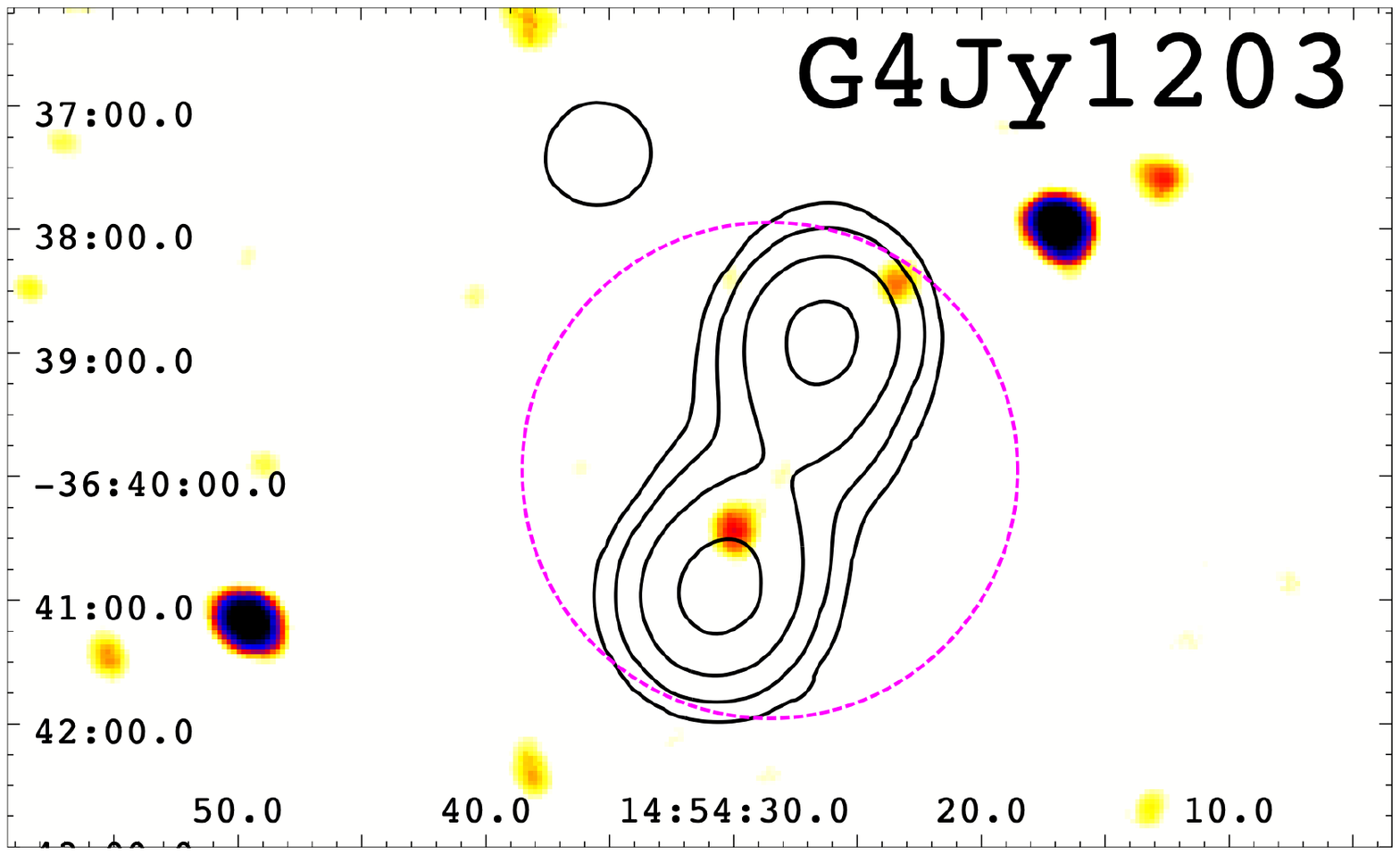}
\includegraphics[width=3.8cm,height=6.4cm,angle=-90]{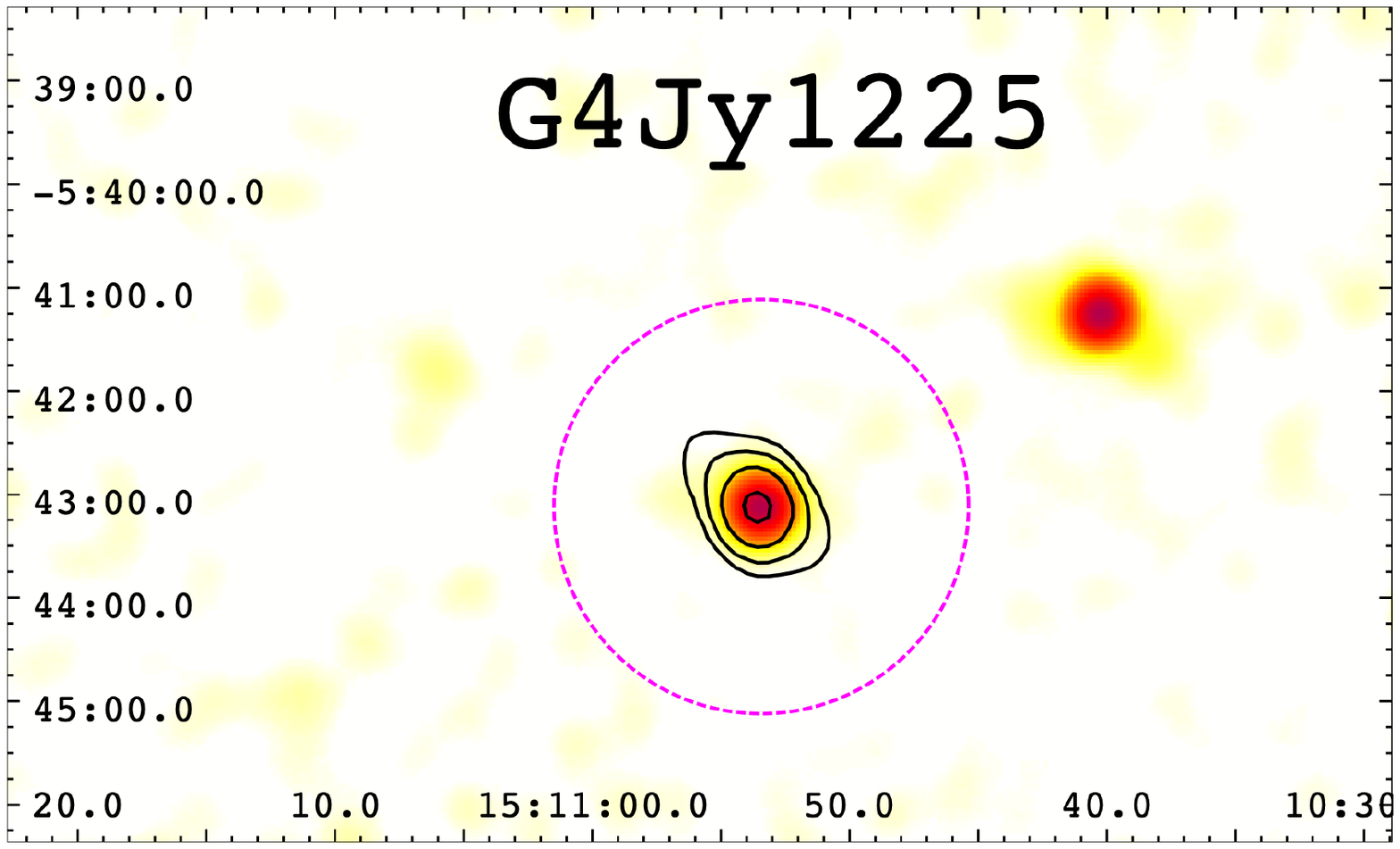}
\includegraphics[width=3.8cm,height=6.4cm,angle=-90]{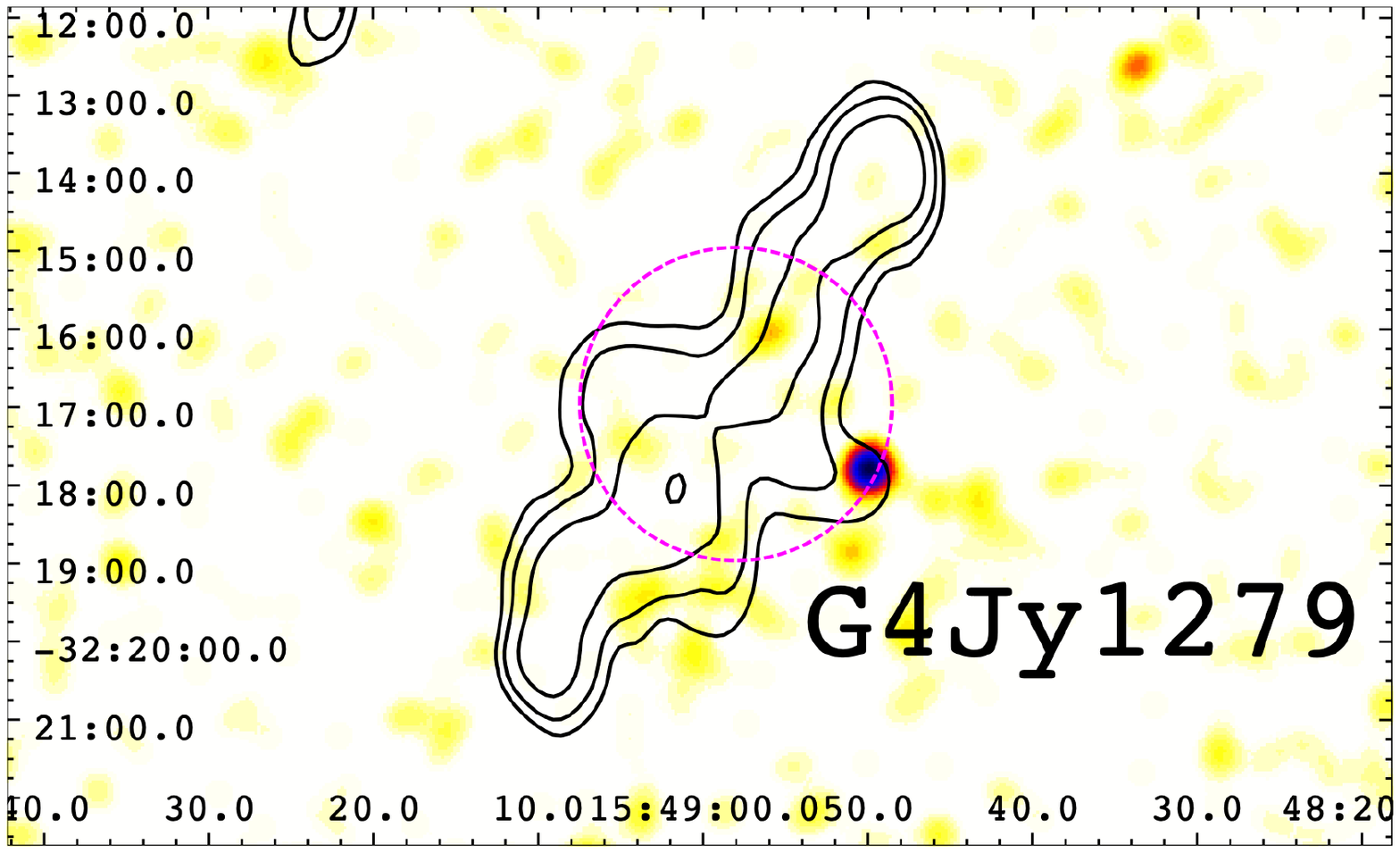}
\includegraphics[width=3.8cm,height=6.4cm,angle=-90]{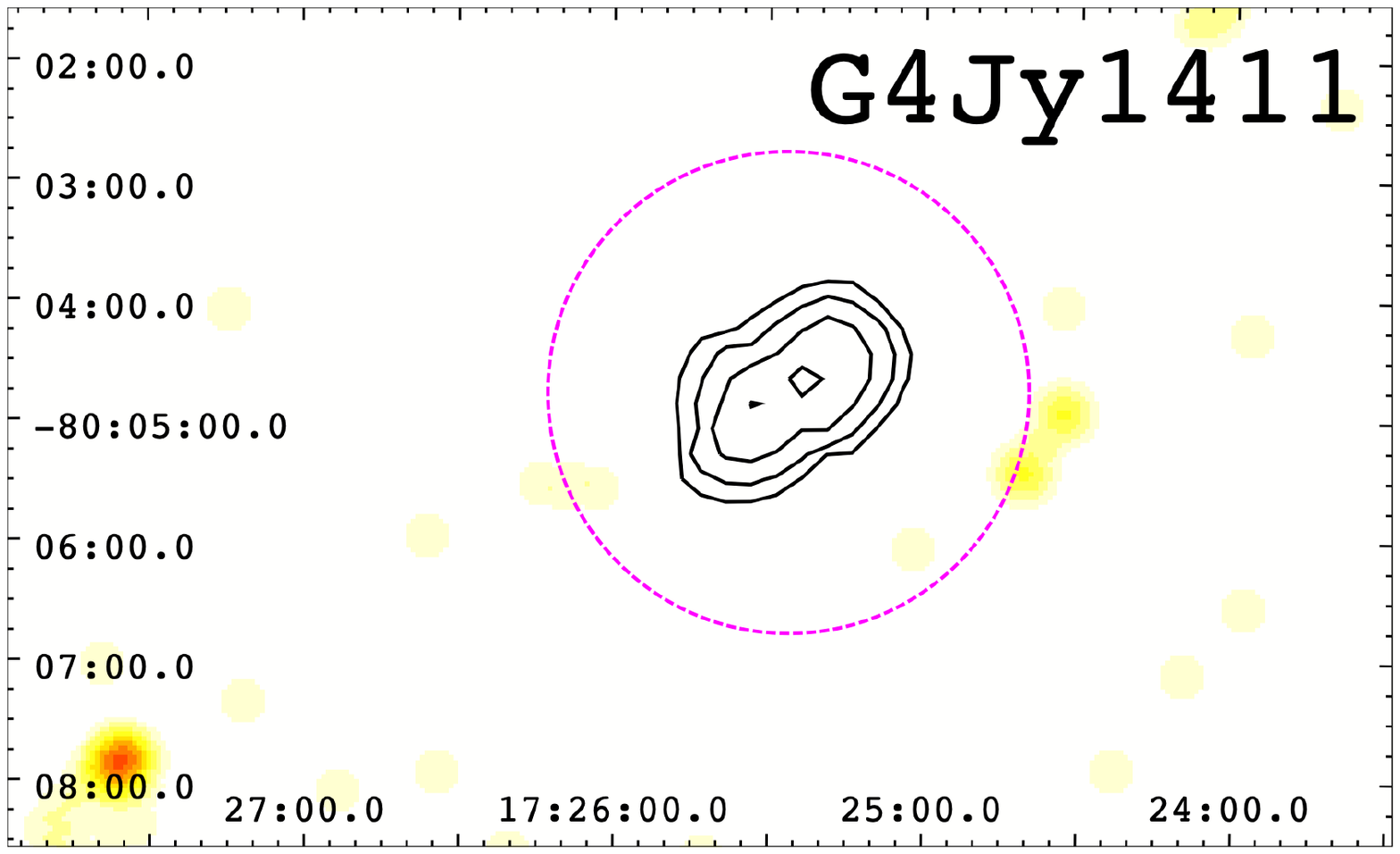}
\includegraphics[width=3.8cm,height=6.4cm,angle=-90]{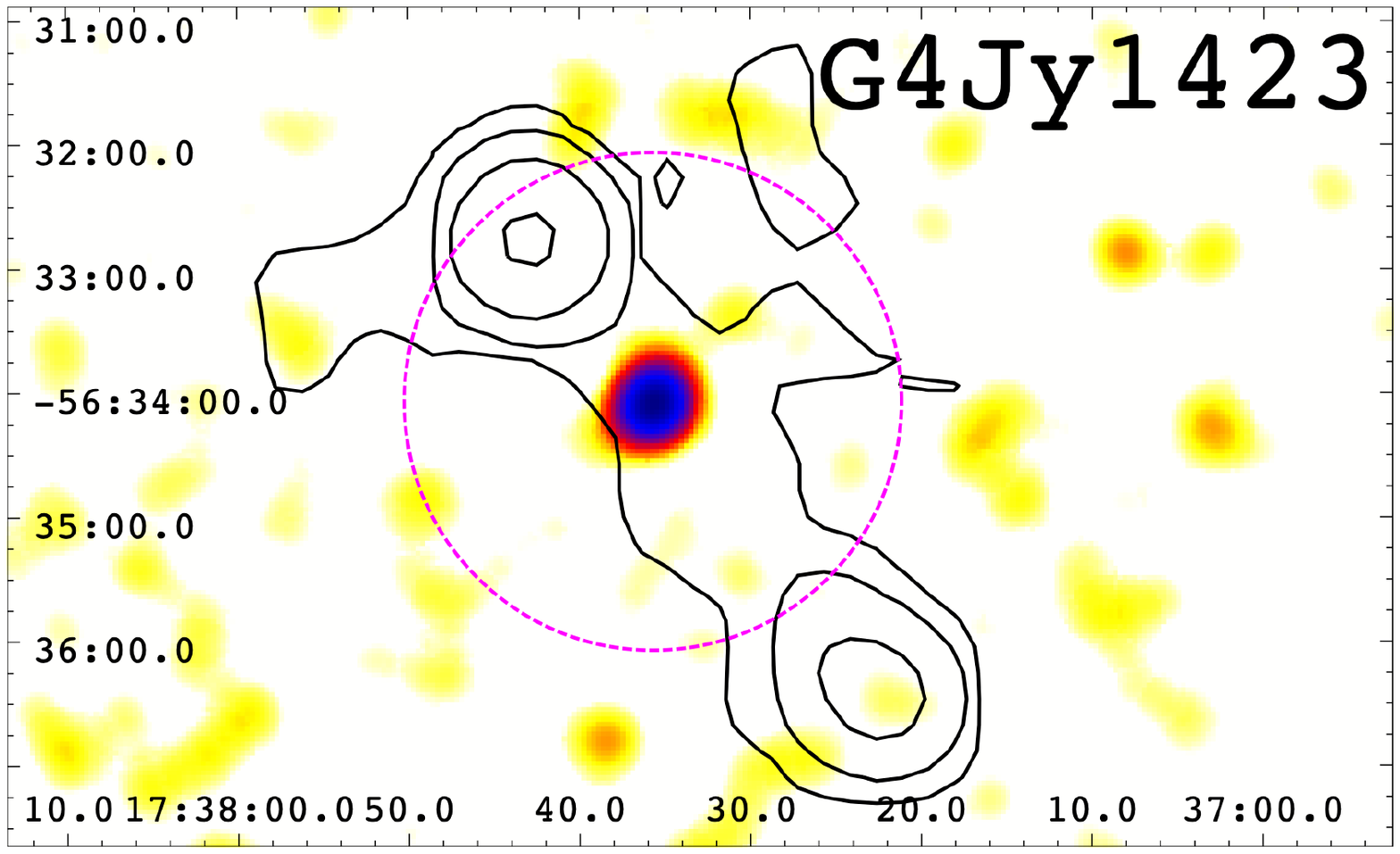}
\includegraphics[width=3.8cm,height=6.4cm,angle=-90]{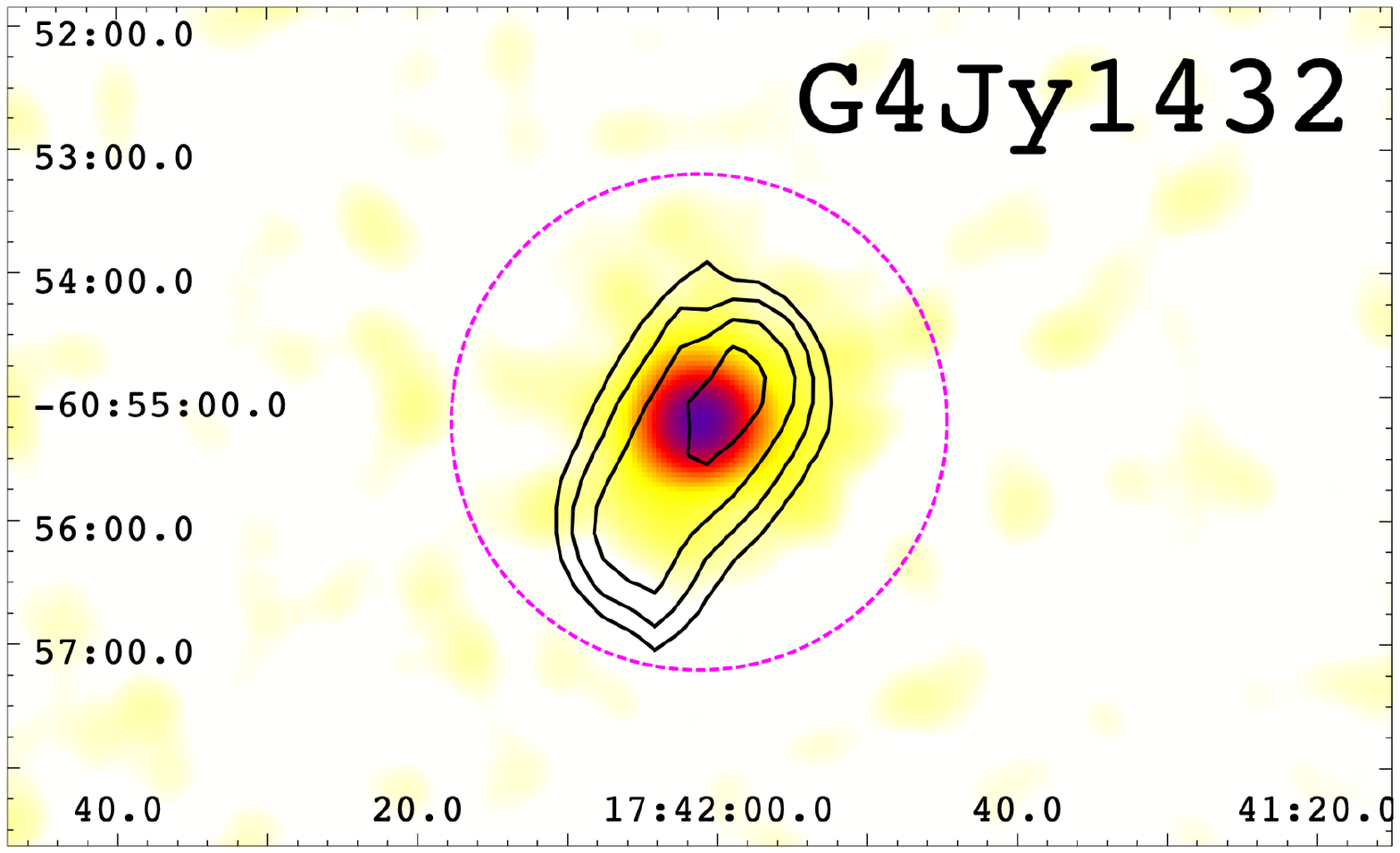}
\includegraphics[width=3.8cm,height=6.4cm,angle=-90]{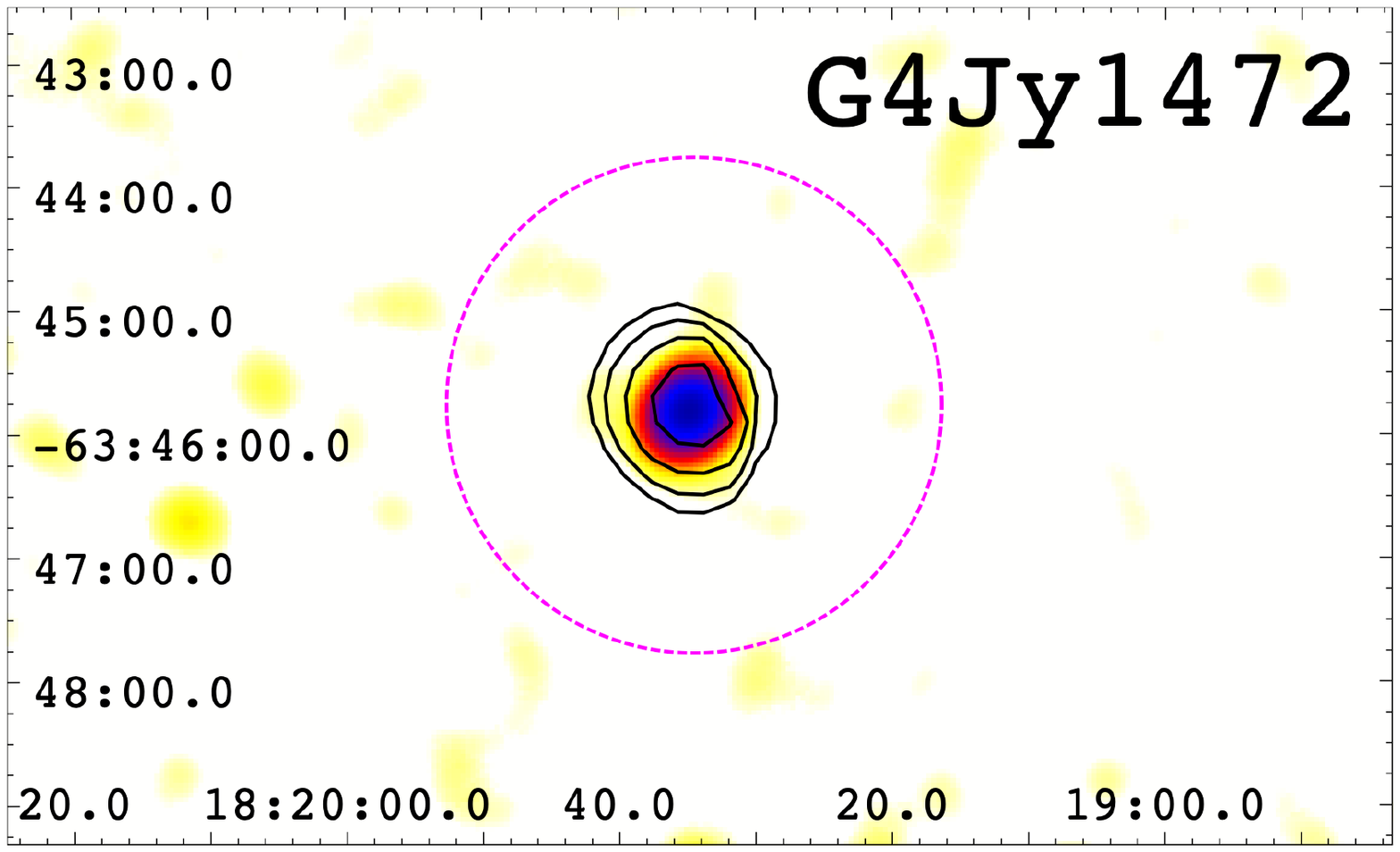}
\includegraphics[width=3.8cm,height=6.4cm,angle=-90]{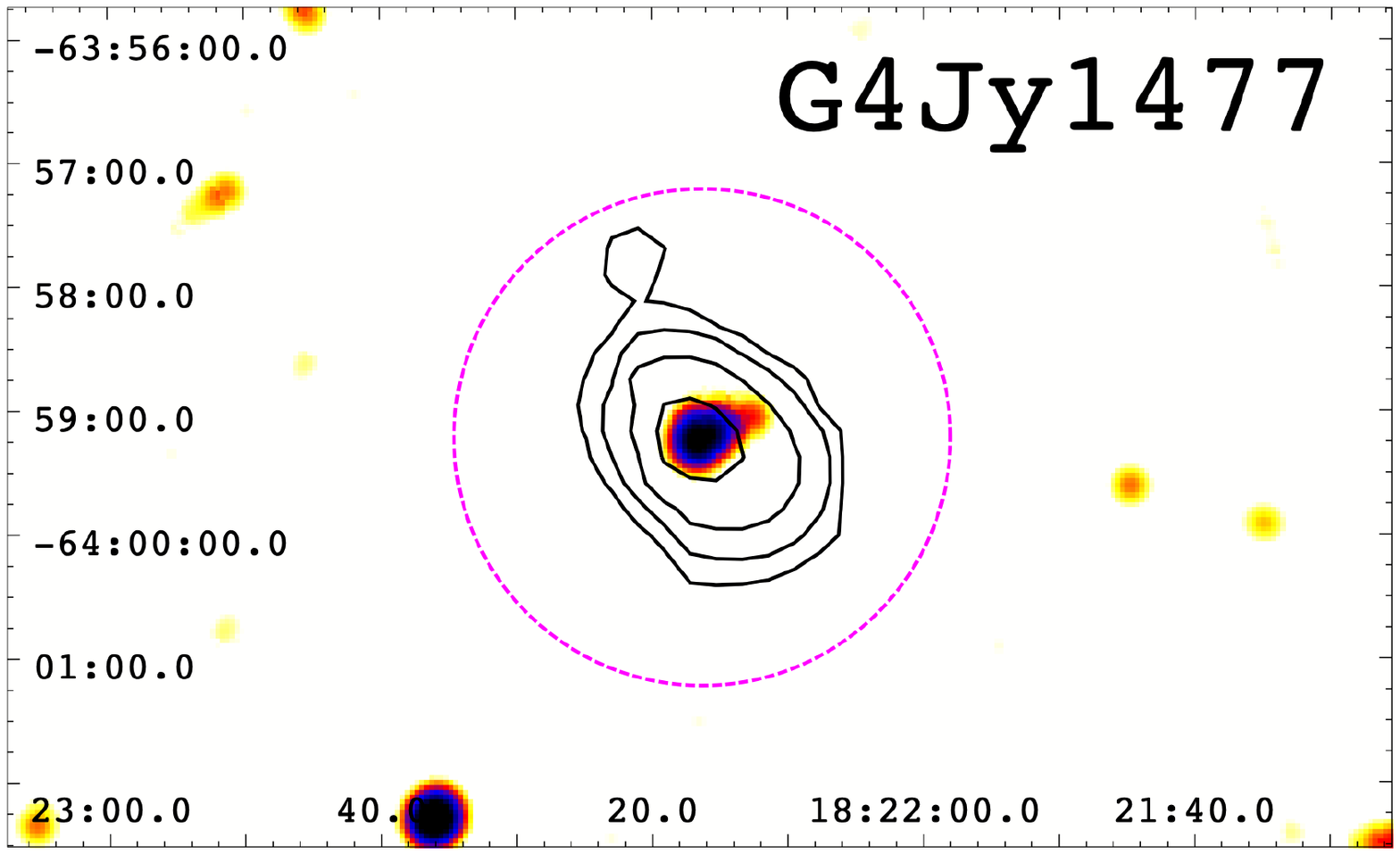}
\caption{Same as Figure~\ref{fig:example} for the following \cs\ radio sources:
G4Jy\,1145, G4Jy\,1148, G4Jy\,1158, G4Jy\,1192, G4Jy\,1203, G4Jy\,1225, G4Jy\,1279, G4Jy\,1411, G4Jy\,1423, G4Jy\,1432, G4Jy\,1472, G4Jy\,1477.}
\end{center}
\end{figure*}

\begin{figure*}[!th]
\begin{center}
\includegraphics[width=3.8cm,height=6.4cm,angle=-90]{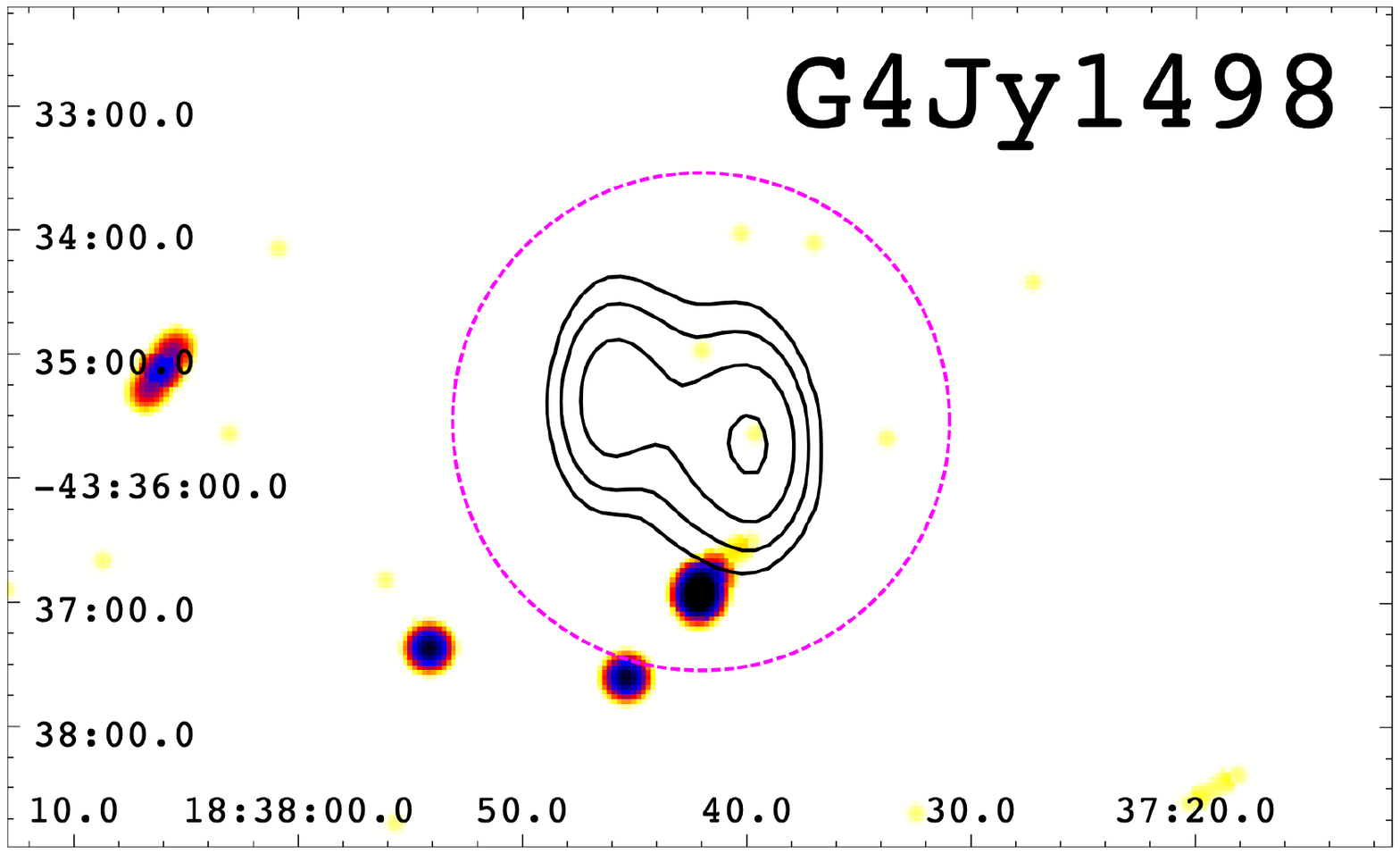}
\includegraphics[width=3.8cm,height=6.4cm,angle=-90]{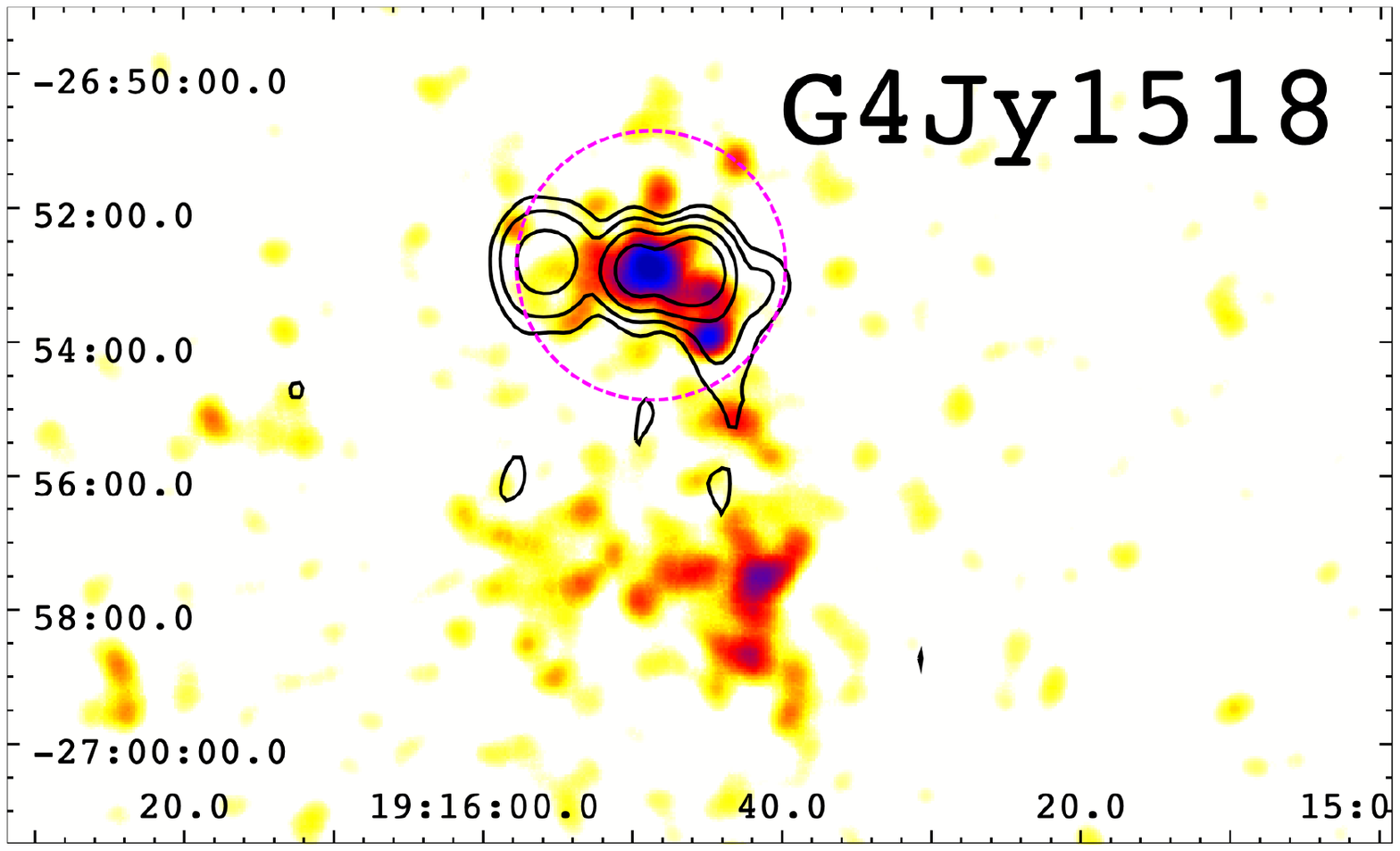}
\includegraphics[width=3.8cm,height=6.4cm,angle=-90]{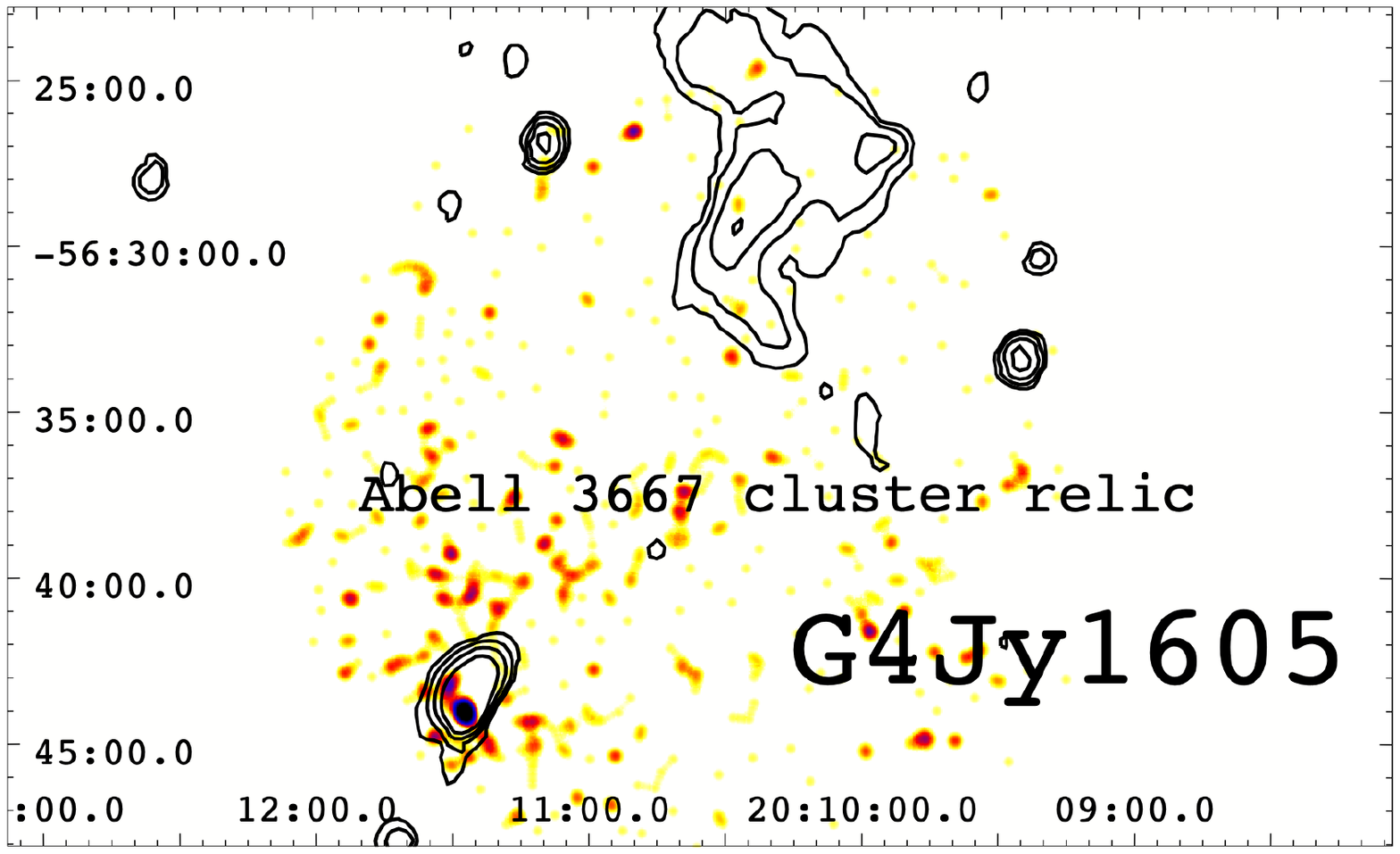}
\includegraphics[width=3.8cm,height=6.4cm,angle=-90]{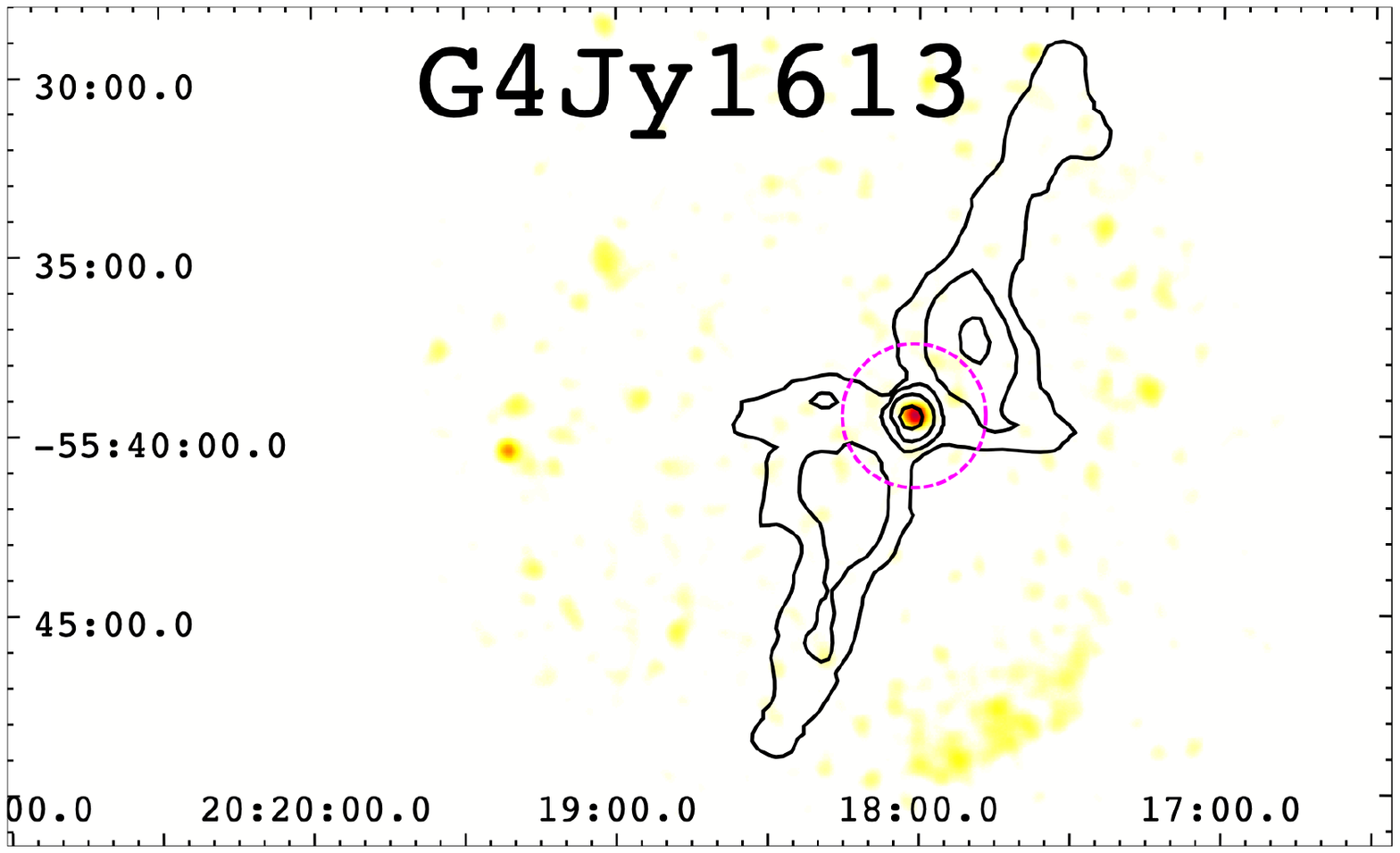}
\includegraphics[width=3.8cm,height=6.4cm,angle=-90]{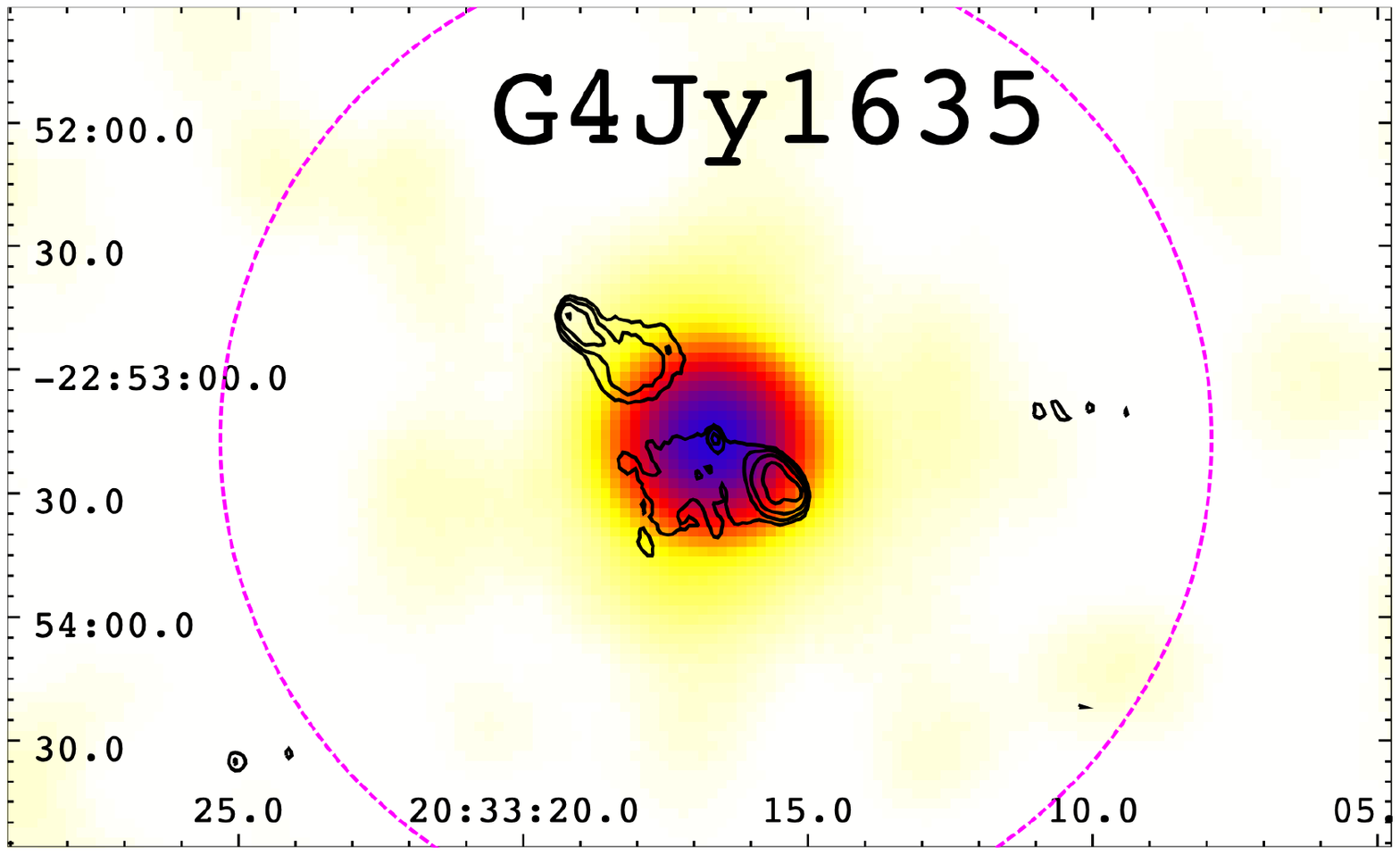}
\includegraphics[width=3.8cm,height=6.4cm,angle=-90]{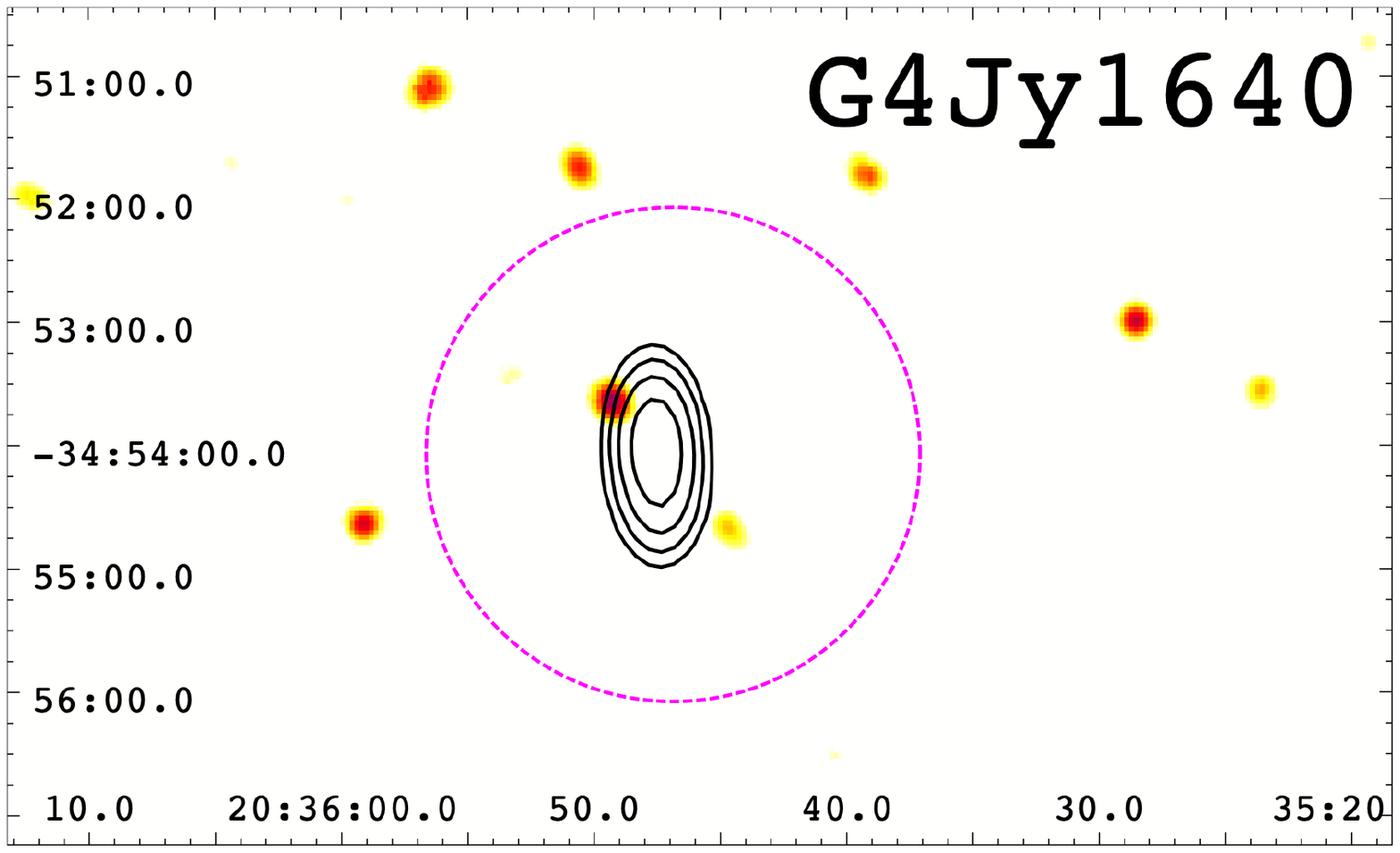}
\includegraphics[width=3.8cm,height=6.4cm,angle=-90]{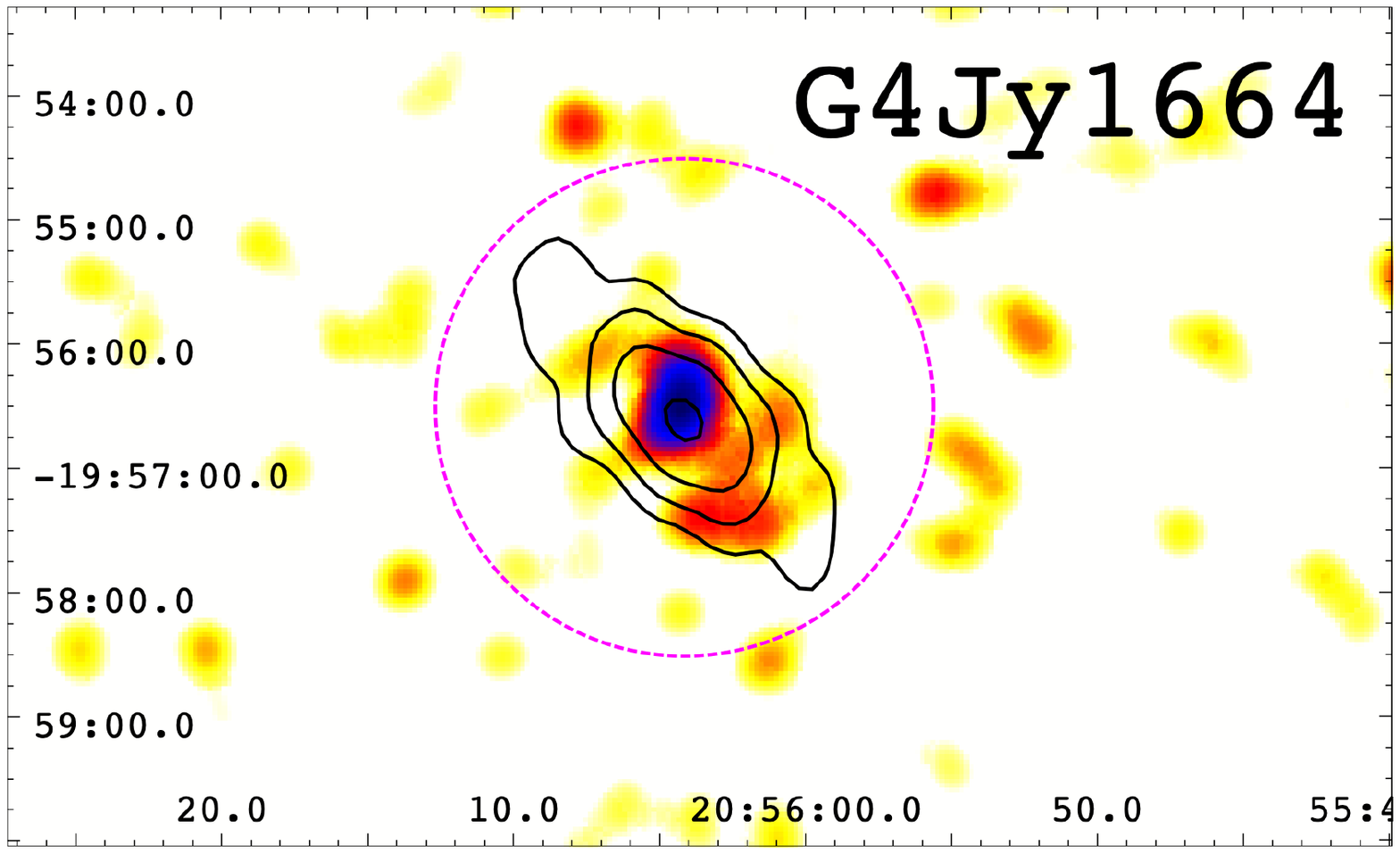}
\includegraphics[width=3.8cm,height=6.4cm,angle=-90]{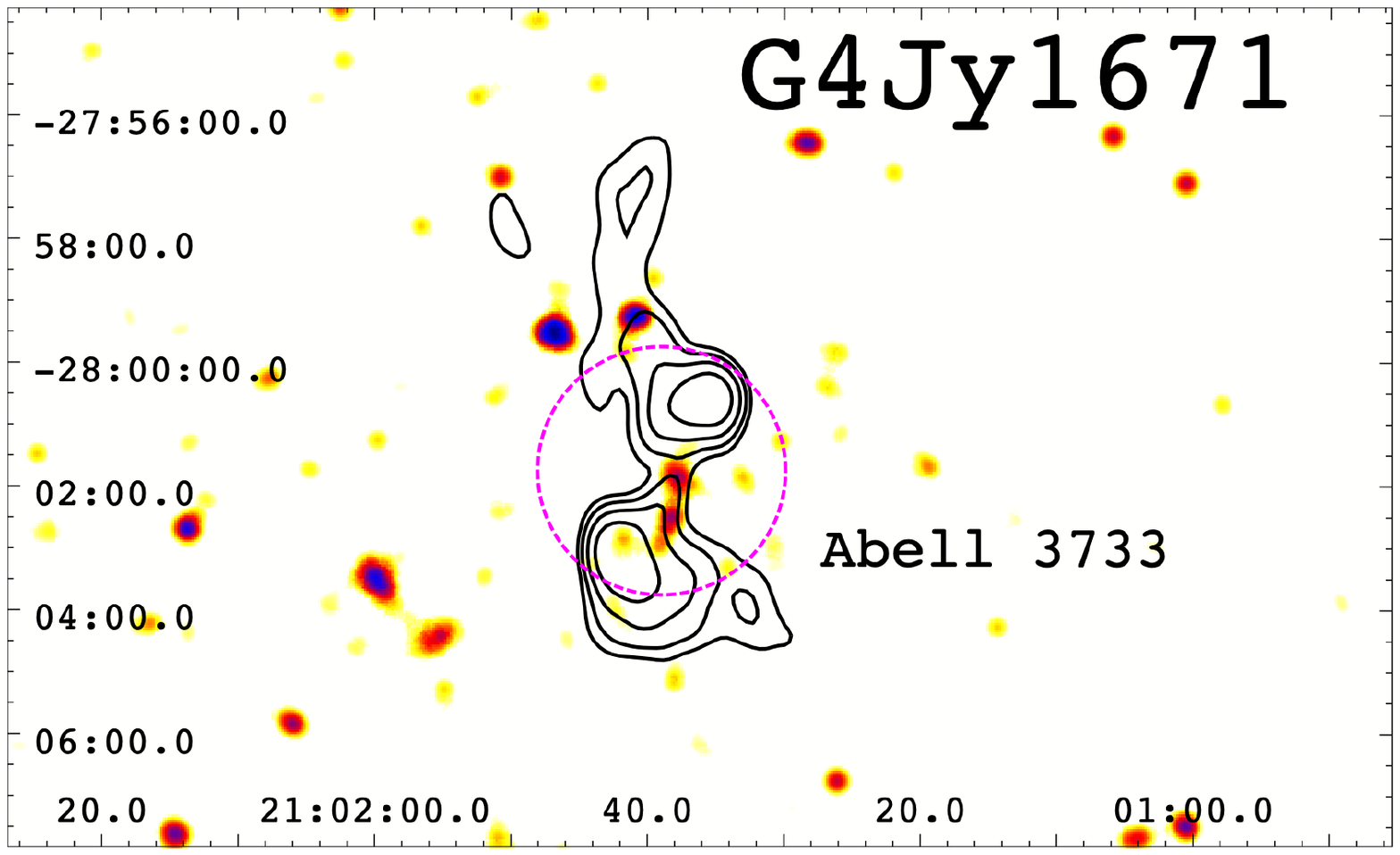}
\includegraphics[width=3.8cm,height=6.4cm,angle=-90]{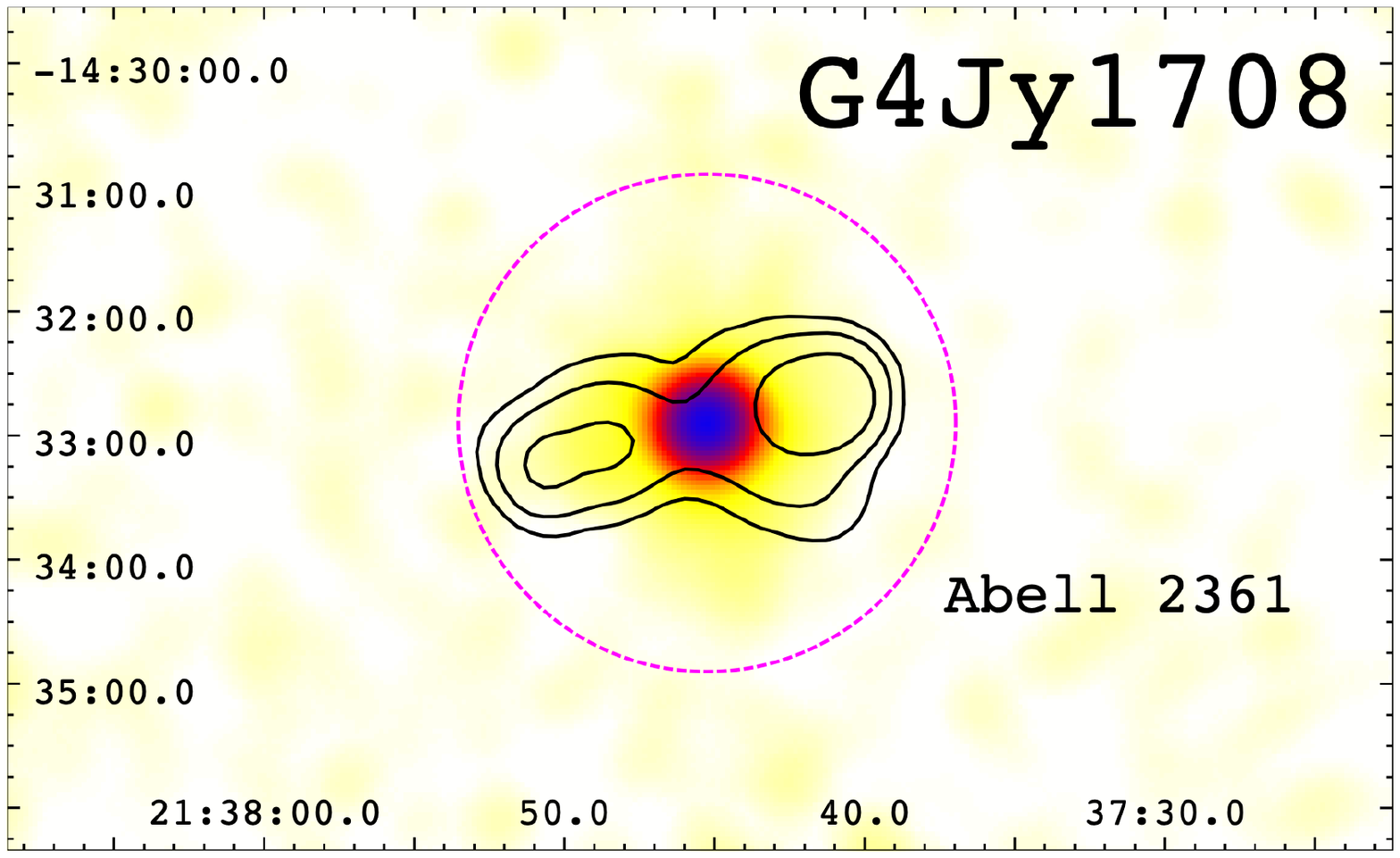}
\includegraphics[width=3.8cm,height=6.4cm,angle=-90]{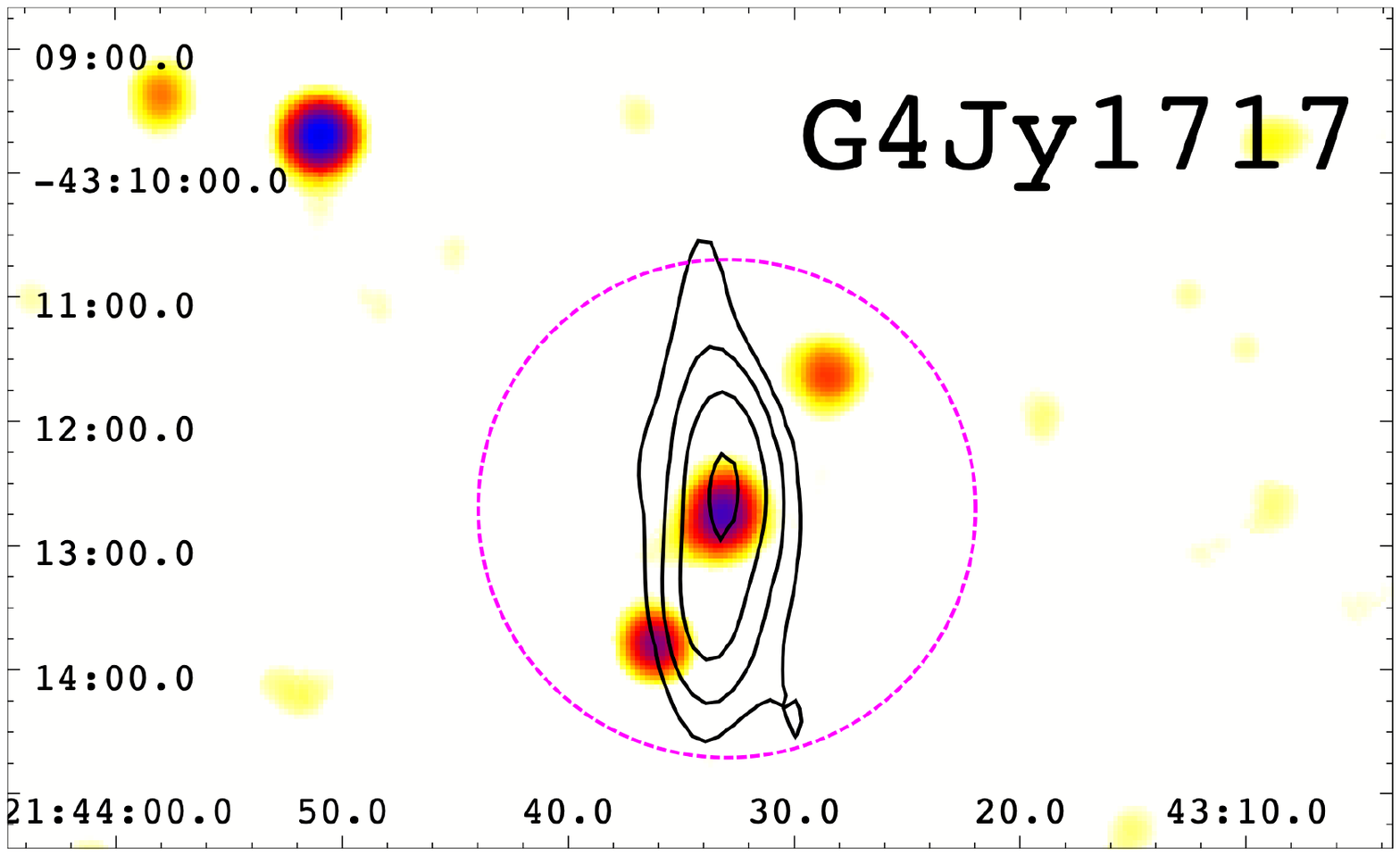}
\includegraphics[width=3.8cm,height=6.4cm,angle=-90]{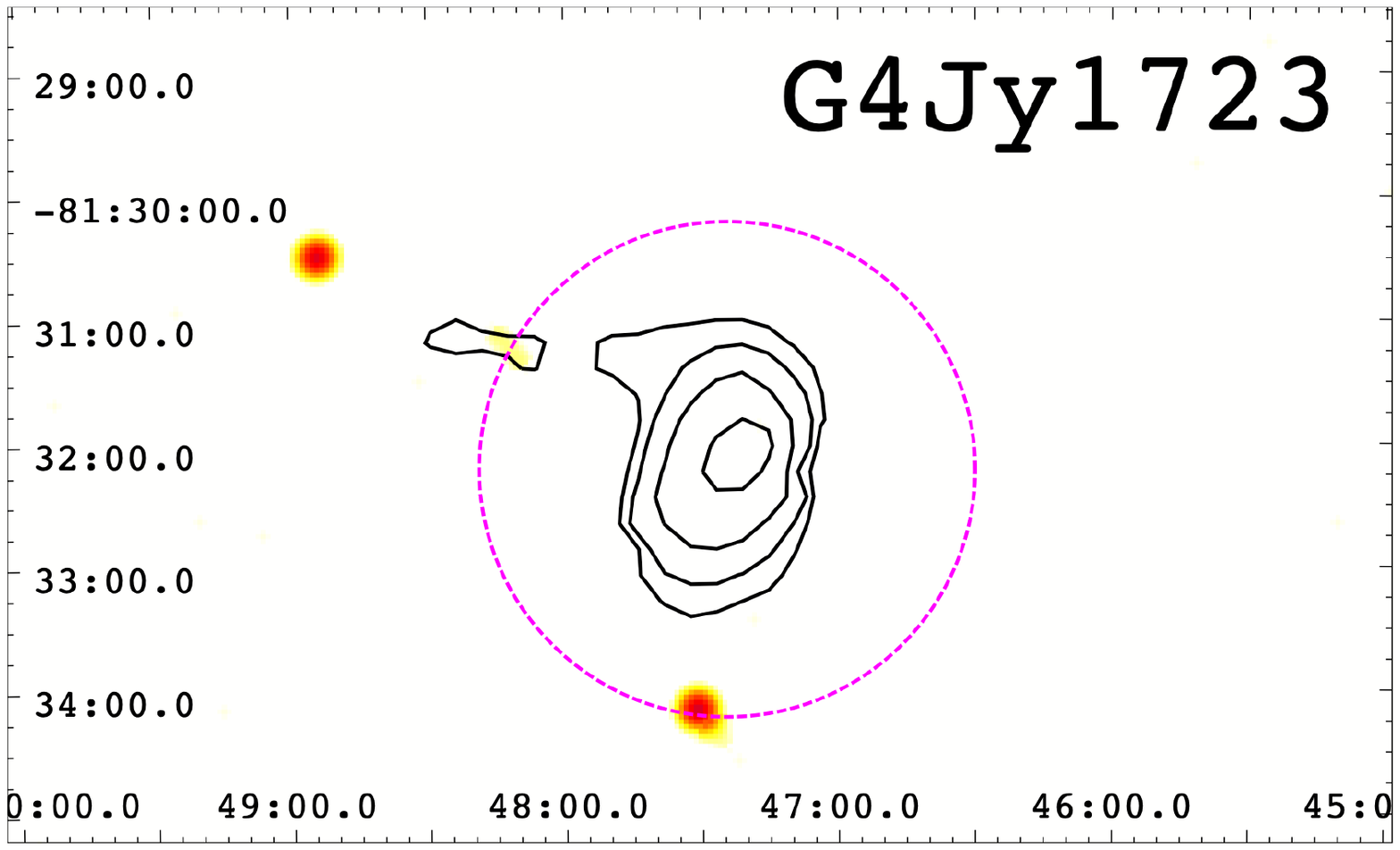}
\includegraphics[width=3.8cm,height=6.4cm,angle=-90]{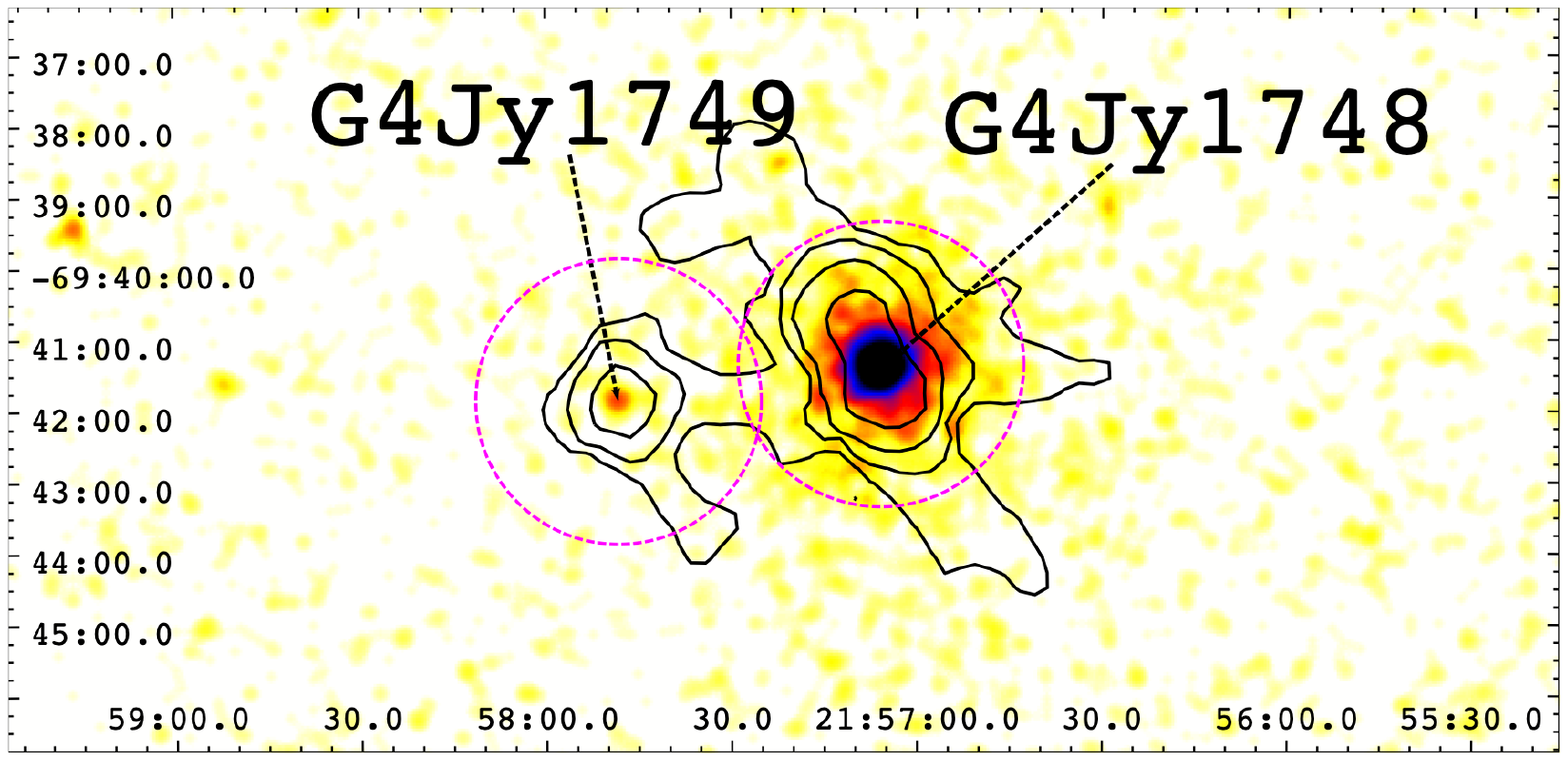}
\caption{Same as Figure~\ref{fig:example} for the following \cs\ radio sources:
G4Jy\,1498, G4Jy\,1518, G4Jy\,1605, G4Jy\,1613, G4Jy\,1635, G4Jy\,1640, G4Jy\,1664, G4Jy\,1671, G4Jy\,1708, G4Jy\,1717, G4Jy\,1723, G4Jy\,1748 \& 1749.}
\end{center}
\end{figure*}

\begin{figure*}[!th]
\begin{center}
\includegraphics[width=3.8cm,height=6.4cm,angle=-90]{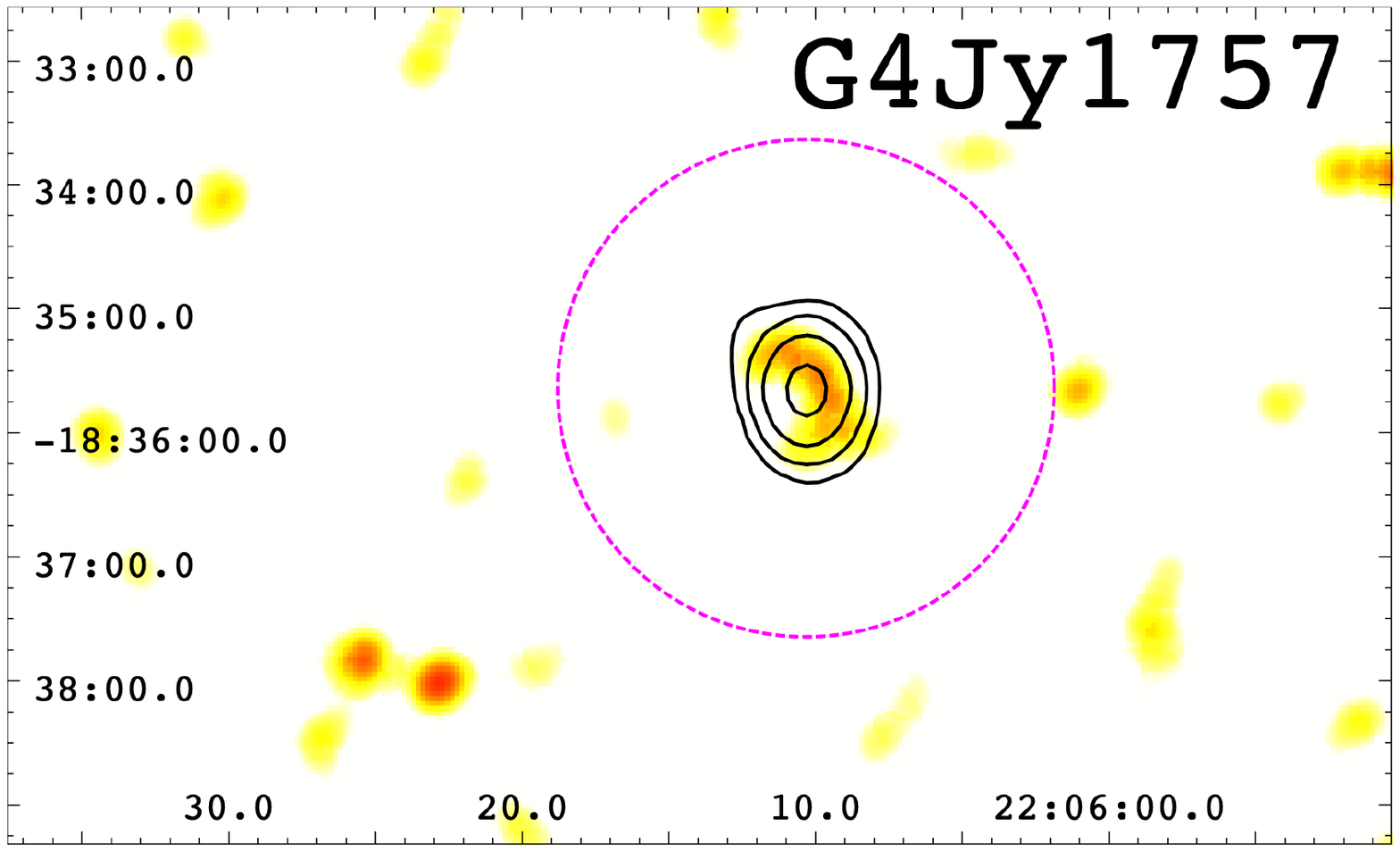}
\includegraphics[width=3.8cm,height=6.4cm,angle=-90]{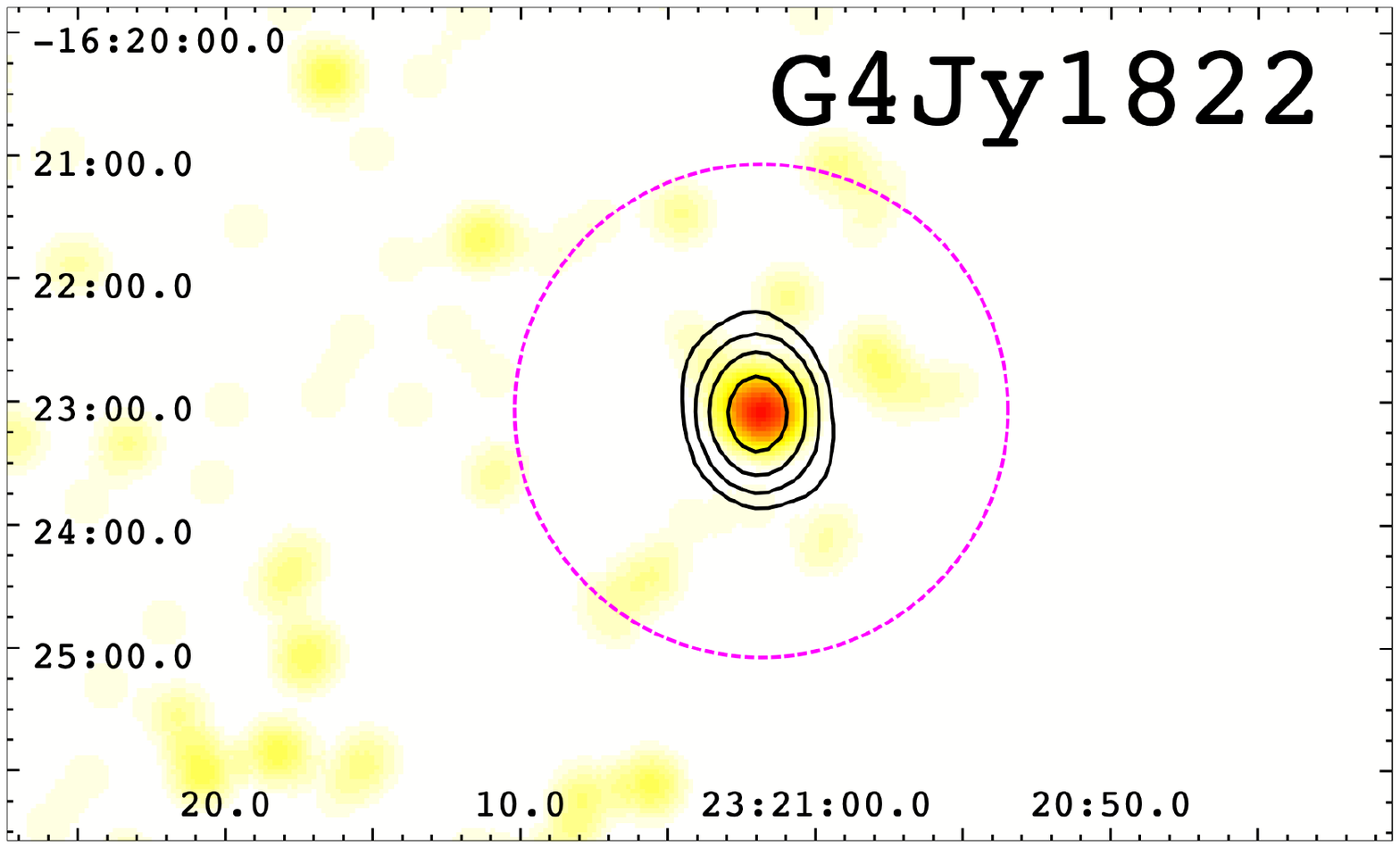}
\includegraphics[width=3.8cm,height=6.4cm,angle=-90]{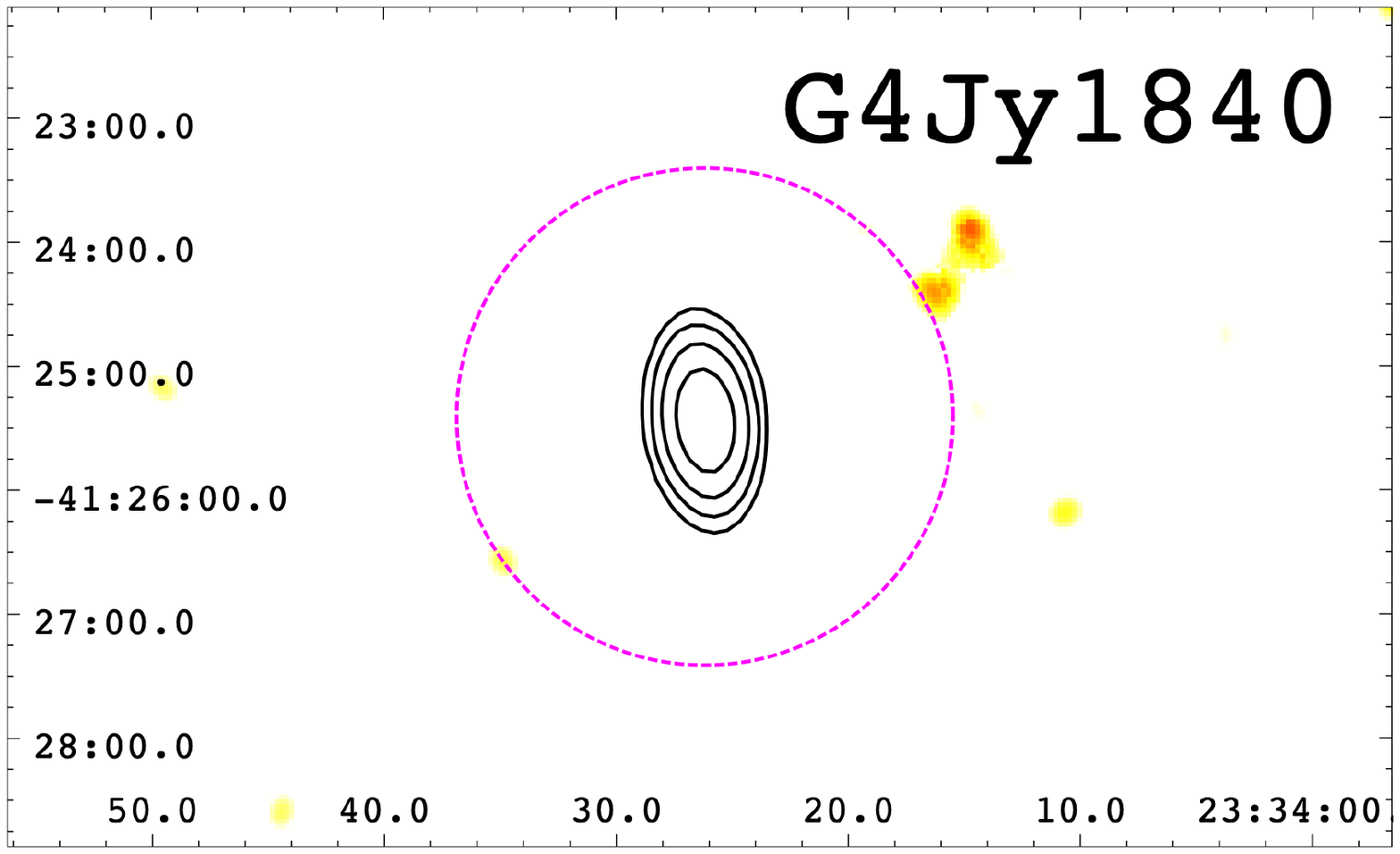}
\includegraphics[width=3.8cm,height=6.4cm,angle=-90]{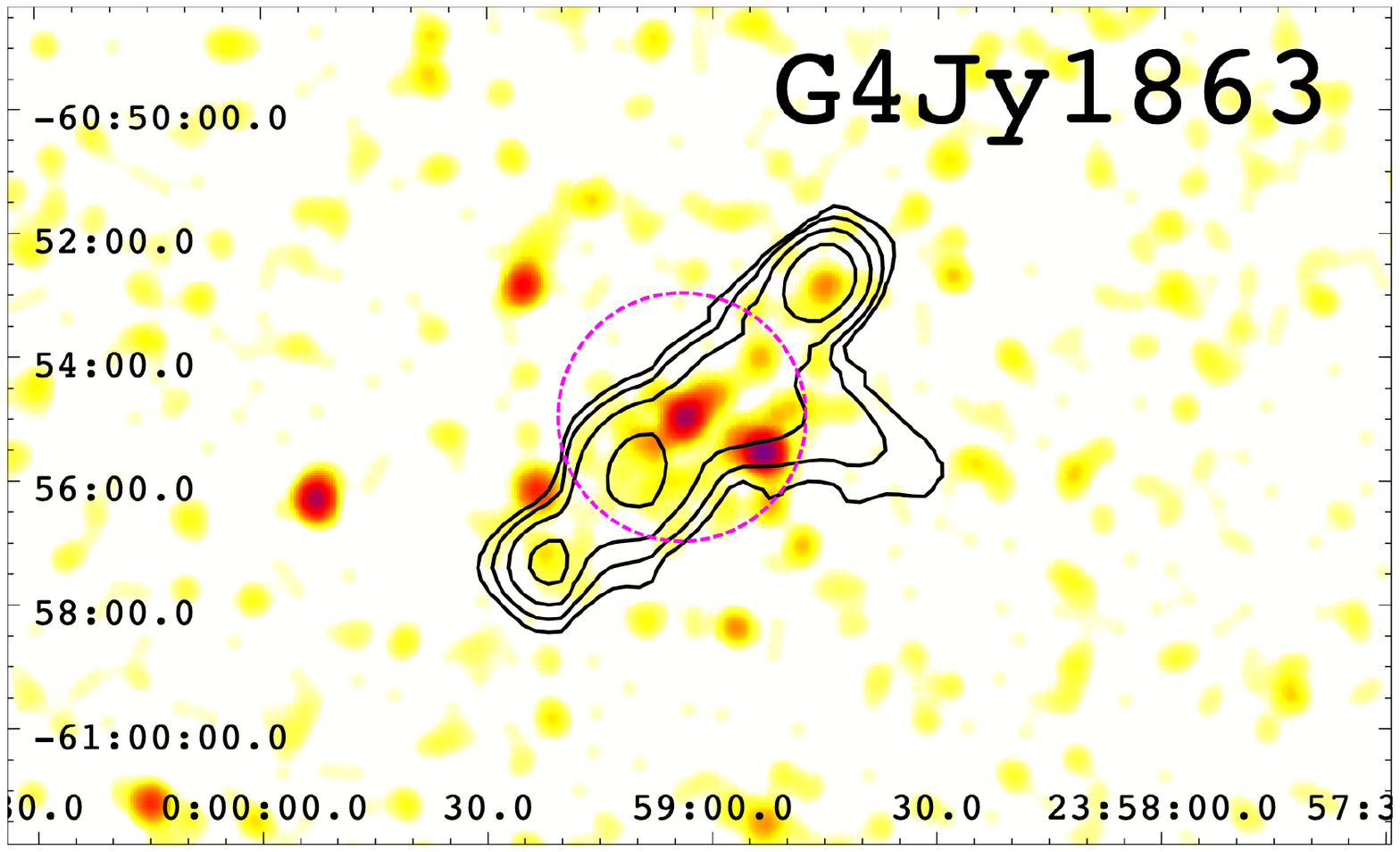}
\caption{Same as Figure~\ref{fig:example} for the following \cs\ radio sources: 
G4Jy\,1757, G4Jy\,1822, G4Jy\,1840 and G4Jy\,1863.}
\end{center}
\end{figure*}

\newpage
\startlongtable
\begin{deluxetable}{rrrc}
\tabletypesize{\scriptsize}
\rotate
\tablecaption{Parameters for radio contours overlaid on X-ray images.}
\label{tab:radcon}
\small
\tablehead{\colhead{G4Jy} & \colhead{$\nu$} & \colhead{$l_{min}$} & \colhead{binning} \\ 
\colhead{name} & \colhead{(MHz)} & \colhead{(Jy/beam)} & \colhead{factor}} 
\startdata
  20 &  150 & 0.1   & x4 \\
  27 &  843 & 0.3   & x2 \\
  77 &  150 & 0.05  & x2 \\
  78 &  150 & 0.3   & x3 \\
  85 &  150 & 0.3   & x3 \\
  93 &  150 & 0.3   & x3 \\
 120 &  150 & 0.1   & x3 \\
 162 &  150 & 0.1   & x4 \\
 168 &  150 & 0.025 & x4 \\
 171 &  150 & 0.05  & x3 \\
 213 & 1400 & 0.4   & x2 \\
 217 &  843 & 0.1   & x3 \\
 247 &  150 & 0.1   & x3 \\
 249 &  843 & 0.3   & x2 \\
 257 &  843 & 0.2   & x2 \\
 260 &  150 & 0.1   & x4 \\
 290 &  150 & 1.6   & x2 \\
 293 &  843 & 0.1   & x2 \\
 843 &  843 & 0.3   & x3 \\
 373 & 3000 & 0.002 & x3 \\
 392 & 3000 & 0.01  & x3 \\
 404 &  150 & 0.1   & x4 \\
 411 &  150 & 0.1   & x4 \\
 415 &  150 & 0.1   & x4 \\
 416 &  843 & 1.6   & x2 \\
 427 &  843 & 0.3   & x3 \\
 446 &  843 & 0.6   & x2 \\
 464 &  150 & 0.1   & x4 \\
 492 & 3000 & 0.005 & x2 \\
 506 & 3000 & 0.002 & x2 \\
 510 &  150 & 0.1   & x3 \\
 518 &  843 & 0.3   & x2 \\
 540 &  150 & 0.1   & x6 \\
 563 &  150 & 0.1   & x4 \\
 580 &  150 & 0.1   & x4 \\
 613 &  843 & 0.15  & x2 \\
 614 &  150 & 0.1   & x3 \\
 618 &  843 & 0.6   & x2 \\
 619 &  150 & 0.1   & x3 \\
 644 & 1400 & 0.005 & x3 \\
 651 &  150 & 0.1   & x2 \\
 672 &  843 & 0.1   & x4 \\
 718 &  843 & 0.2   & x2 \\
 721 &  150 & 0.1   & x4 \\
 723 &  843 & 0.1   & x3 \\
 835 &  150 & 0.1   & x4 \\
 836 &  150 & 0.1   & x4 \\
 854 &  150 & 0.1   & x4 \\
 876 &  150 & 0.05  & x4 \\
 917 & 3000 & 0.006 & x2 \\
 927 &  150 & 0.1   & x4 \\
 933 &  150 & 0.1   & x4 \\
 950 &  150 & 0.1   & x4 \\
1034 & 3000 & 0.002 & x2 \\
1071 &  150 & 0.1   & x4 \\
1079 &  150 & 0.1   & x4 \\
1080 & 1400 & 0.04  & x2 \\
1135 &  150 & 0.1   & x2 \\
1136 &  150 & 0.1   & x4 \\
1145 &  150 & 0.2   & x2 \\
1148 &  150 & 0.2   & x3 \\
1158 &  150 & 0.1   & x4 \\
1192 &  150 & 0.1   & x4 \\
1203 & 1400 & 0.001 & x4 \\
1225 &  150 & 0.1   & x4 \\
1279 &  150 & 0.02  & x3 \\
1411 &  843 & 0.2   & x2 \\
1423 &  843 & 0.1   & x4 \\
1432 &  843 & 0.16  & x2 \\
1472 &  843 & 0.16  & x2 \\
1477 &  843 & 0.1   & x3 \\
1498 &  150 & 0.1   & x3 \\
1518 &  150 & 0.02  & x4 \\
1605 &  843 & 0.01  & x2 \\
1613 &  843 & 0.01  & x3 \\
1635 & 3000 & 0.001 & x3 \\
1640 &  150 & 1.6   & x2 \\
1664 &  150 & 0.1   & x3 \\
1671 &  150 & 0.1   & x2 \\
1708 &  150 & 0.1   & x4 \\
1717 &  150 & 0.3   & x3 \\
1723 &  843 & 0.1   & x3 \\
1748 &  843 & 0.1   & x4 \\
1757 &  150 & 0.1   & x4 \\
1822 &  150 & 0.1   & x4 \\
1840 &  150 & 0.16  & x2 \\
1863 &  843 & 0.1   & x3 \\ 
\enddata
\tablecomments{
Col. (1) the G4Jy name of the radio source, also adopted in the \cs\ catalog; col. (2) frequency of the radio map used to draw contours in MHz; col. (3) minimum level of the radio contours $l_{min}$; col. (4) binning factor for the radio contours.
}
\end{deluxetable}

\newpage
\section{\swf\ - \chn\ comparison}
\label{app:xraycompare}
Four \cs\ radio sources, in the selected sample, have X-ray observations in the \chn\ archive that are still unpublished but publicly available. Thus we retrieved those datasets and we compare \swf-XRT and \chn\ event files. There are other \cs\ radio sources observed with \chn\ but these datasets were all already discussed in the literature (see e.g., individual notes for the \cs\ catalog in the Appendix B of paper I for more details).

Data reduction procedures adopted for \chn\ observations are extensively described in our previous papers \citep[see e.g.,][and references therein for more details]{massaro09a,massaro09b,massaro11a,jimenez21}. Here we followed the standard procedure described in the \chn\ Interactive Analysis of Observations (CIAO) threads\footnote{\underline{http://cxc.harvard.edu/ciao/guides/index.html}} and we used CIAO v4.14 with the \chn\ Calibration Database (CALDB) version 4.9.6. to generate level 2 event files. These were obtained using the $acis\_process\_events$ task of $chandra\_repro$ script and events were filtered for grades 0,2,3,4,6.  Light curves were also extracted and inspected for every dataset to verify the absence of high background intervals. Such time intervals were never found in the selected observations. The four radio sources analyzed here, namely, G4Jy\,171, G4Jy\,260, G4Jy\,411 and G4Jy\,1613, were observed with ACIS-S detector in VFAINT mode. In particular G4Jy\,171 was observed twice (obs.ID number 16099 and 17577, respectively) but we only reduced the longer observation ($25\,\mathrm{ksec}$ nominal $T_{exp}$) carried out on 2015-02-02, while G4Jy\,260 (obs.ID 21488, 2020-10-17, $10\,\mathrm{ksec}$), G4Jy\,411 (obs.ID 23098, 2019-12-12, $10\,\mathrm{ksec}$) and G4Jy\,1613 (obs.ID 23177, 2020-02-25, $10\,\mathrm{ksec}$) where all pointed one time with \chn\, the last two being selected as cool attitude targets\footnote{https://cxc.harvard.edu/proposer/CCTs.html}.

\begin{figure*}[!th]
\begin{center}
\includegraphics[width=7.cm,height=8.4cm,angle=-90]{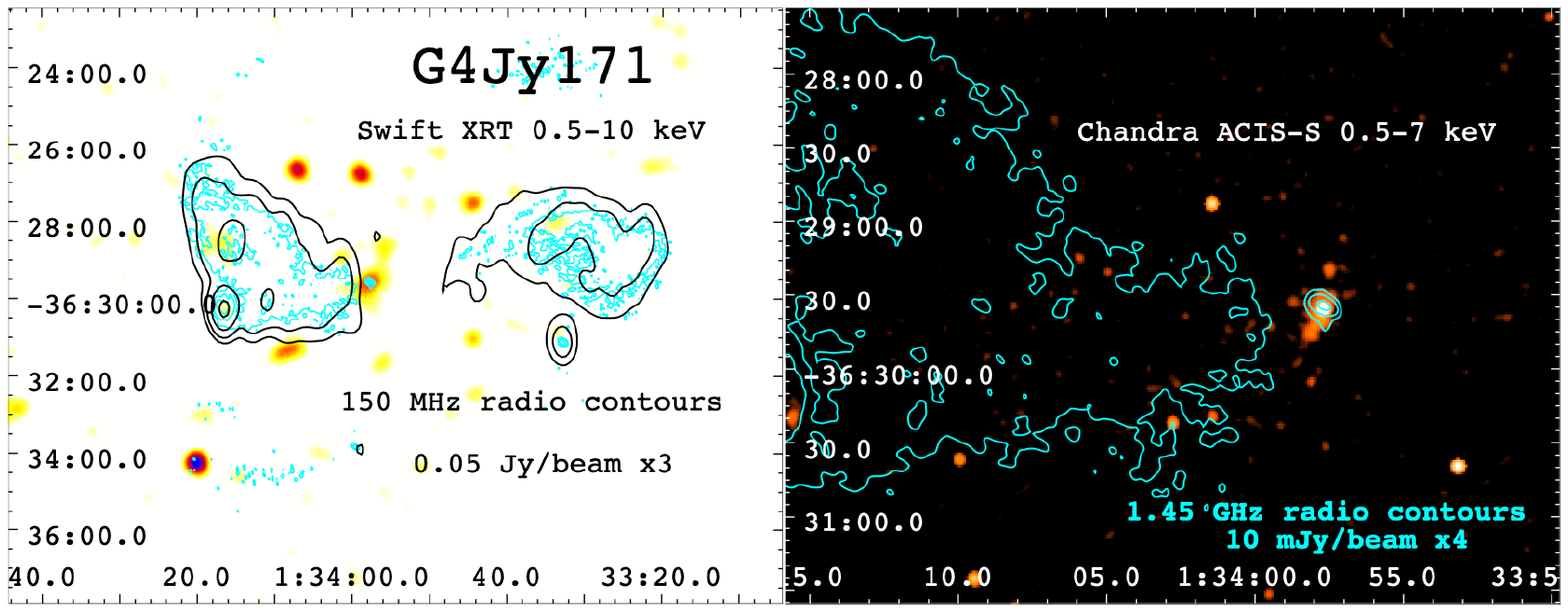}
\includegraphics[width=7.cm,height=8.4cm,angle=-90]{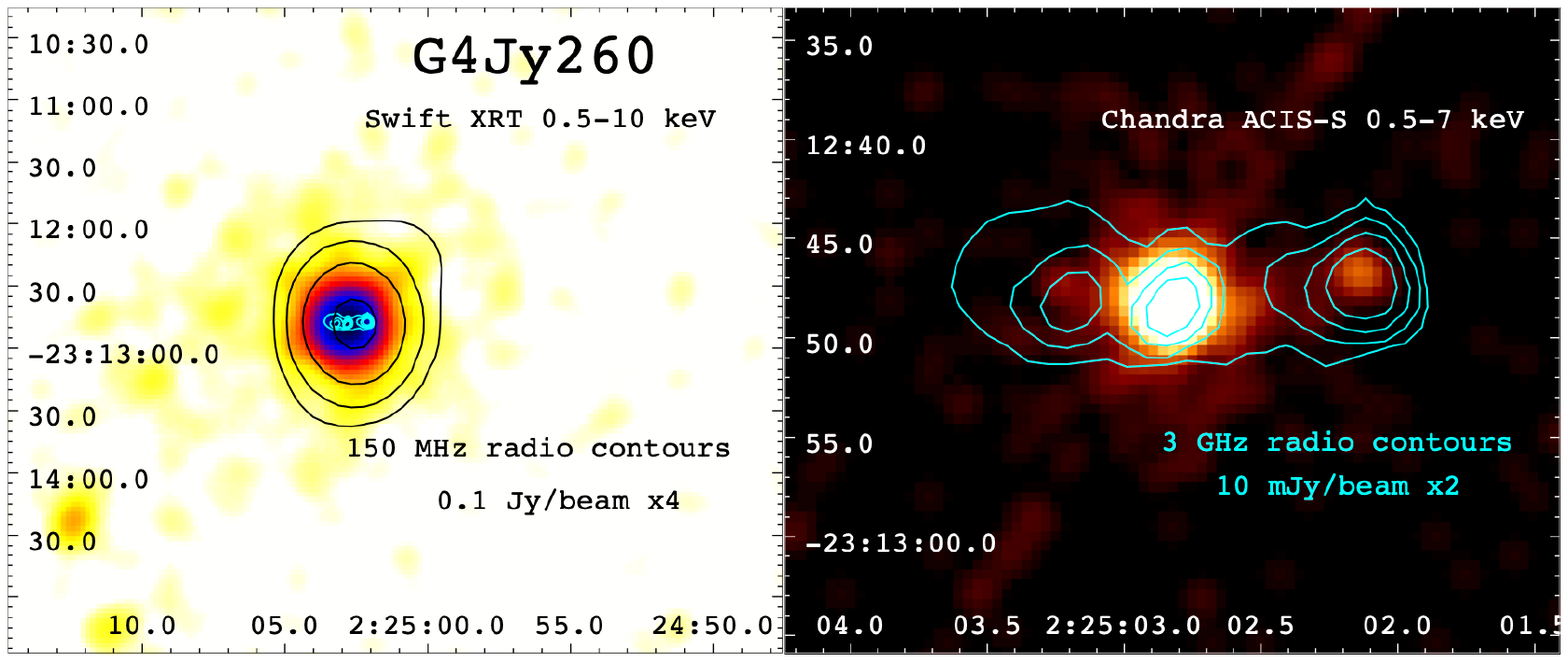}
\includegraphics[width=7.cm,height=8.4cm,angle=-90]{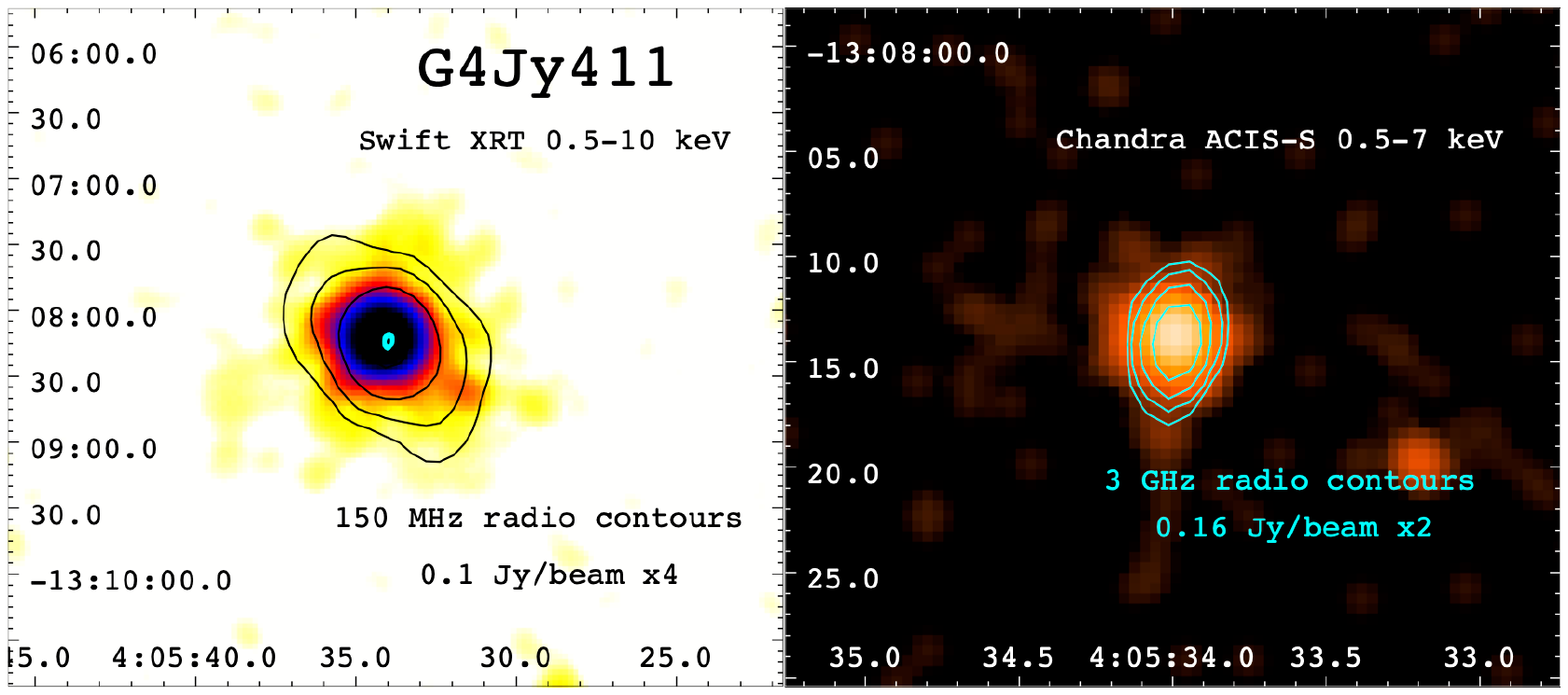}
\includegraphics[width=7.cm,height=8.4cm,angle=-90]{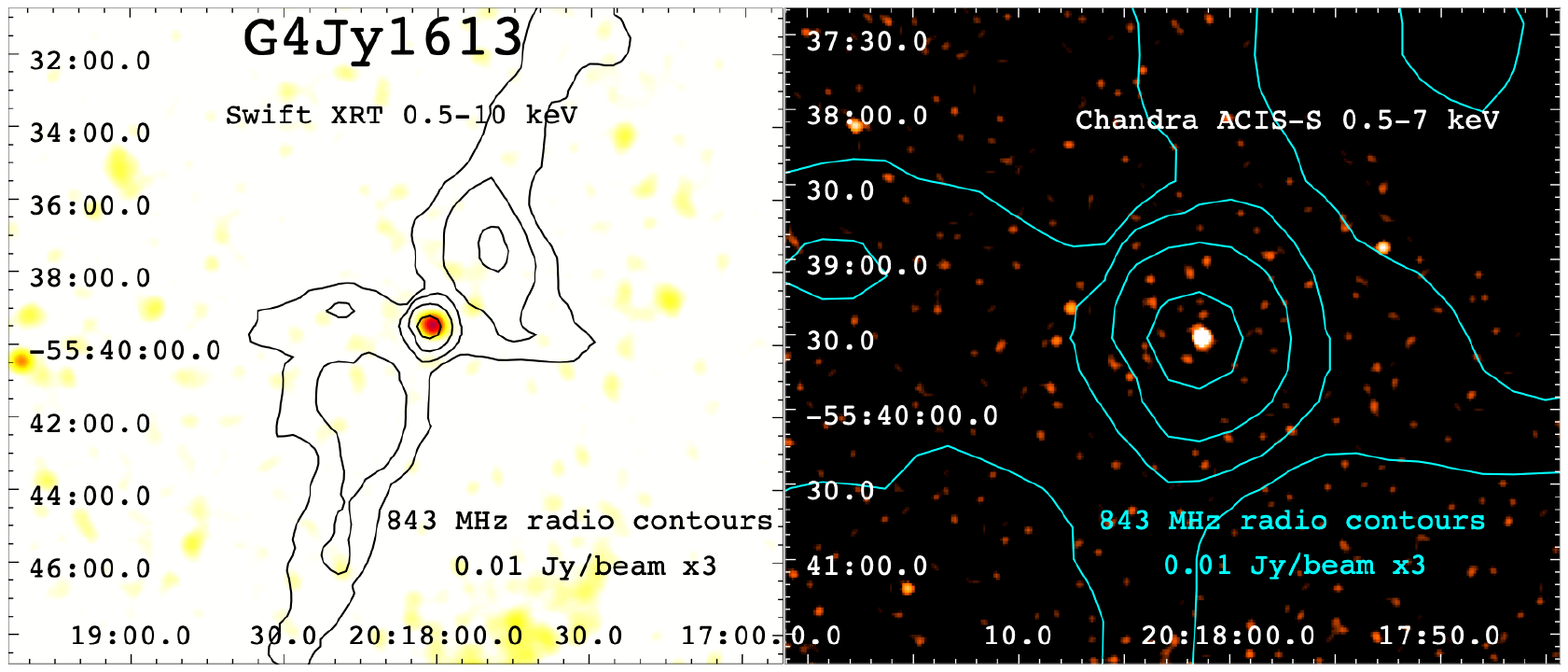}
\caption{{\it Left panels)} \swf-XRT images, as shown in Appendix A. {\it Right panels)} \chn\ images obtained from event files restricted in the 0.5-7\,keV energy range. Radio contours at different frequencies are overlaid in black on the \swf\ images and in cyan on both panels. The frequency of the radio maps from which radio contours were drawn are reported in each panel together with the intensity of the first level and the a binning factor.}
\end{center}
\label{fig:chandra}
\end{figure*}

In Figure~\ref{fig:chandra} we show the comparison between the \swf-XRT event file, as shown in Figure~\ref{fig:example}, in comparison with the \chn\ level 2 event file restricted in the 0.5-7 keV\,band. We found that the \chn\ X-ray image of G4Jy\,260 clearly show the readout streak (north-south direction) due to the presence of pileup \citep[see e.g.,][and references therein]{davis01,mccollough05} but we discovered that the radio hotspot, located on the western side of the nucleus, has an X-ray counterpart. Then we detected the X-ray counterpart of all radio cores for these four \cs\ sources thus confirming the \swf-XRT results. Finally, no clear signatures of extended X-ray emission was found in these \chn\ datasets with the only exception of a marginal detection of diffuse X-ray radiation around G4\,Jy171.


\begin{thebibliography}{} 
\bibitem[Abbott et al. (2018)]{abbott18} Abbott, T. M. C., Abdalla, F. B., Allam, S. et al. 2018 ApJS, 239, 25 
\bibitem[Abell (1958)]{abell58} Abell, George O. 1958 ApJS, 3, 211 
\bibitem[Abell et al. (1989)]{abell89} Abell, G. O., Corwin, H. G. Jr., Olowin, O. P. 1989 ApJS, 70, 1 
\bibitem[Abdollahi et al. (2020)]{abdollahi20} Abdollahi, S.; Acero, F.; Ackermann, M.; Ajello, M.; Atwood et al. 2020 ApJS ,247, 33 
\bibitem[Ahn et al. (2012)]{ahn12} Ahn, C. P., Alexandroff, R., Allende Prieto, C., Anderson, Scott F., Anderton, T. et al. 2012 ApJS, 203, 21 
\bibitem[Aldinucci et al. (2017)]{aldinucci17} M. Aldinucci, S. Bagnasco, S. Lusso, P. Pasteris, S. Rabellino, and S. Vallero. OCCAM: a flexible, multi-purpose and extendable HPC cluster. Journal of Physics: Conference Series, 898(8):082039, 2017. 
\bibitem[Bacon et al. (2010)]{bacon10} Bacon, R., Accardo, M., Adjali, L. et al. 2010 in Society of Photo-Optical Instrumentation Engineers (SPIE) Conference Series, Vol. 7735, Proc. SPIE, 773508 
\bibitem[Bagchi et al. (1998)]{bagchi98} Bagchi, J., Pislar, V., Lima Neto, G. B. 1998 MNRAS, 296L, 23 
\bibitem[Baldi et al. (2010)]{baldi10} Baldi, R. D., Chiaberge, M., Capetti, A. et al. 2010 ApJ, 725, 2426 
\bibitem[Baldwin (1975)]{baldwin75} Baldwin A. J. 1975 ApJ, 201, 26 
\bibitem[Balmaverde et al. (2012)]{balmaverde12} Balmaverde, B., Capetti, A., Grandi, P., Torresi, E., Chiaberge, M. et al. 2012 A\&A, 545A, 143 
\bibitem[Balmaverde et al. (2018a)]{balmaverde18a} Balmaverde, B., Capetti, A., Marconi, A., Venturi, G. 2018 A\&A, 612, A19 
\bibitem[Balmaverde et al. (2018b)]{balmaverde18b} Balmaverde, B., Capetti, A., Marconi, A., Venturi, G., Chiaberge, M. et al. 2018 A\&A, 619, 83 
\bibitem[Balmaverde et al. (2019)]{balmaverde19} Balmaverde, B., Capetti, A., Marconi, A., Venturi, G., Chiaberge, M. et al. 2019 A\&A, 632, 124 
\bibitem[Balmaverde et al. (2021)]{balmaverde21} Balmaverde, B., Capetti, A., Marconi, A., Venturi, G., Chiaberge, M. et al. 2021 A\&A, 645, 12 
\bibitem[Balmaverde et al. (2022)]{balmaverde22} Balmaverde, B., Capetti, A., Baldi, R. D., Baum, S., Chiaberge, M. et al. 2022 A\&A, 662, 23 
\bibitem[Begelman et al. (1984)]{begelman84} Begelman, M. C., Blandford, R. D. \& Rees, M. J. 1984, Reviews of Modern Physics, 56, 255 
\bibitem[Bergeron \& Kunth (1984)]{begeron84} Bergeron J., Kunth D., 1984, MNRAS, 207, 263 
\bibitem[Ben Bekhti et al. (2016)]{hi4pi} HI4PI Collaboration, N. Ben Bekhti, L. Floer, et al., 2016, A\&A, 594, A116 
\bibitem[Bennett (1962)]{bennett62} Bennett, A. S. 1962 MmRAS, 68, 163 
\bibitem[Bennett et al. (2014)]{bennett14} Bennett, C. L., Larson, D., Weiland, J. L., Hinshaw, G. 2014 ApJ 794, 135 
\bibitem[Best et al. (1999)]{best99} Best, P. N., R\''{o}ttgering, H. J. A., Lehnert, M. D. 1999 MNRAS 310, 223 
\bibitem[Best et al. (2000)]{best00} Best, P. N., R\''{o}ttgering, H. J. A., Lehnert, M. D. 2000 MNRAS, 315, 21
\bibitem[Birkinshaw et al. (2002)]{birkinshaw02} Birkinshaw, M., Worrall, D.M. \& Hardcastle, M.J. 2002 MNRAS, 335, 142 
\bibitem[Borne \& Hoessel (1984)]{borne84} Borne, K. D.; Hoessel, J. G. 1984 BAAS, 16, 881 
\bibitem[Browne \& Savage (1977)]{browne77} Browne, L. W. A.; Savage, Ann 1977 MNRAS, 179, 65 
\bibitem[Bruni et al. (2020)]{bruni20} Bruni, G.; Panessa, F.; Bassani, L.; Dallacasa, D.; Venturi, et al. 2020 MNRAS, 494, 902 
\bibitem[Burbidge \& Kinman (1966)]{burbidge66} Burbidge, E. M., Kinman, T. D. 1966 ApJ, 145, 654 
\bibitem[Burbidge (1967)]{burbidge67} Burbidge, E. M. 1967 ApJ, 149L, 51 
\bibitem[Burbidge \& Burbidge (1972)]{burbidge72} Burbidge, E. M., Burbidge, G. R. 1972 ApJ, 172, 37 
\bibitem[Burgess \& Hunstead (2006a)]{burgess06a} Burgess, A. M. \& Hunstead, R. W. 2006a AJ, 131, 100 
\bibitem[Burgess \& Hunstead (2006b)]{burgess06b} Burgess, A. M. \& Hunstead, R. W. 2006b AJ, 131, 114 
\bibitem[Burns et al. (1981)]{burns81} Burns, J. O. 1981, MNRAS, 195, 523 
\bibitem[Burrows et al. (2000)]{burrows00} Burrows, David N., Hill, Joanne E., Nousek, John A., Wells, Alan A., Short, Alexander D., Willingale, Richard, Citterio, Oberto, Chincarini, G., Tagliaferri, G. 2000 SPIE, 4140, 64 
\bibitem[Burrows et al. (2005)]{burrows05} Burrows, D., Hill, J. E., Nousek, J. A., et al. 2005, SSRv, 120, 165 
\bibitem[Buttiglione et al. (2009)]{buttiglione09} Buttiglione, S., Capetti, A., Celotti, A., Axon, D.J., Chiaberge, M., Macchetto, F.D., Sparks, W.B., 2009 A\&A 495, 1033 
\bibitem[Buttiglione et al. (2010)]{buttiglione10} Buttiglione, S., Capetti, A., Celotti, A., Axon, D.J., Chiaberge, M., Macchetto, F.D., Sparks, W.B., 2010 A\&A 509, A6 
\bibitem[Buttiglione et al. (2011)]{buttiglione11} Buttiglione, S., Capetti, A., Celotti, A., Axon, D. J., Chiaberge, M., Macchetto, F. D., Sparks, W. B. 2011 A\&A, 525A, 28 
\bibitem[Capalbi et al. (2005)]{capalbi05} Capalbi, M., Perri, M., Saija, B. \& Tamburelli, F. 2005, ASI Science Data Center, 1 
\bibitem[Carter et al. (1983)]{carter83} Carter, D.; Malin, D. F. 1983 MNRAS, 203, 49 
\bibitem[Cava et al. (2009)]{cava09} Cava, A.; Bettoni, D.; Poggianti, B. M.; Couch, W. J.; Moles, M. et al 2009 A\&A, 495, 707 
\bibitem[Chiaberge et al. (2000)]{chiaberge00} Chiaberge, M., Capetti, A. \& Celotti, A. 2000 A \& A, 355, 873 
\bibitem[Chiaberge et al. (2015)]{chiaberge15} Chiaberge, M., Gilli, R., Lotz, J. M., Norman, C. 2015 ApJ, 806, 147 
\bibitem[Cohen et al. (2007)]{cohen07} Cohen, A. S., Lane, W. M., Cotton, W. D. et al. 2007 AJ, 134, 1245 
\bibitem[Collmar et al. (2010)]{collmar10} Collmar, W.,  B\''{o}ttcher, M.,  Krichbaum, T. P., Agudo, I., Bottacini, E. et al. 2010 A\&A, 522, 66 
\bibitem[Condon et al. (1998)]{condon98} Condon, J. J., Cotton, W. D., Greisen, E. W. et al. 1998 AJ, 115, 1693 
\bibitem[Condon et al. (2021)]{condon21} Condon, J. J., Cotton, W. D., White, S. V., Legodi, S., Goedhart, S., McAlpine, K., Ratcliffe, S. M., Camilo, F. 2021 ApJ, 917, 18 
\bibitem[Croston et al. (2005)]{croston05} Croston, J. H. et al. 2005 ApJ, 626, 733 
\bibitem[Cusumano et al. (2010)]{cusumano10} Cusumano, G., La Parola, V., Segreto, A., Ferrigno, C., Maselli, A. et al. 2010 A\&A, 524, 64 
\bibitem[D’Elia et al. (2013)]{delia13} D’Elia, V., Perri, M., Puccetti, S., et al. 2013, A\&A, 551, A142 
\bibitem[Dai et al. (2015)]{dai15} Dai, Xinyu, Griffin, Rhiannon D.,  Kochanek, Christopher S.,  Nugent, Jenna M., Bregman, Joel N. 2015 ApJS, 218, 8 
\bibitem[Danziger \& Goss (1983)]{danziger83} Danziger, I. J., Goss, W. M. 1983 MNRAS, 202, 703 
\bibitem[Davis (2001)]{davis01} Davis, J. E. 2001 ApJ, 562, 575 
\bibitem[de Koff et al. (1996)]{dekoff96} de Koff, Sigrid, Baum, Stefi A., Sparks, William B., Biretta, John, Golombek, Daniel et al. 1996 ApJS, 107, 621 
\bibitem[Dicken et al. (2014)]{dicken14} Dicken, D., Tadhunter, C., Morganti, R. et al. 2014 ApJ, 788, 98 
\bibitem[di Serego et al. (1994)]{diserego94} di Serego Alighieri, S., Danziger, I. J., Morganti, R., Tadhunter, C. N. 1994 MNRAS 269, 998  
\bibitem[Djorgovski et al. (1988)]{djorgovski88} Djorgovski, S., Spinrad, H., McCarthy, P., Dickinson, M., van Breugel, W. et al. 1988 AJ, 96, 836 
\bibitem[Ebeling et al. (1996)]{ebeling96} Ebeling, H.; Voges, W.; Bohringer, H.; Edge, A. C.; Huchra, J. P.; Briel, U. G. 1996 MNRAS, 281, 799 
\bibitem[Edge et al. (1959)]{edge59} Edge, D. O., Shakeshaft, J. R., McAdam, W. B., Baldwin, J. E., Archer, S. 1959 MmRAS, 68, 37 
\bibitem[Efstathiou et al. (1980)]{efstathiou80} Efstathiou, G.; Ellis, R. S.; Carter, D. 1980 MNRAS, 193, 931 
\bibitem[Ellingson et al. (1987)]{ellingson87} Ellingson, E.; Green, R. F.; Yee, H. K. C. 1987 BAAS, 19, 685 
\bibitem[Ellingson (1988)]{ellingson88} Ellingson, E. 1988 BAAS, 20, 1025 
\bibitem[Ellingson et al. (1989)]{ellingson89} Ellingson, E.; Yee, H. K. C.; Green, Richard F.; Kinman, T. D. 1989 AJ, 97, 1539 
\bibitem[Ellingson et al. (1991)]{ellingson91}  Ellingson, E.; Green, R. F.; Yee, H. K. C. 1991ApJ, 378, 476 
\bibitem[Evans et al. (2006)]{evans06} Evans, D. A., Worrall, D. M., Hardcastle, M. J. et al. 2006 ApJ, 642, 3796 
\bibitem[Evans et al. (2014)]{evans14} Evans, P. A., Osborne, J. P. ;  Beardmore, A. P., Page, K. L., Willingale, R. et al. 2014 ApJS, 210, 8 
\bibitem[Evans et al. (2020)]{evans20} Evans, P. A., Page, K. L., Osborne, J. P., Beardmore, A. P.,  Willingale, R. et al. 2020 ApJS, 247, 54 
\bibitem[Fabian (2012)]{fabian12} Fabian, A. C. 2012, ARA\&A, 50, 455 
\bibitem[Flewelling et al. (2020)]{flewelling20} Flewelling, H. A., Magnier, E. A., Chambers, K. C. et al. 2020, ApJS, 251, 7 
\bibitem[Garc\'ia-P\'erez et al. (2022)]{garcia22} Garc\'ia-P\'erez, A., Jimenez-Gallardo, A., Pe\~na-Herazo, H. A., F. Massaro, White, S. V. et al. 2022 ApJS in prep. 
\bibitem[Gehrels et al. (2004)]{gehrels04} Gehrels, N., Chincarini, G., Giommi, P., Mason, K. O., Nousek, J. A. et al. 2004 ApJ, 611, 1005 
\bibitem[Giommi et al. (1992)]{giommi92} Giommi, P., Angelini, L., Jacobs, P., Tagliaferri, G. 1992 ASPC, 25, 100 
\bibitem[Giovannini et al. (2005)]{giovannini05} Giovannini, G., Taylor, G. B., Feretti, L., Cotton, W. D., Lara, L., Venturi, T. 2005, ApJ 618, 635 
\bibitem[Grossov\'a et al. (2019)]{grossova19} Grossov\'a, R.; Werner, N.; Rajpurohit, K.; Mernier, F.; Lakhchaura, K.; Gab\'anyi, K. et al. 2019 MNRAS, 488, 1917
\bibitem[Grossov\'a et al. (2022)]{grossova22} Grossov\'a, Romana; Werner, Norbert; Massaro, Francesco; Lakhchaura, Kiran et al. 2022 ApJS, 258, 30
\bibitem[Hardcastle \& Worrall (2000)]{hardcastle00} Hardcastle, M. J., Worrall, D. M. 2000 MNRAS, 314, 359 
\bibitem[Hardcastle et al. (2002)]{hardcastle02} Hardcastle, M. J. et al. 2002 ApJ, 581, 948 
\bibitem[Hardcastle et al. (2006)]{hardcastle06} Hardcastle, M. J., Evans, D. A., Croston, J. H. 2006 MNRAS, 370, 1893 
\bibitem[Hardcastle \& Croston (2020)]{hardcastle20} Hardcastle, M. J. \& Croston J. H. 2020 NewAR, 8801539 
\bibitem[Harris \& Grindlay (1979)]{harris79} Harris, D. E. \& Grindlay, J. E. 1979 MNRAS, 188, 25 
\bibitem[Harris \& Krawczynski (2006)]{harris06} Harris, D. E. \& Krawczynski, H. 2006 ARA\&A, 44, 463 
\bibitem[Harvaneck et al. (2001)]{harvaneck01} Harvanek, M., Ellingson, E., Stocke, J. T. et al. 2001 AJ, 122, 2874 
\bibitem[Hayashida et al. (2015)]{hayashida15} Hayashida, M., Nalewajko, K., Madejski, G. M., Sikora, M., Itoh, R. et al. 2015 ApJ, 807, 79 
\bibitem[Hilbert et al. (2016)]{hilbert16} Hilbert, B., Chiaberge, M., Kotyla, J. P., Tremblay, G. R., Stanghellini, C. et al. 2016 ApJS, 225, 12 
\bibitem[Hilker et al. (1999)]{hilker99} Hilker, M., Infante, L., Vieira, G., Kissler-Patig, M., Richtler, T. 1999 A\&AS, 134, 75
\bibitem[Hill et al. (2004)]{hill04} Hill, Joanne E., Burrows, David N., Nousek, John A., Abbey, Anthony F., Ambrosi, Richard M. et al. 2004SPIE.5165..217 
\bibitem[Hiltner \& Roeser (1991)]{hiltner91} Hiltner, P. R. \& Roeser, H. J. 1991 A \& A, 244, 37 
\bibitem[Hintzen et al. (1983)]{hintzen83} Hintzen, P.; Ulvestad, J.; Owen, F. 1983 AJ, 88, 709 
\bibitem[Ho \& Kim (2009)]{ho09} Ho, Luis C., Kim, Minjin 2009 ApJS 184, 398 
\bibitem[Hoopes et al. (1996)]{hoopes96} Hoopes, Charles G., Walterbos, Rene A. M., Greenwalt, Bruce E. 1996 AJ 112, 1429 
\bibitem[Hoyle (1965)]{hoyle65} Hoyle, F. 1965 Natur, 208, 111 
\bibitem[Hunstead et al. (1978)]{hunstead78} Hunstead, R. W., Murdoch, H. S., Shobbrook, R. R. 1978 MNRAS, 185, 149
\bibitem[Hurley-Walker et al. (2017)]{hurley17} Hurley-Walker N., Callingham J.R., Hancock P.J., Franzen T.M.O., Hindson L., Kapinska A.D., Morgan J., et al. 2017 MNRAS, 464, 1146 
\bibitem[Hutchings et al. (1996)]{hutchings96} Hutchings, J. B., Gower, A. C., Ryneveld, S., Dewey, A. 1996 AJ, 111, 2167 
\bibitem[Kempner et al. (2004)]{kempner04} Kempner J. C., Blanton E. L., Clarke T. E., Enßlin T. A., Johnston-Hollitt M., Rudnick L., 2004, in Reiprich T., Kempner J., Soker N., eds, The Riddle of Cooling Flows in Galaxies and Clusters of galaxies. p. 335 (arXiv:astro-ph/0310263) 
\bibitem[Kenworthy et al. (2018)]{kenworthy18} Kenworthy, Matthew A., Snik, Frans, Keller, Christoph U., Doelman, David, Por, Emiel H. et al. 2018 SPIE, 10702E, 46 
\bibitem[Kinman \& Burbidge (1967)]{kinman67} Kinman, T. D., Burbidge, E. M. 1967 ApJ, 148L, 59 
\bibitem[Killeen et al. (1988)]{killeen88a} Killeen, N. E. B.; Bicknell, G. V.; Ekers, R. D. 1988 ApJ, 325, 180
\bibitem[Killeen et al. (1986)]{killeen86} Killeen, N. E. B.; Bicknell, G. V.; Carter, D. 1986 ApJ, 309, 45 
\bibitem[Kosiba et al. (2022)]{kosiba22} Kosiba, M., Pe\~na-Herazo, H. A., Massaro, F., Masetti, N., Paggi, A. et al. 2022 A\&A in press 
\bibitem[Kuraszkiewicz et al. (2021)]{kuraszkiewicz21} Kuraszkiewicz, J., Wilkes, B. J., Atanas, A. et al. 2021 ApJ, 913, 134 
\bibitem[Jauncey et al. (1978)]{jauncey78} Jauncey, D. L., Wright, A. E., Peterson, B. A., Condon, J. J. 1978 ApJ, 219L, 1 
\bibitem[Johnson et al. (2018)]{johnson18} Johnson, Sean D.; Chen, Hsiao-Wen; Straka, Lorrie A.; Schaye, Joop; Cantalupo, Sebastiano 2018 ApJ, 869L, 1
\bibitem[Jones et al. (2004)]{jones04} Jones, D. H., Saunders, W., Colless, M. et al. 2004 MNRAS, 355, 747 
\bibitem[Jones et al. (2009)]{jones09} Jones, D. H., Read, M. A., Saunders, W.  et al. 2009 MNRAS, 399, 683 
\bibitem[Ichinohe et al. (2015)]{ichinohe15} Ichinohe, Y., Werner, N., Simionescu, A., Allen, S. W., Canning, R. E. A., Ehlert, S., Mernier, F., Takahashi, T. 2015 MNRAS 448, 2971 
\bibitem[Intema et al. (2017)]{intema17} Intema, H. T., Jagannathan, P., Mooley, K. P. et al. 2017 A \& A, 598, A78 
\bibitem[Ivezi{\'c} et al. (2019)]{ivezic19} Ivezi{\'c}, {\v{Z}}., Kahn, S. M., Tyson, J. A.  et al. 2019 ApJ, 873, 111 
\bibitem[Jimenez-Gallardo et al. (2020)]{jimenez20} Jimenez-Gallardo, A., Massaro, F., Prieto, M. A., Missaglia, V., Stuardi, C. et al. 2020 ApJS, 250, 7 
\bibitem[Jimenez-Gallardo et al. (2021)]{jimenez21} Jimenez-Gallardo, A., Massaro, F., Paggi, A. et al. 2021 ApJS, 252, 23 
\bibitem[Johnston-Hollitt et al. (2008)]{johnston08} Johnston-Hollitt M., Hunstead R. W., Corbett E. 2008, A\&A, 479, 1 
\bibitem[Joye \& Mandel (2003)]{joye03} Joye, W. A. \& Mandel, E. 2003 adass, 295, 489 
\bibitem[Labiano et al. (2007)]{labiano07} Labiano, A.; Barthel, P. D.; O'Dea, C. P.; de Vries, W. H.; P\'erez, I.; Baum, S. A. 2007 A\&A, 463, 97 
\bibitem[Lacy et al. (2020)]{lacy20}  Lacy, M., Baum, S. A., Chandler, C. J. et al. 2020 PASP 132, 035001 
\bibitem[Laing et al. (1983)]{laing83} Laing, R. A., Riley, J. M., Longair, M. S. 1983 MNRAS 204, 151 
\bibitem[Large et al. (1981)]{large81} Large, M. I., Mills, B. Y.,  Little, A. G., Crawford, D. F., Sutton, J. M. 1981 MNRAS, 194, 693 
\bibitem[Larionov et al. (2020)]{larionov20} Larionov, V. M., Jorstad, S. G., Marscher, A. P. ;  Villata, M., Raiteri, C. M. 2020 MNRAS, 492, 3829 
\bibitem[Law-Green et al. (1995)]{law95} Law-Green, J. D. B., Leahy, J. P., Alexander, P. et al.  1995 MNRAS 274, 939 
\bibitem[Lehnert et al. (1999)]{lehnert99} Lehnert, M. D., Miley, G. K., Sparks, W. B., Baum, S. A., Biretta, J. et al. 1999 ApJS, 123, 351 
\bibitem[Liu et al. (2013)]{liu13} Liu, T., Tozzi, P., Tundo, E., Moretti, A., Wang, J.-X. et al. 2013 A\&A, 549, 143 
\bibitem[Liu et al. (2015)]{liu15} Liu, Teng, Tozzi, Paolo, Tundo, Elena, Moretti, Alberto, Rosati, Piero et al. 2015 ApJS, 216, 28 
\bibitem[Longair (1970)]{longair70} Longair, M. S. 1970 MNRAS, 150, 155 
\bibitem[Longair et al. (1995)]{longair95} Longair, M. S., Best, P. N., Rottgering, H. J. A. 1995 MNRAS, 275L, 47 
\bibitem[Ly et al. (2005)]{ly05} Ly, C., De Young, D. \& Bechtold, J. 2005, ApJ, 618, 609 
\bibitem[Lynds et al. (1965)]{lynds65} Lynds, C. R., Stockton, A. N. \& Livingston, W. C. 1965, ApJ, 142, 1667 
\bibitem[Lynds (1967)]{lynds67} Lynds, C. R. 1967 ApJ, 147, 837 
\bibitem[Madrid et al. (2006)]{madrid06} Madrid, J. P., Chiaberge, M., Floyd, D., Sparks, W. B., Macchetto, D., 2006 ApJS, 164, 307 
\bibitem[Marchesini et al. (2019)]{marchesini19} Marchesini, E. J., Paggi, A., Massaro, F.,  Masetti, N., D'Abrusco, R.,  Andruchow, I., de Menezes, R. 2019 A\&A, 631, 150 
\bibitem[Marchesini et al. (2020)]{marchesini20} Marchesini, E. J., Paggi, A., Massaro, F.,  Masetti, N., D'Abrusco, R., Andruchow, I. 2020 A\&A, 638 ,128 
\bibitem[Marenbach \& Appenzeller (1982)]{marenbach82} Marenbach, G., Appenzeller, I. 1982 A\&A108, 95 
\bibitem[Marshall et al. (2011)]{marshall11} Marshall, H. L.; Gelbord, J. M.; Schwartz, D. A.; Murphy, D. W.; Lovell, J. E. J. et al. 2011 ApJS, 193, 15 
\bibitem[Martel et al. (1999)]{martel99} Martel, Andr\'e R., Baum, Stefi A., Sparks, William B., Wyckoff, Eric, Biretta, John A. et al. 1999 ApJS, 122, 81 
\bibitem[Maselli et al. (2022)]{maselli22} Maselli, Alessandro, Forman, William R. , Jones, Christine, Kraft, Ralph P., Perri, Matteo 2022 ApJS, 262, 51 
\bibitem[Massaro et al. (2008a)]{massaro08a} Massaro, F., Tramacere, A., Cavaliere, A., Perri, M. \& Giommi, P. 2008a, A\&A, 478, 395 
\bibitem[Massaro et al. (2008b)]{massaro08b} Massaro, F., Giommi, P., Tosti, G., et al. 2008b, A\&A, 489, 1047 
\bibitem[Massaro et al. (2009a)]{massaro09a} Massaro, F., Harris, D. E., Chiaberge, M., Grandi, P., Macchetto, F. D., Baum, S. A., O'Dea, C. P., Capetti, A. 2009a ApJ, 696, 980
\bibitem[Massaro et al. (2009b)]{massaro09b} Massaro, F., Chiaberge, M., Grandi, P., Giovannini, G., O'Dea, C. P., Macchetto, F. D., Baum, S. A., Gilli, R., Capetti, A., Bonafede, A., Liuzzo, E. 2009b ApJL, 692, 123 
\bibitem[Massaro et al. (2010)]{massaro10} Massaro, F., Harris, D. E., Tremblay, G. R., Axon, D., Baum, S. A. et al. 2010 ApJ, 714, 589 
\bibitem[Massaro et al. (2011a)]{massaro11a} Massaro, F., Harris, D. E., Cheung, C. C. 2011 ApJS, 197, 24 
\bibitem[Massaro et al. (2012a)]{massaro12a} Massaro, F., Tremblay, G. R., Harris, D. E., Kharb, P., Axon, D. et al. 2012a ApJS, 203 ,31 
\bibitem[Massaro et al. (2013a)]{massaro13a} Massaro, F., Harris, D. E., Tremblay, G. R., Liuzzo, E., Bonafede, A., Paggi, A. 2013a ApJS, 206, 7 
\bibitem[Massaro et al. (2016)]{massaro16} Massaro, F., \'Alvarez Crespo, N.,  D'Abrusco, R., Landoni, M., Masetti, N. et al. 2016 Ap\&SS, 361, 337 
\bibitem[Massaro et al. (2018)]{massaro18} Massaro, F., Missaglia, V., Stuardi, C., Harris, D. E., Kraft, R. P. et al. 2018 ApJS, 234, 7 
\bibitem[Massaro et al. (2022)]{massaro22} Massaro, F. et al. 2022 ApJS submitted 
\bibitem[Mauch et al. (2003)]{mauch03} Mauch, T., Murphy, T., Buttery, H. J. et al. 2003 MNRAS, 342, 1117 
\bibitem[McCarthy et al. (1995)]{mccarthy95} McCarthy, Patrick J., Spinrad, Hyron, van Breugel, Wil 1995 ApJS, 99, 27 
\bibitem[McCarthy et al. (1996a)]{mccarthy96a} McCarthy, Patrick J., Baum, Stefi A., Spinrad, Hyron 1996 ApJS, 106, 281 
\bibitem[McCarthy et al. (1997)]{mccarthy97} McCarthy, Patrick J., Miley, George K., de Koff, Sigrid, Baum, Stefi A., Sparks, William B. et al. 1997 ApJS, 112, 415 
\bibitem[McCollough \& Rots (2005)]{mccollough05} McCollough, M. L. \& Rots, A. H. 2005 ASPC, 347, 478 
\bibitem[McMullin et al. (2020)]{mcmullin20} McMullin, J., Diamond, P., McPherson, A. et al. 2020 in Proceedings of the SPIE, 11445, 1144512 
\bibitem[McNamara \& Nulsen (2007)]{mcnamara07} McNamara B. R. \& Nulsen, P. E. J. 2007 ARA \& A, 45, 117 
\bibitem[McNamara \& Nulsen (2012)]{mcnamara12} McNamara B. R. \& Nulsen, P. E. J. 2012 New Journal of Physics, 14, 5 
\bibitem[Mingo et al. (2017)]{mingo17} Mingo, B., Hardcastle, M. J., Ineson, J., Mahatma, V., Croston, J. H., Dicken, D., Evans, D. A., Morganti, R., Tadhunter, C. 2017 MNRAS, 470, 2762 
\bibitem[Missaglia et al. (2019)]{missaglia19} Missaglia, V., Massaro, F., Capetti, A., Paolillo, M., Kraft, R. P. et al. 2019 A\&A, 626, 8 
\bibitem[Moretti et al. (2004)]{moretti04} Moretti, A., Campana, S., Tagliaferri, G., et al. 2004, in Proc. SPIE, Vol. 5165, X-Ray and Gamma-Ray Instrumentation for Astronomy XIII, ed. K. A. Flanagan \& O. H. W. Siegmund, 232–240 
\bibitem[Morganti (2017)]{morganti17} Morganti, R. 2017 Frontiers in Astronomy and Space Sciences, 4, 42 
\bibitem[Morton \& Tritton (1982)]{morton82} Morton, D. C.; Tritton, K. P. 1982 MNRAS, 198, 669 
\bibitem[O'Dea et al. (1991)]{odea91} O'Dea, C. P., Baum, S. A., Stanghellini, C. 1991 ApJ, 380, 66 
\bibitem[Oh et al. (2018)]{oh18} Oh, Kyuseok; Koss, Michael; Markwardt, Craig B.; Schawinski, Kevin; Baumgartner, Wayne H. et al. 2018 ApJS, 235, 4 
\bibitem[Oke (1974)]{oke74} Oke, J. B.1974 ApJS, 27, 21 
\bibitem[Oke \& Gunn (1983)]{oke83} Oke, J. B. \& Gunn, J. E. 1983 ApJ 266, 713 
\bibitem[Owen \& Rudnick (1976)]{owen76} Owen, F. N. \& Rudnick, L. 1976, ApJ, 205, L1 
\bibitem[Owers et al. (2009)]{owers09} Owers M. S., Couch W. J., Nulsen P. E. J. 2009, ApJ, 693, 901 
\bibitem[Paggi et al. (2013)]{paggi13} Paggi, A., Massaro, F., D’Abrusco, R., et al. 2013, ApJS, 209, 9 
\bibitem[Parisi et al. (2014)]{parisi14} Parisi, P.; Masetti, N.; Rojas, A. F.; Jim\'enez-Bail\'on, E.; Chavushyan, V. et al. 2014 A\&A, 561, 67 
\bibitem[Pe\~na-Herazo et al. (2020)]{herazo20} Pe\~na-Herazo, H. A., Amaya-Almaz\'an, R. A., Massaro, F., de Menezes, R., Marchesini, E. J. et al. 2020 A\&A, 643, 103 
\bibitem[Pe\~na-Herazo et al. (2022)]{herazo22} Pe\~na-Herazo, H. A., Massaro, F., Chavushyan, V., Masetti, N., Paggi, A., Capetti, A. 2022 A\&A, 659, 32 
\bibitem[Peterson \& Bolton (1972)]{peterson72} Peterson, Bruce A.; Bolton, J. G. 1972 ApJ, 173L, 19 
\bibitem[Privon et al. (2008)]{privon08} Privon, G. C., O'Dea, C. P., Baum, S. A., Axon, D. J., Kharb, P., 2008 ApJS, 175, 423 
\bibitem[Quintana \& Ramirez (1995)]{quintana95a} Quintana, H. ;  Ramirez, A. 1995 ApJS, 96, 343 
\bibitem[Ramos Almeida et al. (2011)]{ramos11} Ramos Almeida, C., Tadhunter, C. N., Inskip, K. J. et al. 2011 MNRAS, 410, 1550 
\bibitem[Roming et al. (2005)]{roming05} Roming, Peter W. A., Kennedy, Thomas E., Mason, Keith O., Nousek, John A., Ahr, Lindy et al. 2005 SSRv, 120, 95 
\bibitem[Sandage (1978)]{sandage78} Sandage, A. 1978 AJ, 83, 904 
\bibitem[Sbarufatti et al. (2006)]{sbarufatti06} Sbarufatti, B.; Falomo, R.; Treves, A.; Kotilainen, J. 2006 A\&A, 457, 35 
\bibitem[Sejake et al. (2023)]{sejake23} Sejake et al. 2023 MNRAS, 518, 4290 
\bibitem[Simpson et al. (1993)]{simpson93} Simpson, C.; Clements, D. L.; Rawlings, S.; Ward, M. 1993 MNRAS, 262, 88 
\bibitem[Simpson et al. (1999)]{simpson99} Simpson, Chris, Rawlings, Steve, Lacy, Mark 1999 MNRAS, 306, 828 
\bibitem[Spinrad et al. (1985)]{spinrad85} Spinrad, H., Marr, J., Aguilar, L., Djorgovski, S. 1985 PASP, 97, 932 
\bibitem[Springob et al. (2005)]{springob05} Springob, Christopher M., Haynes, Martha P., Giovanelli, Riccardo, Kent, Brian R. 2005 ApJS, 160, 149 
\bibitem[Stickel \& Kuehr (1994)]{stickel94} Stickel, M.; Kuehr, H. 1994 A\&AS, 105, 67 
\bibitem[Stocke et al. (1998)]{stocke98} Stocke, John T., Penton, Steve, Harvanek, Michael, Neely, W. A., Blades, J. Chris 1998 AJ, 115, 451 
\bibitem[Storchi-Bergmann et al. (1996)]{storchi96} Storchi-Bergmann, T.; Wilson, A. S.; Mulchaey, J. S.; Binette, L. 1996 A\&A, 312, 357 
\bibitem[Stuardi et al. (2018)]{stuardi18} Stuardi, C., Missaglia, V., Massaro, F., Ricci, F., Liuzzo, E. et al. 2018 ApJS, 235, 32 
\bibitem[Sun (2009)]{sun09} Sun, M. 2009 ApJ, 704, 158 
\bibitem[Tadhunter et al. (1993)]{tadhunter93} Tadhunter, C. N., Morganti, R., di Serego-Alighieri, S., Fosbury, R. A. E., Danziger, I. J. 1993, MNRAS, 263, 999 
\bibitem[Taylor (2005)]{taylor05} Taylor M. B., 2005, ASPC, 347, 29 
\bibitem[Teague et al. (1990)]{teague90} Teague, Peter F.; Carter, David; Gray, Peter M. 1990 ApJS, 72, 715 
\bibitem[Thompson et al. (1990)]{thompson90} Thompson, D. J.,  Djorgovski, S., de Carvalho, R. 1990 PASP, 102, 1235 
\bibitem[Tingay et al. (2013)]{tingay13} Tingay, S. J.,  Goeke, R.,  Bowman, J. D.,  Emrich, D.,  Ord, S. M. et al. 2013 PASA, 30, 7 
\bibitem[Tozzi et al. (2014)]{tozzi14} Tozzi, P., Moretti, A., Tundo, E., Liu, T., Rosati, P. et al. 2014 A\&A, 567, 89 
\bibitem[Tritton (1972)]{tritton72} Tritton, K. P. 1972 MNRAS, 158, 277 
\bibitem[Tundo et al. (2012)]{tundo12} Tundo, E., Moretti, A., Tozzi, P., Teng, L., Rosati, P.,  Tagliaferri, G., Campana, S. 2012 A\&A, 547, 57 
\bibitem[Urry \& Padovani (1995)]{urry95} Urry \& Padovani 1995 PASP 107, 803 
\bibitem[Wayth et al. (2015)]{wayth15} Wayth, R. B., Lenc, E., Bell, M. E. et al. 2015 PASP 32, 12 
\bibitem[Wegner et al. (2003)]{wegner03} Wegner, G.; Bernardi, M.; Willmer, C. N. A.; da Costa, L. N.; Alonso, M. V. et al. 2003 AJ, 126, 2268 
\bibitem[Wenger et al. (2000)]{wenger00} Wenger et al. 2000,A\&AS,143, 9 
\bibitem[Werner et al. (2012)]{werner12} Werner, M. W., Murphy, D. W., Livingston, J. H., Gorjian, V., Jones, D. L.,  2012 ApJ, 759, 86 
\bibitem[White et al. (2020a)]{white20a} White, S. V.,  Franzen, T. M. O.,  Riseley, C. J.,  Wong, O. I., Kapi\'nska, A. D. et al. 2020a PASA, 37, 17  
\bibitem[White et al. (2020b)]{white20b} White, S. V.,  Franzen, T. M. O.,  Riseley, C. J.,  Wong, O. I., Kapi\'nska, A. D. et al. 2020b PASA, 37, 18  
\bibitem[Whiteoak (1972)]{whiteoak72} Whiteoak J. B., 1972, Aust. J. Phys., 25,233 
\bibitem[Wilkes et al. (2013)]{wilkes13} Wilkes, B. J., Kuraszkiewicz, J., Haas, M. et al. 2013 ApJ, 773, 15 
\bibitem[Willis \& Parker (1966)]{willis66} Wills, D. \& Parker, E. A. 1966 MNRAS 131, 503  
\bibitem[Worrall (2009)]{worrall09} Worrall, D. M. 2009 A\&ARv, 17, 1 
\bibitem[Wright et al. (1977)]{wright77} Wright, A. E.; Jauncey, D. L.; Peterson, B. A.; Condon, J. J. 1977 ApJ, 211L, 115 
\bibitem[Wright et al. (1979)]{wright79} Wright, A. E.; Peterson, B. A.; Jauncey, D. L.; Condon, J. J. 1979 ApJ, 229, 73 
\bibitem[Wright et al. (2010)]{wright10} Wright E. L., et al. 2010 AJ, 140, 1868 
\bibitem[Worrall \& Birkinshaw (2017)]{worrall17} Worrall, D. M.; Birkinshaw, M. 2017 MNRAS, 467, 2903 
\bibitem[Yates et al. (1989)]{yates89} Yates, M. G.; Miller, L.; Peacock, J. A. 1989 MNRAS, 240, 129 
\bibitem[Yee \& Green (1983)]{yee83} Yee, H. K. C.; Green, R. F. 1983 BAAS, 15, 957 
\bibitem[Young et al. (2005)]{young05} Young, A.J., Wilson, A.S., Tingay, S.J. \& Heinz, S. 2005, ApJ, 622, 830 
\bibitem[Younis et al. (1985)]{younis85} Younis, S.; Meaburn, J.; Stewart, P. 1985 A\&A, 147, 178 
\end{thebibliography}
\end{document}